\documentclass[aps,prd,twocolumn,superscriptaddress,nofootinbib]{revtex4-2}

\usepackage{color}
\usepackage[dvipsnames]{xcolor}

\newcommand{\tododoublecheck}[1]{}

\usepackage{siunitx}
\DeclareSIUnit \h {\ensuremath{\mathit{h}}}
\DeclareSIUnit \Msun {M_\odot}
\DeclareSIUnit \parsec {pc}
\DeclareSIUnit \deg {deg}

\usepackage{graphicx}	%
\usepackage{amsmath}	%
\usepackage{amssymb}	%

\usepackage{colortbl}
\usepackage{physics}
\usepackage[normalem]{ulem}

\def\aj{AJ}%
\def\apj{ApJ}%
\def\apjs{ApJS}%
\def\aap{A\&A}%
\def\mnras{MNRAS}%
\def\prd{Phys.~Rev.~D}%
\def\prl{Phys.~Rev.~Lett.}%
\def\nat{Nature}%
\def\jcap{JCAP}%
\def\physrep{Phys.~Rep.}%
\usepackage[colorlinks=true,allcolors=blue]{hyperref}

\usepackage[capitalise]{cleveref}

\usepackage{xspace}
\newcommand{\getdist}{\textsc{getdist}\xspace}

\newcommand{\camb}{\textsc{CAMB}\xspace}
\newcommand{\ccl}{\textsc{CCL}\xspace}

\newcommand{\pymangle}{\textsc{pymangle}\xspace}
\newcommand{\healpy}{\textsc{healpy}\xspace}
\newcommand{\healpix}{\textsc{healpix}\xspace}
\newcommand{\namaster}{\textsc{NaMaster}\xspace}

\begin{document}

\title{Constraining gravity with a new precision \texorpdfstring{E\textsubscript{G}}{EG} estimator using Planck + SDSS BOSS}

\author{Lukas Wenzl}
\email{ljw232@cornell.edu}
\affiliation{Department of Astronomy, Cornell University, Ithaca, NY, 14853, USA}

\author{Rachel Bean}
\affiliation{Department of Astronomy, Cornell University, Ithaca, NY, 14853, USA}

\author{Shi-Fan Chen}
\affiliation{School of Natural Sciences, Institute for Advanced Study, 1 Einstein Drive, Princeton NJ 08540}

\author{Gerrit~S.~Farren}
\affiliation{DAMTP, Centre for Mathematical Sciences, University of Cambridge, Wilberforce Road, Cambridge CB3 OWA, UK}
\affiliation{Kavli Institute for Cosmology Cambridge, Madingley Road, Cambridge CB3 0HA, UK}

\author{Mathew S.~Madhavacheril}
\affiliation{Department of Physics and Astronomy, University of Pennsylvania, Philadelphia, PA, 19104, USA}

\author{Gabriela A. Marques}
\affiliation{Fermi National Accelerator Laboratory, P. O. Box 500, Batavia, IL 60510, USA}
\affiliation{Kavli Institute for Cosmological Physics, University of Chicago, Chicago, IL 60637, USA}

\author{Frank J. Qu}
\affiliation{DAMTP, Centre for Mathematical Sciences, University of Cambridge, Wilberforce Road, Cambridge CB3 OWA, UK}
\affiliation{Kavli Institute for Cosmology Cambridge, Madingley Road, Cambridge CB3 0HA, UK}

\author{Neelima Sehgal}
\affiliation{Physics and Astronomy Department, Stony Brook University, Stony Brook, NY 11794}

\author{Blake D.~Sherwin}
\affiliation{DAMTP, Centre for Mathematical Sciences, University of Cambridge, Wilberforce Road, Cambridge CB3 OWA, UK}
\affiliation{Kavli Institute for Cosmology Cambridge, Madingley Road, Cambridge CB3 0HA, UK}

\author{Alexander van Engelen}
\affiliation{School of Earth and Space Exploration, Arizona State University, 781 Terrace Mall, Tempe, AZ 85287, U.S.A.}

\date{\today}

\begin{abstract}
The $E_G$ statistic is a discriminating probe of gravity developed to test the prediction of general relativity (GR) for the relation between gravitational potential and clustering on the largest scales in the observable universe. We present a novel high-precision estimator for the $E_G$ statistic using CMB lensing and galaxy clustering correlations that carefully matches the effective redshifts across the different measurement components to minimize corrections. A suite of detailed tests is performed to characterize the estimator's accuracy, its sensitivity to assumptions and analysis choices, and the non-Gaussianity of the estimator's uncertainty is characterized. After finalization of the estimator, it is applied to \textsl{Planck} CMB lensing and SDSS CMASS and LOWZ galaxy data. We report the first harmonic space measurement of $E_G$ using the LOWZ sample and CMB lensing and also updated constraints using the final CMASS sample and the latest \textsl{Planck} CMB lensing map. We find $\hat E_G^{\rm \textsl{Planck}+CMASS} = 0.36^{+0.06}_{-0.05}  (\textrm{68.27\%}) $ and $\hat E_G^{\rm \textsl{Planck}+LOWZ} = 0.40^{+0.11}_{-0.09} (\textrm{68.27\%}) $, with additional subdominant systematic error budget estimates of 2\% and 3\% respectively. Using $\Omega_{\rm m,0}$ constraints from \textsl{Planck} and SDSS BAO observations, $\Lambda$CDM-GR predicts $E_G^{\rm GR} (z = 0.555) = 0.401 \pm 0.005$ and $E_G^{\rm GR} (z = 0.316) = 0.452 \pm 0.005$ at the effective redshifts of the CMASS and LOWZ based measurements. We report the measurement to be in good statistical agreement with the $\Lambda$CDM-GR prediction and report that the measurement is also consistent with the more general GR prediction of scale-independence for $E_G$. This work provides a carefully constructed and calibrated statistic with which $E_G$ measurements can be confidently and accurately obtained with upcoming survey data.
\end{abstract}

\maketitle

\section{Introduction} \label{sec:introduction}

Twenty-five years ago, the late-time accelerated expansion of the universe was discovered \citep{Riess1998,Perlmutter1999}, implying that the universe is currently dominated by a component distinct from the clustering dark matter and Standard Model particles, coined ``dark energy" that is driving this expansion. To date, the physical reason for dark energy remains unresolved \citep{Peebles2003}. Current observational constraints on the expansion history of the universe can be well explained by a cosmological constant $\Lambda$. Together with cold dark matter (CDM) and baryonic matter this $\Lambda$CDM model forms the current standard model of cosmology due to its simplicity and also its ability to fit a wide range of cosmological data (e.g. \citep{Alam2017_DR12_cosmo_analysis,PlanckCollaboration2018,Freedman2019,Madhavacheril2023}). This however is not the only theoretical model of gravity that can fit the data. While simple in its inherently constant energy density, the fine-tuning and coincidence problems related to explaining the observed magnitude of $\Lambda$ \citep{Weinberg1989} have led to the consideration of other gravitational models. These include, for example, $f(R)$ gravity \citep{Carroll2004}, Chameleon gravity \citep{Khoury2004}, DGP \citep{Dvali2000}, TeVeS \citep{Bekenstein2004} and others (see \citet{Clifton2012} for an in-depth overview). These models generally contain screening mechanisms that hide their effects in high-density environments allowing them to pass gravity tests on solar system scales and they can be fine-tuned to match the expansion history of the universe including the late time accelerated expansion \citep{Jain2010}. This degeneracy makes these modified gravity theories difficult to distinguish from distance measurements alone. 

However, these alternative theories of gravity also have implications for the growth of large-scale structures (LSS) in the universe differing from what is expected for general relativity. In particular, the Poisson equation which relates the gravitational potentials to overdensities is altered, and the potential experienced by relativistic particles like photons can be different from the potential experienced by non-relativistic particles \citep{Huterer2015}. Similar to how comparing lensing around the sun with the gravitational dynamics of the planets can be used to test general relativity (GR) on solar system scales, one can devise tests on the scale of a significant fraction of the observable universe to constrain gravity further \citep{Uzan2010}. A wide range of tests have been proposed that leverage the distinct growth of structure in modified gravity models to differentiate them from GR generally or $\Lambda$CDM specifically (see \citep{Jain2008,Jain2010,Ishak2019,Hou2023} for overviews).

One promising pathway to constrain gravity in this regime has been proposed in \citet{Zhang2007}: the so-called $E_G$ statistic, that combines a measurement of the divergence of the peculiar velocity field $\theta$ with a measurement of gravitational lensing $\nabla^2 (\psi -\phi)$. A comparison between these two measurements is sensitive to changes in the Poisson equations and differences between the two gravitational potentials enabling it to differentiate between GR and modified gravity models with degenerate expansion histories \citep{Pullen2015,Leonard2015}. The velocity field is measured by exploiting matter conservation so that on linear scales it can be found by measuring the linear growth rate $f$ and matter overdensities $\delta$. The underlying matter field dominated by dark matter cannot be observed directly so one generally instead leverages biased tracers like galaxies that are related to the underlying matter field by a bias factor $b_g$. The $E_G$ statistic is specifically chosen as the ratio of two measurements that have the same dependence on the galaxy bias leading to a cancellation on linear scales. Spectroscopic redshift information is ideal to accurately determine the divergence of the peculiar velocity field,  given it measures the 3D clustering of galaxies. An analogous quantity for photometric survey data is the $D_G$ statistic, introduced by \citep{Giannantonio2016} and applied in several studies, e.g., \citep{Bianchini2018,Omori2019a,Marques2020a}. $D_G$ shares a conceptual resemblance to $E_G$, since it also combines lensing and clustering information to remove the dependence on galaxy bias. However, $E_G$ has the advantage of being scale-independent in GR and is more directly related to deviations in the Poisson equation.

The initial proposal of the $E_G$ statistic suggested the use of weak lensing measurements of background galaxies as the tracer of lensing \citep{Zhang2007}. This was used for the first measurement of $E_G$ presented in \citet{Reyes2010} and has been done in a range of further analyses \citep{Blake2016,delaTorre2017,Alam2017,Amon2018,Singh2019,Blake2020}. The authors of \citet{Pullen2015} proposed to use Cosmic Microwave Background (CMB) lensing measurements as the lensing tracer allowing constraints on larger scales. Their proposed estimator (herein referred to as $\hat E_G^{\rm Pullen}$) has been applied to CMB lensing measurements from \textsl{Planck} and spectroscopic galaxy samples from the Sloan Digital Sky Survey (SDSS) \citep{Pullen2016,Singh2019,Zhang2021}.

Upcoming next-generation observatories offer the opportunity to significantly tighten constraints on the $E_G$ statistic. These include for CMB lensing the Atacama Cosmology Telescope (ACT) \citep{Qu2023}, the South Pole Telescope (SPT) \citep{Story2015,Pan2023} Simons Observatory (SO) \citep{SimonsObservatory2019} and CMB-S4 \citep{Abazajian2016} and upcoming spectroscopic galaxy samples from the Dark Energy Spectroscopic Instrument (DESI) \citep{DESICollaboration2016} and the Spectro-Photometer for the History of the Universe, Epoch of Reionization, and Ices Explorer (SPHEREx) \citep{Dore2014}.

However, to fully realize the potential of larger statistical constraining power with upcoming datasets careful approaches are required. Precision estimators for which systematic errors are well within the smaller statistical uncertainties are needed. There are significant analysis challenges that need to be carefully modeled. In this work, we aim to prepare for upcoming data, creating a precision pipeline that allows accurate estimation of the $E_G$ statistic using CMB lensing and spectroscopic galaxy data. We present a newly revised estimator for the $E_G$ statistic building on previous work and making key advances.

We apply this new estimator for the $E_G$ statistic to well-tested datasets, namely CMB lensing from \textsl{Planck} and galaxy clustering from SDSS BOSS, and report new constraints. The data used has been previously extensively tested for systematics (see e.g. \citep{Pullen2016,Doux2018,Singh2019,Chen2022}). We do not reproduce these tests on the data here but note that such tests will also be a key part of future work with new upcoming datasets, to carefully characterize the datasets prior to any $E_G$ analysis to ensure constraints can be confidently attributed to cosmological, and not instrumental or astrophysical, effects.

The structure of this paper is as follows: in \cref{sec:results_summary} we summarize the key results from the paper, the definition and motivation of the $E_G$ statistic are discussed in \cref{sec:E_G_statistic}, and the datasets used for the analysis are described in \cref{sec:data}. Then, in \cref{sec:E_G-estimator}, the new and updated $E_G$ estimator applicable to precision CMB lensing and spectroscopic galaxy clustering data is motivated and presented. 
In \cref{sec:Measuring_observables} the measurement of the observables from data is discussed. In \cref{sec:ratio_distribution} the non-Gaussian uncertainty of the $E_G$ estimator is investigated and statistical tests to compare with GR predictions are developed. The results of the measurements are presented in \cref{sec:results} and the results are discussed in \cref{sec:conclusion}.

\section{Summary of key results}\label{sec:result_summary} \label{sec:results_summary}

In this work, we introduce a novel precision estimator for the $E_G$ statistic to test a key prediction of GR about the structure formation of the universe that is distinct from predictions for alternative gravity models. We apply this new estimator to well-established data in the form of CMB lensing measurements from \textsl{Planck} PR4 \citep{Carron2022} and galaxy clustering measurements from SDSS BOSS DR12 \citep{Reid2016}, to present new constraints on the $E_G$ statistic. GR predictions used in this work are based on the $\Lambda$CDM constraints given in the Planck 2018 results \citep{PlanckCollaboration2018} using CMB anisotropy and baryon acoustic oscillations (BAO) from SDSS BOSS DR12 \citep{Alam2017_DR12_cosmo_analysis}.

\begin{figure}
\includegraphics[width=\columnwidth]{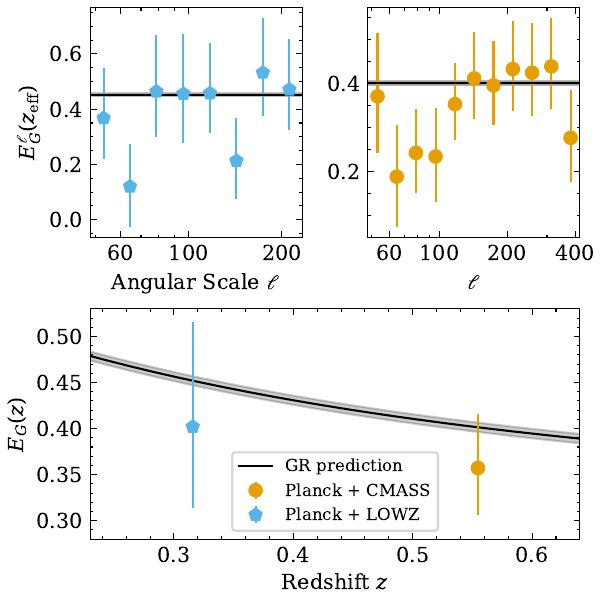}
\caption{Summary of the $E_G$ statistic measurements presented in this work. Shown is $E_G$ as a function of redshift [lower] and for each of the two measurements at their effective redshift as a function of angular scale $\ell$ [upper]. We present a measurement using \textsl{Planck} CMB lensing combined with SDSS CMASS galaxies [orange circles] and combined with SDSS LOWZ galaxies [blue pentagons] with 68.27\% confidence ranges for the error. %
The $\Lambda$CDM-GR expectation for the fiducial cosmology (assuming $\Omega_{\rm m,0}=0.3111\pm 0.0056$) is shown as a black line with a grey one sigma uncertainty band. }
\label{fig:main_result}
\end{figure}

The presented estimator for the $E_G$ statistics uses CMB lensing and spectroscopic galaxy clustering information. Key to the derivation is to establish the effective redshift of the observables used for the measurement and the introduction of a reweighting scheme that allows matching the effective redshift of the angular cross-power spectrum to the angular auto-power spectrum and the redshift space distortions (RSD) analysis which are all combined for the estimator. Additionally, we minimize the approximations during the derivation, resulting in a precise estimator that is applicable to upcoming precision data and mainly limited by astrophysical systematics. The impact of astrophysical systematics can be minimized in upcoming precision data by calculating the $E_G$ statistic in narrower bins in redshift. 

The estimator, all applied corrections, and analysis choices were finalized and frozen before applying the estimator to the \textsl{Planck} and SDSS BOSS data. There were no major changes to the result after the initial calculation of the $\hat E_G$ and $\hat E_G^{\ell}$ estimates. We briefly summarize key insights from the measurements:

\begin{itemize}
    \item GR predicts the $E_G$ statistic to be scale-independent on linear scales while in many alternative theories of gravity, $E_G$ is scale-dependent. We find our observations to be statistically consistent with scale-independence, in line with GR predictions. Our tests are based on constraints from two measurements with effective redshifts of $z_{\rm eff}^{\textsl{Planck} \rm + CMASS} = 0.555$ and  $z_{\rm eff}^{\textsl{Planck} \rm + LOWZ} = 0.316 $ sensitive to scales in the approximate range $k \in [0.02, 0.2] \textrm{Mpc}^{-1}$. The binned measurements as a function of scale are shown in the top panels of \cref{fig:main_result}.
    
    \item Since our data is consistent with scale-independence we combine the constraints across the angular scales assuming $E_G$ as a constant. We find overall constraints on $E_G$ of 
    $\hat E_G^{\rm \textsl{Planck}+CMASS} = 0.36^{+0.06}_{-0.05} (\textrm{stat}) $ and $\hat E_G^{\rm \textsl{Planck}+LOWZ} = 0.40^{+0.11}_{-0.09} (\textrm{stat}) $ which are visualized in the bottom panel of \cref{fig:main_result}. 
    
    \item Within the framework of GR the well-determined expansion history of the universe makes direct predictions for the growth history and thereby $E_G$. 
    We compare our measurement to predictions for $E_G$ based on a $\Lambda$CDM fit, specifically a measurement of $\Omega_{\rm m, 0}$ from \citep{PlanckCollaboration2018} using \textsl{Planck} CMB anisotropy and SDSS BOSS BAO measurements. We find that each of our measurements \textsl{Planck}+CMASS and \textsl{Planck}+LOWZ is statistically consistent with the respective $\Lambda$CDM-GR prediction. In addition to a consistency test of the data with GR the presented results can also be compared to predictions for alternative models of gravity, enabling constraints on their degrees of freedom.
\end{itemize}

\section{The \texorpdfstring{E\textsubscript{G}}{EG} statistic }\label{sec:background} \label{sec:E_G_statistic}

Modified gravity models, such as $f(R)$ gravity \citep{Carroll2004} and Chameleon gravity \citep{Khoury2004}, offer an alternative framework to understand the observed accelerated expansion of the universe. However, they typically have implications for the formation of cosmic structures that differ from GR.

For GR, the Einstein field equations are solved, under the assumption of isotropy and homogeneity on large scales, with the Friedmann-Lema\^itre-Robertson-Walker (FLRW) metric and the Friedmann equations. The dynamics of such a universe are described as perturbation fields in this background metric: $\psi$ for the time component and $\phi$ for the spatial component. The perturbed FRW metric for a flat universe is given by $\dd s^2 = (1+ 2\psi) \dd t^2 - a^2 (1+2\phi) \dd \textbf{x}^2$ where $a(t)$ is the scale factor. With this metric, the dynamics of the universe are described by \citep{Ma1995} 
\begin{align}
    \nabla^2 \psi &= - 4 \pi G a^2 \rho(a) \delta, \\
    \phi &= - \psi,
\end{align}
where $\rho$ is the background matter density and $\delta$ the matter density pertubations. In GR the perturbation fields of time and space are equal (up to a sign convention), giving the generalized Poisson equation relating the lensing of light $\nabla^2 \phi$ and the fractional overdensity $\delta$. 
This represents a prediction of GR about how gravitational lensing relates to the clustering in our Universe. 

This relation changes for many alternative models of gravity in which the expansion history $a(t)$ is tuned to match GR well within observational uncertainties, however, the dynamics of the universe change. In general, for a large class of models, the modified Poisson equations can be described as \citep{Pullen2015}
\begin{align}
    \nabla^2 \psi &= - 4 \pi G a^2 \mu(k, a) \rho(a) \delta, \\
    \phi &= - \gamma(k,a) \psi.
\end{align}
Here, $\gamma$ and $\mu$ are general functions that can be time and scale-dependent for a given modified gravity scenario. The $\mu$ function parametrizes the effective strength of gravity, and $\gamma$ is the gravitational slip that quantifies the difference in the perturbation fields. We can also use the common re-parametrization of $\gamma$ to $\Sigma \equiv \frac{1}{2} \mu (1+\gamma)$, the lensing parameter, which is what the $E_G$ statistic, defined below, more directly measures. In the GR limit $\mu, \gamma$ and $\Sigma$ are $1$. A large class of modified gravity scenarios can be mapped to these general functions \citep{Silvestri2013,Baker2014,Pullen2015,Wenzl2022}.

The $E_G$ statistic is constructed as a quantity to test the relation between lensing and matter clustering as predicted by the Poisson equations. For observations, the clustering is difficult to measure directly since the observable galaxies are a biased tracer of the underlying matter overdensities. The $E_G$ statistic instead investigates the divergence of the peculiar velocity field $\theta$ which, on linear scales, is related to the underlying matter perturbations as $\theta = - f\delta$ where $f=d\ln \delta/d\ln a$ is the growth rate. The definition of the $E_G$ statistic in harmonic space, as introduced in \citet{Zhang2007}, is given by 
\begin{align}
    E_G (k, z) \equiv \left[ \frac{\nabla^2 (\psi -\phi ) }{3 H_0^2 (1+z) \theta} \right]_{k}. \label{eq:EG}
\end{align}
The definition is chosen so that one can build an estimator from Power spectra involving galaxy tracers where the effect of galaxy bias cancels on linear scales and therefore the measured $E_G$ statistic can conveniently be compared to the expectation value for GR. We discuss this in detail later in this section when discussing the 3D estimator. 

By plugging in the Poisson equation and using $\theta = - f\delta$ we can find the expectation value for GR to be
\begin{equation}
    E_G^{\rm GR} (z) = \frac{\Omega_{\rm m,0}}{f(z)}, \label{eq_EG_GR}
\end{equation}
where $\Omega_{\rm m, 0} = \frac{\rho}{\rho_{\rm crit}}$ is the energy density in the form of matter today, in units of the critical density $\rho_{\rm crit} = \frac{3 H_0^2}{8\pi G}$. The value for $E_G$ is predicted to be scale-independent for GR in the linear regime. More specifically we can quantify this for the current standard model of cosmology, $\Lambda$CDM, which represents a model of general relativity where the dominant components of the universe today are a cosmological constant $\Lambda$, CDM, baryonic matter, and no spacial curvature. %
Within the $\Lambda$CDM model the growth rate can be well approximated by $f(z) = \Omega_{\rm m}(z)^{0.55}$ \citep{Linder2008_growth} and therefore $E_G$ is predicted from the background expansion using a constraint on $\Omega_{\rm m, 0}$ alone. Based on measurements of the expansion history and early universe in the form of BAO and CMB anisotropy observations, the 6-parameter $\Lambda$CDM model is tightly constrained. In particular, for our fiducial cosmology based on \textsl{Planck} CMB anisotropies and SDSS BAO \citep{PlanckCollaboration2018} we have $\Omega_{\rm m, 0} = 0.3111\pm 0.0056$. This gives us a percent-level $\Lambda$CDM-GR prediction for the $E_G$ statistic based on geometric measurements of the expansion history of the universe against which we can compare the growth of large-scale structure-derived $E_G$ estimates at the effective redshifts of the surveys used in this work.

Furthermore, for a given alternative gravity scenario with modifications $\gamma$ and $\Sigma$, the expression for $E_G$ is \citep{Pullen2015}
\begin{equation}
    E_G^{\rm ModGrav} (k, z) = \frac{\Omega_{\rm m,0} \Sigma(k, z)}{f(k, z)},
\end{equation}
where the growth rate $f(k, z)$ and $E_G (k, z)$ can now be scale dependent. The $E_G$ statistic thus directly scales with modifications to the gravitational lensing potential quantified by $\Sigma$ and is also sensitive to changes in the growth rate. 
Results on $E_G$ can be compared to a given model of modified gravity by deriving the expected $E_G$ value based on the modified Poisson equations of the model. 
We refer the interested reader to \citet{Pullen2015} where $E_G$ is derived for the example cases of $f(R)$ gravity and Chameleon gravity.

Based on the definition in \cref{eq:EG}, an effective 3D estimator with expectation value equal to $E_G$ is given by \citep{Zhang2007,Pullen2015}
\begin{equation}
    \hat{E}_G(k, z)=\frac{c^2 \hat{P}_{\nabla^2(\psi-\phi) g}(k, z)}{3 H_0^2(1+z) \hat{P}_{\theta g}(k, z)}, \label{estimator_3D}
\end{equation}
where $H_0$ is the Hubble constant and the hats indicate estimation from observables. The $\hat{P}_{\theta g}(k, z)$ is the cross-power spectrum between the divergence of peculiar velocities and galaxy clustering. On linear scales, the divergence of the peculiar velocities relates to overdensities, $\delta_g$, of a galaxy tracer, as $\theta = - \beta \delta_{g} $ with $\beta = f / b_{g}$, where $b_{g}$ is the bias of the tracer relative to the underlying matter field. Therefore we can approximate the correlation in the denominator as $\beta \hat{P}_{gg}(k, z)$ where we have the galaxy power spectrum as well as $\beta$ which can be constrained from an RSD measurement. In total the effect of galaxy bias cancels on linear scales for this estimator since $\beta \propto 1/b_{g}$, $\hat{P}_{gg} \propto b_{g}^2$ gives $\hat{P}_{\theta g} \propto b_{g}$ and we also have $\hat{P}_{\nabla^2(\psi-\phi) g}\propto b_{g}$. This allows a direct comparison of the measurement to the theory prediction. 

The $\hat{P}_{\nabla^2(\psi-\phi) g}$ is the 3D power spectrum between lensing and galaxy clustering. While the $E_G$ statistic is defined in 3D, the lensing information needed for the estimate is only available as a projected quantity on the sky. Therefore estimators of the $E_G$ statistic are generally built for the projected statistics \citep{Zhang2007,Reyes2010,Pullen2015}.

When combining projected quantities, as is explicit in the $E_G$ statistic, one has to consider carefully the effective redshift of each observed angular power spectrum for the datasets employed to ensure a consistent apples-to-apples comparison.

\section{Data} 
\label{sec:data}
\subsection{\texorpdfstring{\textsl{Planck}}{Planck} CMB lensing data}
\label{sec:Planckdata}

We use CMB lensing measurements based on CMB observations with the \textsl{Planck} Satellite\footnote{\url{https://www.cosmos.esa.int/web/planck}}. The official \textsl{Planck} 2018 public release 3 (PR3) was presented in \citep{PlanckCollaboration2018_overview} and the corresponding analysis of CMB lensing was presented in \citep{PlanckCollaboration2018_lensing}. For this work, we use the latest reprocessing of the \textsl{Planck} data using \textsc{NPIPE}, referred to as PR4 \citep{PlanckCollaboration_PR4_NPIPE}. PR4 contains approximately 8\% more data and represents an improved processing pipeline compared to PR3, resulting in smaller errors overall. The \textsl{Planck} PR4 CMB power spectra give statistically consistent constraints for $\Lambda$CDM compared to \textsl{Planck} PR3 \citep{Rosenberg2022,Tristram2023}. 

We use the CMB lensing map reconstruction based on the \textsl{Planck} PR4 data products presented in \citet{Carron2022}\footnote{\url{https://github.com/carronj/planck_PR4_ lensing}}. A quadratic estimator with additional optimal filtering compared to the PR3 lensing map which in combination with the additional CMB data, results in approximately 20\% stronger constraints on the amplitude of the power spectrum. The amplitude of this \textsl{Planck} PR4 lensing map is statistically consistent with \textsl{Planck} PR3 lensing measurements, with the relative amplitude reported as $1.004 \pm 0.024$ (68\% limits) \citep{Carron2022}. The \textsl{Planck} CMB lensing maps have been extensively tested for systematic effects like foreground contamination in the context of cross-correlation analyses (e.g.  \citep{Pullen2016,Krolewski2020,Chen2022}).

The \textsl{Planck} PR4 lensing map is estimated for the scales $8 \leq \ell \leq 2048$.
We apply a low pass filter $\exp(-(\ell / \ell_{\rm max})^{20})$ to the data with $l_{\rm max} = 1800$ to avoid small-scale noise bleeding into our measurement, for which we only consider scales $\ell < 420$, and we also apodize the mask to avoid sharp edges as done in \citet{White2022}. We perform all calculations in Equatorial coordinates and therefore rotate the \textsl{Planck} maps from the Galactic coordinates.

In addition to the \textsl{Planck} data, we also leverage in this analysis a set of  480 CMB simulated lensing maps used in \citet{Carron2022}. 
They are CMB lensing reconstructions on the CMB map simulations originally presented in \citet{PlanckCollaboration_PR4_NPIPE}. These simulated CMB maps contain a fiducial cosmology signal, realistic noise, and systematics of the analysis pipeline.

\subsection{SDSS BOSS data}

For our analysis, we use the Baryon Oscillation Spectroscopic Survey (BOSS) catalogs of galaxies with spectroscopic redshift information \citep{Dawson2013}. BOSS was part of the Sloan Digital Sky Survey III \citep{Eisenstein2011}, it covers 10,252 square degrees of sky and contains 1,198,006 galaxies. We use the final data release 12 (DR12) which was presented in \citet{Reid2016}\footnote{\url{https://data.sdss.org/sas/dr12/boss/lss/}} and whose cosmological analyses are summarized in \citet{Alam2017_DR12_cosmo_analysis}. The data consists of two large-scale structure galaxy catalogs for cosmological analysis: CMASS and LOWZ. For cosmological analysis, we use the CMASS galaxy catalog in the redshift range of $ 0.43 < z < 0.7 $ and the LOWZ galaxy catalog in the range $0.15 < z < 0.43$.

In the following, we describe in detail how we construct the overdensity maps, used for the cosmological analysis, from the galaxy catalogs. Throughout the analysis \healpix maps with $\mathit{nside}=1024$ are used. 

A mask is constructed containing the fractional coverage ($f_i$) of the survey for each pixel $i$ of the overdensity map. SDSS provides completeness information per sector and veto areas as mangle polygon files. In \Cref{SDSS_mask_making} we describe in detail how we convert these to a pixelized map. We apply a minimum cutoff coverage fraction of $0.6$ to the mask to avoid pixels with low coverage. This cutoff removes 5908 galaxies (0.76\%) for CMASS and 3169 galaxies (0.88\%) for LOWZ. %
Based on the mask the sky fraction of the galaxy sample is determined as 
\begin{equation}
    f_{\rm sky, gal} \equiv \sum_i f_i / N_{\rm pix}. \label{eq:fskygal}
\end{equation}
We also estimate $f_{\rm sky, gal, Planck PR4 overlap}$, the sky fraction for only the pixels that overlap with the Planck PR4 CMB lensing data, by only summing the pixels where the CMB lensing mask is non-zero. After accounting for the redshift cuts and cutoff on the mask the CMASS and LOWZ samples consist of 771294 and 358593 galaxies respectively and cover fractions $f_{\rm sky, gal}^{\rm CMASS} = 0.225$ and $f_{\rm sky, gal}^{\rm LOWZ} = 0.200$ of the sky. The Planck PR4 CMB lensing map covers 96\% of both the CMASS map and LOWZ maps, resulting in $f_{\rm sky, gal, Planck PR4 overlap}^{\rm CMASS } = 0.216$ and $f_{\rm sky, gal, Planck PR4 overlap }^{\rm LOWZ} = 0.192$.
 
To account for systematic effects in the data, weights are applied when constructing the overdensity maps. These weights account for trends in seeing, galactic latitude, redshift failures, and close pairs. The SDSS team provides these as inverse variance weights to correct for the effects. The weight for each CMASS or LOWZ galaxy $g$ is given by \citep{Reid2016}
\begin{equation}
w_{g} = (w_{\rm NOZ} + w_{\rm CP} -1)\cdot  w_{\rm SEEING} \cdot w_{\rm STAR}
\end{equation}
accounting in order for redshift failures, close pairs with fiber collisions, effects of seeing, and bright star contamination. 
Additionally, the FKP weights $w_{\rm FKP}$ given in \citet{Reid2016} are applied. The FKP weights are redshift dependent and therefore need to be accounted for when calculating the galaxy redshift distribution $\dd N/\dd z$. Two separate overdensity maps are constructed: one with the weights $w_{\rm tot} = w_{g} w_{\rm FKP}$ for the auto-correlation and one with weights $w_{\rm tot} =w_g w_{\rm FKP} w_\times$ for the cross-correlation where additional weights (\cref{cross_reweighting}) are applied to match the effective redshift of the auto-correlation. These weight choices are made to match the effective redshift of the angular power spectra with the effective redshift of the RSD analysis (see \cref{sec:two_point_correlation_functions_and_effective_redshift}). Including the FKP weights is optimal for the RSD analysis, reducing the error on $\beta$ significantly \citep{GilMarin2016,Chuang2017} and therefore to match the effective redshift of the angular power spectra we also need to include the FKP weights for the galaxy overdensity maps.

The galaxy overdensity maps $\delta_i$ are constructed as
\begin{equation}
\delta_i = \frac{n_i} {f_i \bar n} -  1,
\label{eq:overdensitymap}
\end{equation}
where we used weighted galaxy counts per pixel $n_i = \sum_{g \in i} w_{\rm tot}$ and the average weighted galaxy count $\bar n = 1/N \sum_i n_i / f_i$. $N$ is the number of pixels with coverage above the cutoff of 0.6, and $f_i$ is the mask.

The impact of $w_{g}$ weights on $E_G$ analyses has been investigated in \citep{Pullen2016} where it was reported that only the seeing and star weights significantly affect the results compared to the size of the statistical uncertainty. Both of these weights are well motivated to correct observational systematics and have been shown to be necessary to recover unbiased clustering measurements \citep{Anderson2014}. The other weights ($w_\times$ and $w_{\textrm{FKP}}$) are only applied based on redshift, not position so can not introduce potential contamination to the correlation but they do affect the measured values as they change the effective redshift of the measurements. 

For the auto-correlation of the galaxies, shot noise needs to be accounted for. The coupled shot noise can be estimated analytically as \citep{Nicola2021,Marques2023}:
\begin{align} 
 \tilde N^{\rm shot} &= \frac{f_{\rm sky, gal}}{n_{\rm eff}} \label{eq:shotnoise}, \\
  n_{\rm eff} &= \frac{\left( \sum_{g} w_{\rm tot}\right) ^2}{4\pi f_{\rm sky,gal} \sum_{g} w_{\rm tot}^2}, \label{eq:neff}
\end{align}
where $n_{\rm eff}$ is the effective number density of galaxies. 

\section{New precision \texorpdfstring{E\textsubscript{G}}{EG}  estimator for CMB lensing and galaxy clustering } 
\label{sec:E_G-estimator}

\subsection{Angular power spectra and effective redshift} \label{sec:two_point_correlation_functions_and_effective_redshift}

In harmonic space, we can measure the angular (cross) power spectra of the observed maps. These angular power spectra are sensitive to the underlying clustering quantified by a 3D matter power spectrum $P(k, z)$. This clustering is measured as projected onto the sky where the projection is described by window functions $W$ that characterize the sensitivity of the observed tracers to the underlying clustering as a function of distance. The observed angular power spectra are therefore generally described under the Limber approximation as
\begin{eqnarray}
    C_\ell^{A B} &\equiv \int \dd \chi \frac{W_A(\chi) W_B(\chi)}{\chi^2} P\left( k=\frac{\ell+1/2}{\chi}, z(\chi)\right), \label{2ptcorr_general}
\end{eqnarray}
where $\chi$ is the comoving distance to redshift $z$, and $\ell$ is the multipole moment.

The angular cross-power spectrum between CMB lensing and galaxy clustering as well as the angular auto-power spectrum for galaxy clustering are given by \citep{Pullen2015}
\newcommand\GRequl{\mathrel{\stackrel{\makebox[0pt]{\mbox{\normalfont\tiny (GR)}}}{=}}}
\begin{align}
C^{\kappa g}_\ell &=\int \dd{z}  \frac{\hat W_{\kappa}(z) W_{g}(z)}{\chi^{2}(z)} P_{\nabla^2 (\psi - \phi) g}\qty(k=\frac{\ell+1/2}{\chi(z)}, z)  \\ 
&\GRequl\int \dd{z}  \frac{W_{\kappa}(z) W_{g}(z)}{\chi^{2}(z)} P_{\delta g}\qty(k=\frac{\ell+1/2}{\chi(z)}, z), \\
C^{gg}_\ell &=\int \dd{z} \frac{H(z)}{c} \frac{W_{g}^2(z)}{\chi^{2}(z)} P_{gg}\qty(k=\frac{\ell+1/2}{\chi(z)}, z),
\end{align}
where $g$ refers to galaxy clustering and $\kappa$ refers to CMB lensing, and $P_{\delta \delta}$ to the matter power spectrum. For the cross-correlation in the case of GR, the Poisson equation can be used to express the measurement as a function of the matter power spectrum. Galaxy clustering is a biased tracer of the underlying matter power spectrum which on linear scales can approximately be described as a linear bias factor $b_{g}$: $P_{\delta g} = b_{g} P_{\delta \delta}$,  $P_{gg} = b_{g}^2 P_{\delta \delta}$ and $P_{\nabla^2 (\psi - \phi) g} = b_{g} P_{\nabla^2 (\psi - \phi) \delta}$. 

Each tracer considered has a specific kernel function $W_{A} (z)$ characterizing it. The lensing kernel for a source at redshift $z_{\rm S}$ is given by
\begin{align}
    W_{\kappa}\left(z, z_{\rm S}\right)&=\frac{3 H^2_{0} \Omega_{\rm m,0} }{2c^2}\hat W_{\kappa} (z, z_S),\\
    \hat W_{\kappa} (z, z_{\rm S}) &\equiv (1+z)\chi(z)\left(1- \frac{\chi(z) }{ \chi ( z_{\rm S})}\right).
\label{eq:lensing_kernel_fixed_source}
\end{align}
For CMB lensing the source is the surface of the last scattering $z_{*}$, therefore the CMB lensing kernel is given by $W_{\kappa}(z) = W_{\kappa}\left(z, z_{*}\right)$, $\hat W_{\kappa}(z) = \hat W_{\kappa}\left(z, z_{*}\right)$.

For galaxy clustering, the kernel is given by the weighted and normalized galaxy redshift distribution 
\begin{equation}
    W_{g}\left(z \right)= \frac{\dd N}{\dd z}.
\end{equation}

Note that beyond this description observed angular power spectra have additional subdominant contributions like contamination of the galaxy clustering from foreground gravitational lensing, referred to as magnification bias (discussed in \cref{sec:magnification_bias}). Throughout the analysis, angular power spectra for the fiducial cosmology are calculated using \ccl\footnote{\url{https://github.com/LSSTDESC/CCL}} \citep{Chisari2019} with the underlying 3D power spectra calculated using \camb \citep{Lewis2000,Howlett2012}. 

Angular power spectra between two tracers sample the underlying 3D power spectra at a range of redshift as defined by the combination of the two window functions $W_A(\chi), W_B (\chi)$. The effective redshifts of the angular power spectra are given by \citep{Chen2022}
\begin{align}
    z_{\rm eff}^{\rm AB} = \frac{\int \dd\chi \, \chi^{-2} W_A (\chi) W_B (\chi) z(\chi)}{\int \dd\chi \, \chi^{-2} W_A (\chi) W_B (\chi)},
\end{align}
and more specifically for the angular cross-power spectrum $C_\ell^{\kappa g}$ and angular auto-power spectrum $C_\ell^{gg}$ as
\begin{align}
    z_{\rm eff}^{\rm cross} &= \frac{\int \dd z \, \chi^{-2} \hat W_\kappa(z) W_{g} (z) z}{\int \dd z \, \chi^{-2} \hat W_\kappa(z) W_{g} (z) },  \\
    z_{\rm eff}^{\rm auto} &= \frac{\int \dd z \, \chi^{-2}(z)  H(z) c^{-1} W_{g}^2(z) z}{\int \dd z \, \chi^{-2}(z)  H(z) c^{-1} W_{g}^2(z)}. \\
\end{align}
It is important to note, therefore, that the effective redshift of the auto and cross-correlation for the same galaxy sample do not match in general. Comparing the angular power spectrum measured at different effective redshifts would bias the $E_G$ estimation. This can be avoided by reweighting each galaxy for the cross-correlation by an additional weight
\begin{align}
    w_\times (z) = \frac{\dd N}{\dd z} \frac{1}{\hat W_\kappa(z) I} \label{cross_reweighting},\\
    I = \int \dd z \frac{W_{g}^2 (z)}{\hat W_\kappa (z)},
\end{align}
evaluated at our fiducial cosmology where $I$ is the normalization so that the kernel for the reweighted sample
\begin{align}
    W^*_g \equiv \frac{\dd N^*}{\dd z} = \frac{\dd N}{\dd z} w_\times (z) 
\end{align}
is correctly normalized ($\int \dd z W^*_{g}=1$).
The effective redshift of the angular cross-power spectrum with the reweighting, $C_\ell^{\kappa g*}$, then matches the effective redshift of the angular auto-power spectrum. Additionally, the effective redshift for an RSD analysis based on the 3D clustering of the same galaxy data matches the effective redshift of the 2D clustering analysis \citep{Chen2022} which in total gives
\begin{align}
    z_{\rm eff} \equiv z_{\rm eff}^{\rm cross*} = z_{\rm eff}^{\rm auto} = z_{\rm eff}^{\beta}.
    \label{eq:zeff}
\end{align} 
These three measurements at the same effective redshift can be combined to build an estimator for the $E_G$ statistic. Note for this to hold the galaxy sample needs to be treated the same for all three measurements, including the same weights except for the additional reweighting for the cross-correlation.

In the original SDSS BOSS analyses, which were referenced for the redshift of the measurement in previous $E_G$ analyses \citep{Pullen2016,Singh2019}, the weighted mean redshift of the galaxy sample was used to estimate the redshift of the measurement. This is less accurate than the above prescription, especially for angular power spectra \citep{Chen2022}. 

The difference in effective redshift between auto and cross-correlation has been noted in earlier $E_G$ estimations using CMB lensing and galaxy clustering. It was addressed in these works by a correction that relies on accurate simulations and a HOD \citep{Pullen2016,Singh2019,Zhang2021}. In this work, by reweighting, we remove the need for such a correction.

\subsection{Derivation of the estimator}
\label{sec:EGderiv}
In this section, we present a novel estimator for $E_G$ using CMB lensing and galaxy clustering. We build on previous analyses \citep{Pullen2015,Pullen2016,Singh2019,Zhang2021} but make key improvements by carefully matching the effective redshifts of the observables, reducing the corrections needed, and minimizing the error introduced from approximations. %

To define a consistent projected $E_G$ estimator the 3D power spectra in \cref{estimator_3D} need to be projected with the same set of window functions. Based on the discussion of effective redshift a convenient choice is using the galaxy kernel resulting in an estimator as a function of scale $\ell$ at the redshift $z_{\rm eff}$
\begin{equation}
    \hat{E}_G^{\ell} (z_{\rm eff})=\frac{2c^2 \hat{C}_\ell^{\nabla^2(\psi-\phi) g}}{3 H_0^2 \beta \hat{C}_\ell^{gg}},
\end{equation}
where $\hat C_\ell^{gg}$ is the angular auto-power spectrum of the galaxy sample, $\beta$ is measured from an RSD analysis at the same effective redshift as the auto-correlation and 
\begin{align}
    \hat{C}_\ell^{\nabla^2(\psi-\phi) g} &= \frac{1}{2}\int \dd z \frac{H(z)}{c (1+z)} \frac{W_{g}^2(z)}{\chi^2 (z)} \hat{P}_{\nabla^2(\psi-\phi) g}(k, z). 
\end{align}
This helper function is closely related to the observable cross-correlation $C_\ell^{\kappa g*}$. We, under careful consideration of the approximations needed, equate this expression with $C_\ell^{\kappa g*}$, minimizing the overall error of the estimator.

By applying the reweighting scheme (\cref{cross_reweighting}) to the cross-correlation analyses in order to match the effective redshift of the cross-correlation to the auto-correlation, the galaxy kernel for the cross-correlation is
\begin{align}
    W_{g}^*(z) = W_{g}(z) w_\times (z).
\end{align} 
Plugging in $w_\times (z)$ we have that the projection kernel $W_{g}^2(z)$ can be expressed as
\begin{equation}
    W_{g}^2(z) =  \hat W_\kappa(z) W_{g}^*(z) I . \label{eq:proj_kernel_eqality_to_lensing}
\end{equation}
This reweighting of the galaxy sample removes the mismatch in effective redshift that would otherwise be present between the auto and cross-correlations. Without this reweighting a scale-dependent bias of multiple percent, depending on $\dd N/\dd z$,  would be present in the estimator (see \cref{comparingwithPullenEGestimator}). %
We note that the re-weighting also affects the variance of the weights which affects the shot noise in the galaxy data. In practice for our particular case, where we also apply the redshift-dependent FKP weights the redshift-dependent weights partially cancel so that the overall effect of the re-weighting is a decrease in variance compared to the sample for the auto-correlation increasing the constraining power of the cross-correlation. 

Using \cref{eq:proj_kernel_eqality_to_lensing} the helper function can be well approximated with a cross-correlation measurement as 
\begin{align}
    \hat{C}_\ell^{\nabla^2(\psi-\phi) g} &\approx  \frac{H(z_{\rm eff}) I}{c} \int \dd z \frac{\hat W_\kappa(z) W_{g}^*(z)}{\chi^2(z)} \hat{P}_{\nabla^2(\psi-\phi) g} (k,z) \\
    &=  \frac{H( z_{\rm eff}) I}{c} C_\ell^{\kappa g*}, \label{eq:numerator}
\end{align}
where $ C_\ell^{\kappa g*}$ is our cross-correlation measurement with the reweighted galaxy sample and we approximate $H(z)$ as slowly varying within the redshift sample.

In summary, the new $E_G$ estimator based on CMB lensing and a galaxy sample is given by 
\begin{equation}
    \hat E_G^\ell (z_{\rm eff}) \approx \Gamma(z_{\rm eff}) \frac{C_\ell^{\kappa g*}}{\beta C_\ell^{gg}} \label{E_G_estimator}
\end{equation}
where 
\begin{equation}
    \Gamma (z_{\rm eff}) \equiv \frac{2 c H(z_{\rm eff}) }{3 H_0^2} \int \dd z \frac{W_{g}^2(z)}{\hat W_\kappa (z)}
\end{equation}
and $z_{\rm eff}$ is the effective redshift of the observables given by \cref{eq:zeff}.

\subsection{Accuracy of the \texorpdfstring{E\textsubscript{G}}{EG} estimator}
\label{sec:E_G_estimator_accuracy}

Throughout the derivation of the $E_G$ estimator presented in \cref{sec:EGderiv}, some approximations were needed. While the overall bias is minimized by carefully matching the effective redshift of each measurement through the use of a reweighting scheme, $\beta$, and $H$ are assumed as slowly varying within the redshift range of the galaxy sample and evaluated at the effective redshift of the measurement. Astrophysical effects like an evolution of the galaxy bias as a function of redshift and the effect of magnification bias can affect the result. 
We now characterize the accuracy of the estimator as defined in \cref{E_G_estimator} and its sensitivity to astrophysical effects for the \textsl{Planck} CMB lensing maps and SDSS BOSS galaxy samples considered in this work. In this work, we neglect errors in the redshift estimation of the galaxies since they are based on spectroscopic observations and are expected to be small. For the reweighting, the co-moving distance across the redshifts covered by the galaxy sample at the fiducial cosmology is used (see \cref{cross_reweighting,eq:lensing_kernel_fixed_source}). This is not a concern since the $E_G$ statistic tests models that reproduce the expansion history as measured for our universe but predict deviations in the growth of structure. We neglect the small uncertainty in the fiducial cosmology for the reweighting in this analysis.

First, in \cref{sec:noastrocomp}, we consider the numerical accuracy of the estimator when ignoring the astrophysical complications of dependencies in the galaxy bias and the effect of magnification bias. Then, in \cref{sec:biasevol} we quantify the systematic bias from the redshift evolution of the galaxy bias. Finally, in \cref{sec:magnification_bias}, we consider that the observables are affected by magnification bias for which we quantify the effect on our estimator and find a correction in the case of CMASS. 

\subsubsection{Systematic without astrophysical complications}
\label{sec:noastrocomp}

The overall systematic bias of the estimator as a result of the approximations made during the derivation can be quantified by comparing the estimator calculated from the analytic observables as defined in \cref{E_G_estimator} for the fiducial cosmology with the $E_G^{\rm GR}$ value at the fiducial cosmology. The overall systematic bias is given by 
\begin{align}
    S^{\ell}_\Gamma  \equiv \frac{E_G^{\rm GR} (z_{\rm eff})} {\hat E_G^\ell (z_{\rm eff})}. \label{eq:error_in_EG}
\end{align}

\begin{figure}
\includegraphics[width=\columnwidth]{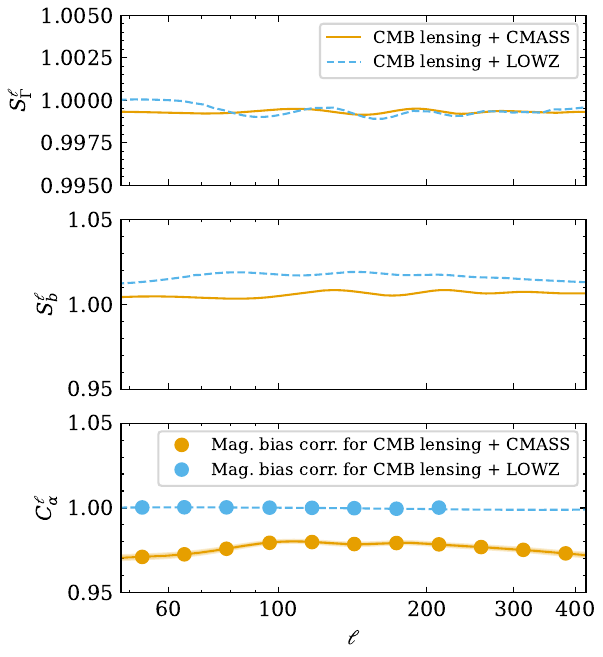}
\caption{Accuracy of the $E_G$ estimator presented in this work for CMB lensing combined with the CMASS [orange, solid line] and LOWZ [blue, dashed line] galaxy samples with one indicating consistency between estimator and input. [Upper panel] The overall systematic error of the estimator, $S_\Gamma^\ell$, before considering astrophysical systematics. [Middle] Our estimate of the systematic error from bias redshift evolution within the galaxy samples of CMASS and LOWZ, $S_b^\ell $. [Lower] The overall impact of magnification bias, $C_\alpha^{\ell}$, which we correct for with the binned correction shown as dots.}
\label{fig:systeamtic_biases}
\end{figure}

The upper panel of \cref{fig:systeamtic_biases} shows the result for $S^\ell_\Gamma$ at our fiducial cosmology. The $E_G^{\rm GR}$ value is calculated using \cref{eq_EG_GR} for our fiducial cosmology. The $\hat E_G^\ell (z_{\rm eff})$ value is calculated by plugging into \cref{E_G_estimator} the angular power spectra and $\beta$ calculated for our fiducial cosmology with a constant galaxy bias of $b_{g}=2.0$ and no magnification bias using \ccl (see \cref{sec:two_point_correlation_functions_and_effective_redshift}). 
For the estimator presented in this work, we find that the $S^{\ell}_\Gamma$ systematic is negligible compared to the observational uncertainty, with $|S^{\ell}_\Gamma - 1|< 0.3\% $ on all scales in our cosmological ranges and for both CMB lensing + CMASS and CMB lensing + LOWZ. We conservatively include the upper limit of $0.3\%$ in our systematic error budget. 

For a previous harmonic space estimator for $E_G$ presented in \citet{Pullen2016} a similar quantity referred to as $C_\Gamma$ has been investigated; the quantity $S_\Gamma^{\ell}$ also considers the redshift evolution of the growth $f$ but is otherwise analogous (not considering the redshift evolution of $f$ leads to an amplitude difference of around 0.1\% so they are equivalent within the accuracy relevant to this discussion). Therefore we can compare the overall systematic of the two estimators. For the previous estimator, the systematic was at the level of a few percent, depending on $\dd N(z) /\dd z$, %
and required a correction based on N-body simulations and assuming a HOD for the galaxy samples \citep{Pullen2015,Pullen2016,Zhang2021}. In \cref{comparingwithPullenEGestimator} we compare the estimator presented in this work with the one presented in \citet{Pullen2015} in detail and we discuss the impact of each difference. For the estimator presented in this work, no correction based on simulations needs to be applied before considering astrophysical effects since the bias $S_\Gamma$ is negligible, and therefore the results do not rely on the accuracy of simulations or choice of HOD. 

\subsubsection{Systematic from bias evolution with redshift}
\label{sec:biasevol}

Next, one can consider the effect of redshift evolution of the galaxy bias. So far a constant galaxy bias has been assumed on linear scales. To separate $\beta$ and $C_\ell^{gg}$ there is an implicit assumption that the galaxy bias $b_{g}$ is not only scale independent but also constant in redshift. If there is a significant evolution of the galaxy bias with redshift within a given galaxy sample this can introduce a systematic bias. Given a bias evolution $b_{g}(z)$ we can analytically estimate the systematic bias to the estimator from this redshift evolution as

\begin{align}
    S_b^\ell &= \frac{C_\ell^{gg}}{b_{g} C_{\ell}^{\delta g}},\\
    C_\ell^{\delta g} &\equiv \int \dd z \frac{H(z)}{c} \frac{W_{g}^2(z)}{\chi^2(z)} P_{\delta g}(k,z), \label{eq_corr_with_one_bias_removed}
\end{align}
where $b_{g}$ is the constant galaxy bias at the effective redshift as before but the power spectra projected in the angular power spectra are given by $P_{gg} = b_{g}^2(z) P_{\delta \delta}$ and $P_{\delta g} = b_{g}(z) P_{\delta \delta}$ with $b_{g}(z_{\rm eff}) = b_{g}$. We implement this calculation using \ccl. The implementation of \texttt{ccl.NumberCountsTracer} is already set up to handle a redshift-dependent bias as an input which we use to calculate $C_\ell^{gg}$ here. For $C_\ell^{\delta g}$ we correlate one \texttt{ccl.NumberCountsTracer} with the redshift-dependent bias and one with a bias equal to one resulting in the quantity of \cref{eq_corr_with_one_bias_removed}.

This characterization of systematic error due to the deviation of galaxy bias from a constant is identical to $"C_b (\ell)"$ in previous analyses \citep{Pullen2016,Zhang2021}. In \citet{Pullen2016} this was studied for CMASS based on power spectra from N-body simulations and assuming an HOD testing for the impact of bias evolution and also scale dependence of the bias. They report systematics of between 0 and 1\% over their cosmological range.

We can also study this systematic from measurements of $b_{g}(z)$ based on the data. 
To do this measurement one has to assume GR, so we can not directly differentiate an astrophysical bias evolution within the sample from a deviation from GR that affects the clustering amplitude. Therefore we can not easily correct this systematic by constraining the biased evolution of the galaxy samples. However, we can study how large the effect on our estimator is to inform if a measured disagreement with GR could be caused by the bias evolution informing our systematic bias estimate. If it is found that the systematic bias from bias evolution is relevant compared to the observational uncertainty, narrower bins in redshift would need to be used to reduce the systematic bias. We note that in principle there could be a scenario where there is a strong bias evolution with a redshift that is largely canceled by an opposite deviation from GR in the growth history. In such a case, the systematic bias would not be fully captured by this estimate. We neglect this in this work but note that this could be investigated in future work with higher constraining power by comparing $E_G$ results for narrower bins in redshift with results for the full samples. 

In \citet{Farren2023} the galaxy bias was constrained for CMASS and LOWZ in redshift increments of $\Delta z = 0.05$ assuming a $\Lambda$CDM growth history. Using the values reported in their Table 13 and performing a linear least squared fit results in slopes $s$ of approximately $s = [1.2, 1.6, 1.8,  1.8]$ for CMASS North, CMASS South, LOWZ North, and LOWZ South. We use $b_{g, \rm LOWZ}(z) = 2.0 + 1.8 (z - 0.316)$ and conservatively assume the steeper evolution for CMASS $b_{g, \rm CMASS}(z) = 2.0 + 1.6 (z - 0.555)$. Note that for this test the focus is the estimate of the evolution with redshift, the test is only weakly sensitive to small differences in the overall amplitude from the fiducial value of 2.0. We note that the best-fit for the values from \citet{Farren2023} for the LOWZ amplitude is approximately 2.0 for both North and South and for the CMASS amplitude is 2.2 and 2.1 for the North and South samples respectively, all consistent with the overall $b_{g}$ fitted to data in this work (see \cref{sec:galaxy_sims}).

The middle panel in \cref{fig:systeamtic_biases} shows the constraints on $S_b^\ell$. Given the estimated bias redshift evolution, the systematic bias in $E_G$ when using CMASS is at most 1\% and when using LOWZ at most 2\%. These systematic biases are much smaller than the statistical uncertainty of the measurement and therefore bias evolution with redshift is not of significant concern for the analysis. We account for the effect in our systematic error budget with 1\% uncertainty on $E_G$ for CMASS and 2\% for LOWZ.

\subsubsection{Correcting for Magnification bias} 
\label{sec:magnification_bias}

So far in the discussion, we have neglected the effect of magnification bias on the observable cross and auto-correlation.
The brightness of observed galaxies, similarly to their shape, gets modulated by lensing along the line of sight from intervening gravitational potential \citep{Hildebrandt2009,Schmidt2012,Duncan2014,Hildebrandt2016,Thiele2020,Unruh2020}. For surveys with a detection threshold, this magnification affects the local observed number density of galaxies and thereby introduces a systematic effect known as the magnification bias. 

In the literature, this is known to be a significant systematic for current-generation surveys, especially on the cross-correlation signal, and therefore needs to be accounted for in cosmological analyses \citep{Krolewski2020,Maartens2021,vonWietersheimKramsta2021,Duncan2022,EuclidCollaboration2022_magnification_bias,ElvinPoole2023, Wenzl2023magbias}. Specifically for the $E_G$ statistic it has been highlighted as an important systematic that should be accounted for \citep{MoradinezhadDizgah2016,Yang2018,Ghosh2019} and has been included in a recent $E_G$ constraint using SDSS eBOSS \citep{Zhang2021}.
However, the magnification bias effect has been neglected by previous $E_G$ estimates using SDSS BOSS and Planck CMB lensing \citep{Pullen2016,Singh2019}.

The magnification bias effect can be quantified via the function $\alpha(z)$ that describes the sensitivity of the sample selection to lensing. For a review of the functional form of the impact of the magnification bias on the observable angular auto- and cross-power spectra, we refer the interested reader to \citet{Wenzl2023magbias}. 
We leverage the recent measurements of $\alpha$ for the LOWZ and CMASS samples from \citep{Wenzl2023magbias} for our specific set of analysis choices and accounting for the full photometric selections of the samples. 

For LOWZ we use $\alpha_{\rm LOWZ} = 2.47 \pm 0.11$ where we use the mean value between the best-fit value for the galaxy sample with and without reweighting for the cross-correlation. The uncertainty conservatively combines statistical uncertainty ($0.02$), systematic uncertainty from redshift evolution ($0.07$) as well as half the difference between with and without reweighting for the cross-correlation ($0.02$). 

For CMASS we account for the observed redshift evolution by using the best-fit function 
\begin{equation}
    \alpha_{\rm CMASS}(z) = 2.71 + 8.78 (z - 0.55),
    \label{eq:cmass_zdep}
\end{equation}
with statistical uncertainties for the constant and slope of 0.08 and 1.26. We conservatively widen the uncertainty of the constant to 0.10 to account for the reported systematic uncertainty from the light profile choice of 0.02.

In the estimate of $\hat E_G$, we account for the effect of magnification bias by including a correction term $C^\ell_\alpha$ throughout our analysis by replacing the $\Gamma$ factor in \cref{E_G_estimator} as
\begin{align}
    C^\ell_\alpha &\equiv \frac{C_\ell^{\kappa g*}}{C_\ell^{\kappa g*,\rm mag }} \frac{C_\ell^{gg,\rm mag  }}{C_\ell^{gg}}, \\
    \Gamma &\rightarrow \Gamma_\ell \equiv \Gamma  C^\ell_\alpha,
\end{align}
where the $C_\ell^{\kappa g*, \rm mag }$ and $C_\ell^{gg, \rm mag}$ are the angular cross- and auto-power spectra additionally including the effect of magnification bias calculated as described in \citep{Wenzl2023magbias}. All angular power spectra are calculated at the fiducial cosmology and binned using the mode coupling matrix as described in \cref{sec:namaster}. The correction factors for LOWZ and CMASS are shown in \cref{fig:systeamtic_biases}. For LOWZ the correction is negligible, with corrections of $<0.1\%$ in the cosmological range of this analysis. For the case of \textsl{Planck} + CMASS, the magnification bias impact is 2-3\% and we correct this effect in our analysis with the binned corrections shown as dots in the figure. Note that this correction is such that it reduces the amplitude of the $E_G$ measurement for both SDSS BOSS samples and any other galaxy sample with $\alpha > 1$. 

When plotting the measurement of angular power spectra compared to theory spectra we also include the magnification bias contribution in the comparison theory curves. The magnification bias contribution is also consistently included in the simulated galaxy maps and covariance estimation. From the uncertainty of the magnification bias, we estimate the uncertainty of the correction applied. Like the correction itself, this is negligible for LOWZ. For CMASS the uncertainty of $\alpha(z)$ translates to at most $0.2\%$ uncertainty in $C_\alpha^{\ell}$, shown in \cref{fig:systeamtic_biases} as a narrow band around the estimate. Based on this we account for a $0.2\%$ error for our systematic budget of the estimate involving CMASS.

\section{Measuring the observables from data} 
\label{sec:Measuring_observables}

In this section, we describe the signal extraction approach and discuss how we characterize the uncertainty on our observables and validate our pipeline. The angular power spectra estimation approach is outlined in \cref{sec:namaster}. \cref{sec:simulated_datavec,sec:covariance} describe the use of simulations for covariance estimation and to perform pipeline validation and \cref{sec:shot_noise_test} describes a test of the shot noise estimate. The power spectrum measurements from the \textsl{Planck} PR4 CMB lensing map and the SDSS BOSS galaxy samples are summarized in \cref{sec:angular_power_spectra_measurements} and the RSD parameter $\beta$ in \cref{sec:beta}. 

\subsection{Angular power spectrum estimation} \label{sec:namaster}

To measure the angular auto- and cross-power spectra from our observed maps we use a pseudo-Cl estimator. Given two maps (A, B) expanded in spherical harmonics ($a_{\ell m}$, $b_{\ell m}$) one can estimate the auto- or cross-power spectrum at each multipole $\ell$ as
\begin{align}
\hat D_\ell^{A B} = \frac{1}{2\ell +1 } \sum_{m=-\ell}^{\ell} \operatorname{Re} (a_{\ell m} b_{\ell m}^{*}). \label{angular_power_anafast}
\end{align}
In general, our observations only cover a fraction of the sky so that the overall amplitude of this simple estimate $\hat D_\ell$ is approximately a factor $f_{\rm sky}$, the sky fraction of the overlap between the maps, lower than the expectation for the full sky. Additionally, complex mask geometries couple originally independent modes at neighboring scales. To reduce the correlation between the measurements, the large noise of the individual multiples, and for computational efficiency, we measure the angular power spectra in bandpowers $i$\footnote{In this section we use the subscript $i$ for clarity but then simplify to using $\ell$ for the rest of the work referring to the effective $\ell$ of each bandpower which we also use in plots.}. We use the unbiased pseudo-$C_\ell$ estimator \namaster\footnote{\url{https://github.com/LSSTDESC/NaMaster}} presented in \citet{Alonso2019}. The estimates for the angular power spectra are given by
\begin{align}
\hat C_i^{A B} =  \sum_{i'} [M^{-1}]_{i i'}  \frac{1}{N_{i'}} \sum_{\ell \in i'} \hat D_\ell^{A B}, \label{eq_binned_angular_power_spectrum_estimate}
\end{align}
where $N_i$ is the number of multipoles $\ell$ in bandpower $i$ and $M$ is the so called binned coupling matrix. This binned coupling matrix is defined by the geometry of the observational masks of the two maps being correlated as well as the bandpowers and it accounts for both the loss of overall amplitude from partial sky coverage and the coupling between modes (See \citep{Alonso2019} for further details).

Additionally, for the galaxy auto correlation $C_\ell^{gg}$ the shot noise in the map correlates with itself. This contribution needs to be subtracted to estimate the underlying cosmological correlation. We do this before binning as
\begin{align}
\hat D_\ell \rightarrow  \hat D_\ell - \tilde N^{\rm shot} \label{eq:shotnoisesubtraction}
\end{align}
where the coupled shot noise estimate $N^{\rm shot}$ is from \cref{eq:shotnoise}.

We use a map resolution parametrized by $\mathit{nside}=1024$ throughout our analysis\footnote{This represents a map with $N_{\rm pix} = 12 \cdot \mathit{nside}^2 = 12,582,912$ pixel.}. Since the galaxy map is constructed in real space we need to account for the loss of power due to the finite pixelization used \citep{Marques2023}. To do this we apply the pixel window function specific to our map resolution as an effective beam function for the galaxy field objects in \namaster. For the auto-correlation, this is done after shot-noise subtraction.

We also use the mode coupling matrix to find the expectation value of the angular power spectrum for each band power given by 
\begin{align}
\langle  \hat C_\ell \rangle = \sum_{\ell'} M_{\ell \ell'} C_{\ell'} . \label{eq:binned_theory_spectra}
\end{align}
where $C_\ell$ is the unbinned theory spectrum. We apply this throughout our analysis when comparing theory curves to the measured $\hat C_i^{A B}$and to bin scale-dependent correction factors into the band powers.

We use a log-spaced binning scheme for our analysis. For computational accuracy and to avoid leakage at the edges we perform all computations in the full range given by our map resolution ($\ell \in [2, 3071]$). 

For our cosmological analysis, however, we only use linearly dominated scales up to $k= \SI{0.2}{Mpc^{-1}}$, a real space cutoff of around $\SI{30}{Mpc}$ (comoving). For our logarithmic binning, these scale cuts limit us to $\ell_{\rm max} = 420$ and $\ell_{\rm max} = 233$ for the effective redshifts of CMASS and LOWZ respectively. We also limit the cosmological analysis on large scales to $\ell_{\rm min} = 48$ following the recommendation in \citet{Chen2022} to conservatively avoid large scales where foregrounds in the galaxy maps could introduce spurious correlations with CMB lensing. %

The 11 log-spaced bins in the cosmological range used throughout the analysis have edges $[48, 59,   71,   87, 106,  129,  157,  191,  233,$ $283, 345,  420 ]$ where each bin is inclusive of the lower edge and exclusive of the upper edge as per \namaster convention. In the following, we will label each bin by their effective $\ell$ which are given by $[ 53, 64.5, 78.5, 96, 117,$ $142.5, 173.5,$ $211.5, 257.5, 313.5, 382]$ with the analysis for LOWZ only using the first 8 bins.

\subsection{Simulated data}
\label{sec:simulated_datavec}

To validate our estimates of the angular correlation functions, and to estimate the covariance of the measurements, we consider simulated realizations of the observables. 
The CMB lensing reconstruction has a known transfer function, in \cref{sec:CMBsims} we use the simulated realizations of the CMB lensing maps, described in \cref{sec:Planckdata}, to estimate a norm correction to account for this. In \cref{sec:galaxy_sims}, we create realizations of the galaxy overdensity maps with and without reweighting for the cross-correlation. We create them with the same level of shot noise as our galaxy samples and give them the expected correlation with the CMB lensing realizations and to each other. Finally, in \cref{sec:input_recovery} we use our simulated observables to test our measurement code against the input.

For the theory curves to draw our simulated realizations from, we use our fiducial cosmology discussed in \cref{sec:result_summary}.

\subsubsection{Simulated CMB lensing maps and norm correction} 
\label{sec:CMBsims}

We use the set of 480 simulations used in \citet{Carron2022} to estimate the norm correction of the \textsl{Planck} PR4 CMB lensing maps and to estimate the covariances of our cross-correlations with the CMASS and LOWZ galaxy samples. 
For all of these, we know the Gaussian realization of the CMB lensing angular power spectrum used as an input for the simulation ($a_{\ell m}^{\kappa}$).

It is well known that lensing reconstruction with quadratic estimators in the presence of a mask can lead to offsets of the overall normalization \citep{BenoitLevy2013,Carron2023,Farren2023}. This normalization offset was not accounted for in previous $E_G$ measurements with \textsl{Planck} and SDSS \citep{Pullen2016,Singh2019,Zhang2021}.

In the context of cross-correlation with galaxy samples, \citet{Farren2023} showed that this overall normalization can be corrected based on end-to-end simulations of the pipeline. Using the approach in \citep{Farren2023} and \citet{Krolewski2023}, we calculate the Monte Carlo correction based on the set of 480 simulations as the ratio between the cross-correlation of appropriately masked CMB lensing input realizations with the cross-correlation of the CMB lensing input realization to the map reconstruction. The functional form is given by
\begin{align} 
\textrm{mc-corr}(\ell) &\equiv \frac{\bar C_{\ell}^{\kappa_{g \rm -mask}, \kappa_{\rm CMB-mask,apodized}}  }{\bar C_{\ell}^{\kappa_{g\rm -mask}, \hat \kappa_{\rm apodized}} } , \\
 \hat a_{\ell m}^{\kappa} &\rightarrow \textrm{mc-corr}(\ell )\cdot \hat a_{\ell m}^{\kappa}, \label{eq:mc_corr}
\end{align}
where $\kappa_{g\rm -mask}$ are the input $\kappa$ maps $a_{\ell m}$ with the mask of the galaxy sample to be correlated with applied, $\kappa_{\rm apodized}$ are the reconstructed maps $\hat a_{\ell m}^{\kappa}$ with the same apodization and low pass filter we apply to the data and $\kappa_{\rm CMB-mask, apodized}$ are the input $\kappa$ maps $a_{\ell m}$ with the CMB lensing mask applied including the same apodization that is applied to the data. We here use a hat to indicate maps estimated with our pipeline and no hat for the input maps. Note that the \textsl{Planck} PR4 CMB lensing mask is binary. We take the average angular power spectrum $\bar C_\ell$ over all 480 simulations. The correlation is calculated without accounting for the mask coupling in (see \cref{angular_power_anafast}) since we do not want to account for the coupling matrix in our norm correction to avoid correcting it twice. We use the implementation \texttt{healpy.sphtfunc.anafast} \citep{Gorski2005,Zonca2019}. 

For the \textsl{Planck} PR4 $\kappa$ map, anisotropic filtering was used which can make the norm correction dependent on the specific area of overlap of the galaxy sample that we correlate with. While for the full map, the norm correction is approximately 2\%, we find that for the overlap with CMASS and LOWZ, the mc-correction for each is approximately 4\% up in the cosmological range. We apply the mc-correction throughout our analysis. We re-calculate the mc-correction for each galaxy sample and the North and Southern patches separately when considering them separately in our analysis. The norm correction affects the amplitude of the angular cross-correlation measurements and thereby directly shifts our $E_G$ measurement. We will discuss the level of impact the norm correction has on our measured result in \cref{sec:consistency_tests}.

\subsubsection{Simulated correlated galaxy maps} 
\label{sec:galaxy_sims}

While the $E_G$ statistic is designed to optimally cancel the galaxy bias to create simulated realizations of the observation we need to estimate the galaxy bias for each galaxy sample. We iteratively update the galaxy bias to ensure the amplitudes of the theory curves for the galaxy auto-correlation used to make our simulations match well with the clustering amplitude in our galaxy samples. For this purpose, we measured the $C_\ell^{gg}$ and then fitted our theory curve to the data with only the galaxy bias as a free parameter. We performed the fit in the cosmological ranges for CMASS and LOWZ and used relative weights for each bin informed by the analytic error (see \cref{sec:covariance} for the description of our analytic error calculation). 

For the initial analytic error, we assumed our fiducial cosmology and an initial guess of 2.0 for the galaxy bias for both the CMASS and LOWZ samples, consistent with the literature \citep{White2011,Parejko2013}. We also include magnification bias in the theory curves since the effect is in the measurement and to ensure that our end-to-end pipeline test covers the magnification bias correction. For our full CMASS (LOWZ) galaxy map and fiducial cosmology, we find a best-fit value of $b_g = 2.14$ ($b_g=2.02$). We use these best-fit galaxy bias values, the magnification bias measurement, and our fiducial cosmology to define the theory curves used to create the simulated galaxy maps. 

With the theory curves established, we now leverage them to create the galaxy maps with the expected cross-correlation with our set of CMB lensing simulations and also the expected auto-correlation including the shot noise contribution. Given the need to consider two different weighting schemes for the auto- and cross-correlation analysis (see \cref{E_G_estimator}), two versions of each galaxy map simulations are created to effectively match the respective redshift distributions of the two weighting schemes while also being correctly correlated to each other. 
Given our theory spectra, we build the $a_{\ell m}$ of the simulated maps by combining (building on the approaches in \citep{Kamionkowski1997,Farren2023})
\begin{align}
    a_{\ell m}^{g^X} = \frac{C_\ell^{\kappa g^X}}{C_\ell^{\kappa \kappa}} a_{\ell m}^{\kappa} + a_{\ell m}^{g^X, \rm uncorr.} + a_{\ell m}^{g, \rm noise},
\end{align}
where $g^X \in [g, g^*]$ and the contributions are drawn so that
\begin{align}
    &\langle a_{\ell m}^{g^X, \rm uncorr.}, (a_{\ell m}^{g^Y, \rm uncorr.})^* \rangle = \nonumber\\
    &\delta_{\ell \ell'} \delta_{m m'} \left( C_\ell^{g^X g^Y} -\frac{C_\ell^{\kappa g^X} C_\ell^{\kappa g^Y}}{C_\ell^{\kappa \kappa}} \right) \\
    &\langle a_{\ell m}^{g^X, \rm noise}, (a_{\ell m}^{g^X, \rm noise})^* \rangle = \delta_{\ell \ell'} \delta_{m m'} n_{\rm eff}^{g^X}.
\end{align}
Here $n_{\rm eff}^{g^X}$ is the effective noise for each galaxy weighting scheme as defined in \cref{eq:neff}. 

For the implementation, we use the \texttt{synalm} function from \healpy and then we apply the respective galaxy mask to the sims. The $a_{\ell m}^{\kappa}$ are the set of Gaussian realizations of the CMB lensing field used for the generation of the simulated CMB lensing maps. We note that the $a_{\ell m}^\kappa$ were generated for a slightly different input $C_\ell^{\kappa \kappa}$ than our fiducial cosmology. To account for this and avoid a slight bias in the simulated maps, we use the exact $C_\ell^{\kappa \kappa}$ used to generate the $a_{\ell m}^{\kappa}$. 

For the shot noise in the galaxy maps, we assume Gaussianity and sample the noise for each pixel according to our effective noise as well as the coverage fraction for each pixel. The shot noise between the galaxy maps with and without reweighting is correlated, to account for this we use the same realization of the standard normal for both and scale them according to the shot noise in each. This approach ensures that the simulated maps are correctly correlated between them so we can accurately estimate the full covariance including cross-covariances.

\subsection{Covariance estimation}
\label{sec:covariance}

We need to carefully characterize the covariance matrix of our measurement. Due to the complicated mask geometry of the galaxy samples, measurements at different scales become correlated. As discussed in \cref{sec:namaster} we are using wide bins which reduce the correlation between bins, but neighboring bins will still have some amount of correlation. Additionally, the auto- and cross-correlation at similar scales are also expected to be correlated with each other as both scale with the amount of clustering in the galaxy data. 

We use multiple approaches to estimate the full covariance to check for consistency and validate our approach. We use our set of simulated maps to estimate the covariance for our baseline. We then compare the simulated covariance to an analytic estimate where we compare both the marginalized errors and the off-diagonals. Finally, we perform an internal error estimate on the data using a jackknifing technique to validate the marginalized errors.

As our baseline, we estimate the sample covariance from our set of $N_{\rm sims} = 480$ measurements of the angular power spectra on our simulated maps as 
\begin{align} 
 \widehat {\textrm{Cov}}(X_\ell, Y_{\ell'}) = \frac{1}{N_{\rm sims}-1} \sum_{k=1}^{N_{\rm sims}} \left( X_\ell^{(k) }- \bar X_\ell\right) \left( Y_{\ell'}^{(k) }- \bar Y_{\ell'}\right),  %
\end{align}
where $X_\ell, Y_\ell \in \left[ \hat C_\ell^{\kappa g*},\hat C_\ell^{gg}\right] $ are the angular power spectra for our simulated maps discussed in \cref{sec:simulated_datavec} and $\bar X_\ell, \bar Y_\ell$ are the means over all simulations. %
With this, we can estimate the covariance of each observable as well as the cross-covariance between the auto and cross-correlations. We note that here it is crucial that the galaxy sample with and without reweighting are correctly correlated in our simulated maps so that the cross-covariance is captured in the estimate. While the estimate of the covariance from simulations is unbiased, the inverse is a biased estimate. We correct for this by applying a Hartlap correction when inverting the covariance matrix given by \citep{Hartlap2007}
\begin{align}
    \widehat {\textrm{Cov}}^{-1} \rightarrow \widehat {\textrm{Cov}}^{-1} \left(1 -\frac{N_d +1 }{N_{\rm sims}}\right) \label{eq:Hartlap}
\end{align}
where $N_d$ is the length of the data vector considered\footnote{For considerations of each 2pt correlation function this is the number of bins in the cosmological range. For the $\hat E_G$ estimate we invert the full covariance of both 2pt correlation functions doubling $N_d$ and for the bin-wise estimate $\hat E_G^{\ell}$ we have $N_d = 2$.}.

We compare our estimate from simulations with an analytic estimate as well as jackknife errors to test for consistency.
We analytically estimate the full covariance including the cross covariances assuming Gaussianity and using our theory power spectra based on our fiducial cosmology and measured galaxy bias (see \cref{sec:simulated_datavec}). In the absence of a mask, the off diagonals would be zero and the error for each multipole would be given by  \citep{Marques2023}
\begin{align}
(\sigma_\ell^{XY})^2 = \frac{1}{2\ell +1} [C_{\ell}^{XX} C_{\ell}^{YY} + (C_{\ell}^{XY})^2 ],
\end{align}
where $X,Y \in [\kappa, g, g*]$. Here $C_{\ell}^{XX}$ and $C_{\ell}^{YY}$ refer to the auto-correlation inclusive of the noise in the map. For the galaxy auto-correlation, we use the theory auto correlation plus our shot noise estimate $N_\ell$. For the CMB lensing auto-correlation we use the theory auto-correlation plus the noise curve of the \textsl{Planck} PR4 CMB lensing map. The single noise curve does not account for the anisotropy of the noise in the CMB lensing map which can make the analytic covariance less accurate overall \citep{Carron2022}. The complex mask geometry leads to coupling between different modes which can approximately be analytically modeled \citep{Efstathiou2004,Couchot2017}. We use the implementation presented in \citet{GarciaGarcia2019} which introduced the \texttt{NmtCovarianceWorkspace} class within \namaster. The implementation takes as inputs the angular power spectra inclusive of noise as well as the mask geometry in the form of the coupling matrices introduced in \cref{eq_binned_angular_power_spectrum_estimate}. We note that we include magnification bias for the theory spectra here but it has very little impact. For the cross-correlation of \textsl{Planck} CMB lensing and CMASS, the covariance increases by about half a percent compared to not including magnification bias.

\begin{figure}
\includegraphics[width=\columnwidth]{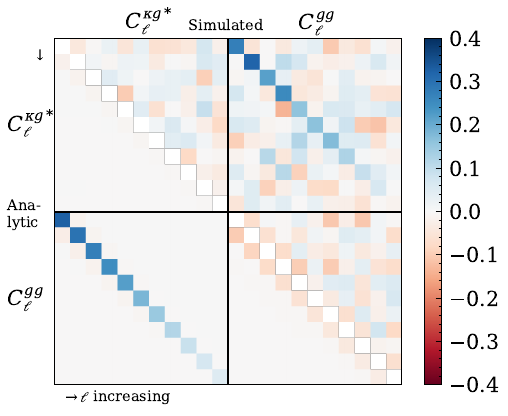}
\caption{Visualizing the off-diagonal correlation matrix between $C_\ell^{\kappa g*}$ and $C_\ell^{gg}$ for \textsl{Planck} and CMASS. Shown is the correlation matrix defined as $\textrm{Cov} (C_\ell^{XY}, C_\ell^{AB}) / (\sigma_\ell^{XY} \sigma_\ell^{AB}) $ for the bins in the cosmological range. The upper triangle shows the simulated covariance which is the baseline for the results presented in this work and the lower triangle shows the analytic covariance. An arrow indicates the direction of increasing $\ell$ for each axis. Correlations between $C_\ell^{gg}$ and $C_\ell^{\kappa g*}$ are up to around 0.35 for our largest-scale bin. %
The off-diagonals of the simulated covariance match the analytic covariance well within statistical uncertainty. For a comparison of the diagonals of the covariances see \cref{fig:cross_auto_correlation_Planck_CMASS}. For \textsl{Planck} + LOWZ see \cref{appendix_LOWZ_plots}.}
\label{fig:correlation_matrix}
\end{figure}

The off diagonals of the simulated and analytic covariance for \textsl{Planck} + CMASS are visualized as a correlation matrix in \cref{fig:correlation_matrix}. The correlation matrix is calculated as $\textrm{Cov} (C_\ell^{XY}, C_\ell^{AB}) / (\sigma_\ell^{XY} \sigma_\ell^{AB})$. We find good agreement between the two techniques up to the statistical uncertainty of the simulated covariance. Additionally, we find, as expected, a significant correlation between the auto- and cross-angular power spectra with correlation factors up to 0.35 on our largest scales in the cosmological analysis range. We also compare the marginalized errors of the covariances in the middle panels of \cref{fig:cross_auto_correlation_Planck_CMASS} where we show that in the cosmological analysis range the errors using our simulated covariance and analytic covariance agree at the percent level. 
We also show the same consistency of the covariance estimates for \textsl{Planck} + LOWZ in \cref{appendix_LOWZ_plots}.

Estimating the covariance from simulations or analytic modeling assumes that we include all sources of noise present in the data. We can validate these error estimates by comparing them to an internal estimate from the data using a jackknife approach. For jackknife resampling, we split the overlapping area into $N_{\rm jack}$ patches\footnote{We recursively bisect the overlapping area along lines of RA and DEC. We split so that each patch contains equal $\sum f_i$ based on our galaxy mask and split along the more spread out direction.}. From these patches, we create $N_{\rm jack}$ maps with one patch left out in each. For each of these, we calculate the auto and cross-correlations. We can then estimate the covariance as \citep{Norberg2009}
\begin{align} 
 \widehat {\textrm{Cov}}(X_\ell, Y_{\ell'}) = \frac{N_{\rm jack}-1}{N_{\rm jack}} \sum_{k=1}^{N_{\rm jack}} \left( X_\ell^{(k) }- \bar X_\ell\right) \left( Y_{\ell'}^{(k) }- \bar Y_{\ell'}\right), 
\end{align}
where $X_\ell, Y_\ell \in \left[ \hat C_\ell^{\kappa g*}, \hat C_\ell^{gg}\right] $. For the auto-correlations, we need to account for the change in shot noise in the galaxy map when removing a patch. We adapt the analytic shot noise estimate (\cref{eq:shotnoise}) for the subsample as
\begin{equation}
    \tilde N^{\rm shot, (k)} = N^{\rm shot} \frac{N_{\rm jack-1}}{N_{\rm jack}}.
\end{equation}
For the cross-covariance between $\hat C_\ell^{\kappa g*}, \hat C_\ell^{gg}$ we restrict the auto-correlation to the same footprint as the cross-correlation and use the same splits to capture the correlation correctly.

The smallest patch is chosen to still contain the largest scales of the cosmological range following \citet{Pullen2016}. Based on this we decided to use 32 patches. Because of this small number, the statistical uncertainty of the marginalized jackknife error for each $X_\ell$ is large compared to the other methods above. Additionally, jackknife resampling overestimates the covariance of the measurement systematically \citep{Marques2020,Mohammad2022}. This is easiest to see in real space where 2pt functions are based on pairs of objects. Pairs across patches are doubly removed from the sampling and therefore the variance is over-counted \citep{Mohammad2022}.

We use the jackknife errors to validate the marginalized bin-wise errors of our angular power spectrum measurements. They are a conservative overestimate of 10-20\% with large statistical uncertainty \citep{Marques2020} and therefore we interpret them as validating our error estimation if they are on average 10-20\% larger than our baseline error estimate and there are no outliers more than 50\% larger. In the middle panels of \cref{fig:cross_auto_correlation_Planck_CMASS} we compare the marginalized jackknife errors to the other techniques finding that they are on average around 20\% higher with statistical variation between 0 and 50\% higher, consistent with our expectations. From this reasonable agreement, we conclude that in the cosmological range, the noise in the data is well represented in the simulations we use to estimate our baseline covariance.

We do not have sufficient samples to accurately estimate the full covariance including the off-diagonal elements with the jackknifing technique and therefore do not give a constraint on $\hat E_G$ marginalized across scales for the jackknife errors. We, however, do use them to estimate the binwise $\hat E_G^{\ell}$ where we only need the marginalized error per bin and the cross-covariance between the two. We again apply Hartlap corrections when inverting the covariance where in \cref{eq:Hartlap} we replace $N_{\rm sims}$ with $N_{\rm jack}$.

\subsection{Input recovery on Simulations} 
\label{sec:input_recovery}

\begin{figure}
\includegraphics[width=\columnwidth]{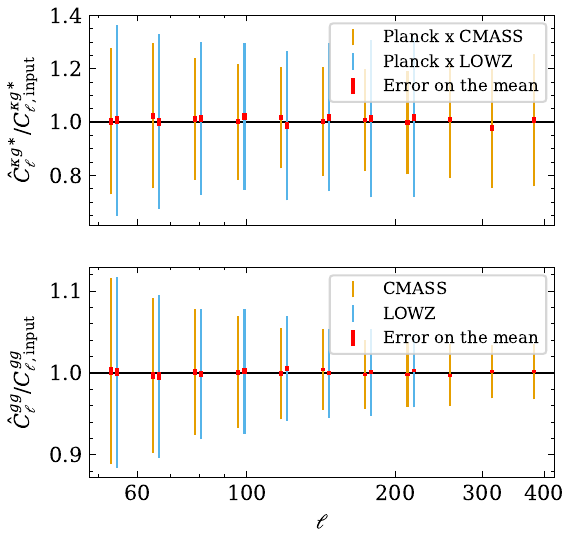}
\caption{
Demonstration of pipeline accuracy on simulated data. The ratio of the mean estimated cross-correlation, $\hat C_\ell^{\kappa g}$, [upper] and auto-correlation, $\hat C_\ell^{gg}$, [lower] with the input theory spectrum is shown for the cases involving CMASS [orange] and LOWZ [blue, slightly shifted to the right for readability] with the statistical uncertainty of a single measurement, $\sigma$, and of the mean for $N_{\rm sims}=480$ simulations,  $\sigma/\sqrt{N_{\rm sims}}$ [red].}
\label{fig:sim_recovery}
\end{figure}

Based on our set of 480 simulated map realizations we test if our analysis pipeline accurately recovers the input angular power spectra for the simulations. 
In \cref{fig:sim_recovery} we show the ratio of the mean recovered cross and auto-correlation signals, $\bar {\hat C}_\ell^{\kappa g*}$ and $\bar{\hat C}_\ell^{gg}$, for our set of 480 simulations to the input angular power spectra. Shown is the uncertainty of a single measurement $\sigma$ [orange/blue] and the uncertainty on the mean $\sigma/ \sqrt{N_{\rm sims}}$ [red].

The level of agreement with the input is quantified by calculating the probability to exceed the given $\chi^2$ value (PTE) over the set of scales in the range of cosmological scales to be used to analyze the experimental data. For this,  the covariance on the mean is the full simulation-based covariance in the cosmological range divided by $N_{\rm sims}$. The $\chi^2$  is for the difference between the mean measurement $\bar {\hat C}_\ell^{XY}$ and the input $C_\ell^{XY}$ given by
\begin{align}
    \chi^2_{\rm recovery} = \sum_{\ell \ell'}  d_{\ell}^{\rm recovery} \left(\frac{1}{N_{\rm sims}}\textrm{Cov}\right) ^{-1}_{\ell \ell'} d_{\ell'}^{\rm recovery},
\end{align}
where $d_{\ell}^{\rm recovery} = \bar {\hat C}_\ell^{XY} - C_\ell^{XY}$ and $XY$ is either $\kappa g*$ or $gg$. The PTE value for this $\chi^2_{\rm recovery} $ is used to indicate if our measurements on the simulations are statistically consistent with the input. Since we already confirmed the accuracy of the covariance estimation a low PTE value here would indicate a bias in the analysis pipeline. Given the number of simulations $N_{\rm sims}$ this test is sensitive to systematic biases of $\sigma/\sqrt{N_{\rm sims}}$ where $\sigma$ is the measurement uncertainty for an individual bin. To complement the PTE test, we also fit the input theory curve to the mean measurement on the sims with a free amplitude $A$ and calculate the difference in amplitude $\Delta A  = A -1$ to report the accuracy of the test in absolute terms.

\begin{table}
\begin{tabular}{l|p{30mm}|p{30mm}}
Sim. recovery & \textsl{Planck} + CMASS & \textsl{Planck} + LOWZ \\ \hline
$C_\ell^{\kappa g*}$ &  PTE = 0.10 \newline $\Delta A$ = $( 0.5 \pm 0.3)\% $&  PTE = 0.38 \newline $\Delta A$ =$( 0.9 \pm 0.5)\%$ \\\hline
$C_\ell^{gg}$ &  PTE = 0.49 \newline $\Delta A$ = $(0.0 \pm 0.1)\%$ &  PTE = 0.59 \newline $\Delta A$ =$( 0.1 \pm 0.1)\%$
\end{tabular}
\caption{Comparison of the mean signal recovered from 480 simulated CMB lensing and galaxy clustering maps using the analysis pipeline compared to the input statistic. The PTE values test if the mean signal is statistically consistent with the input. In addition, the theory curve is fitted to the mean estimated signal with a free amplitude $A$. Values for $\Delta A = A-1$ are listed. \label{tab:sim_recovery}}
\end{table}

The resulting PTE values and the difference in fitted amplitude to the input are summarized in \cref{tab:sim_recovery}. For each data set and statistic, the mean measurement is statistically consistent with the input, showing no significant disagreement. 
We note that for the galaxy auto-correlation simulation recovery test for the LOWZ sample, we originally observed a marginal failure with a PTE value of 0.03. The amplitude of this mean recovered auto-correlation was still only around $(-0.2\pm 0.1)\%$ low. We do enough PTE tests that a marginal failure can be a statistical outlier. Since this is a simulation-based test we reran the simulation generation to do another independent test. The second run passed without issue (PTE 0.59) therefore we concluded the failure was just statistical and we proceeded with the second run. We discuss the summary of our PTE statistics in \cref{sec:conclusion}.

We conclude that our analysis pipeline accurately estimates the angular power spectra to at least within $\sigma/\sqrt{N_{\rm sims}}  \approx 0.05 \sigma$ where $\sigma$ is the measurement uncertainty for individual bins. In absolute terms, the cross-correlation measurement amplitudes are shown to be accurate to at least the percent level and the galaxy auto-correlation measurement amplitudes are accurate to at least the 0.1 percent level. We conservatively account for the level at which we validated the pipeline in the systematic error budget for the $E_G$ estimate with 1\% for \textsl{Planck} + CMASS and 2\% for \textsl{Planck} + LOWZ. 

Our test covers the entire angular power spectrum estimation pipeline as described in \cref{sec:namaster} and the norm correction of the CMB lensing map as described in \cref{sec:CMBsims}.

\subsection{Shot noise test} \label{sec:shot_noise_test}

Measuring the angular auto-power spectrum of the galaxy samples with a pseudo-Cl estimator relies on accurate subtraction of the shot noise as described in \cref{eq:shotnoisesubtraction}. We use an analytic estimate of the shot noise that accounts for the variance of the weights applied to the galaxy samples (see \cref{eq:shotnoise}). We can check our analytic estimate of the shot noise by comparing it to a measurement from data. To estimate the shot noise from data we split the galaxy sample into two equal parts and construct the galaxy overdensity for both. Each overdensity map $\delta_i$ contains the cosmological signal $\delta$ as well as shot noise $n_i$. The shot noise in each map is independent since the two sub-samples of the data are distinct. By taking the difference between the two maps, the cosmological contribution cancels and we are left with only shot noise in the map when taking the auto-correlation \citep{Nicola2020}
\begin{equation}
    \langle |\delta_1 - \delta_2|^2\rangle = \langle |n_1|^2\rangle + \langle |n_2|^2\rangle.
\end{equation}
Each sample has half the number density of the full galaxy sample, so the total shot noise is 4 times higher than our full galaxy sample. We can therefore get an observational estimate of the shot noise contained in the sample by calculating the auto-correlation of the difference map (without subtracting any shot noise) and dividing it by 4. Since the map contains shot noise and no signal this measurement is not affected by a pixel window function and we therefore do not apply a correction for it. To compare our analytic estimate of the coupled shot noise with the measurement from data we need to apply the inverse of the binned coupling matrix as done for the theory spectra (see \cref{eq:binned_theory_spectra})
\begin{equation}
       N_\ell^{\rm shot} = \sum_{\ell'} \left[ \mathcal{M}^{-1} \right]_{\ell \ell'} \tilde N^{\rm shot}.
\end{equation}
\begin{figure}
\includegraphics[width=\columnwidth]{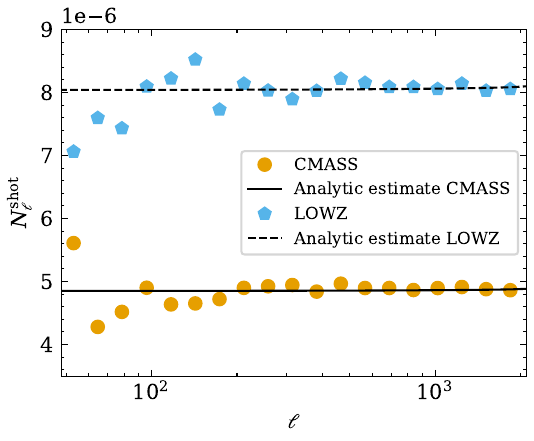}
\caption{Comparing our analytic estimate of the shot noise in the galaxy sample [solid black line for CMASS, dashed black line for LOWZ] with an estimate from data for both CMASS [orange circles] and LOWZ [blue pentagons]. The data estimate has large statistical uncertainty on large scales and gets more accurate for small scales (high $\ell$). The analytic estimate matches the constraint from the data well.}
\label{fig:shotnoisetest}
\end{figure}
In \cref{fig:shotnoisetest} we compare the analytic estimates to the estimates from data for both the CMASS and LOWZ samples. On large scales the statistical uncertainty from the observational estimate is large but especially on small scales we see that it matches our analytic estimate very well visually. To quantify this we divide the difference map estimate by the analytic estimate. In the range $757 \leq \ell < 2024$  the mean and standard deviation of the ratios for CMASS are $1.003 \pm 0.004$ and for LOWZ are $1.001\pm 0.005$. Therefore the analytic estimate agrees with the difference map estimate well within a percent on small scales and the two are consistent with each other within the statistical scatter of the difference map estimate. 

We note that this shows that \cref{eq:shotnoise} correctly accounts for the variance in weights across the sky. A naive approach of assuming equal weights would simplify the shot noise estimate to $n_{\rm eff, naive} = \frac{N_{\rm gal}}{4\pi f_{\rm sky,gal}}$ with $N_{\rm gal}$ the number count of galaxies in the sample. However, SDSS has significant variance in the weights. Using the naive formula would underestimate the shot noise for the CMASS sample by around 21\% and for the LOWZ sample by around 7\% (both averaged over $757 \leq \ell < 2024$). 

Overall this test gives credibility to our auto-correlation result as the subtraction of the shot noise is a crucial step and we have shown that the analytic estimate fits the data accurately. The shot noise estimate is also important for getting an accurate estimate of the analytic covariance for both cross- and auto-correlation.

\subsection{Angular power spectrum measurements}
\label{sec:angular_power_spectra_measurements}

\begin{figure*}
\includegraphics[height=10cm]{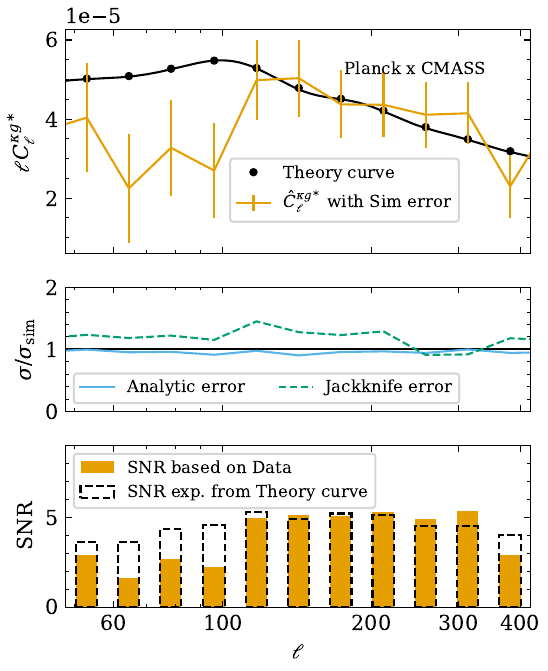}
\includegraphics[height=10cm]{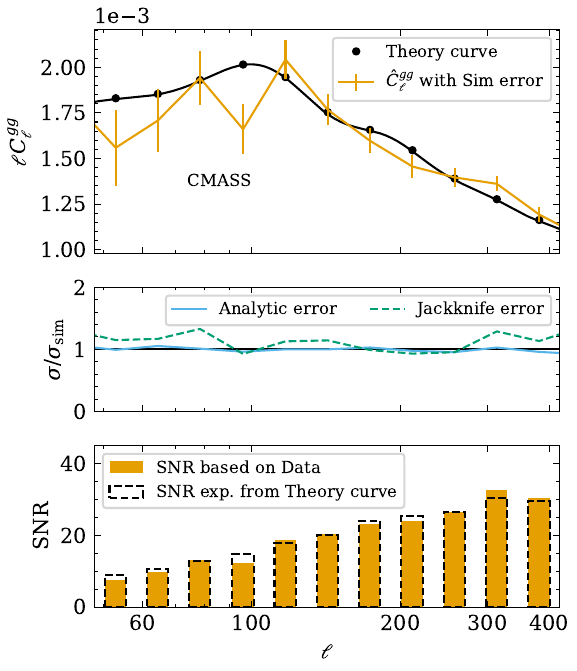}
\caption{Measurement of [Left] the angular cross-power spectrum, $C_\ell^{\kappa g*}$, for the CMB lensing map based on \textsl{Planck} PR4 and the reweighted CMASS galaxy sample and [Right] the angular auto CMASS galaxy power spectrum. [Top] The angular power spectrum with simulation-based errors as well as our theory curve. [Middle] Comparison of the analytic [solid blue line] and jackknife error estimates [dashed green] to the baseline marginalized errors from the simulations. [Lower] The signal-to-noise ratio, calculated as the ratio of measurement and measurement uncertainty is shown [filled orange]. For reference, the ratio of the theory expectation and the measurement uncertainty is also shown [dashed black]. For \textsl{Planck} + LOWZ see \cref{appendix_LOWZ_plots}.}
\label{fig:cross_auto_correlation_Planck_CMASS}
\end{figure*}

In \cref{fig:cross_auto_correlation_Planck_CMASS} we present the angular cross-correlation measurement $\hat C_{\ell}^{\kappa g*}$ for the \textsl{Planck} CMB lensing map and the CMASS galaxy sample, based on our pipeline and covariance estimate and, for reference only, the theory curve based on the fiducial cosmology and a fit of the galaxy bias to the amplitude of the galaxy auto-correlation measurement. 
We also present the galaxy auto-correlation of the CMASS galaxy sample. The same plot for \textsl{Planck} + LOWZ can be found in \cref{appendix_LOWZ_plots}.

Visually the measurements are consistent with the shape of the theory curves. For the cross-correlation between \textsl{Planck} and CMASS, we recover the known result in the literature that the measurement tends to be low compared to theory expectation on large scales ($\ell < 100$) although to a less significant degree than reported in e.g. \citep{Pullen2016}. 

In the bottom panels of the angular power spectrum measurement plots (\cref{fig:cross_auto_correlation_Planck_CMASS,appendix_LOWZ_plots}) the signal-to-noise ratios (SNR) are shown for each case. We also show the ratio of the theory curves with the measurement which can be more directly compared with the constraints of other surveys. For the log-spaced binning used in this work the expected SNR of the cross-correlation, which carries much larger uncertainty than the auto-correlation, is roughly constant between bins. Therefore, the uncertainties on the $E_G$ estimates between bins are comparable, making them simpler to compare and combine.

\subsection{\texorpdfstring{$\beta$}{beta} measurement from RSD analysis}
\label{sec:beta}

We redo the RSD analysis on the original BOSS DR12 data vectors from \citet{GilMarin2016} to measure $\beta$ for CMASS and LOWZ. We do this to compare the constraints with updated theoretical modeling to the original results and also to be able to account for the cross-covariance between $f \sigma_8$ and $b \sigma_8$ when calculating the derived parameter $\beta$ which is not included in the published results. 
The data vectors from \citet{GilMarin2016} are combined measurements of the power spectra for the full samples obtained by sky-averaging the NGC and SGC measurements, along with window matrices connecting the translation-invariant theoretical three-dimensional power spectrum $P_\ell$ of the galaxies to that of the windowed observed field
\begin{equation}
    P_{\ell}^{\rm obs}(k_i) = \sum_{\ell' = 0, 2} W^{\ell \ell'}_{ij} P_{\ell'}(k_j).
\end{equation}
These three-dimensional measurements were computed in \citet{GilMarin2016} assuming a fiducial cosmology with $\Omega_{m}^{\rm fid} = 0.31$.\footnote{We thank H{\'e}ctor Gil-Mar{\'i}n for providing us with the official BOSS measurements including window matrices.}

The ratio $\beta$ compares the amplitude of the real and redshift-space clustering signal in galaxies, which become contaminated by the nonlinearities of structure formation towards smaller scales (higher $k$). As in the official BOSS results \citep{GilMarin2016} we fit the linear theory clustering-weighted bias and growth rate, $b \sigma_8(z)$ and $f(z) \sigma_8(z)$ ($\beta = f/b$), assuming a fiducial linear power spectrum $P_{\rm lin}(k)$ and clustering amplitude $\sigma_8(z)$ given by the best-fit \textsl{Planck} value \cite{PlanckCollaboration2018}. The latter assumptions are necessary to compute the contributions of next-to-leading-order nonlinearities; we note that the signal-to-noise of the BOSS power spectra are not sufficient to break the $b-\sigma_8$ and $f-\sigma_8$ degeneracies, such that freeing the value of $\sigma_8$ in fits can lead to undesirable parameter-volume effects \cite{Maus2023}. Finally, as in the official BOSS fits, and to be agnostic to anisotropies due to deviations of the expansion history from the fiducial cosmology, we perform our fits with the Alcock-Paczynski parameters $\alpha_{\parallel,\perp}$ as free parameters.

In order to marginalize over the nonlinearities of structure formation in our analysis we adopt the Lagrangian perturbation theory (LPT) model of \citet{Chen2020,Chen2021}, implemented in the publicly available code \texttt{velocileptors}\footnote{\url{https://github.com/sfschen/velocileptors}}. This model contains both higher-order nonlinear bias terms as well as effective-theory corrections, e.g., isotropic and anisotropic $k^2 P_{\rm lin}(k)$ counterterms and stochastic noise terms including a scale-dependent $k^2 \mu^2$ contribution. It represents a more careful treatment of nonlinear galaxy clustering and redshift-space distortions compared to the perturbation theory models used in the official BOSS results, particularly in the consistent effective-theory treatment of small-scale physics beyond the reach of perturbation theory. As a result, our RSD constraints are slightly less tight than those from the official analysis, though we note that this is not a weakness of the measurement but rather a result of including the full theoretical uncertainty in our model to this order. We use the publicly available pipeline developed for \citet{Chen2022} and adopt the parameters and priors given in Table 2 of that work.

\begin{figure}
\includegraphics[width=\columnwidth]{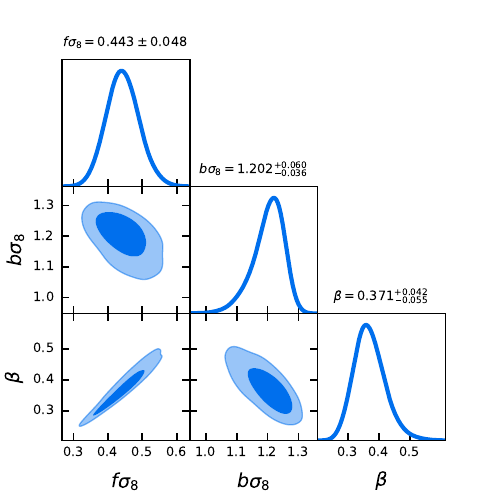}
\caption{RSD fit to the clustering signal of the CMASS data set. We show the corner-plot of the fitting parameters $f \sigma_8, b\sigma_8$ as well as the derived parameter $\beta = f \sigma_8 /(b\sigma_8)$ that we use as part of our $E_G$ estimate. For each parameter, we also show the mean value and 68.27\% error. For LOWZ see \cref{appendix_LOWZ_plots}.}
\label{fig:cornerplot_beta_CMASS}
\end{figure}

We use $k_{\rm max} = 0.2 h/\textrm{Mpc}$. We show the cornerplot for the constraints on $f \sigma_8, b\sigma_8$ and the derived parameter $\beta = f \sigma_8/ (b \sigma_8)$ for CMASS in \cref{fig:cornerplot_beta_CMASS} using \getdist \citep{Lewis2019getdist}\footnote{\url{https://github.com/cmbant/getdist}}. We also report the plot of the weighted mean value and one sigma errors for each parameter. For the same plot for the LOWZ sample see \cref{appendix_LOWZ_plots}.

For the CMASS sample we find $f \sigma_8 = 0.443\pm 0.048$, $b\sigma_8 = 1.202^{+0.060}_{-0.0336}$. These are in good agreement with the values reported in \citet{GilMarin2016} where for the original perturbation theory models of the original BOSS results and a slightly higher $k_{\rm max} = 0.24 h/\textrm{Mpc} $ they found $f \sigma_8 = 0.444\pm 0.038$, $b\sigma_8 = 1.222\pm 0.021$ for the CMASS sample.

For the LOWZ sample we find $f \sigma_8 = 0.426^{+0.071}_{-0.082}$, $b\sigma_8 = 1.271^{+0.080}_{-0.049}$ which is also in good agreement with the values from \citet{GilMarin2016} which are $f \sigma_8 = 0.394\pm 0.052$, $b\sigma_8 = 1.281\pm 0.035$ for the LOWZ sample.

For the derived parameter $\beta$ we find $\beta = 0.371^{+0.042}_{-0.055} $ for the CMASS sample and $\beta = 0.338^{+0.058}_{-0.081}$ for the LOWZ sample. While less directly comparable our $\beta$ results are also broadly consistent within one sigma with other RSD analyses of the SDSS BOSS data: \citet{Chuang2017} do not use the optimal RSD weights and a slightly less restrictive upper redshift cut on CMASS finding $\beta = 0.435\pm0.070$ for CMASS and $\beta = 0.301\pm 0.066$ for LOWZ; For the earlier BOSS DR11 data \citep{Pullen2016} report $\beta = 0.368\pm 0.046$ for the CMASS sample. For our $E_G$ measurement we use the full probability density function (pdf) of the $\beta$ measurement as shown in the corner plots the process for which we describe in detail next.

\section{Non-Gaussian Uncertainty of \texorpdfstring{E\textsubscript{G}}{EG} measurement} 
\label{sec:ratio_distribution}

To estimate the significance of any $E_G$ measurement, we need to be able to correctly calculate the probability distribution function (pdf) for the $E_G$ statistic given the measurement covariances of the three constituent observables: $C_{\ell}^{\kappa g*}, C_{\ell}^{gg}$ and $\beta$. Even when these observables are Gaussian, since $E_G$ involves a ratio of them, its uncertainty can be significantly skewed, and non-Gaussian \citep{Sun2023}. In using the data across multiple bins, one also has to carefully consider how to combine them and the potential correlations between bandpowers.

We investigate the probability distribution function (pdf) of our estimator without assuming Gaussianity, describe how we report our constraints, and compare to assuming Gaussianity in \cref{sec:EGpdf}. Additionally, we build PTE tests to test our measurement for consistency with the $\Lambda$CDM-GR prediction (\cref{sec:PTE_GR}) and for consistency with scale independence (\cref{sec:scale_independence_test}).

\subsection{pdf of our \texorpdfstring{\^{E}\textsubscript{G}}{EG} estimator}
\label{sec:EGpdf}

Let us first consider the ratio of the measured angular power spectra $\hat C_\ell^{\kappa g}$ and $ \hat C_\ell^{gg}$. %
We include the $\Gamma_\ell$ factor and define $R\equiv \Gamma_\ell \hat C_\ell^{\kappa g*}/ \hat C_\ell^{gg} $. 
The probability density for this ratio is given by the ratio distribution \citep{Curtiss1941}
\begin{align}
    p_R = \int \dd y |y| p_{XY}(Ry \hat C_\ell^{gg},y \hat C_\ell^{gg}) , \label{p_R_integral}
\end{align}
where $R$ is the ratio we want to find and $p_{XY}$ is the pdf of our observed angular power spectra which are correlated Gaussian random variables with our measurement as mean and the covariance of our observation  
\begin{align}
    &XY \sim \mathcal{N}\left( \mu = \left( \begin{matrix}
           \Gamma_\ell \hat C_\ell^{\kappa g*} \\
           \hat C_\ell^{gg} \\
        \end{matrix}\right), \right. \nonumber\\ 
        &\left. \textrm{Cov}_{\ell \ell'} = \begin{bmatrix}
           \Gamma_\ell \Gamma_{\ell'} \, \textrm{Cov}(C_\ell^{\kappa g*}, C_{\ell'}^{\kappa g*}), \Gamma_{\ell'} \, 
           \textrm{Cov} (C_\ell^{\kappa g*}, C_{\ell'}^{gg}) \\
           \Gamma_{\ell}  \, \textrm{Cov} (C_{\ell}^{gg}, C_{\ell'}^{\kappa g*}), \textrm{Cov} ( C_{\ell}^{gg}, C_{\ell'}^{gg}) \\
        \end{bmatrix} \right).
\end{align}
With this approach, we can find the pdf for the ratio both for each bin individually ($R_\ell$) and one overall constraint $R$ over our cosmological range in $\ell$. For the latter, we implicitly assume that $E_G$ is scale-invariant as predicted for GR. We also separately evaluate this prediction by testing for evidence of scale dependence (see \cref{sec:scale_independence_test}).

Given the ratio of the angular power spectra, we can then complete our calculation of our estimator ($\hat E_G$, \cref{E_G_estimator}) by dividing $R$ by $\beta$. To obtain the probability density function for our estimator $\hat E_G$ given the data we calculate the ratio distribution
\begin{align}
    p_{\hat E_G} (\hat E_G | \hat C_\ell^{\kappa g}, \hat C_\ell^{gg}, \beta ) = \int \dd \beta' |\beta'| p_R (\hat E_G \cdot \beta' ) p_\beta(\beta'),
\end{align}
where $p_\beta$ is the pdf of the $\beta$ measurement. %
Here, and throughout the analysis, we assume no correlation between $\beta$ and the angular power spectra. This correlation is expected to be negligible since the $\beta$ measurement is based on the full 3D clustering of the galaxies while the angular power spectra only use a single projection onto the sky \citep{Pullen2016}.

Crucially, this approach does not assume Gaussianity for multiplying or dividing our measurements. This ensures that we capture the asymmetric pdf for large uncertainties where the Gaussian approximation breaks down. We can also perform this approach per bin to measure the statistic for each bin $\hat E_G^\ell$. 

\begin{figure}
\includegraphics[width=\columnwidth]{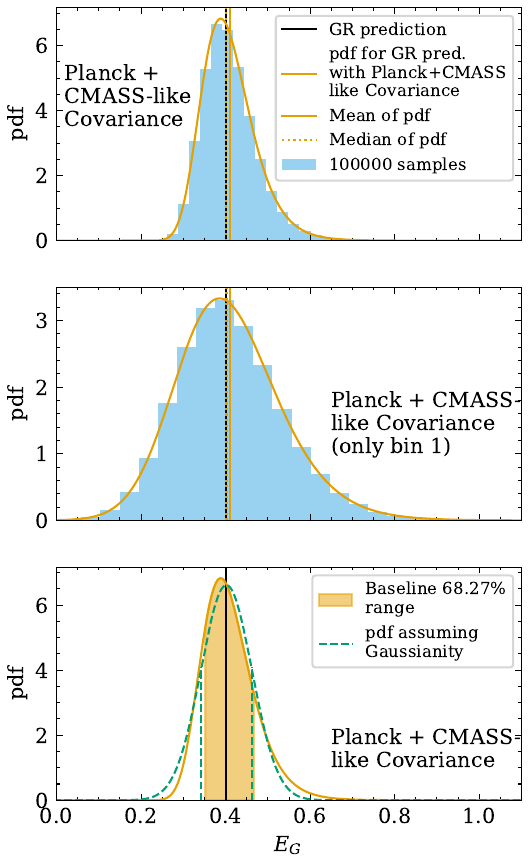}
\caption{
Characterization of the non-Gaussian pdf for the $\hat E_G$ estimator for the fiducial cosmology and \textsl{Planck} + CMASS data covariance [orange curve]. The fiducial value for $E_G$ [vertical black line] is shown along with the mean of the pdf [vertical orange full line] and median [vertical orange dotted line]. The distribution of 100,000 samples from the theory model of the data vector and covariance are also shown [blue shading]. The top panel shows the case for combined measurement using all bins and the middle panel the case of only using the first bin. In the bottom panel, we show for comparison the result when assuming Gaussianity [green dashed] for the combined measurement.}
\label{fig:pdf_EG_GR_value}
\end{figure}

\cref{fig:pdf_EG_GR_value} shows the $E_G$ pdf distribution for the fiducial cosmology and observational covariances when combining the correlation and $\beta$ measurements across our cosmological range. For this test, $\beta$ is treated simply as a Gaussian with a standard deviation equal to the mean of the upper and lower 68.27\% errors. 

We also show the distribution of $\hat E_G$ estimates for a large number of samples (100,000) from the fiducial cosmology and our observational covariance, showing that the pdf describes the distribution accurately. Note our pdf weights each binned measurement by their relative uncertainty but for the sampled points we take the mean of the binned estimates assuming equal weights since it is unclear how to estimate and define the relative uncertainty for each bin without assuming Gaussianity. Since the uncertainty of each of the binned measurements, for our specific log-spaced binning scheme, is of comparable size, the difference is negligible. We also confirmed that the estimated pdf for each of the individual bins accurately matches the distribution of a large number of samples. 

The pdf for the $\hat E_G$ estimator is significantly skewed. The mean of the distribution is higher than the GR input value, the mode (peak) of the distribution is lower than the input value and the median of the distribution recovers the input value accurately. We quantify this for the example of a \textsl{Planck} + CMASS-like covariance at our fiducial cosmology shown in the top panel of \cref{fig:pdf_EG_GR_value}: Compared to the GR input value of $E_G = 0.401$, the mean is $\Delta E_G = 0.008$ ($2.0\%$) high and the mode $\Delta E_G = -0.014$ ($3.6\%$) low, however the median matches well with $|\Delta E_G| < 0.0005$ ($<0.2\%$).
Given this, we use the median of the distribution of the pdf as the best estimate in the analysis. Consistently we use the middle 68.27\% of the area under the pdf as the one sigma range.

To facilitate comparison with other results in the literature, which assumed Gaussian errors, we can compare our pdf with the result one would get when assuming Gaussianity and performing Gaussian error propagation. For reference, the Gaussian error propagation for the estimator is summarized in \cref{error_propagation}. To get the covariance $\textrm{Cov}(\hat E_G^\ell, \hat E_G^{\ell'})$ from the fractional covariance we get from the error propagation we scale by our estimate $\hat E_G$\footnote{We use the weighted average over the bins. To find this we first take the mean of the $\hat E^{\ell}_G$ and use it to calculate an initial covariance. Then use this initial covariance to calculate the weighted mean across the bins and use the weighted mean to find the covariance. In practice, the relative weighting makes little difference.}. We can then under the assumption of Gaussianity describe the pdf for $\hat E_G$ as
\begin{align}
    p_{\hat E_G}^{\rm Gaussian} (\hat E_G) \propto \textrm{exp} \left(\frac{- \chi_{\rm Gaus}^2}{2} \right) 
    \end{align}
with
\begin{align}
    \chi_{\rm Gaus}^2 = \sum_{\ell, \ell'} (\hat E^{\ell}_G - \hat E_G) \textrm{Cov}^{-1}(\hat E_G^\ell, \hat E_G^{\ell'})  (\hat E^{\ell'}_G - \hat E_G)
\end{align}
where we apply Hartlap corrections to the inverse covariance as done throughout our analysis. This $p_{\hat E_G}^{\rm Gaussian} $ is equivalent to the approach in \citet{Pullen2016} and \citet{Zhang2021}. The best estimate given in their approaches is the peak of the distribution where $\chi^2_{\rm Gaus}$ is minimized and the one sigma errors can be found from the $\chi_{\rm Gaus}^2 = \chi^2_{\rm Gaus, min} + 1$ surface.

For the fiducial cosmology, this Gaussian approach gives the same best estimate as our pdf but the errors are different. When applying to measured data, the best-fit assuming Gaussianity for a single bin is the same as in our pdf, but the relative weighting of the measurements is different so the marginalized result across scales is not the same. The best-fit in the Gaussian case is equal to the weighted mean of the bin-wise measurements $E_G^{\ell}$, weighted according to the inverse variance of each bin given by the Gaussian covariance. 
Since the bin-wise $E_G^{\ell}$ is high on average, the marginalized result will also be biased high. 

The comparison of the Gaussian and full pdf in \cref{fig:pdf_EG_GR_value} shows that the upper error gets underestimated and the lower error gets overestimated. To estimate comparable uncertainty levels for the skewed distribution we use the central 68.27\% confidence range which matches the one sigma error in the Gaussian limit. For the fiducial cosmology and for the \textsl{Planck} + CMASS covariance, the Gaussian estimate underestimates the upper error by 12\% (0.060 for Gaussian vs. 0.068 for full pdf) and overestimates the lower error by 11\% (0.060 vs. 0.054) compared to our pdf. For the $95.45\%$ (2$\sigma$ equivalent) confidence range, the difference between Gaussian and non-Gaussian pdf is even more pronounced: the Gaussian estimate of the upper error is 24\% low (0.121 vs. 0.16) and the lower error is 23\% high (0.121 vs. 0.098). This is relevant when reporting a low $E_G$ measurement compared to expectation as done in \citet{Pullen2016} since the underestimated upper error leads to an overestimation of tension with the $\Lambda$CDM-GR prediction. The exact difference depends on the measurement since the pdf shape depends on the measurement and the best-fit values for the two approaches are generally not identical. As a simplified example to make the point, we can reduce the $C_\ell^{\kappa g}$ from our theory curve by a factor 0.7 so that according to Gaussian statistics the $\Lambda$CDM-GR prediction is outside of the central 99.07\% range (equivalent of 2.6$\sigma$) matching the value reported in \citet{Pullen2016}, where we tuned the factor manually to get a similar disagreement. Fully accounting for the asymmetric pdf, the best-fit value is the same in this example, but the $\Lambda$CDM-GR value is only outside the central 95\% range due to the asymmetric errors (based on $\textrm{PTE}_{\rm GR}$ as defined in \cref{eq:PTEGR}; equivalent to 2.0$\sigma$ for Gaussian statistics), which would only be a marginally significant result.
We argue, therefore, that it is crucial to correctly account for the asymmetric pdf of the estimator in any interpretation of the measurement results.

\subsection{PTE test for consistency with GR value} 
\label{sec:PTE_GR}

Given the pdf, we can test a measurement against the expectation value from $\Lambda$CDM-GR, $E_G^{\rm GR}$. We calculate the PTE value for consistency with the $\Lambda$CDM-GR prediction as the distance from the median in the cumulative distribution given by
\begin{align}
    \textrm{PTE}_{\rm GR} = 1 - 2 |0.5 - P(E_G < E_G^{\rm GR}) | \label{eq:PTEGR}
\end{align}
The theoretical $\Lambda$CDM-GR prediction itself carries uncertainty $\sigma_{E_G^{\rm GR}}$ because it relies on a measurement of $\Omega_{m,0}$ based on the expansion history. We account for the theory uncertainty by convolving our pdf for $E_G$ with the normal distribution $\mathcal{N}(\mu=0,\,\sigma^2_{E_G^{\rm GR}})$ but we will see that this makes very little difference in our case since $\sigma_{E_G^{\rm GR}}$ is much smaller than the measurement uncertainty for our data. %

\begin{figure}
\includegraphics[width=\columnwidth]{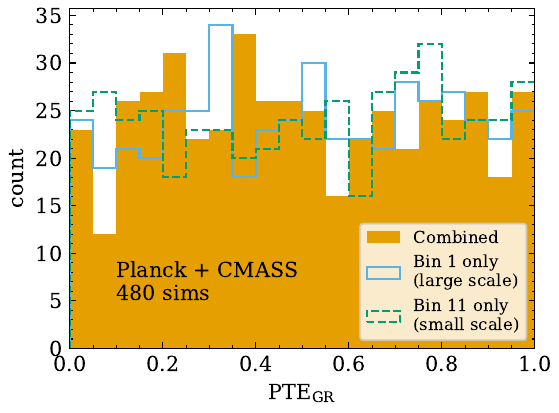}
\caption{Characterizing the statistical consistency test between the $E_G$ measurement and the $\Lambda$CDM-GR prediction by applying it to simulations. For each of the set of 480 simulated \textsl{Planck} + CMASS-like data vectors drawn around the fiducial cosmology, we calculate the $\textrm{PTE}_{\rm GR}$ for consistency with the GR value. Results are shown when combining all multipole bins in the cosmological range [orange solid] and for two single bandpowers at the largest ($\ell=53$) [blue line] and smallest scale ($\ell=382$) [green dashed] of the cosmological analysis range. The $\textrm{PTE}_{\rm GR}$ distributions are consistent with the expected flat distribution for a valid test.
}
\label{fig:PTE_GR_Planck_CMASS}
\end{figure}

In \cref{fig:PTE_GR_Planck_CMASS} we show the $\textrm{PTE}_{\rm GR}$ values for $\hat E_G$ measured on our 480 simulated data vectors compared to the input $E_G^{\rm GR}$. We here do not need to marginalize over the uncertainty of $E_G^{\rm GR}$ as we use our fiducial cosmology as an input to our simulations. We find that for all multipole bins combined, and individual bins, a flat distribution within the Poisson uncertainty showing that $\textrm{PTE}_{\rm GR}$ accurately tests for consistency with GR while accounting for the asymmetric pdf of our measurement.

\subsection{PTE test for consistency with scale independence} 
\label{sec:scale_independence_test}

Another key prediction for $E_G$ in GR is scale independence. Above we implicitly assumed scale independence when measuring the combined constraint on $E_G$ across multiple bins. To study whether the data are consistent with the assumption of scale independence we consider a distance measure, building on \citet{Miyatake2017}, defined as
\footnote{We note that the $E_G$ value that minimizes this $\chi^2$ distance measure is a biased estimator of the $E_G$ statistic. We use this distance measure only to test for scale dependence, not to get an unbiased estimate of the ratio.}
\begin{equation}
    \chi^2 (E_G)= \sum_{\ell \ell'} d_\ell \textrm{Cov'}_{\ell\ell'}^{-1.} d_\ell' \label{estimator_of_ell},
\end{equation}
where for the $E_G$ statistic we have $d_\ell \equiv \Gamma_\ell \hat C_\ell^{\kappa g*} - E_G \beta \hat C_\ell^{gg}$. %
The covariance is given by
\begin{align}
    \textrm{Cov'}_{\ell\ell'} &\equiv \Gamma_\ell \Gamma_{\ell'} \textrm{Cov}(C_\ell^{\kappa g*}, C_{\ell'}^{\kappa g*}) \nonumber \\
    &-  E_G \Gamma_\ell \beta  \, \left[ \textrm{Cov} (C_\ell^{\kappa g*}, C_{\ell'}^{gg}) + \textrm{Cov} (C_{\ell}^{gg}, C_{\ell'}^{\kappa g*}) \right] \nonumber \\
    &+ \left(  E_G  \right)^2 \left[ \beta^2 \textrm{Cov} (C_{\ell}^{gg},  C_{\ell'}^{gg})  + \hat C_\ell^{gg} \hat C_{\ell'}^{gg}\sigma_\beta^2\right].
\end{align}
As before, we neglect the correlation between $\beta$ and the angular power spectra and assume the $\beta$ measurement is Gaussian. Note that the covariance $\textrm{Cov'}$ depends on $E_G$. 

\begin{figure}
\includegraphics[width=\columnwidth]{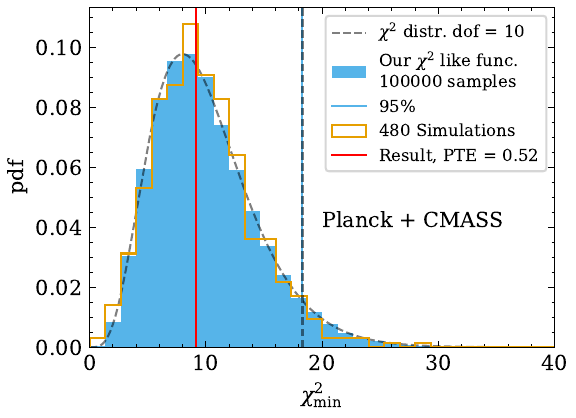}
\includegraphics[width=\columnwidth]{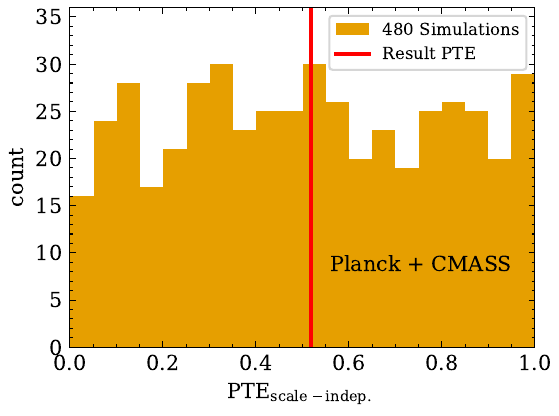}
\caption{Characterizing the statistical consistency test of the $E_G$ measurement for scale independence for the case of \textsl{Planck} + CMASS. [Upper] The minimum value for the $ \chi^2$ distance measure. The $\chi^2$ distribution for the degrees of freedom [grey dashed curve] is shown to be consistent with the results from the 480 simulations [orange] and the distribution for 100,000 samples around the fiducial cosmology and baseline (simulation-derived) covariance. The value of significance for the $\chi^2$ distribution [grey dashed vertical] and for the 100,000 samples (95\% of sampled distance measures fall below) [blue full vertical] are shown to be in good agreement. [Lower] The resulting $\textrm{PTE}_{\rm scale-indep.}$ values for the 480 simulations. The $\textrm{PTE}_{\rm scale-indep.}$ distribution is consistent with the expected flat distribution for a valid test. Our result for the \textsl{Planck} + CMASS $\hat E_G$ estimate, $\textrm{PTE}_{\rm scale-indep.} = 0.52$, obtained in \cref{sec:results} is shown as a vertical red line. For \textsl{Planck} + LOWZ see \cref{appendix_LOWZ_plots}.}
\label{fig:ScaledependenceTest}
\end{figure}

The distribution of minima for this distance measure closely follows a $\chi^2$ function with degrees of freedom equal to the number of bins minus one (since we find the minimum). 
\cref{fig:ScaledependenceTest} shows the $\chi^2$ distribution for the case of \textsl{Planck} + CMASS over the full cosmological range consisting of 11 bins spanning $48 \leq \ell<420$. 
For comparison, we show a histogram of $\chi^2_{\rm min}$ values for 100,000 samples around the fiducial cosmology using the baseline (simulation-derived)  covariance. Since the covariance is exact for these samples we do not apply Hartlap corrections. 
The plot demonstrates that the distribution of the samples and the line of significance, for which 95\% of the $\chi^2_{\rm min}$ samples are below, are consistent with the analytical distribution. 

The figure also shows the histogram of the $\chi^2_{\rm min}$ for the 480 simulated data vectors. 
In the case of the simulated data vectors as well as the observed data vector, the baseline covariance carries uncertainty. For an unbiased estimate of the inverse covariance used in the $\chi^2$ calculation, we apply Hartlap corrections. We only have 480 simulations, therefore the Poisson scatter in the histogram is larger but overall the distribution matches well with the $\chi^2$ distribution as well.

We use a Pearson’s chi-square test to test our observation for consistency with scale independence, given as
\begin{align}
    \textrm{PTE}_{\rm scale-indep.} = P_{\chi^2} (\chi^2 > \chi^2_{\rm min} ) ,
\end{align}
the probability of getting a higher $\chi^2$ than for our measurement as based on the analytic $\chi^2$ pdf for our degrees of freedom. 
A significant result ($\textrm{PTE}_{\rm scale-indep.}<0.05$) would indicate that the data is not well described by a constant ratio and therefore would be evidence for scale dependence. We however note a significant result could also be caused by underestimating the covariance of the measurement or from underlying scale-dependent systematic biases. %

To test our approach we calculated the $\textrm{PTE}_{\rm scale-indep.}$ value for the 480 simulated data vectors. As shown in \cref{fig:ScaledependenceTest}, the distribution of PTE values is consistent with a flat distribution within the Poisson uncertainties of the counts indicating that our scale dependence test works correctly.

\section{Results for \texorpdfstring{E\textsubscript{G}}{EG} with \texorpdfstring{\textsl{Planck}}{Planck} and SDSS BOSS} 
\label{sec:results}

\subsection{Baseline results for the \texorpdfstring{E\textsubscript{G}}{EG} statistic} \label{sec:baseline_results}
We now apply the $E_G$ estimator presented in this work to the \textsl{Planck} PR4 CMB lensing data and SDSS BOSS galaxy samples using our measured angular power spectra and RSD parameter $\beta$. 

\begin{figure}
\includegraphics[width=\columnwidth]{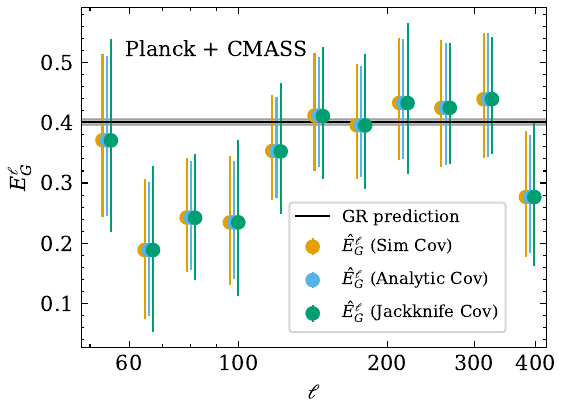}
\includegraphics[width=\columnwidth]{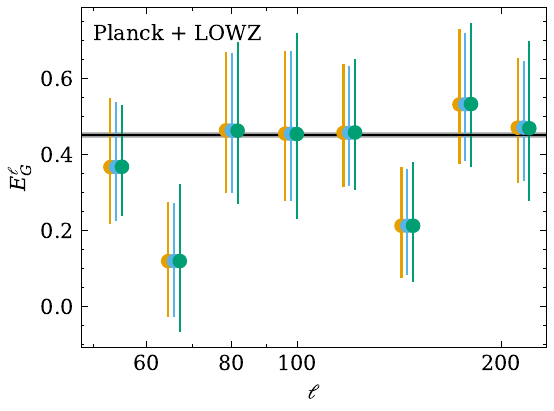}
\caption{%
Measurements of $\hat E_G^\ell$ for [Upper] \textsl{Planck} + CMASS and [Lower] \textsl{Planck} + LOWZ for each multipole bin. The median value is indicated as dots and the errors show 68.27\% confidence ranges. The marginalized errors for all three of the covariance estimate approaches are given: simulations [black], analytic [blue], and jackknife [green] with the latter two shifted to the right for visibility. 
The GR prediction is shown [solid line] for reference with the one sigma uncertainty [grey shaded]. 
}
\label{fig:E_G_ell_Planck_CMASS}
\end{figure}

\cref{fig:E_G_ell_Planck_CMASS} shows the binwise estimate $E_G^\ell$ for \textsl{Planck} PR4 + CMASS and for \textsl{Planck} PR4 + LOWZ for the cosmological range in scales, with the $\Lambda$CDM-GR expectation value as a reference. For each, three different estimates of the 68.27\% confidence ranges (the equivalent of one sigma error for Gaussian statistics) are shown: the baseline simulation-based covariance as well as the analytic and jackknife covariances. For individual bins, the estimator gives the same best-fit value independent of the error. Similar to the individual angular power spectra the different error techniques are also consistent with each other within expectation for the binwise $E_G^\ell$ constraints. Simulation-based errors and analytic errors match closely and the jackknife errors are a systematic overestimate of 20-30\% and larger statistical variation consistent with expectations (see \cref{sec:covariance}). 

The statistical test for scale-independence (\cref{sec:scale_independence_test}) evaluates if the measurement is statistically consistent with a constant across the cosmological range for each of the measurements. Scale independence is a direct prediction of GR that is violated in many models of modified gravity. Finding the data inconsistent with a constant could indicate tension with GR, but could also be caused by scale-dependent systematic effects. This test is insensitive to an overall normalization systematic in the analysis. 
We find that both the \textsl{Planck} + CMASS and \textsl{Planck} + LOWZ measurements are statistically consistent with a constant across the scales in the cosmological range. We find $\textrm{PTE}_{\rm scale-indep} = 0.52$ for \textsl{Planck} + CMASS and also show the result in the context of the results on the set of simulations in \cref{fig:ScaledependenceTest}. For \textsl{Planck} + LOWZ we find $\textrm{PTE}_{\rm scale-indep} = 0.37$. Both cases indicate that the binwise measurements are within measurement uncertainty fully consistent with scale independence as predicted by GR. Since there is no evidence for scale dependence one can combine the measurements across the cosmological range to find an overall constraint on $E_G$.

It is worth noting that while GR predicts scale-independent growth in linear theory, we expect structure formation to introduce scale dependence on small scales due to non-linearities. The transition between linear and nonlinear growth can be rigorously studied within perturbation theory, which should be valid over the range of scales and observables we fit \citep{Chen2022}. In particular, our cumulative constraint on $E_G$ in either sample is roughly $20\%$, which is wider than the expected correction due to non-linearities, some of which further cancels in the ratio of the auto and cross-correlation, on the scales ($k < 0.2$ Mpc$^{-1}$) that we use in this work. However, we caution that this systematic will be more significant for future, more constraining data. These non-linearities can also be taken into account by marginalizing over the effects of nonlinear bias as computed in perturbation theory; we leave such considerations to future work.

\begin{table*}
\begin{tabular}{l|l|l|l|l|l}
 & $z_{\rm eff}$  & GR prediction $E_G^{\rm GR}$ & Measurement $\hat E_G$ & $\textrm{PTE}_{\rm scale-indep}$ & $\textrm{PTE}_{\rm GR}$ \\ \hline
Planck PR4 + CMASS & $ 0.555$ & $0.401\pm 0.005$ & $0.36^{+0.06}_{-0.05}  (\textrm{stat})$ & 0.52 & 0.46 \\
Planck PR4 + LOWZ & $ 0.316$ & $0.452\pm0.005$ & $0.40^{+0.11}_{-0.09} (\textrm{stat}) $ & 0.37 & 0.64
\end{tabular}
\caption{Summary of the results for the $\hat E_G$ estimator presented in this work applied to \textsl{Planck} PR4 and SDSS BOSS DR12 CMASS and LOWZ data. Shown is the effective redshift of the measurement, $z_{\rm eff}$, the $\Lambda$CDM-GR prediction at the effective redshift based on our fiducial cosmology, $E_{G}^{\rm GR}$, the measurement result, $\hat{E}_{G}$, and the PTE values for consistency of the measurement with scale independence as well as for consistency with the $\Lambda$CDM-GR prediction. Measurements for both datasets are statistically consistent with scale independence as predicted by GR and statistically consistent with the value predicted by $\Lambda$CDM-GR for our fiducial cosmology, they match within the 68.27\% confidence range. \label{table:main_results}}
\end{table*}

\begin{figure}
\includegraphics[width=\columnwidth]{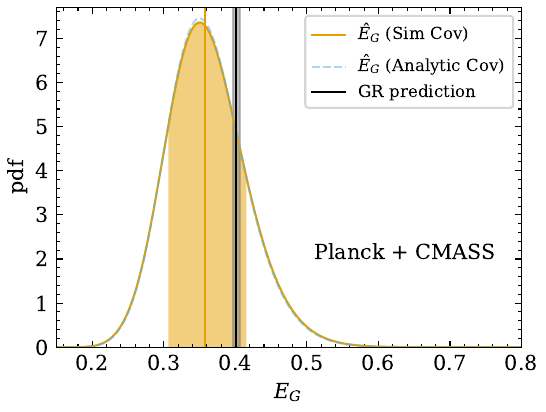}
\includegraphics[width=\columnwidth]{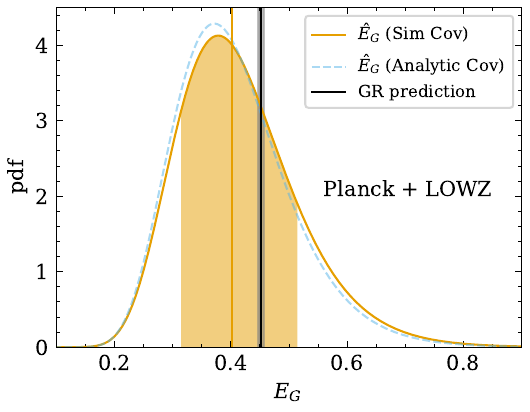}
\caption{The pdf of the measurement $\hat E_G$ marginalized over the cosmological range of scales for \textsl{Planck} + CMASS [upper panel] and \textsl{Planck} + LOWZ [lower panel]. Shown is the pdf of the $\hat E_G$ measurement using the baseline simulation-based covariance [solid orange line] with the median [vertical orange] and central 68.27\% confidence range [shaded orange] highlighted. The resulting pdf for the measurement when using the analytic covariance is also shown [dashed blue line]. The $\Lambda$CDM-GR prediction is shown as a vertical black line with the one sigma uncertainty shaded. Our measurement is in good agreement with the GR prediction with $\textrm{PTE}_{\rm GR} = 0.46$ for \textsl{Planck} + CMASS and $\textrm{PTE}_{\rm GR} = 0.64$ for \textsl{Planck} + LOWZ.}
\label{fig:E_G_Planck_CMASS}
\end{figure}

In \cref{fig:E_G_Planck_CMASS} the overall constraints on $\hat E_G$ for \textsl{Planck} + CMASS and \textsl{Planck} + LOWZ are shown, giving the full pdf, along with the median value and the central 68.27\% confidence range for the $\hat E_G$ estimator combining the constraints across the scales in the cosmological ranges. The figure also shows a vertical line indicating the $\Lambda$CDM-GR prediction for our fiducial cosmology with the one sigma uncertainty shown as a grey band. %

For \textsl{Planck} + CMASS we find $E_G^{\rm \textsl{Planck}+CMASS} = 0.36^{+0.06}_{-0.05}  (\textrm{stat})$ where the statistical error is showing the 68.27\% confidence range. The 95.45\% confidence ranges are given by $ _{-0.10}^{+0.13}$. At the effective redshift of the measurement the $\Lambda$CDM-GR prediction based on our fiducial cosmology is $E_G^{\rm GR} (z = 0.555) = 0.401 \pm 0.005$.

For \textsl{Planck} + LOWZ we find $\hat E_G^{\rm \textsl{Planck}+LOWZ} = 0.40^{+0.11}_{-0.09} (\textrm{stat})$ where the statistical error is showing the 68.27\% confidence interval. The 95.45\% confidence ranges are given by $_{-0.16}^{+0.26}$. Note that the 95.45\% confidence range is more than twice for the upper error and less than twice for the lower error compared to the 68.27\% uncertainties due to the non-Gaussian shape of the uncertainty. At the effective redshift of the measurement, the $\Lambda$CDM-GR prediction based on the fiducial cosmology is $E_G^{\rm GR} (z = 0.316) = 0.452 \pm 0.005$

For both the \textsl{Planck} + CMASS and \textsl{Planck} + LOWZ cases the $\Lambda$CDM-GR prediction lies within the 68.27\% confidence range of the measurement. This agreement is quantified with the statistical test for consistency with GR (\cref{sec:PTE_GR}). The test accounts for the statistical uncertainty of the measurement and the uncertainty of the $\Lambda$CDM-GR prediction. We find $\textrm{PTE}_{\rm GR} = 0.46$ for \textsl{Planck} + CMASS and  $\textrm{PTE}_{\rm GR} = 0.64$ for \textsl{Planck} + LOWZ. Therefore we find the marginalized $E_G$ measurements in good agreement with the $\Lambda$CDM-GR prediction.
The uncertainty of the theory prediction is largely negligible in comparison to the measurement uncertainty. Not accounting for the theory uncertainty by convolving our measurement pdf with the theory uncertainty before calculating the $\textrm{PTE}_{\rm GR}$ gives 0.44 and 0.64 for \textsl{Planck} + CMASS and \textsl{Planck} + LOWZ respectively which is only a marginal difference.

On top of the statistical uncertainty, the analysis can also be affected by systematic errors. Throughout the analysis, conservative estimates of the systematic error budget for the estimate are given. These include approximations made to derive the estimator, bias evolution, magnification bias (\cref{sec:E_G_estimator_accuracy}), and the level to which the pipeline is validated (\cref{fig:sim_recovery}). 
Overall, combining the estimates in quadrature and rounding up we have systematic error budgets of 2\% for \textsl{Planck} + CMASS and 3\% for \textsl{Planck} + LOWZ, well within the statistical uncertainties of the measurements, which are approximately $^{+17\%}_{-14\%}$ and $^{+28\%}_{-23\%}$ respectively. 
The subdominant systematic uncertainty is not accounted for in the test for consistency with the $\Lambda$CDM-GR value, since it has unknown statistical properties and is an upper estimate that can artificially inflate the level of agreement with GR. We can, for reference, conservatively assume the systematic error as one sigma of a Gaussian and convolve it with the statistical pdf as done for the $\Lambda$CDM-GR prediction uncertainty. The sigmas from the systematic error and theory error are added in quadrature. This widens the pdf, increasing the $\textrm{PTE}_{\rm GR}$ values to 0.46 for \textsl{Planck} + CMASS and 0.65 for \textsl{Planck} + LOWZ, an increase of 0.5 (hidden below the rounding) and 1 percentage point respectively. Even without accounting for the systematic uncertainty, the result agrees with $\Lambda$CDM-GR and a conservative upper estimate of the total systematic error only increases the level of the agreement and we find the difference is only marginal justifying neglecting the systematic error budget for our statistical test.

Here, we disclose how data was handled in this analysis to avoid biases in the statistical tests from fine-tuning the analysis choices. During the development of the angular power spectrum estimation, simulations at the fiducial cosmology and a preliminary galaxy bias of 2.0 were used, first Gaussian simulations and then the simulations discussed in \cref{sec:simulated_datavec}. The $\hat E_G$ estimator, all the corrections applied, the non-Gaussian error estimation, and all analysis choices including for example the fiducial cosmological range in scales and the use of the simulation-based covariance were finalized before revealing the results on data. After the analysis approach was frozen, we refined the galaxy bias, to make the covariance matrix more accurate, as described in \cref{sec:galaxy_sims}. We then calculated the pdfs for $E_G$ and $E_G^{\ell}$ on data and reported the result. There were no significant changes to the results after the initial calculation. After the results were finalized we added the theory curves to the plots of the angular power spectra for comparison (e.g. \cref{fig:cross_auto_correlation_Planck_CMASS}) and investigated the sensitivity of our results to a range of analysis choices (\cref{sec:consistency_tests}). 
The data used in this analysis has been public and used in numerous analyses before this work, therefore the general insight that on large scales the cross-correlation between \textsl{Planck} CMB lensing and SDSS BOSS tends low is well-known. Furthermore, the $E_G$ statistic has been estimated for previous iterations of the \textsl{Planck} CMB lensing and SDSS BOSS datasets with different approaches which were finalized and published before this analysis. After finalizing our results we compared them to previous analyses in detail (\cref{sec:resultsincontext}).

A significant value, defined in this work as PTE $<0.05$ will naturally occur on average 5\% of the time even when the underlying data is statistically consistent. Therefore to fully contextualize any significant tests we summarize the total number of PTE tests. 
In this work, we present a total of eight statistical PTE value tests for the data and pipeline. Given this number of tests, we could potentially find a significant value by chance (with a probability of around 33\% when assuming independence). %
Indeed for our initial set of LOWZ simulations, the galaxy auto correlation recovery test failed marginally as we disclose in \cref{sec:input_recovery}. Since we can create an independent set of simulations we reran the test which passed and we concluded that this was a chance failure. The other PTE value tests presented and in particular none of the four PTE tests that evaluate consistency with GR show a significant result. In summary, we find the values of the PTE tests presented consistent with a flat PTE distribution, i.e., no statistical tension that would indicate inaccuracies in our pipeline or disagreement with GR predictions. %

The baseline results and tests are summarized in \cref{table:main_results}. Next, we investigate the sensitivity of the results to analysis choices.

\subsection{Consistency tests} \label{sec:consistency_tests}

\begin{figure*}
\includegraphics[width=1.4\columnwidth]{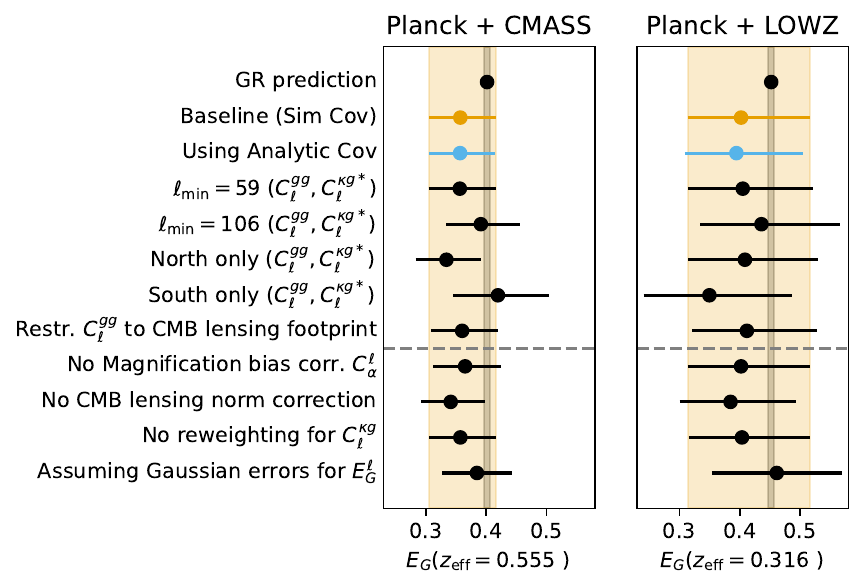}
\caption{The sensitivity of the baseline $\hat E_G$ results presented in this work to analysis choices and corrections applied. The $\Lambda$CDM-GR predicted value and one sigma error based on the fiducial cosmology [grey band] and the 68.27\% confidence range for the baseline measurements from this work [orange band] are used as references. These are compared with the results using an analytic covariance instead of simulation-based, restricting to smaller scales with $\ell_{\rm min} = 59$ and $\ell_{\rm min} = 106$, analyzing the angular power spectra for the BOSS North and South patches only and restricting the auto-correlation to the same footprint as the cross-correlation. Additionally, below the grey dashed line, the impact of removing well-motivated analysis steps is shown: removing the magnification bias correction (negligible for \textsl{Planck} + LOWZ), not correcting the CMB lensing normalization, not reweighting the galaxy sample for the cross-correlation and not accounting for the non-Gaussianity of the estimator uncertainty. }
\label{fig:analysis_choices}
\end{figure*}

The $E_G$ estimate presented in this work consists of a complex analysis that has a range of steps where multiple potential approaches are valid. We investigate some of these analysis choices for their impact on the final constraints. We also quantify the impact of improved corrections and consideration of the uncertainty. The tests are summarized in \cref{fig:analysis_choices}.

Apart from the simulation-based approach we use for our baseline covariance, we could also use an analytically derived covariance. We find them to be in full agreement with each other (third row of \cref{fig:analysis_choices}).

We can consider a more conservative cut of $\ell_{\rm min} = 59$ (rather than the baseline $\ell_{\rm min} = 49)$, cutting the first bin from the analysis, to more closely match the cosmological range used in \citet{Pullen2016}. We find that the results with the restricted range agree well with the full range (fourth row). Additionally, we also present the result when cutting all the large-scale bins that show a low result, applying $\ell_{\rm min} = 106$. This shows that the statistically insignificant tendency of the measurement to be low compared to the $\Lambda$CDM-GR prediction is driven by the largest scales for both the CMASS and LOWZ measurements.

\subsubsection{North vs South sample}
In this work, we present combined constraints from the SDSS BOSS North and South maps. Previous analyses have noted and investigated differences between the North and South observations of the SDSS datasets (e.g. \citep{Chen2022}). 

We here investigate the difference in our results between North and South. While we only have the $\beta$ measurement available for the full sample since we rely on the published data vector that combines the samples for the RSD analysis, we can split the angular power spectra and calculate them for the North and Southern maps separately. We recalculate the norm correction for the CMB lensing maps specific to the North and South samples and derive the covariance matrix from the same simulations split into the two patches. The results for $E_G$ calculated using only the Northern and Southern patches are shown in \cref{fig:analysis_choices}. The separate results are in statistical agreement with each other, justifying combining them. The Northern patch contains the majority of the objects which results in a smaller uncertainty for that result, which also agrees closer with the combined sample.

\subsubsection{Investigating \texorpdfstring{\textsl{Planck}}{Planck} overlap vs full map}

The \textsl{Planck} CMB lensing map covers most (about 96\%) but not all of the SDSS data. We use the full galaxy map to calculate the auto-correlations. 
If the effective galaxy bias for the overlap region is different from the one for the full sample this could introduce a bias, since while we always use the full SDSS overdensity map the cross-correlation is additionally restricted by the overlap region with the \textsl{Planck} data. The LSS samples are designed to be uniform across the footprint through the use of weights. 
We do not expect the small difference to matter in the analysis, but we can investigate whether our results are sensitive to this difference by recalculating the $E_G$ statistic with the auto-correlation restricted to the overlap region with the \textsl{Planck} CMB lensing map. Again we only have the $\beta$ measurement for the full sample. The results are shown in \cref{fig:analysis_choices}, indicating that our results are insensitive to the choice of restricting the auto-correlation to the overlap with the CMB lensing map.  

\subsubsection{Impact of analysis steps}

In the analysis, we carefully account for the effect of magnification bias, correct the normalization of the CMB lensing maps, match the effective redshift of all observables, and account for the non-Gaussianity of the $\hat E_G$ estimator uncertainty. We can investigate the impact of these by removing them individually and comparing the results. The results for removing each of them individually are shown in \cref{fig:analysis_choices}. They all affect the result by only a fraction of the $68.27\%$ ($1\sigma$) uncertainty. Removing any one of them, the results are still statistically consistent with GR and they do not all shift in the result in the same direction. The effect of magnification bias and the normalization of the CMB lensing maps approximately cancel each other out for the \textsl{Planck} + CMASS measurement. When not reweighting the galaxy sample for the angular cross-correlation the expected value of the estimator has a scale-dependent bias at the percent level due to the mismatch in effective redshift as we show in \cref{comparingwithPullenEGestimator}. This is especially relevant when testing for scale dependence. When averaging over our cosmological range in scales the bias cancels partially and we find that the difference is small compared to the statistical uncertainty for both measurements.

We have motivated these analysis steps in this work and argue that they should be applied for an accurate estimate of $E_G$, nonetheless, these checks show that given the statistical uncertainty of the measurement, these analysis steps are not decisive for our overall conclusion that the measurements are statistically consistent with GR. As statistical constraining power increases with upcoming datasets, these analysis steps will become increasingly impactful.

\subsection{Results in context} \label{sec:resultsincontext}

\begin{figure}
\includegraphics[width=\columnwidth]{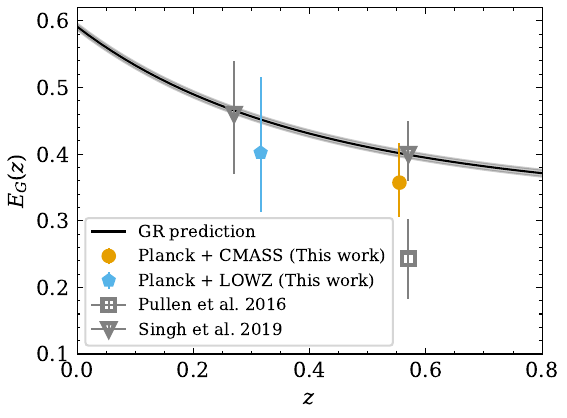}
\caption{
Overview of the cosmological constraints based on the $\hat E_G$ estimator presented in this work applied to the \textsl{Planck} PR4 CMB lensing map and the SDSS BOSS DR12 CMASS and LOWZ galaxy samples. Shown are the measurements of $E_G$ at their respective effective redshift for \textsl{Planck} + CMASS [orange dots] and \textsl{Planck} + LOWZ [blue pentagons]. The $\Lambda$CDM-GR expectation for the fiducial cosmology (assuming $\Omega_{\rm m,0}=0.3111\pm 0.0056$) is shown as a black line with a grey one sigma uncertainty band. Literature results that use earlier versions of the \textsl{Planck} and SDSS data and different approaches to estimate the $E_G$ statistic are shown in grey. Note that the redshifts of the different measurements are not shifted for visibility, other results using \textsl{Planck} and SDSS BOSS data quote different effective redshifts. The statistical 68.27\% confidence ranges for each measurement are shown, systematic error budgets are not plotted, they are reported as subdominant for all cases.}
\label{fig:results_literature_comparison}
\end{figure}

In \cref{fig:results_literature_comparison} we plot the baseline marginalized $E_G$ constraints presented in this work compared to other constraints from the literature using \textsl{Planck} CMB lensing and SDSS BOSS. The plot shows $E_G$ as a function of redshift. The curve for $\Lambda$CDM-GR with our fiducial cosmology is shown as a solid black line with one sigma uncertainty shown as a grey band. The $E_G$ measurements presented in this work are shown with an orange circle showing the result for \textsl{Planck} PR4 + CMASS, $E_G(z=0.555) = 0.36^{+0.06}_{-0.05} (\textrm{stat})$, and a blue pentagon showing the result for \textsl{Planck} PR4 + LOWZ, $\hat E_G(z=0.316)= 0.40^{+0.11}_{-0.09} (\textrm{stat})$. 

Additional literature results using \textsl{Planck} and SDSS BOSS to measure $E_G$ are also shown, specifically from \citet{Pullen2016} who find $E_G(z = 0.57) = 0.243 \pm 0.060 (\textrm{stat}) \pm 0.013 (\textrm{sys})$ for \textsl{Planck} + CMASS and \citet{Singh2019} who find $E_G(z = 0.27) = 0.46_{-0.088}^{+0.082}$ for \textsl{Planck} + CMASS and $E_G(z = 0.57) = 0.4_{-0.051}^{+0.049}$ for \textsl{Planck} + LOWZ with $~3\%$ systematic error budgets for both. Note the plot only shows the statistical uncertainty in each case. For an overview of additional $E_G$ results based on BOSS quasars or with galaxy weak lensing as the lensing tracer, we refer the interested reader to Fig. 11 in \citet{Zhang2021}.

Compared to the results in \citet{Pullen2016}, we find similar measurement uncertainty, however, the best-fit value in this previous result was $\Delta E_G = 0.12$ lower, leading to a reported significant tension with $\Lambda$CDM-GR. Their estimate is also using a harmonic space estimator \citep{Pullen2015}. We use the same pseudo-Cl approach for the calculation of the auto-correlation as for the cross-correlation with careful consideration of the shot noise while the previous analysis used a quadratic minimum variance estimator for the auto-correlation. The \citet{Pullen2016} measurement uses \textsl{Planck} PR2 \citep{PlanckCollaboration2016} and SDSS CMASS DR11 while  \textsl{Planck} PR4 and SDSS CMASS DR12 are used for our analysis. 
The CMASS DR12 catalog contains approximately 12\% more galaxies compared to DR11. The catalog construction calculation of systematics weights is largely the same, with only minor improvements, including a redefinition of the close pair and redshift failure weight calculation affecting only small fractions of the sample and small changes to the stellar density map and seeing estimation for weights \citep{Reid2016}.
\citet{Pullen2016} report the mean redshift of the galaxy sample instead of the effective redshift as defined in this work.  Using the more accurate estimation of effective redshift presented in this work would assign a lower redshift to their result leading to an even stronger tension with $\Lambda$CDM-GR in their findings than was reported. 

In the present work, we motivate, develop, and apply a significantly updated $\hat E_G$ estimator, that carefully considers and matches the effective redshift of all observables and does not require corrections based on simulations and an HOD, in contrast to the estimator used in the previous analysis \citep{Pullen2015}. We compare the estimators in detail in \cref{comparingwithPullenEGestimator}, finding that under ideal assumptions the results with the previous estimator together with its significant correction terms match the estimator presented in this work. We argue that the removal of the reliance on large simulation-based corrections makes our approach more robust against potential mismatches between the simulations and the data. We find that the corrections required for the previous estimator are very sensitive to the redshift distribution of galaxies and the analysis assumptions, for example, whether redshift-dependent FKP weights are used or not.
Additionally, we add corrections for magnification bias and the CMB lensing normalization that were not accounted for in \citet{Pullen2016}.  We find in our analysis that these two effects approximately cancel for \textsl{Planck} + CMASS and also find that restricting our analysis to $\ell \geq 59$, which is closer to $\ell \geq 62$ used in \citep{Pullen2016}, does not significantly change our results (see \cref{fig:analysis_choices}). The \citet{Pullen2016} analysis also assumes Gaussian errors for the binwise and marginalized estimates. As shown in \cref{sec:EGpdf}, this can significantly underestimate the upper error which for a low measurement overestimates the tension with $\Lambda$CDM-GR. 
Our result for $\beta$ for CMASS DR12 of $\beta = 0.371^{+0.042}_{-0.055} $ is wholly consistent with the value found in \citet{Pullen2016} using CMASS DR11 of $\beta = 0.386 \pm 0.046$. 

In summary, we have outlined and assessed the sensitivity of the differences in the present analysis relative to those used in \citet{Pullen2016} to different analysis approaches and differences in the datasets used. The additional corrections for magnification bias and transfer function as well as the slightly less conservative large-scale cutoff are not significant drivers for the differences in results. 
We further discussed in detail the consistency of the updated data products used in this analysis compared to the previous investigation. There are a range of small differences in the data that could induce differences in the results. %
We consider reprocessing previous data products, and the reconstruction of results in prior published work, as out of scope for this analysis. 
The consideration of the non-Gaussian error of the estimator in the present analysis is another contributor to the difference in result.

We can also compare to the measurements presented in \citet{Singh2019}. The reported measurements are slightly higher than the results we presented in this work but the disagreement is not as substantial as with \citet{Pullen2016}, and both results lead to the same overall conclusion of consistency with the $\Lambda$CDM-GR prediction. We discuss potential sources for the small difference. %
The \citet{Singh2019} estimates are based on \textsl{Planck} PR2 and SDSS BOSS DR12 and use a real space estimator. The use of a real space estimator makes them less directly comparable to the harmonic estimator presented in this work. %
We quote the result values given in their Table 1 for the comparison. %
The analysis uses different redshift cuts for the galaxy samples. Most notably the upper cutoff for LOWZ is $z < 0.36$ instead of the $z< 0.43$ used in this work, explaining the lower effective redshift of their measurement. Since $E_G$ decreases with redshift, this brings the two results to closer agreement (see \cref{fig:results_literature_comparison}). Additionally, the lower redshift cutoffs for both samples differ slightly, but this is unlikely to have a relevant effect. 
\citet{Singh2019} note that for combining harmonic space constraints on $E_G^\ell$ one needs to account for the full covariance making the marginalized result susceptible to noisy covariances. They argue that the large Hartlap corrections needed when inverting the covariance could bias the result for harmonic space measurements of $E_G$ when using jackknife errors. 
In this work, we do not use jackknife samples to estimate the full covariance since we argue the number of samples available is not sufficient to invert the covariance reliably. For our simulation-based baseline covariance, we have 480 samples, so for our marginalization requiring the full covariance the Hartlap corrections are only 4.8\% for \textsl{Planck}+CMASS and 3.5\% for \textsl{Planck}+LOWZ. Furthermore, we also confirmed that we recover consistent results when using an analytic covariance that does not suffer from these issues (see \cref{fig:E_G_Planck_CMASS,fig:analysis_choices}). 
\citet{Singh2019} do discuss magnification bias but estimate it based on an approach that assumes that the SDSS BOSS samples are magnitude-limited, which significantly underestimates the effect (as demonstrated in \citet{Wenzl2023magbias}). 
Another difference is that we additionally correct for the CMB lensing norm correction. These two differences partially cancel for \textsl{Planck} + CMASS and are not large enough to explain the difference overall. 
In \citet{Singh2019} the non-Gaussian uncertainty of the $E_G$ estimator was not considered, therefore their approach is more closely comparable to an inference assuming Gaussianity. Indeed when assuming Gaussianity the results we find (see \cref{fig:analysis_choices}) match the results of \citet{Singh2019} relatively closely.

\section{Conclusion}\label{sec:conclusion}

In this work, we have developed a significantly updated $\hat E_G$ estimator to constrain the properties of gravity using galaxy and CMB lensing data. 
We have applied this novel estimator to the \textsl{Planck} PR4 CMB lensing map and the SDSS BOSS DR12 CMASS and LOWZ galaxy catalogs, spanning two redshift ranges, and find that the measurements are statistically consistent with GR predictions. 

The novel $\hat E_G$ estimator combines a reweighted angular cross-power spectrum between CMB lensing and galaxy clustering, the angular auto-power spectrum of the galaxy sample, and the $\beta$ parameter from an RSD analysis. The estimator is carefully derived so that the effective redshifts of all the measurements match, to minimize the overall error introduced by approximations and to avoid the need for simulation+HOD-based corrections. 

A suite of detailed tests is performed to characterize the estimator's accuracy and its sensitivity to assumptions and choices in the analysis. The findings demonstrate that the estimator serves as an unbiased one for $E_G$ in that statistical uncertainties are dominant over the major sources of potential systematic bias considered. For the CMASS and LOWZ samples, the overall bias of the estimator is within the 0.2\% level before considering astrophysical systematics.

Galaxy bias evolution causes a percent level systematic to the estimator for these galaxy samples. It is difficult to correct for this redshift evolution of the galaxy bias since it can only be constrained by assuming GR. However, for the statistical uncertainty of the given sample, the effect is marginal, not restricting comparisons to the GR predictions. Furthermore for future measurements with larger statistical constraining power this systematic can be reduced by measuring $E_G$ in narrower redshift bins. Another systematic effect is from magnification bias. This bias can be constrained from the underlying photometric selection of the galaxy sample \citep{Wenzl2023magbias}, allowing careful corrections to account for the effect. The relevance of magnification bias grows with redshift: while it is negligible for a LOWZ-like sample, for CMASS we find a 2-3\% correction and expect the impact to increase further with redshift.

In \cref{sec:ratio_distribution} we studied the non-Gaussian uncertainty of the $E_G$ estimator. Especially for individual measurements with large statistical uncertainty, either current datasets or future datasets split into narrow redshift bins, we found that the pdf for the $E_G$ statistic is significantly asymmetric. This makes the $68.27\%$ (the equivalent of $1\sigma$ for a Gaussian) confidence limits asymmetric around the median of the distribution and it can affect the best-fit value when combining multiple bins. We also derive two statistical tests that we use to test the measurement for consistency with scale independence and the $E_G$ value predicted by GR.

We reported the constraints from the new $\hat E_G$ estimator with the \textsl{Planck} PR4 CMB lensing map and SDSS BOSS DR12 CMASS and LOWZ galaxy samples in \cref{table:main_results}. Constraints from \textsl{Planck}+CMASS sample an effective redshift of $z_{\rm eff}^{\textsl{Planck} \rm + CMASS} = 0.555$ and constraints from \textsl{Planck}+CMASS sample $z_{\rm eff}^{\textsl{Planck} \rm + LOWZ} = 0.316 $. 
We measured $\hat E_G^\ell$ as a function of angular scale covering a range of approximately $k \in [0.02, 0.2] \textrm{Mpc}^{-1}$. We tested the null hypothesis of GR that the $E_G$ statistic is scale-independent, carefully accounting for the cross-covariance between different angular scales. We found $\textrm{PTE}_{\rm scale-indep}$ values of $52\%$ for \textsl{Planck} + CMASS and $37\%$ for \textsl{Planck} + LOWZ, indicating full consistency of the measurement with scale independence as predicted by GR. This test only weakly depends on $\Lambda$CDM assumptions through the reweighting scheme and the magnification bias correction which each impacts the result well below the statistical uncertainty. 

We combined the constraints across angular scales in the cosmological analysis range, assuming $E_G$ is a constant as a function of scale. We found  68.27\% confidence interval values of $\hat E_G^{\rm \textsl{Planck}+CMASS} = 0.36^{+0.06}_{-0.05}  (\textrm{stat}) $ and $\hat E_G^{\rm \textsl{Planck}+LOWZ} = 0.40^{+0.11}_{-0.09} (\textrm{stat}) $ for \textsl{Planck} + CMASS and \textsl{Planck} + LOWZ respectively, 
with additional subdominant systematic error budgets of 2\% and 3\% respectively. 

The $E_G$ statistic theory predictions for the $\Lambda$CDM cosmology at the effective redshifts for CMASS and LOWZ are $E_G^{\rm GR} (z = 0.555) = 0.401 \pm 0.005$ and $E_G^{\rm GR} (z = 0.316) = 0.452 \pm 0.005$. Comparing these to our measurements under careful consideration of the non-Gaussian uncertainty we found $\textrm{PTE}_{\rm GR}$ values of $44\%$ and $64\%$ for Planck+CMASS and Planck+LOWZ respectively, indicating that the measured $\hat E_G$ values are in good agreement with $\Lambda$CDM-GR predictions.

We overall found the datasets to be statistically fully consistent with GR predictions. We showed in detail in \cref{fig:analysis_choices} that this conclusion is insensitive to a range of analysis choices and that the corrections derived in this work are not decisive for the overall conclusion of agreement with the GR prediction. In \cref{sec:baseline_results} we additionally described how the data was handled to avoid unconscious bias in the results and discuss the full set of statistical tests performed.

In the context of previous $E_G$ analysis on the \textsl{Planck} and SDSS BOSS datasets, we found that our harmonic space results are reasonably consistent with the real-space-based results presented in \citet{Singh2019}, especially when considering differences in the redshift cuts applied to the data and if we assume a Gaussian likelihood for the measurement. However, we do emphasize that the non-Gaussian uncertainty needs to be used to accurately determine constraints. We also discussed that we found a higher result than the one presented in \citet{Pullen2016} for \textsl{Planck} + CMASS which had found results in significant tension with $\Lambda$CDM-GR. We discussed differences in the estimator and analysis in detail.

The estimator for the $E_G$ statistic presented in this work includes redshift-dependent reweighing for the cross-correlation, to reduce the scale-dependent bias of the estimator, which requires accurate galaxy redshifts as provided in spectroscopic surveys. Applying this to potential future $E_G$ measurements using photometric samples, such as from LSST \citep{Ivezic2019}, \textsl{Euclid} \citep{Laureijs2011} and \textsl{Roman} \citep{Spergel2015} as suggested in \citet{Pullen2015}, will require careful consideration in future work. 

In addition to testing consistency with GR, the $E_G$ scale independent and dependent measurements presented here also open up the potential for comparisons with predictions from alternative gravity models. This includes evidence of deviations of  $E_G$ from the GR-predicted magnitude and redshift evolution, as in the Chameleon model, and scale dependence, as in, for example, $f(R)$ models \citep{Zhang2007,Pullen2015}.

The $E_G$ estimator presented in this work is readily applicable to upcoming cosmological datasets like CMB lensing measurements from ACT \citep{Qu2023}, SPT \citep{Pan2023}, SO \citep{SimonsObservatory2019} and CMB-S4 \citep{Abazajian2016} as well as spectroscopic galaxy samples from DESI \citep{DESICollaboration2016} and SPHEREx \citep{Dore2014} which will allow interesting new constraints on gravity with this statistic. %

\begin{acknowledgments}

We thank Anthony Pullen for providing further information about the work presented in \citet{Pullen2016} and helpful comments on a preliminary draft of this work.
We thank Alex Krolewski for helpful discussions about the bias evolution with redshift for SDSS BOSS. 
We thank H{\'e}ctor Gil-Mar{\'i}n for providing us with the official BOSS measurements including window matrices from \citep{GilMarin2016} for the RSD analysis.
We thank Zeyang Sun for helpful clarifications and discussions on ratio distributions presented in \citet{Sun2023}.

The work of LW and RB is supported by NSF grant AST-2206088, NASA ATP grant 80NSSC18K0695, and NASA ROSES grant 12-EUCLID12-0004. MM acknowledges support from NSF grants AST-2307727 and AST-2153201 and NASA grant 21-ATP21-0145. 
GSF acknowledges support through the Isaac Newton Studentship and the Helen Stone Scholarship at the University of Cambridge. 
GSF and BS acknowledge support from the European Research Council (ERC) under the European Union’s Horizon 2020 research and innovation programme (Grant agreement No. 851274). 
SC acknowledges the support of the National Science Foundation at the Institute for Advanced Study. 
GM is part of Fermi Research Alliance, LLC under Contract No. DE-AC02-07CH11359 with the U.S. Department of Energy, Office of Science, Office of High Energy Physics. 
NS acknowledges support from DOE award number DE-SC0020441. 
AvE acknowledges support from NASA grants 80NSSC23K0747 and 80NSSC23K0464. 

This research used resources of the National Energy Research Scientific Computing Center (NERSC), a U.S. Department of Energy Office of Science User Facility located at Lawrence Berkeley National Laboratory, operated under Contract No. DE-AC02-05CH11231 using NERSC award HEP-ERCAPmp107

SDSS-III is managed by the Astrophysical Research Consortium for the Participating Institutions of the SDSS-III Collaboration including the University of Arizona, the Brazilian Participation Group, Brookhaven National Laboratory, Carnegie Mellon University, University of Florida, the French Participation Group, the German Participation Group, Harvard University, the Instituto de Astrofisica de Canarias, the Michigan State/Notre Dame/JINA Participation Group, Johns Hopkins University, Lawrence Berkeley National Laboratory, Max Planck Institute for Astrophysics, Max Planck Institute for Extraterrestrial Physics, New Mexico State University, New York University, Ohio State University, Pennsylvania State University, University of Portsmouth, Princeton University, the Spanish Participation Group, University of Tokyo, University of Utah, Vanderbilt University, University of Virginia, University of Washington, and Yale University.

\end{acknowledgments}

\appendix

\section{SDSS Mask making} \label{SDSS_mask_making}

\begin{figure}
\includegraphics[width=1\columnwidth]{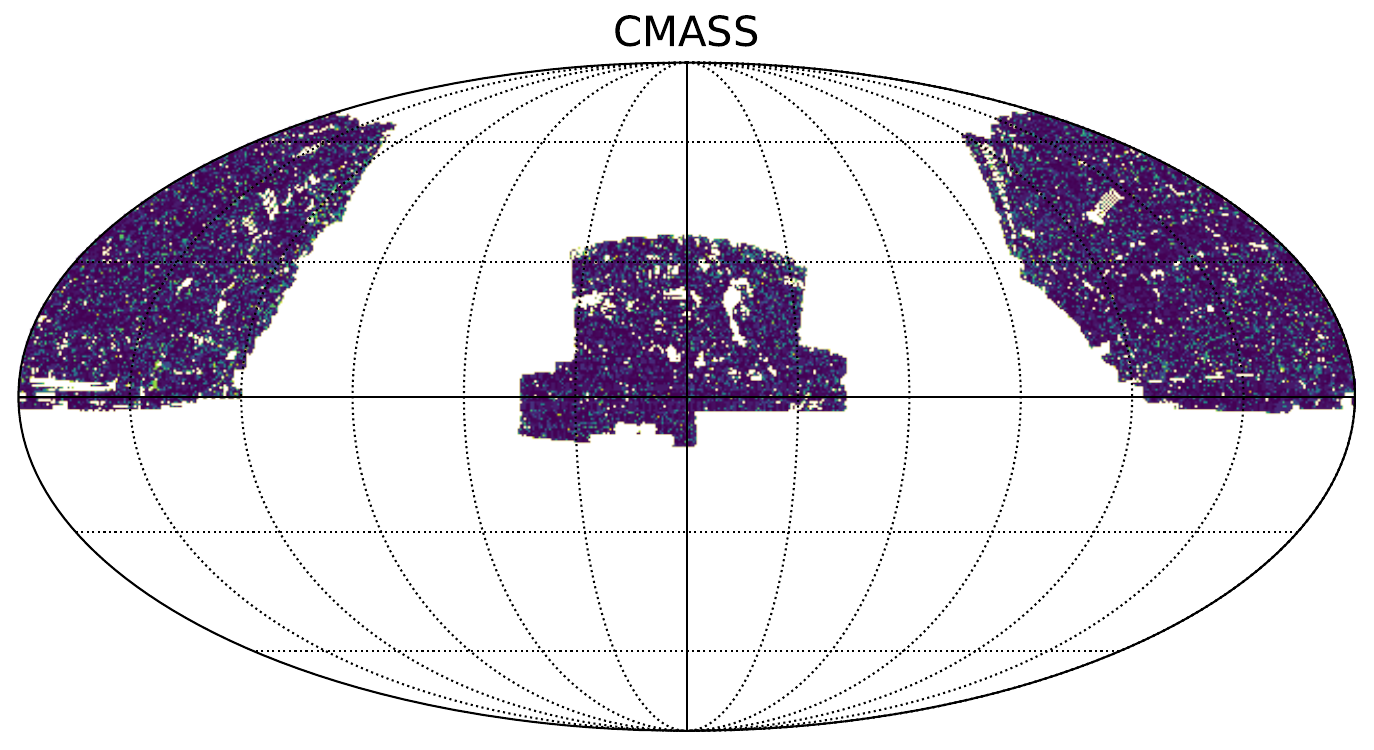}
\includegraphics[width=1\columnwidth]{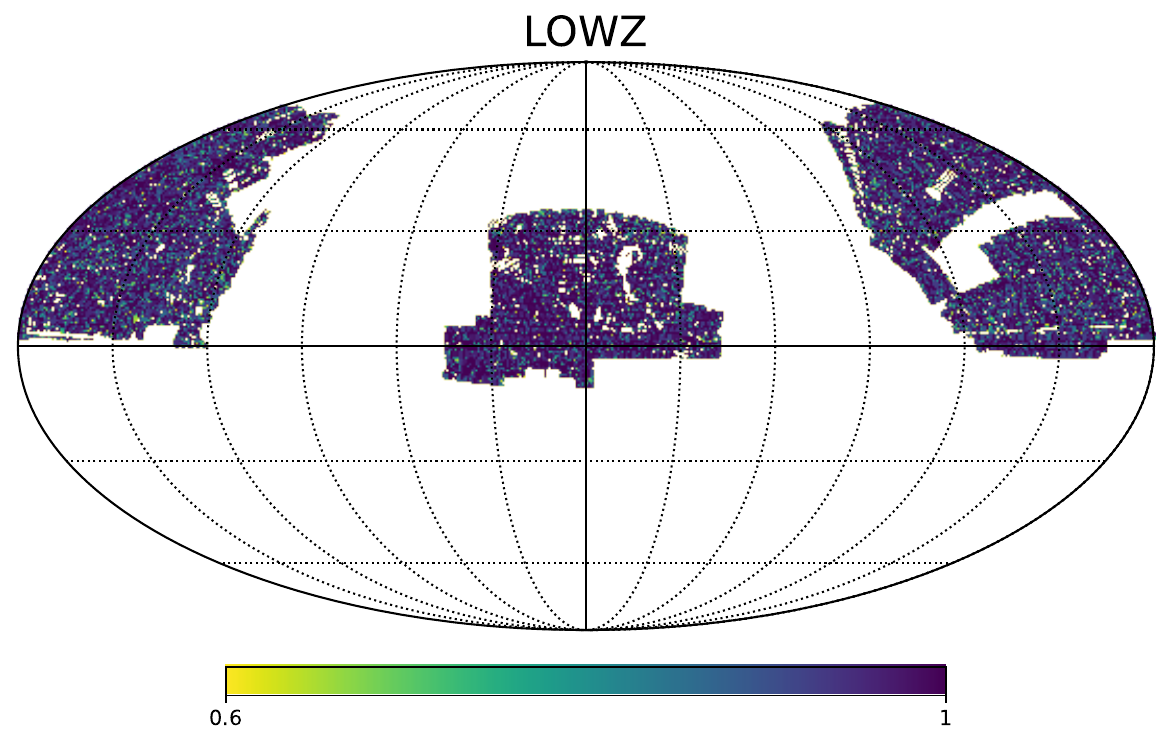}
\caption{Observational masks of the CMASS and LOWZ samples in Equatorial coordinates and Mollweide projection. The values show the fractional coverage of each pixel combining survey footprint, local completeness, and accounting for veto regions. Shown are the pixels with coverage fraction above 60\% which is the lower cutoff used in the analysis.}
\label{fig:masks}
\end{figure}

In this section, we describe how we construct the masks for the galaxy samples LOWZ and CMASS shown in \cref{fig:masks}. Accurate masks are crucial for an unbiased estimate of the normalization of the 2pt correlation functions in harmonic space, while the original SDSS analysis primarily focused on real-space approaches. We need to estimate the fractional coverage of each pixel for our \healpix maps using $\mathit{nside}=1024$. To find the fractional coverage we need to combine information about the survey footprint and the local completeness, as well as account for veto regions. Veto regions were not included as part of the SDSS completeness calculation and therefore need to be accounted for separately.

The SDSS team provides completeness information per sector and veto areas as mangle polygon files\footnote{https://data.sdss.org/sas/dr12/boss/lss/geometry/}. Sectors are defined as unique combinations of points. 
We use \pymangle\footnote{https://github.com/esheldon/pymangle} to check individual sky locations against these polygon files. To accurately estimate the fractional coverage of each pixel we check 4096 sample points per pixel. Each point that is inside a veto region or outside the survey footprint is set to 0 and the other points are set to the local completeness value. The average of all points is then the coverage fraction of the pixel considered. This is equivalent to checking the centers of pixels in a $\mathit{nside}=65,536$ map and then down-sampling via averaging. For optimal memory efficiency, we do this in patches and for speed in parallel.   
With this, we estimate the coverage fraction accurate to $<0.5\%$ (1 sigma). Note that this quoted accuracy neglects uncertainty on the original estimates of completeness and the definition of the veto regions presented in \citep{Reid2016} which we presume to be accurate. %

This avoids potential inaccuracies from representing \healpix pixels as polygons and is much more computationally efficient. It also adds less additional shot noise than using randoms to construct the masks, although, for the scales of interest in this analysis, the difference is mostly negligible. %

\section{Comparing our \texorpdfstring{E\textsubscript{G}}{EG} estimator to a previous estimator} \label{comparingwithPullenEGestimator}

The new estimator $\hat E_G$ we present in \cref{E_G_estimator} contains notable changes compared to a previous harmonic space $E_G$ estimator for CMB lensing and galaxy clustering presented in \citet{Pullen2015,Pullen2016} given by
\begin{align}
    \hat E_G^{\rm Pullen} (\bar z) &= \Gamma^{\rm Pullen} (\bar z) \frac{C_\ell^{\kappa g}}{\beta C_\ell^{gg}}, \\
    \Gamma^{\rm Pullen} (\bar z) &= \frac{2c H(\bar z)}{3 H_0^2} \frac{W_{g}(\bar z)}{\hat W_\kappa(\bar z)},
    \label{eq:PullenEG}
\end{align}
where $\bar z^{\rm Pullen} = 0.57$ is their quoted redshift of the measurement. The key improvements of our estimator are:
\begin{enumerate}
  \item Our final expression for $\Gamma$ has a factor $\int \dd z \frac{W_{g}^2(z)}{\hat W_\kappa (z)}$ instead of the $W_{g}(\bar z) / \hat W_\kappa(\bar z)$ factor in $\hat E_G^{\rm Pullen}$.
  \item We reweigh the galaxy samples for the cross-correlation to match the effective redshift of the measurement to the auto-correlation. So our $\hat E_G$ uses $C_\ell^{\kappa g*}$ compared to $C_\ell^{\kappa g}$ in $\hat E_G^{\rm Pullen2015}$.
  \item We choose to project our angular power spectra with the clustering kernel which defines our effective redshift. This choice is convenient since then the effective redshift of the RSD measurement matches that of the angular power spectra.
  \item These changes make the overall systematics of the estimator negligible when inputting a simulation with exactly linear bias and no other astrophysical systematics, $S_\Gamma^\ell$ (\cref{eq:error_in_EG}), as opposed to in \citet{Pullen2016} where a correction of approximately $5\%$ based on simulations and assuming a HOD is used.
  \item We add a correction for the magnification bias of the galaxy samples (see \cref{sec:magnification_bias}).
  \item We account for the amplitude error in the CMB lensing reconstruction due to the mask (see \cref{sec:CMBsims}).
  \item We account for the non-Gaussian uncertainty of the estimator (see \cref{sec:ratio_distribution}).  %
\end{enumerate}
The first change could be directly applied to the $\hat E_G^{\rm Pullen2016}$ estimator without any other changes to partially resolve the large overall offset of the estimator. We first discuss the underlying reason for this large offset. The approximation $W_{g}(\chi) \approx \frac{W_{g}^2(\chi)}{W_{g}(\chi (\bar z))}$ used in \citet{Pullen2015,Pullen2016} would be close for galaxy samples where $\dd N(z)/\dd z$ is close to a constant in the observed redshift range. However, for realistic galaxy samples, this approximation can introduce a large offset, depending sensitively on the shape of the $\dd N(z)/\dd z$ distribution and the choice of redshift $\bar z$. We can quantify this with the overall error of the estimator $S_\Gamma^\ell$ for the case of SDSS CMASS. 
In \citet{Pullen2016} FKP weights were not considered so in the following, to allow for comparison to their results, we also removed FKP weights for this discussion, to have a comparable galaxy sample. We include FKP weights in our analysis since they optimally weight the data for the RSD analysis minimizing the uncertainty on $\beta$. We note that the accuracy of the estimator presented in this work is similar with or without the use of FKP weights, but since these weights are redshift dependent they do sensitively affect the accuracy of the $\hat E_G^{\rm Pullen2016}$ estimator. 
Using our well-motivated effective redshift calculation, we find $z_{\rm eff} = 0.531$ for the CMASS sample without FKP weights, the offset $S_\Gamma^\ell$ in the final estimate using $\hat E_G^{\rm Pullen} ( z_{\rm eff})$ is roughly 20\%. For their choice of redshift of $\bar z^{\rm Pullen} =0.57$ and using $\hat E_G^{\rm Pullen} ( \bar z^{\rm Pullen} =0.57)$ the offset is roughly 5\% as discussed in their paper. This higher value of redshift was used based on the value reported in the BOSS DR11 release paper \citep{Anderson2014}. %
Note that one could also choose a lower redshift to similarly reduce the amount of correction needed. The correction only reduces because the approximation above is more accurate for redshifts off the peak for a CMASS-like redshift distribution. %

Even after this change a scale-dependent (ringing) bias remains in the estimator that is caused by the mismatch of the effective redshift of the auto and cross-correlations sample. We resolve this by reweighing the galaxy sample for the cross-correlation and shifting the effective redshift to match the one for the auto-correlation. 

\begin{figure}
\includegraphics[width=\columnwidth]{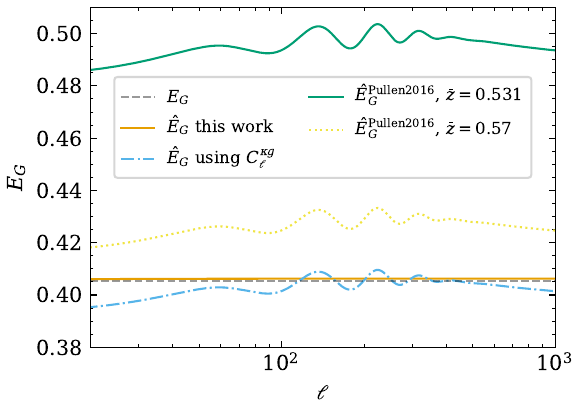}
\caption{A summary of how the improvements to the $\hat E_G$ estimator proposed in this work impact the overall accuracy of the estimator for our fiducial cosmology. Shown is the case of CMB lensing combined with a CMASS-like galaxy sample without considering FKP weights with $z_{\rm eff} = 0.531$. Shown are the analytic $E_G$ value for the fiducial cosmology [black dashed], $\hat E_G$ using the estimator in this work [orange, full], our estimator but using $C_\ell^{\kappa g}$ rather than the reweighted $C_\ell^{\kappa g*}$ [blue, dot-dashed] and the estimator used in \citet{Pullen2016} for both the redshift assumed in that work, $z=0.57$, [yellow, dotted] and at the effective redshift of the sample, $z=0.531$ [green, full]. See \cref{comparingwithPullenEGestimator} for details.}
\label{fig:E_G_estimator_performance}
\end{figure}

\begin{figure}
\includegraphics[width=\columnwidth]{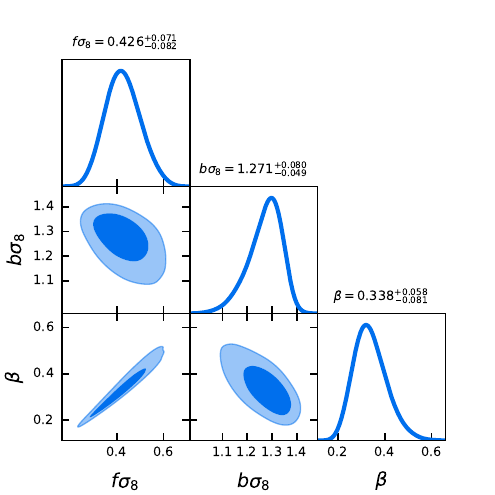}
\caption{
RSD fit to the clustering signal of the LOWZ data set. The corner plot shows the fitting parameters $f \sigma_8, b\sigma_8$ as well as the derived parameter $\beta = f \sigma_8 /(b\sigma_8)$ used as part of our $E_G$ estimate. For each parameter, we also show the mean value and one sigma error.
}
\label{fig:cornerplot_beta_LOWZ}
\end{figure}

\begin{figure}
\includegraphics[width=\columnwidth]{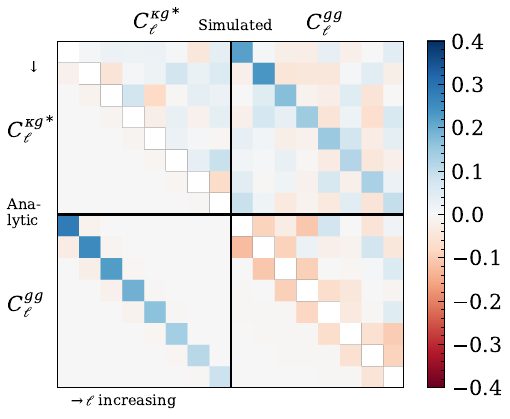}
\caption{Visualizing the off-diagonal correlation matrix between $C_\ell^{gg}$ and $C_\ell^{\kappa g*}$ for \textsl{Planck} and LOWZ. Shown is the correlation matrix defined as $\textrm{Cov} (C_\ell^{XY}, C_\ell^{AB}) / (\sigma_\ell^{XY} \sigma_\ell^{AB}) $ with the diagonal removed for readability. The upper triangle shows the simulated covariance and the lower triangle the analytic covariance. Correlations between $C_\ell^{gg}$ and $C_\ell^{\kappa g*}$ are up to around 0.35 for our narrowest large-scale bins. The off-diagonals of the simulated covariance match the analytic covariance well within statistical uncertainty. For a comparison of the diagonals of the covariances see \cref{fig:cross_auto_correlation_Planck_LOWZ}.}
\label{fig:correlation_matrix_plot_PlanckPR4_LOWZ}
\end{figure}

In \cref{fig:E_G_estimator_performance} we show the expectation value of the estimator for a CMASS-like sample\footnote{Without FKP weights to allow for direct comparison with \citet{Pullen2016} but otherwise as described in \cref{sec:data}. Note that the choice of using or not using FKP weights has no significant effect on the accuracy of the new estimator presented in this work.}, when plugging in the analytic expressions for our fiducial cosmology and using the effective redshift of the sample as defined in this work ($z_{\rm eff} = 0.531$). This test does not model the magnification bias and uses a linear galaxy bias. Our final $\hat E_G$ estimator (solid orange line) matches the theory prediction $E_G$ (black dashed line) very well ($|S_\Gamma^\ell -1|< 0.3\% $). If we use $C_\ell^{\kappa g}$ instead of our reweighted $C_\ell^{\kappa g*}$, shown as a blue dot-dashed line, we see a scale-dependent bias of a few percent. If we additionally use $W_{g}(\bar z) / \hat W_\kappa(\bar z)$ instead of $\int \dd z \frac{W_{g}^2(z)}{\hat W_\kappa (z)}$ we recover the functional form of the previous estimator $\hat E_G^{\rm Pullen2016}$. For our effective redshift of the CMASS sample (shown as a green solid line), this estimator has an overall 20\% bias that needs to be corrected for, as we already discussed by considering $S_\Gamma^\ell$. For $\bar z = 0.57$ as assumed in \citet{Pullen2016} (shown as a yellow dotted line) the bias is around 5\% matching what is shown in their Fig. 3. Note that for the different $\bar z = 0.57$, the theory prediction and the other factors in $\Gamma$ also shift slightly.

We want to stress that together with their $C^\Gamma$ correction the estimator for $E_G$ presented in \citet{Pullen2015,Pullen2016} is unbiased. However, the large correction relies on the accuracy of N-body simulations and the choice of HOD. %
We argue it is preferable to adapt our improved estimator to not require this correction for an unbiased result.

\section{Additional plots for LOWZ sample} \label{appendix_LOWZ_plots}

\begin{figure*}
\includegraphics[width=1.0155\columnwidth]{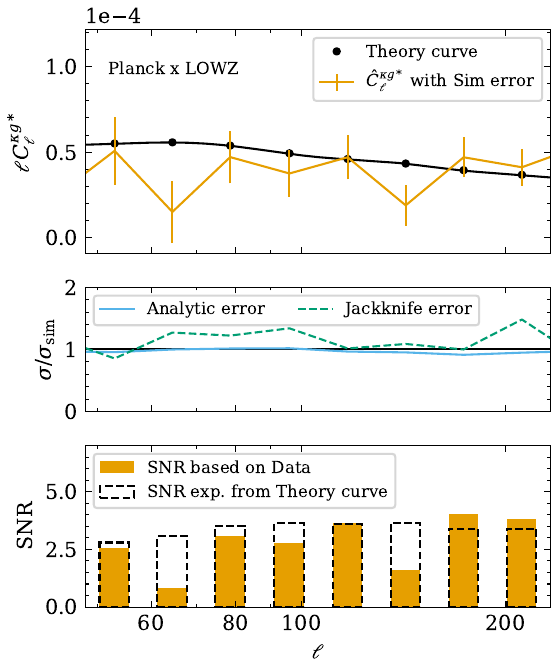}
\includegraphics[width=0.9845\columnwidth]{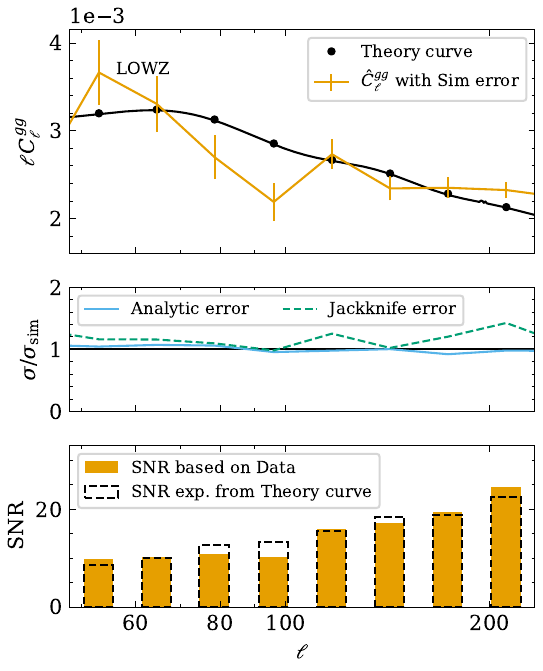}
\caption{Measurement of [Left] the angular cross-power spectrum, $C_\ell^{\kappa g*}$, for CMB lensing map based on \textsl{Planck} PR4 and the reweighted LOWZ galaxy sample and [Right] the angular auto LOWZ galaxy power power spectrum (equivalent figure for CMASS shown in \cref{fig:cross_auto_correlation_Planck_CMASS}). [Top] The angular power spectrum with simulation-based errors as well as our theory curve. [Middle] Comparison of the analytic [solid blue line] and jackknife error estimates [dashed green] to the baseline marginalized errors from the simulations. [Lower] The signal-to-noise ratio, calculated as the ratio of measurement and measurement uncertainty is shown [filled orange]. For reference, the ratio of the theory expectation and the measurement uncertainty is also shown [dashed black]. }
\label{fig:cross_auto_correlation_Planck_LOWZ}
\end{figure*}

\begin{figure}
\includegraphics[width=\columnwidth]{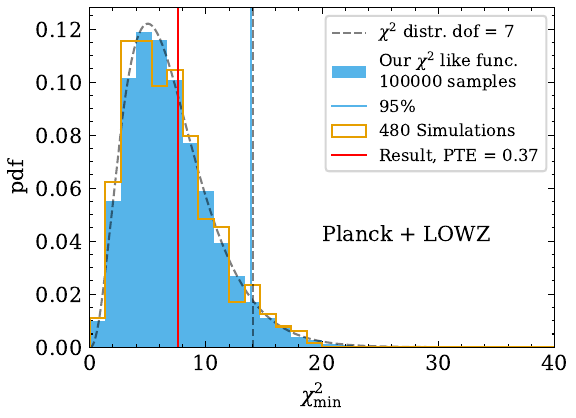}
\includegraphics[width=\columnwidth]{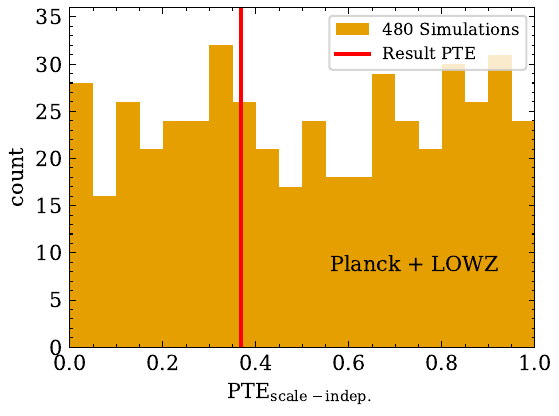}
\caption{Characterizing the statistical test for consistency of our measurement with scale independence for \textsl{Planck} + LOWZ. Content as described for CMASS in \cref{fig:ScaledependenceTest}. }
\label{fig:ScaledependenceTest_LOWZ}
\end{figure}

Here we show additional plots involving the LOWZ dataset. In \cref{fig:cornerplot_beta_LOWZ} we show the RSD fit to measure beta for the LOWZ sample. 
In \cref{fig:correlation_matrix_plot_PlanckPR4_LOWZ} we show the off-diagonal covariance matrix as a correlation matrix for the angular power spectrum measurements with \textsl{Planck} and LOWZ.

\cref{fig:cross_auto_correlation_Planck_LOWZ} shows our cross-correlation measurement between \textsl{Planck} and LOWZ as well as the auto-correlation measurement of LOWZ.

In \cref{fig:ScaledependenceTest_LOWZ} we characterize the statistical test for scale independence of our measurement for the case of \textsl{Planck} + LOWZ.

\section{Error propagation for covariances}\label{error_propagation}

We, for convenience, discuss the error propagation for the covariances when combining multiple $C_\ell$ when assuming Gaussianity. %
If the two measurements are significantly correlated we need to account for the cross-covariance.

When multiplying or taking the ratio of two measurements, as is done for $E_G$ it is convenient to work in fractional errors, defining 
\begin{align}
    \textrm{CovN}(A_\ell, B_{\ell'}) \equiv \frac{\textrm{Cov}(A_\ell, B_{\ell'}) }{ A_\ell B_{\ell'}}.
\end{align}
For multiplication or division $M_\ell \equiv  C_\ell^X \cdot  (C_\ell^Y)^{\pm 1}$, Gaussian error propagation gives
\begin{multline}
\textrm{CovN}(M_\ell, M_{\ell'}) = \\\textrm{CovN}(C_\ell^X, C_{\ell'}^X)  + \textrm{CovN}(C_\ell^Y, C_{\ell'}^Y) \\ 
\pm \textrm{CovN}(C_\ell^X, C_{\ell'}^Y) \pm \textrm{CovN}(C_{\ell}^Y, C_{\ell'}^X).
\end{multline}
For the full $E_G$ estimator as defined in \cref{E_G_estimator} which depends on $C_\ell^{\kappa g},C_\ell^{gg}$ and $\beta$ one obtains
\begin{align}
\begin{aligned}
\textrm{CovN}&(E_G^\ell, E_G^{\ell'}) = \\&\textrm{CovN}(C_\ell^{\kappa g}, C_{\ell'}^{\kappa g})  + \textrm{CovN}(C_\ell^{gg}, C_{\ell'}^{gg}) + \left( \frac{\sigma_\beta}{\beta}\right)^2 \\ 
&- \textrm{CovN}(C_\ell^{\kappa g}, C_{\ell'}^{gg}) - \textrm{CovN}(C_{\ell}^{gg}, C_{\ell'}^{\kappa g}) \\
&- \textrm{CovN}(\beta, C_{\ell'}^{\kappa g}) + \textrm{CovN}(\beta, C_{\ell'}^{gg}) \\
&- \textrm{CovN}(C_\ell^{\kappa g}, \beta) + \textrm{CovN}(C_\ell^{gg}, \beta).
\end{aligned}
\end{align}

\bibliography{main}

\begin{thebibliography}{112}%
\makeatletter
\providecommand \@ifxundefined [1]{%
 \@ifx{#1\undefined}
}%
\providecommand \@ifnum [1]{%
 \ifnum #1\expandafter \@firstoftwo
 \else \expandafter \@secondoftwo
 \fi
}%
\providecommand \@ifx [1]{%
 \ifx #1\expandafter \@firstoftwo
 \else \expandafter \@secondoftwo
 \fi
}%
\providecommand \natexlab [1]{#1}%
\providecommand \enquote  [1]{``#1''}%
\providecommand \bibnamefont  [1]{#1}%
\providecommand \bibfnamefont [1]{#1}%
\providecommand \citenamefont [1]{#1}%
\providecommand \href@noop [0]{\@secondoftwo}%
\providecommand \href [0]{\begingroup \@sanitize@url \@href}%
\providecommand \@href[1]{\@@startlink{#1}\@@href}%
\providecommand \@@href[1]{\endgroup#1\@@endlink}%
\providecommand \@sanitize@url [0]{\catcode `\\12\catcode `\$12\catcode
  `\&12\catcode `\#12\catcode `\^12\catcode `\_12\catcode `\%12\relax}%
\providecommand \@@startlink[1]{}%
\providecommand \@@endlink[0]{}%
\providecommand \url  [0]{\begingroup\@sanitize@url \@url }%
\providecommand \@url [1]{\endgroup\@href {#1}{\urlprefix }}%
\providecommand \urlprefix  [0]{URL }%
\providecommand \Eprint [0]{\href }%
\providecommand \doibase [0]{https://doi.org/}%
\providecommand \selectlanguage [0]{\@gobble}%
\providecommand \bibinfo  [0]{\@secondoftwo}%
\providecommand \bibfield  [0]{\@secondoftwo}%
\providecommand \translation [1]{[#1]}%
\providecommand \BibitemOpen [0]{}%
\providecommand \bibitemStop [0]{}%
\providecommand \bibitemNoStop [0]{.\EOS\space}%
\providecommand \EOS [0]{\spacefactor3000\relax}%
\providecommand \BibitemShut  [1]{\csname bibitem#1\endcsname}%
\let\auto@bib@innerbib\@empty
\bibitem [{\citenamefont {{Riess}}\ \emph {et~al.}(1998)\citenamefont
  {{Riess}}, \citenamefont {{Filippenko}}, \citenamefont {{Challis}},
  \citenamefont {{Clocchiatti}}, \citenamefont {{Diercks}}, \citenamefont
  {{Garnavich}}, \citenamefont {{Gilliland}}, \citenamefont {{Hogan}},
  \citenamefont {{Jha}}, \citenamefont {{Kirshner}}, \citenamefont
  {{Leibundgut}}, \citenamefont {{Phillips}}, \citenamefont {{Reiss}},
  \citenamefont {{Schmidt}}, \citenamefont {{Schommer}}, \citenamefont
  {{Smith}}, \citenamefont {{Spyromilio}}, \citenamefont {{Stubbs}},
  \citenamefont {{Suntzeff}},\ and\ \citenamefont {{Tonry}}}]{Riess1998}%
  \BibitemOpen
  \bibfield  {author} {\bibinfo {author} {A.~G. {Riess}}, \bibinfo {author}
  {A.~V. {Filippenko}}, \bibinfo {author} {P.~{Challis}}, \bibinfo {author}
  {A.~{Clocchiatti}}, \bibinfo {author} {A.~{Diercks}}, \bibinfo {author}
  {P.~M. {Garnavich}}, \bibinfo {author} {R.~L. {Gilliland}}, \bibinfo {author}
  {C.~J. {Hogan}}, \bibinfo {author} {S.~{Jha}}, \bibinfo {author} {R.~P.
  {Kirshner}}, et~al.,\ }\href {https://doi.org/10.1086/300499} {\bibfield
  {journal} {\bibinfo  {journal} {\aj}\ }\textbf {\bibinfo {volume} {116}},\
  \bibinfo {pages} {1009} (\bibinfo {year} {1998})},\ \Eprint
  {https://arxiv.org/abs/astro-ph/9805201} {arXiv:astro-ph/9805201 [astro-ph]}
  \BibitemShut {NoStop}%
\bibitem [{\citenamefont {{Perlmutter}}\ \emph {et~al.}(1999)\citenamefont
  {{Perlmutter}}, \citenamefont {{Aldering}}, \citenamefont {{Goldhaber}},
  \citenamefont {{Knop}}, \citenamefont {{Nugent}}, \citenamefont {{Castro}},
  \citenamefont {{Deustua}}, \citenamefont {{Fabbro}}, \citenamefont
  {{Goobar}}, \citenamefont {{Groom}}, \citenamefont {{Hook}}, \citenamefont
  {{Kim}}, \citenamefont {{Kim}}, \citenamefont {{Lee}}, \citenamefont
  {{Nunes}}, \citenamefont {{Pain}}, \citenamefont {{Pennypacker}},
  \citenamefont {{Quimby}}, \citenamefont {{Lidman}}, \citenamefont {{Ellis}},
  \citenamefont {{Irwin}}, \citenamefont {{McMahon}}, \citenamefont
  {{Ruiz-Lapuente}}, \citenamefont {{Walton}}, \citenamefont {{Schaefer}},
  \citenamefont {{Boyle}}, \citenamefont {{Filippenko}}, \citenamefont
  {{Matheson}}, \citenamefont {{Fruchter}}, \citenamefont {{Panagia}},
  \citenamefont {{Newberg}}, \citenamefont {{Couch}},\ and\ \citenamefont
  {{Project}}}]{Perlmutter1999}%
  \BibitemOpen
  \bibfield  {author} {\bibinfo {author} {S.~{Perlmutter}}, \bibinfo {author}
  {G.~{Aldering}}, \bibinfo {author} {G.~{Goldhaber}}, \bibinfo {author} {R.~A.
  {Knop}}, \bibinfo {author} {P.~{Nugent}}, \bibinfo {author} {P.~G. {Castro}},
  \bibinfo {author} {S.~{Deustua}}, \bibinfo {author} {S.~{Fabbro}}, \bibinfo
  {author} {A.~{Goobar}}, \bibinfo {author} {D.~E. {Groom}}, et~al.,\ }\href
  {https://doi.org/10.1086/307221} {\bibfield  {journal} {\bibinfo  {journal}
  {\apj}\ }\textbf {\bibinfo {volume} {517}},\ \bibinfo {pages} {565} (\bibinfo
  {year} {1999})},\ \Eprint {https://arxiv.org/abs/astro-ph/9812133}
  {arXiv:astro-ph/9812133 [astro-ph]} \BibitemShut {NoStop}%
\bibitem [{\citenamefont {{Peebles}}\ and\ \citenamefont
  {{Ratra}}(2003)}]{Peebles2003}%
  \BibitemOpen
  \bibfield  {author} {\bibinfo {author} {P.~J. {Peebles}}\ and\ \bibinfo
  {author} {B.~{Ratra}},\ }\href {https://doi.org/10.1103/RevModPhys.75.559}
  {\bibfield  {journal} {\bibinfo  {journal} {Reviews of Modern Physics}\
  }\textbf {\bibinfo {volume} {75}},\ \bibinfo {pages} {559} (\bibinfo {year}
  {2003})},\ \Eprint {https://arxiv.org/abs/astro-ph/0207347}
  {arXiv:astro-ph/0207347 [astro-ph]} \BibitemShut {NoStop}%
\bibitem [{\citenamefont {{Alam}}\ \emph
  {et~al.}(2017{\natexlab{a}})\citenamefont {{Alam}}, \citenamefont {{Ata}},
  \citenamefont {{Bailey}}, \citenamefont {{Beutler}}, \citenamefont
  {{Bizyaev}}, \citenamefont {{Blazek}}, \citenamefont {{Bolton}},
  \citenamefont {{Brownstein}}, \citenamefont {{Burden}}, \citenamefont
  {{Chuang}}, \citenamefont {{Comparat}}, \citenamefont {{Cuesta}},
  \citenamefont {{Dawson}}, \citenamefont {{Eisenstein}}, \citenamefont
  {{Escoffier}}, \citenamefont {{Gil-Mar{\'\i}n}}, \citenamefont {{Grieb}},
  \citenamefont {{Hand}}, \citenamefont {{Ho}}, \citenamefont {{Kinemuchi}},
  \citenamefont {{Kirkby}}, \citenamefont {{Kitaura}}, \citenamefont
  {{Malanushenko}}, \citenamefont {{Malanushenko}}, \citenamefont {{Maraston}},
  \citenamefont {{McBride}}, \citenamefont {{Nichol}}, \citenamefont
  {{Olmstead}}, \citenamefont {{Oravetz}}, \citenamefont {{Padmanabhan}},
  \citenamefont {{Palanque-Delabrouille}}, \citenamefont {{Pan}}, \citenamefont
  {{Pellejero-Ibanez}}, \citenamefont {{Percival}}, \citenamefont
  {{Petitjean}}, \citenamefont {{Prada}}, \citenamefont {{Price-Whelan}},
  \citenamefont {{Reid}}, \citenamefont {{Rodr{\'\i}guez-Torres}},
  \citenamefont {{Roe}}, \citenamefont {{Ross}}, \citenamefont {{Ross}},
  \citenamefont {{Rossi}}, \citenamefont {{Rubi{\~n}o-Mart{\'\i}n}},
  \citenamefont {{Saito}}, \citenamefont {{Salazar-Albornoz}}, \citenamefont
  {{Samushia}}, \citenamefont {{S{\'a}nchez}}, \citenamefont {{Satpathy}},
  \citenamefont {{Schlegel}}, \citenamefont {{Schneider}}, \citenamefont
  {{Sc{\'o}ccola}}, \citenamefont {{Seo}}, \citenamefont {{Sheldon}},
  \citenamefont {{Simmons}}, \citenamefont {{Slosar}}, \citenamefont
  {{Strauss}}, \citenamefont {{Swanson}}, \citenamefont {{Thomas}},
  \citenamefont {{Tinker}}, \citenamefont {{Tojeiro}}, \citenamefont
  {{Maga{\~n}a}}, \citenamefont {{Vazquez}}, \citenamefont {{Verde}},
  \citenamefont {{Wake}}, \citenamefont {{Wang}}, \citenamefont {{Weinberg}},
  \citenamefont {{White}}, \citenamefont {{Wood-Vasey}}, \citenamefont
  {{Y{\`e}che}}, \citenamefont {{Zehavi}}, \citenamefont {{Zhai}},\ and\
  \citenamefont {{Zhao}}}]{Alam2017_DR12_cosmo_analysis}%
  \BibitemOpen
  \bibfield  {author} {\bibinfo {author} {S.~{Alam}}, \bibinfo {author}
  {M.~{Ata}}, \bibinfo {author} {S.~{Bailey}}, \bibinfo {author}
  {F.~{Beutler}}, \bibinfo {author} {D.~{Bizyaev}}, \bibinfo {author} {J.~A.
  {Blazek}}, \bibinfo {author} {A.~S. {Bolton}}, \bibinfo {author} {J.~R.
  {Brownstein}}, \bibinfo {author} {A.~{Burden}}, \bibinfo {author} {C.-H.
  {Chuang}}, et~al.,\ }\href {https://doi.org/10.1093/mnras/stx721} {\bibfield
  {journal} {\bibinfo  {journal} {\mnras}\ }\textbf {\bibinfo {volume} {470}},\
  \bibinfo {pages} {2617} (\bibinfo {year} {2017}{\natexlab{a}})},\ \Eprint
  {https://arxiv.org/abs/1607.03155} {arXiv:1607.03155 [astro-ph.CO]}
  \BibitemShut {NoStop}%
\bibitem [{\citenamefont {{Planck Collaboration}}\ \emph
  {et~al.}(2020{\natexlab{a}})\citenamefont {{Planck Collaboration}},
  \citenamefont {{Aghanim}}, \citenamefont {{Akrami}}, \citenamefont
  {{Ashdown}}, \citenamefont {{Aumont}}, \citenamefont {{Baccigalupi}},
  \citenamefont {{Ballardini}}, \citenamefont {{Banday}}, \citenamefont
  {{Barreiro}}, \citenamefont {{Bartolo}}, \citenamefont {{Basak}},
  \citenamefont {{Battye}}, \citenamefont {{Benabed}}, \citenamefont
  {{Bernard}}, \citenamefont {{Bersanelli}}, \citenamefont {{Bielewicz}},
  \citenamefont {{Bock}}, \citenamefont {{Bond}}, \citenamefont {{Borrill}},
  \citenamefont {{Bouchet}}, \citenamefont {{Boulanger}}, \citenamefont
  {{Bucher}}, \citenamefont {{Burigana}}, \citenamefont {{Butler}},
  \citenamefont {{Calabrese}}, \citenamefont {{Cardoso}}, \citenamefont
  {{Carron}}, \citenamefont {{Challinor}}, \citenamefont {{Chiang}},
  \citenamefont {{Chluba}}, \citenamefont {{Colombo}}, \citenamefont
  {{Combet}}, \citenamefont {{Contreras}}, \citenamefont {{Crill}},
  \citenamefont {{Cuttaia}}, \citenamefont {{de Bernardis}}, \citenamefont {{de
  Zotti}}, \citenamefont {{Delabrouille}}, \citenamefont {{Delouis}},
  \citenamefont {{Di Valentino}}, \citenamefont {{Diego}}, \citenamefont
  {{Dor{\'e}}}, \citenamefont {{Douspis}}, \citenamefont {{Ducout}},
  \citenamefont {{Dupac}}, \citenamefont {{Dusini}}, \citenamefont
  {{Efstathiou}}, \citenamefont {{Elsner}}, \citenamefont {{En{\ss}lin}},
  \citenamefont {{Eriksen}}, \citenamefont {{Fantaye}}, \citenamefont
  {{Farhang}}, \citenamefont {{Fergusson}}, \citenamefont {{Fernandez-Cobos}},
  \citenamefont {{Finelli}}, \citenamefont {{Forastieri}}, \citenamefont
  {{Frailis}}, \citenamefont {{Fraisse}}, \citenamefont {{Franceschi}},
  \citenamefont {{Frolov}}, \citenamefont {{Galeotta}}, \citenamefont
  {{Galli}}, \citenamefont {{Ganga}}, \citenamefont {{G{\'e}nova-Santos}},
  \citenamefont {{Gerbino}}, \citenamefont {{Ghosh}}, \citenamefont
  {{Gonz{\'a}lez-Nuevo}}, \citenamefont {{G{\'o}rski}}, \citenamefont
  {{Gratton}}, \citenamefont {{Gruppuso}}, \citenamefont {{Gudmundsson}},
  \citenamefont {{Hamann}}, \citenamefont {{Handley}}, \citenamefont
  {{Hansen}}, \citenamefont {{Herranz}}, \citenamefont {{Hildebrandt}},
  \citenamefont {{Hivon}}, \citenamefont {{Huang}}, \citenamefont {{Jaffe}},
  \citenamefont {{Jones}}, \citenamefont {{Karakci}}, \citenamefont
  {{Keih{\"a}nen}}, \citenamefont {{Keskitalo}}, \citenamefont {{Kiiveri}},
  \citenamefont {{Kim}}, \citenamefont {{Kisner}}, \citenamefont {{Knox}},
  \citenamefont {{Krachmalnicoff}}, \citenamefont {{Kunz}}, \citenamefont
  {{Kurki-Suonio}}, \citenamefont {{Lagache}}, \citenamefont {{Lamarre}},
  \citenamefont {{Lasenby}}, \citenamefont {{Lattanzi}}, \citenamefont
  {{Lawrence}}, \citenamefont {{Le Jeune}}, \citenamefont {{Lemos}},
  \citenamefont {{Lesgourgues}}, \citenamefont {{Levrier}}, \citenamefont
  {{Lewis}}, \citenamefont {{Liguori}}, \citenamefont {{Lilje}}, \citenamefont
  {{Lilley}}, \citenamefont {{Lindholm}}, \citenamefont {{L{\'o}pez-Caniego}},
  \citenamefont {{Lubin}}, \citenamefont {{Ma}}, \citenamefont
  {{Mac{\'\i}as-P{\'e}rez}}, \citenamefont {{Maggio}}, \citenamefont {{Maino}},
  \citenamefont {{Mandolesi}}, \citenamefont {{Mangilli}}, \citenamefont
  {{Marcos-Caballero}}, \citenamefont {{Maris}}, \citenamefont {{Martin}},
  \citenamefont {{Martinelli}}, \citenamefont {{Mart{\'\i}nez-Gonz{\'a}lez}},
  \citenamefont {{Matarrese}}, \citenamefont {{Mauri}}, \citenamefont
  {{McEwen}}, \citenamefont {{Meinhold}}, \citenamefont {{Melchiorri}},
  \citenamefont {{Mennella}}, \citenamefont {{Migliaccio}}, \citenamefont
  {{Millea}}, \citenamefont {{Mitra}}, \citenamefont {{Miville-Desch{\^e}nes}},
  \citenamefont {{Molinari}}, \citenamefont {{Montier}}, \citenamefont
  {{Morgante}}, \citenamefont {{Moss}}, \citenamefont {{Natoli}}, \citenamefont
  {{N{\o}rgaard-Nielsen}}, \citenamefont {{Pagano}}, \citenamefont
  {{Paoletti}}, \citenamefont {{Partridge}}, \citenamefont {{Patanchon}},
  \citenamefont {{Peiris}}, \citenamefont {{Perrotta}}, \citenamefont
  {{Pettorino}}, \citenamefont {{Piacentini}}, \citenamefont {{Polastri}},
  \citenamefont {{Polenta}}, \citenamefont {{Puget}}, \citenamefont {{Rachen}},
  \citenamefont {{Reinecke}}, \citenamefont {{Remazeilles}}, \citenamefont
  {{Renzi}}, \citenamefont {{Rocha}}, \citenamefont {{Rosset}}, \citenamefont
  {{Roudier}}, \citenamefont {{Rubi{\~n}o-Mart{\'\i}n}}, \citenamefont
  {{Ruiz-Granados}}, \citenamefont {{Salvati}}, \citenamefont {{Sandri}},
  \citenamefont {{Savelainen}}, \citenamefont {{Scott}}, \citenamefont
  {{Shellard}}, \citenamefont {{Sirignano}}, \citenamefont {{Sirri}},
  \citenamefont {{Spencer}}, \citenamefont {{Sunyaev}}, \citenamefont
  {{Suur-Uski}}, \citenamefont {{Tauber}}, \citenamefont {{Tavagnacco}},
  \citenamefont {{Tenti}}, \citenamefont {{Toffolatti}}, \citenamefont
  {{Tomasi}}, \citenamefont {{Trombetti}}, \citenamefont {{Valenziano}},
  \citenamefont {{Valiviita}}, \citenamefont {{Van Tent}}, \citenamefont
  {{Vibert}}, \citenamefont {{Vielva}}, \citenamefont {{Villa}}, \citenamefont
  {{Vittorio}}, \citenamefont {{Wandelt}}, \citenamefont {{Wehus}},
  \citenamefont {{White}}, \citenamefont {{White}}, \citenamefont {{Zacchei}},\
  and\ \citenamefont {{Zonca}}}]{PlanckCollaboration2018}%
  \BibitemOpen
  \bibfield  {author} {\bibinfo {author} {{Planck Collaboration}}, \bibinfo
  {author} {N.~{Aghanim}}, \bibinfo {author} {Y.~{Akrami}}, \bibinfo {author}
  {M.~{Ashdown}}, \bibinfo {author} {J.~{Aumont}}, \bibinfo {author}
  {C.~{Baccigalupi}}, \bibinfo {author} {M.~{Ballardini}}, \bibinfo {author}
  {A.~J. {Banday}}, \bibinfo {author} {R.~B. {Barreiro}}, \bibinfo {author}
  {N.~{Bartolo}}, et~al.,\ }\href {https://doi.org/10.1051/0004-6361/201833910}
  {\bibfield  {journal} {\bibinfo  {journal} {\aap}\ }\textbf {\bibinfo
  {volume} {641}},\ \bibinfo {eid} {A6} (\bibinfo {year}
  {2020}{\natexlab{a}})},\ \Eprint {https://arxiv.org/abs/1807.06209}
  {arXiv:1807.06209 [astro-ph.CO]} \BibitemShut {NoStop}%
\bibitem [{\citenamefont {{Freedman}}\ \emph {et~al.}(2019)\citenamefont
  {{Freedman}}, \citenamefont {{Madore}}, \citenamefont {{Hatt}}, \citenamefont
  {{Hoyt}}, \citenamefont {{Jang}}, \citenamefont {{Beaton}}, \citenamefont
  {{Burns}}, \citenamefont {{Lee}}, \citenamefont {{Monson}}, \citenamefont
  {{Neeley}}, \citenamefont {{Phillips}}, \citenamefont {{Rich}},\ and\
  \citenamefont {{Seibert}}}]{Freedman2019}%
  \BibitemOpen
  \bibfield  {author} {\bibinfo {author} {W.~L. {Freedman}}, \bibinfo {author}
  {B.~F. {Madore}}, \bibinfo {author} {D.~{Hatt}}, \bibinfo {author} {T.~J.
  {Hoyt}}, \bibinfo {author} {I.~S. {Jang}}, \bibinfo {author} {R.~L.
  {Beaton}}, \bibinfo {author} {C.~R. {Burns}}, \bibinfo {author} {M.~G.
  {Lee}}, \bibinfo {author} {A.~J. {Monson}}, \bibinfo {author} {J.~R.
  {Neeley}}, et~al.,\ }\href {https://doi.org/10.3847/1538-4357/ab2f73}
  {\bibfield  {journal} {\bibinfo  {journal} {\apj}\ }\textbf {\bibinfo
  {volume} {882}},\ \bibinfo {eid} {34} (\bibinfo {year} {2019})},\ \Eprint
  {https://arxiv.org/abs/1907.05922} {arXiv:1907.05922 [astro-ph.CO]}
  \BibitemShut {NoStop}%
\bibitem [{\citenamefont {{Madhavacheril}}\ \emph {et~al.}(2023)\citenamefont
  {{Madhavacheril}}, \citenamefont {{Qu}}, \citenamefont {{Sherwin}},
  \citenamefont {{MacCrann}}, \citenamefont {{Li}}, \citenamefont
  {{Abril-Cabezas}}, \citenamefont {{Ade}}, \citenamefont {{Aiola}},
  \citenamefont {{Alford}}, \citenamefont {{Amiri}}, \citenamefont {{Amodeo}},
  \citenamefont {{An}}, \citenamefont {{Atkins}}, \citenamefont {{Austermann}},
  \citenamefont {{Battaglia}}, \citenamefont {{Battistelli}}, \citenamefont
  {{Beall}}, \citenamefont {{Bean}}, \citenamefont {{Beringue}}, \citenamefont
  {{Bhandarkar}}, \citenamefont {{Biermann}}, \citenamefont {{Bolliet}},
  \citenamefont {{Bond}}, \citenamefont {{Cai}}, \citenamefont {{Calabrese}},
  \citenamefont {{Calafut}}, \citenamefont {{Capalbo}}, \citenamefont
  {{Carrero}}, \citenamefont {{Challinor}}, \citenamefont {{Chesmore}},
  \citenamefont {{Cho}}, \citenamefont {{Choi}}, \citenamefont {{Clark}},
  \citenamefont {{C{\'o}rdova Rosado}}, \citenamefont {{Cothard}},
  \citenamefont {{Coughlin}}, \citenamefont {{Coulton}}, \citenamefont
  {{Crowley}}, \citenamefont {{Dalal}}, \citenamefont {{Darwish}},
  \citenamefont {{Devlin}}, \citenamefont {{Dicker}}, \citenamefont {{Doze}},
  \citenamefont {{Duell}}, \citenamefont {{Duff}}, \citenamefont
  {{Duivenvoorden}}, \citenamefont {{Dunkley}}, \citenamefont {{D{\"u}nner}},
  \citenamefont {{Fanfani}}, \citenamefont {{Fankhanel}}, \citenamefont
  {{Farren}}, \citenamefont {{Ferraro}}, \citenamefont {{Freundt}},
  \citenamefont {{Fuzia}}, \citenamefont {{Gallardo}}, \citenamefont
  {{Garrido}}, \citenamefont {{Givans}}, \citenamefont {{Gluscevic}},
  \citenamefont {{Golec}}, \citenamefont {{Guan}}, \citenamefont {{Hall}},
  \citenamefont {{Halpern}}, \citenamefont {{Han}}, \citenamefont {{Harrison}},
  \citenamefont {{Hasselfield}}, \citenamefont {{Healy}}, \citenamefont
  {{Henderson}}, \citenamefont {{Hensley}}, \citenamefont
  {{Herv{\'\i}as-Caimapo}}, \citenamefont {{Hill}}, \citenamefont {{Hilton}},
  \citenamefont {{Hilton}}, \citenamefont {{Hincks}}, \citenamefont
  {{Hlo{\v{z}}ek}}, \citenamefont {{Ho}}, \citenamefont {{Huber}},
  \citenamefont {{Hubmayr}}, \citenamefont {{Huffenberger}}, \citenamefont
  {{Hughes}}, \citenamefont {{Irwin}}, \citenamefont {{Isopi}}, \citenamefont
  {{Jense}}, \citenamefont {{Keller}}, \citenamefont {{Kim}}, \citenamefont
  {{Knowles}}, \citenamefont {{Koopman}}, \citenamefont {{Kosowsky}},
  \citenamefont {{Kramer}}, \citenamefont {{Kusiak}}, \citenamefont {{La
  Posta}}, \citenamefont {{Lague}}, \citenamefont {{Lakey}}, \citenamefont
  {{Lee}}, \citenamefont {{Li}}, \citenamefont {{Limon}}, \citenamefont
  {{Lokken}}, \citenamefont {{Louis}}, \citenamefont {{Lungu}}, \citenamefont
  {{MacInnis}}, \citenamefont {{Maldonado}}, \citenamefont {{Maldonado}},
  \citenamefont {{Mallaby-Kay}}, \citenamefont {{Marques}}, \citenamefont
  {{McMahon}}, \citenamefont {{Mehta}}, \citenamefont {{Menanteau}},
  \citenamefont {{Moodley}}, \citenamefont {{Morris}}, \citenamefont
  {{Mroczkowski}}, \citenamefont {{Naess}}, \citenamefont {{Namikawa}},
  \citenamefont {{Nati}}, \citenamefont {{Newburgh}}, \citenamefont {{Nicola}},
  \citenamefont {{Niemack}}, \citenamefont {{Nolta}}, \citenamefont
  {{Orlowski-Scherer}}, \citenamefont {{Page}}, \citenamefont {{Pandey}},
  \citenamefont {{Partridge}}, \citenamefont {{Prince}}, \citenamefont
  {{Puddu}}, \citenamefont {{Radiconi}}, \citenamefont {{Robertson}},
  \citenamefont {{Rojas}}, \citenamefont {{Sakuma}}, \citenamefont
  {{Salatino}}, \citenamefont {{Schaan}}, \citenamefont {{Schmitt}},
  \citenamefont {{Sehgal}}, \citenamefont {{Shaikh}}, \citenamefont {{Sierra}},
  \citenamefont {{Sievers}}, \citenamefont {{Sif{\'o}n}}, \citenamefont
  {{Simon}}, \citenamefont {{Sonka}}, \citenamefont {{Spergel}}, \citenamefont
  {{Staggs}}, \citenamefont {{Storer}}, \citenamefont {{Switzer}},
  \citenamefont {{Tampier}}, \citenamefont {{Thornton}}, \citenamefont
  {{Trac}}, \citenamefont {{Treu}}, \citenamefont {{Tucker}}, \citenamefont
  {{Ulluom}}, \citenamefont {{Vale}}, \citenamefont {{Van Engelen}},
  \citenamefont {{Van Lanen}}, \citenamefont {{van Marrewijk}}, \citenamefont
  {{Vargas}}, \citenamefont {{Vavagiakis}}, \citenamefont {{Wagoner}},
  \citenamefont {{Wang}}, \citenamefont {{Wenzl}}, \citenamefont {{Wollack}},
  \citenamefont {{Xu}}, \citenamefont {{Zago}},\ and\ \citenamefont
  {{Zhang}}}]{Madhavacheril2023}%
  \BibitemOpen
  \bibfield  {author} {\bibinfo {author} {M.~S. {Madhavacheril}}, \bibinfo
  {author} {F.~J. {Qu}}, \bibinfo {author} {B.~D. {Sherwin}}, \bibinfo {author}
  {N.~{MacCrann}}, \bibinfo {author} {Y.~{Li}}, \bibinfo {author}
  {I.~{Abril-Cabezas}}, \bibinfo {author} {P.~A.~R. {Ade}}, \bibinfo {author}
  {S.~{Aiola}}, \bibinfo {author} {T.~{Alford}}, \bibinfo {author}
  {M.~{Amiri}}, et~al.,\ }\href {https://doi.org/10.48550/arXiv.2304.05203}
  {\bibfield  {journal} {\bibinfo  {journal} {arXiv e-prints}\ ,\ \bibinfo
  {eid} {arXiv:2304.05203}} (\bibinfo {year} {2023})},\ \Eprint
  {https://arxiv.org/abs/2304.05203} {arXiv:2304.05203 [astro-ph.CO]}
  \BibitemShut {NoStop}%
\bibitem [{\citenamefont {{Weinberg}}(1989)}]{Weinberg1989}%
  \BibitemOpen
  \bibfield  {author} {\bibinfo {author} {S.~{Weinberg}},\ }\href
  {https://doi.org/10.1103/RevModPhys.61.1} {\bibfield  {journal} {\bibinfo
  {journal} {Reviews of Modern Physics}\ }\textbf {\bibinfo {volume} {61}},\
  \bibinfo {pages} {1} (\bibinfo {year} {1989})}\BibitemShut {NoStop}%
\bibitem [{\citenamefont {{Carroll}}\ \emph {et~al.}(2004)\citenamefont
  {{Carroll}}, \citenamefont {{Duvvuri}}, \citenamefont {{Trodden}},\ and\
  \citenamefont {{Turner}}}]{Carroll2004}%
  \BibitemOpen
  \bibfield  {author} {\bibinfo {author} {S.~M. {Carroll}}, \bibinfo {author}
  {V.~{Duvvuri}}, \bibinfo {author} {M.~{Trodden}},\ and\ \bibinfo {author}
  {M.~S. {Turner}},\ }\href {https://doi.org/10.1103/PhysRevD.70.043528}
  {\bibfield  {journal} {\bibinfo  {journal} {\prd}\ }\textbf {\bibinfo
  {volume} {70}},\ \bibinfo {eid} {043528} (\bibinfo {year} {2004})},\ \Eprint
  {https://arxiv.org/abs/astro-ph/0306438} {arXiv:astro-ph/0306438 [astro-ph]}
  \BibitemShut {NoStop}%
\bibitem [{\citenamefont {{Khoury}}\ and\ \citenamefont
  {{Weltman}}(2004)}]{Khoury2004}%
  \BibitemOpen
  \bibfield  {author} {\bibinfo {author} {J.~{Khoury}}\ and\ \bibinfo {author}
  {A.~{Weltman}},\ }\href {https://doi.org/10.1103/PhysRevLett.93.171104}
  {\bibfield  {journal} {\bibinfo  {journal} {\prl}\ }\textbf {\bibinfo
  {volume} {93}},\ \bibinfo {eid} {171104} (\bibinfo {year} {2004})},\ \Eprint
  {https://arxiv.org/abs/astro-ph/0309300} {arXiv:astro-ph/0309300 [astro-ph]}
  \BibitemShut {NoStop}%
\bibitem [{\citenamefont {{Dvali}}\ \emph {et~al.}(2000)\citenamefont
  {{Dvali}}, \citenamefont {{Gabadadze}},\ and\ \citenamefont
  {{Porrati}}}]{Dvali2000}%
  \BibitemOpen
  \bibfield  {author} {\bibinfo {author} {G.~{Dvali}}, \bibinfo {author}
  {G.~{Gabadadze}},\ and\ \bibinfo {author} {M.~{Porrati}},\ }\href
  {https://doi.org/10.1016/S0370-2693(00)00669-9} {\bibfield  {journal}
  {\bibinfo  {journal} {Physics Letters B}\ }\textbf {\bibinfo {volume}
  {485}},\ \bibinfo {pages} {208} (\bibinfo {year} {2000})},\ \Eprint
  {https://arxiv.org/abs/hep-th/0005016} {arXiv:hep-th/0005016 [hep-th]}
  \BibitemShut {NoStop}%
\bibitem [{\citenamefont {{Bekenstein}}(2004)}]{Bekenstein2004}%
  \BibitemOpen
  \bibfield  {author} {\bibinfo {author} {J.~D. {Bekenstein}},\ }\href
  {https://doi.org/10.1103/PhysRevD.70.083509} {\bibfield  {journal} {\bibinfo
  {journal} {\prd}\ }\textbf {\bibinfo {volume} {70}},\ \bibinfo {eid} {083509}
  (\bibinfo {year} {2004})},\ \Eprint {https://arxiv.org/abs/astro-ph/0403694}
  {arXiv:astro-ph/0403694 [astro-ph]} \BibitemShut {NoStop}%
\bibitem [{\citenamefont {{Clifton}}\ \emph {et~al.}(2012)\citenamefont
  {{Clifton}}, \citenamefont {{Ferreira}}, \citenamefont {{Padilla}},\ and\
  \citenamefont {{Skordis}}}]{Clifton2012}%
  \BibitemOpen
  \bibfield  {author} {\bibinfo {author} {T.~{Clifton}}, \bibinfo {author}
  {P.~G. {Ferreira}}, \bibinfo {author} {A.~{Padilla}},\ and\ \bibinfo {author}
  {C.~{Skordis}},\ }\href {https://doi.org/10.1016/j.physrep.2012.01.001}
  {\bibfield  {journal} {\bibinfo  {journal} {\physrep}\ }\textbf {\bibinfo
  {volume} {513}},\ \bibinfo {pages} {1} (\bibinfo {year} {2012})},\ \Eprint
  {https://arxiv.org/abs/1106.2476} {arXiv:1106.2476 [astro-ph.CO]}
  \BibitemShut {NoStop}%
\bibitem [{\citenamefont {{Jain}}\ and\ \citenamefont
  {{Khoury}}(2010)}]{Jain2010}%
  \BibitemOpen
  \bibfield  {author} {\bibinfo {author} {B.~{Jain}}\ and\ \bibinfo {author}
  {J.~{Khoury}},\ }\href {https://doi.org/10.1016/j.aop.2010.04.002} {\bibfield
   {journal} {\bibinfo  {journal} {Annals of Physics}\ }\textbf {\bibinfo
  {volume} {325}},\ \bibinfo {pages} {1479} (\bibinfo {year} {2010})},\ \Eprint
  {https://arxiv.org/abs/1004.3294} {arXiv:1004.3294 [astro-ph.CO]}
  \BibitemShut {NoStop}%
\bibitem [{\citenamefont {{Huterer}}\ \emph {et~al.}(2015)\citenamefont
  {{Huterer}}, \citenamefont {{Kirkby}}, \citenamefont {{Bean}}, \citenamefont
  {{Connolly}}, \citenamefont {{Dawson}}, \citenamefont {{Dodelson}},
  \citenamefont {{Evrard}}, \citenamefont {{Jain}}, \citenamefont {{Jarvis}},
  \citenamefont {{Linder}}, \citenamefont {{Mandelbaum}}, \citenamefont
  {{May}}, \citenamefont {{Raccanelli}}, \citenamefont {{Reid}}, \citenamefont
  {{Rozo}}, \citenamefont {{Schmidt}}, \citenamefont {{Sehgal}}, \citenamefont
  {{Slosar}}, \citenamefont {{van Engelen}}, \citenamefont {{Wu}},\ and\
  \citenamefont {{Zhao}}}]{Huterer2015}%
  \BibitemOpen
  \bibfield  {author} {\bibinfo {author} {D.~{Huterer}}, \bibinfo {author}
  {D.~{Kirkby}}, \bibinfo {author} {R.~{Bean}}, \bibinfo {author}
  {A.~{Connolly}}, \bibinfo {author} {K.~{Dawson}}, \bibinfo {author}
  {S.~{Dodelson}}, \bibinfo {author} {A.~{Evrard}}, \bibinfo {author}
  {B.~{Jain}}, \bibinfo {author} {M.~{Jarvis}}, \bibinfo {author}
  {E.~{Linder}}, et~al.,\ }\href
  {https://doi.org/10.1016/j.astropartphys.2014.07.004} {\bibfield  {journal}
  {\bibinfo  {journal} {Astroparticle Physics}\ }\textbf {\bibinfo {volume}
  {63}},\ \bibinfo {pages} {23} (\bibinfo {year} {2015})},\ \Eprint
  {https://arxiv.org/abs/1309.5385} {arXiv:1309.5385 [astro-ph.CO]}
  \BibitemShut {NoStop}%
\bibitem [{\citenamefont {{Uzan}}(2010)}]{Uzan2010}%
  \BibitemOpen
  \bibfield  {author} {\bibinfo {author} {J.-P. {Uzan}},\ }\href
  {https://doi.org/10.1007/s10714-010-1047-8} {\bibfield  {journal} {\bibinfo
  {journal} {General Relativity and Gravitation}\ }\textbf {\bibinfo {volume}
  {42}},\ \bibinfo {pages} {2219} (\bibinfo {year} {2010})},\ \Eprint
  {https://arxiv.org/abs/0908.2243} {arXiv:0908.2243 [astro-ph.CO]}
  \BibitemShut {NoStop}%
\bibitem [{\citenamefont {{Jain}}\ and\ \citenamefont
  {{Zhang}}(2008)}]{Jain2008}%
  \BibitemOpen
  \bibfield  {author} {\bibinfo {author} {B.~{Jain}}\ and\ \bibinfo {author}
  {P.~{Zhang}},\ }\href {https://doi.org/10.1103/PhysRevD.78.063503} {\bibfield
   {journal} {\bibinfo  {journal} {\prd}\ }\textbf {\bibinfo {volume} {78}},\
  \bibinfo {eid} {063503} (\bibinfo {year} {2008})},\ \Eprint
  {https://arxiv.org/abs/0709.2375} {arXiv:0709.2375 [astro-ph]} \BibitemShut
  {NoStop}%
\bibitem [{\citenamefont {{Ishak}}(2019)}]{Ishak2019}%
  \BibitemOpen
  \bibfield  {author} {\bibinfo {author} {M.~{Ishak}},\ }\href
  {https://doi.org/10.1007/s41114-018-0017-4} {\bibfield  {journal} {\bibinfo
  {journal} {Living Reviews in Relativity}\ }\textbf {\bibinfo {volume} {22}},\
  \bibinfo {eid} {1} (\bibinfo {year} {2019})},\ \Eprint
  {https://arxiv.org/abs/1806.10122} {arXiv:1806.10122 [astro-ph.CO]}
  \BibitemShut {NoStop}%
\bibitem [{\citenamefont {{Hou}}\ \emph {et~al.}(2023)\citenamefont {{Hou}},
  \citenamefont {{Bautista}}, \citenamefont {{Berti}}, \citenamefont
  {{Cuesta-Lazaro}}, \citenamefont {{Hern{\'a}ndez-Aguayo}}, \citenamefont
  {{Tr{\"o}ster}},\ and\ \citenamefont {{Zheng}}}]{Hou2023}%
  \BibitemOpen
  \bibfield  {author} {\bibinfo {author} {J.~{Hou}}, \bibinfo {author}
  {J.~{Bautista}}, \bibinfo {author} {M.~{Berti}}, \bibinfo {author}
  {C.~{Cuesta-Lazaro}}, \bibinfo {author} {C.~{Hern{\'a}ndez-Aguayo}}, \bibinfo
  {author} {T.~{Tr{\"o}ster}},\ and\ \bibinfo {author} {J.~{Zheng}},\ }\href
  {https://doi.org/10.3390/universe9070302} {\bibfield  {journal} {\bibinfo
  {journal} {Universe}\ }\textbf {\bibinfo {volume} {9}},\ \bibinfo {eid} {302}
  (\bibinfo {year} {2023})},\ \Eprint {https://arxiv.org/abs/2306.13726}
  {arXiv:2306.13726 [gr-qc]} \BibitemShut {NoStop}%
\bibitem [{\citenamefont {{Zhang}}\ \emph {et~al.}(2007)\citenamefont
  {{Zhang}}, \citenamefont {{Liguori}}, \citenamefont {{Bean}},\ and\
  \citenamefont {{Dodelson}}}]{Zhang2007}%
  \BibitemOpen
  \bibfield  {author} {\bibinfo {author} {P.~{Zhang}}, \bibinfo {author}
  {M.~{Liguori}}, \bibinfo {author} {R.~{Bean}},\ and\ \bibinfo {author}
  {S.~{Dodelson}},\ }\href {https://doi.org/10.1103/PhysRevLett.99.141302}
  {\bibfield  {journal} {\bibinfo  {journal} {\prl}\ }\textbf {\bibinfo
  {volume} {99}},\ \bibinfo {eid} {141302} (\bibinfo {year} {2007})},\ \Eprint
  {https://arxiv.org/abs/0704.1932} {arXiv:0704.1932 [astro-ph]} \BibitemShut
  {NoStop}%
\bibitem [{\citenamefont {{Pullen}}\ \emph {et~al.}(2015)\citenamefont
  {{Pullen}}, \citenamefont {{Alam}},\ and\ \citenamefont {{Ho}}}]{Pullen2015}%
  \BibitemOpen
  \bibfield  {author} {\bibinfo {author} {A.~R. {Pullen}}, \bibinfo {author}
  {S.~{Alam}},\ and\ \bibinfo {author} {S.~{Ho}},\ }\href
  {https://doi.org/10.1093/mnras/stv554} {\bibfield  {journal} {\bibinfo
  {journal} {\mnras}\ }\textbf {\bibinfo {volume} {449}},\ \bibinfo {pages}
  {4326} (\bibinfo {year} {2015})},\ \Eprint {https://arxiv.org/abs/1412.4454}
  {arXiv:1412.4454 [astro-ph.CO]} \BibitemShut {NoStop}%
\bibitem [{\citenamefont {{Leonard}}\ \emph {et~al.}(2015)\citenamefont
  {{Leonard}}, \citenamefont {{Ferreira}},\ and\ \citenamefont
  {{Heymans}}}]{Leonard2015}%
  \BibitemOpen
  \bibfield  {author} {\bibinfo {author} {C.~D. {Leonard}}, \bibinfo {author}
  {P.~G. {Ferreira}},\ and\ \bibinfo {author} {C.~{Heymans}},\ }\href
  {https://doi.org/10.1088/1475-7516/2015/12/051} {\bibfield  {journal}
  {\bibinfo  {journal} {\jcap}\ }\textbf {\bibinfo {volume} {2015}},\ \bibinfo
  {pages} {051} (\bibinfo {year} {2015})},\ \Eprint
  {https://arxiv.org/abs/1510.04287} {arXiv:1510.04287 [astro-ph.CO]}
  \BibitemShut {NoStop}%
\bibitem [{\citenamefont {{Giannantonio}}\ \emph {et~al.}(2016)\citenamefont
  {{Giannantonio}}, \citenamefont {{Fosalba}}, \citenamefont {{Cawthon}},
  \citenamefont {{Omori}}, \citenamefont {{Crocce}}, \citenamefont {{Elsner}},
  \citenamefont {{Leistedt}}, \citenamefont {{Dodelson}}, \citenamefont
  {{Benoit-L{\'e}vy}}, \citenamefont {{Gazta{\~n}aga}}, \citenamefont
  {{Holder}}, \citenamefont {{Peiris}}, \citenamefont {{Percival}},
  \citenamefont {{Kirk}}, \citenamefont {{Bauer}}, \citenamefont {{Benson}},
  \citenamefont {{Bernstein}}, \citenamefont {{Carretero}}, \citenamefont
  {{Crawford}}, \citenamefont {{Crittenden}}, \citenamefont {{Huterer}},
  \citenamefont {{Jain}}, \citenamefont {{Krause}}, \citenamefont
  {{Reichardt}}, \citenamefont {{Ross}}, \citenamefont {{Simard}},
  \citenamefont {{Soergel}}, \citenamefont {{Stark}}, \citenamefont {{Story}},
  \citenamefont {{Vieira}}, \citenamefont {{Weller}}, \citenamefont {{Abbott}},
  \citenamefont {{Abdalla}}, \citenamefont {{Allam}}, \citenamefont
  {{Armstrong}}, \citenamefont {{Banerji}}, \citenamefont {{Bernstein}},
  \citenamefont {{Bertin}}, \citenamefont {{Brooks}}, \citenamefont
  {{Buckley-Geer}}, \citenamefont {{Burke}}, \citenamefont {{Capozzi}},
  \citenamefont {{Carlstrom}}, \citenamefont {{Carnero Rosell}}, \citenamefont
  {{Carrasco Kind}}, \citenamefont {{Castander}}, \citenamefont {{Chang}},
  \citenamefont {{Cunha}}, \citenamefont {{da Costa}}, \citenamefont
  {{D'Andrea}}, \citenamefont {{DePoy}}, \citenamefont {{Desai}}, \citenamefont
  {{Diehl}}, \citenamefont {{Dietrich}}, \citenamefont {{Doel}}, \citenamefont
  {{Eifler}}, \citenamefont {{Evrard}}, \citenamefont {{Neto}}, \citenamefont
  {{Fernandez}}, \citenamefont {{Finley}}, \citenamefont {{Flaugher}},
  \citenamefont {{Frieman}}, \citenamefont {{Gerdes}}, \citenamefont {{Gruen}},
  \citenamefont {{Gruendl}}, \citenamefont {{Gutierrez}}, \citenamefont
  {{Holzapfel}}, \citenamefont {{Honscheid}}, \citenamefont {{James}},
  \citenamefont {{Kuehn}}, \citenamefont {{Kuropatkin}}, \citenamefont
  {{Lahav}}, \citenamefont {{Li}}, \citenamefont {{Lima}}, \citenamefont
  {{March}}, \citenamefont {{Marshall}}, \citenamefont {{Martini}},
  \citenamefont {{Melchior}}, \citenamefont {{Miquel}}, \citenamefont {{Mohr}},
  \citenamefont {{Nichol}}, \citenamefont {{Nord}}, \citenamefont {{Ogando}},
  \citenamefont {{Plazas}}, \citenamefont {{Romer}}, \citenamefont {{Roodman}},
  \citenamefont {{Rykoff}}, \citenamefont {{Sako}}, \citenamefont
  {{Saliwanchik}}, \citenamefont {{Sanchez}}, \citenamefont {{Schubnell}},
  \citenamefont {{Sevilla-Noarbe}}, \citenamefont {{Smith}}, \citenamefont
  {{Soares-Santos}}, \citenamefont {{Sobreira}}, \citenamefont {{Suchyta}},
  \citenamefont {{Swanson}}, \citenamefont {{Tarle}}, \citenamefont {{Thaler}},
  \citenamefont {{Thomas}}, \citenamefont {{Vikram}}, \citenamefont {{Walker}},
  \citenamefont {{Wechsler}},\ and\ \citenamefont
  {{Zuntz}}}]{Giannantonio2016}%
  \BibitemOpen
  \bibfield  {author} {\bibinfo {author} {T.~{Giannantonio}}, \bibinfo {author}
  {P.~{Fosalba}}, \bibinfo {author} {R.~{Cawthon}}, \bibinfo {author}
  {Y.~{Omori}}, \bibinfo {author} {M.~{Crocce}}, \bibinfo {author}
  {F.~{Elsner}}, \bibinfo {author} {B.~{Leistedt}}, \bibinfo {author}
  {S.~{Dodelson}}, \bibinfo {author} {A.~{Benoit-L{\'e}vy}}, \bibinfo {author}
  {E.~{Gazta{\~n}aga}}, et~al.,\ }\href {https://doi.org/10.1093/mnras/stv2678}
  {\bibfield  {journal} {\bibinfo  {journal} {\mnras}\ }\textbf {\bibinfo
  {volume} {456}},\ \bibinfo {pages} {3213} (\bibinfo {year} {2016})},\ \Eprint
  {https://arxiv.org/abs/1507.05551} {arXiv:1507.05551 [astro-ph.CO]}
  \BibitemShut {NoStop}%
\bibitem [{\citenamefont {Bianchini}\ and\ \citenamefont
  {Reichardt}(2018)}]{Bianchini2018}%
  \BibitemOpen
  \bibfield  {author} {\bibinfo {author} {F.~Bianchini}\ and\ \bibinfo {author}
  {C.~L. Reichardt},\ }\href@noop {} {\bibfield  {journal} {\bibinfo  {journal}
  {The Astrophysical Journal}\ }\textbf {\bibinfo {volume} {862}},\ \bibinfo
  {pages} {81} (\bibinfo {year} {2018})}\BibitemShut {NoStop}%
\bibitem [{\citenamefont {Omori}\ \emph {et~al.}(2019)\citenamefont {Omori},
  \citenamefont {Giannantonio}, \citenamefont {Porredon}, \citenamefont
  {Baxter}, \citenamefont {Chang}, \citenamefont {Crocce}, \citenamefont
  {Fosalba}, \citenamefont {Alarcon}, \citenamefont {Banik}, \citenamefont
  {Blazek} \emph {et~al.}}]{Omori2019a}%
  \BibitemOpen
  \bibfield  {author} {\bibinfo {author} {Y.~Omori}, \bibinfo {author}
  {T.~Giannantonio}, \bibinfo {author} {A.~Porredon}, \bibinfo {author}
  {E.~Baxter}, \bibinfo {author} {C.~Chang}, \bibinfo {author} {M.~Crocce},
  \bibinfo {author} {P.~Fosalba}, \bibinfo {author} {A.~Alarcon}, \bibinfo
  {author} {N.~Banik}, \bibinfo {author} {J.~Blazek}, et~al.,\ }\href@noop {}
  {\bibfield  {journal} {\bibinfo  {journal} {Physical Review D}\ }\textbf
  {\bibinfo {volume} {100}},\ \bibinfo {pages} {043501} (\bibinfo {year}
  {2019})}\BibitemShut {NoStop}%
\bibitem [{\citenamefont {Marques}\ and\ \citenamefont
  {Bernui}(2020)}]{Marques2020a}%
  \BibitemOpen
  \bibfield  {author} {\bibinfo {author} {G.~A. Marques}\ and\ \bibinfo
  {author} {A.~Bernui},\ }\href@noop {} {\bibfield  {journal} {\bibinfo
  {journal} {Journal of Cosmology and Astroparticle Physics}\ }\textbf
  {\bibinfo {volume} {2020}}\bibinfo  {number} { (05)},\ \bibinfo {pages}
  {052}}\BibitemShut {NoStop}%
\bibitem [{\citenamefont {{Reyes}}\ \emph {et~al.}(2010)\citenamefont
  {{Reyes}}, \citenamefont {{Mandelbaum}}, \citenamefont {{Seljak}},
  \citenamefont {{Baldauf}}, \citenamefont {{Gunn}}, \citenamefont
  {{Lombriser}},\ and\ \citenamefont {{Smith}}}]{Reyes2010}%
  \BibitemOpen
\bibfield  {number} {  }\bibfield  {author} {\bibinfo {author} {R.~{Reyes}},
  \bibinfo {author} {R.~{Mandelbaum}}, \bibinfo {author} {U.~{Seljak}},
  \bibinfo {author} {T.~{Baldauf}}, \bibinfo {author} {J.~E. {Gunn}}, \bibinfo
  {author} {L.~{Lombriser}},\ and\ \bibinfo {author} {R.~E. {Smith}},\ }\href
  {https://doi.org/10.1038/nature08857} {\bibfield  {journal} {\bibinfo
  {journal} {\nat}\ }\textbf {\bibinfo {volume} {464}},\ \bibinfo {pages} {256}
  (\bibinfo {year} {2010})},\ \Eprint {https://arxiv.org/abs/1003.2185}
  {arXiv:1003.2185 [astro-ph.CO]} \BibitemShut {NoStop}%
\bibitem [{\citenamefont {{Blake}}\ \emph {et~al.}(2016)\citenamefont
  {{Blake}}, \citenamefont {{Joudaki}}, \citenamefont {{Heymans}},
  \citenamefont {{Choi}}, \citenamefont {{Erben}}, \citenamefont
  {{Harnois-Deraps}}, \citenamefont {{Hildebrandt}}, \citenamefont
  {{Joachimi}}, \citenamefont {{Nakajima}}, \citenamefont {{van Waerbeke}},\
  and\ \citenamefont {{Viola}}}]{Blake2016}%
  \BibitemOpen
  \bibfield  {author} {\bibinfo {author} {C.~{Blake}}, \bibinfo {author}
  {S.~{Joudaki}}, \bibinfo {author} {C.~{Heymans}}, \bibinfo {author}
  {A.~{Choi}}, \bibinfo {author} {T.~{Erben}}, \bibinfo {author}
  {J.~{Harnois-Deraps}}, \bibinfo {author} {H.~{Hildebrandt}}, \bibinfo
  {author} {B.~{Joachimi}}, \bibinfo {author} {R.~{Nakajima}}, \bibinfo
  {author} {L.~{van Waerbeke}}, et~al.,\ }\href
  {https://doi.org/10.1093/mnras/stv2875} {\bibfield  {journal} {\bibinfo
  {journal} {\mnras}\ }\textbf {\bibinfo {volume} {456}},\ \bibinfo {pages}
  {2806} (\bibinfo {year} {2016})},\ \Eprint {https://arxiv.org/abs/1507.03086}
  {arXiv:1507.03086 [astro-ph.CO]} \BibitemShut {NoStop}%
\bibitem [{\citenamefont {{de la Torre}}\ \emph {et~al.}(2017)\citenamefont
  {{de la Torre}}, \citenamefont {{Jullo}}, \citenamefont {{Giocoli}},
  \citenamefont {{Pezzotta}}, \citenamefont {{Bel}}, \citenamefont {{Granett}},
  \citenamefont {{Guzzo}}, \citenamefont {{Garilli}}, \citenamefont
  {{Scodeggio}}, \citenamefont {{Bolzonella}}, \citenamefont {{Abbas}},
  \citenamefont {{Adami}}, \citenamefont {{Bottini}}, \citenamefont {{Cappi}},
  \citenamefont {{Cucciati}}, \citenamefont {{Davidzon}}, \citenamefont
  {{Franzetti}}, \citenamefont {{Fritz}}, \citenamefont {{Iovino}},
  \citenamefont {{Krywult}}, \citenamefont {{Le Brun}}, \citenamefont {{Le
  F{\`e}vre}}, \citenamefont {{Maccagni}}, \citenamefont {{Ma{\l}ek}},
  \citenamefont {{Marulli}}, \citenamefont {{Polletta}}, \citenamefont
  {{Pollo}}, \citenamefont {{Tasca}}, \citenamefont {{Tojeiro}}, \citenamefont
  {{Vergani}}, \citenamefont {{Zanichelli}}, \citenamefont {{Arnouts}},
  \citenamefont {{Branchini}}, \citenamefont {{Coupon}}, \citenamefont {{De
  Lucia}}, \citenamefont {{Ilbert}}, \citenamefont {{Moutard}}, \citenamefont
  {{Moscardini}}, \citenamefont {{Peacock}}, \citenamefont {{Metcalf}},
  \citenamefont {{Prada}},\ and\ \citenamefont {{Yepes}}}]{delaTorre2017}%
  \BibitemOpen
  \bibfield  {author} {\bibinfo {author} {S.~{de la Torre}}, \bibinfo {author}
  {E.~{Jullo}}, \bibinfo {author} {C.~{Giocoli}}, \bibinfo {author}
  {A.~{Pezzotta}}, \bibinfo {author} {J.~{Bel}}, \bibinfo {author} {B.~R.
  {Granett}}, \bibinfo {author} {L.~{Guzzo}}, \bibinfo {author} {B.~{Garilli}},
  \bibinfo {author} {M.~{Scodeggio}}, \bibinfo {author} {M.~{Bolzonella}},
  et~al.,\ }\href {https://doi.org/10.1051/0004-6361/201630276} {\bibfield
  {journal} {\bibinfo  {journal} {\aap}\ }\textbf {\bibinfo {volume} {608}},\
  \bibinfo {eid} {A44} (\bibinfo {year} {2017})},\ \Eprint
  {https://arxiv.org/abs/1612.05647} {arXiv:1612.05647 [astro-ph.CO]}
  \BibitemShut {NoStop}%
\bibitem [{\citenamefont {{Alam}}\ \emph
  {et~al.}(2017{\natexlab{b}})\citenamefont {{Alam}}, \citenamefont
  {{Miyatake}}, \citenamefont {{More}}, \citenamefont {{Ho}},\ and\
  \citenamefont {{Mandelbaum}}}]{Alam2017}%
  \BibitemOpen
  \bibfield  {author} {\bibinfo {author} {S.~{Alam}}, \bibinfo {author}
  {H.~{Miyatake}}, \bibinfo {author} {S.~{More}}, \bibinfo {author} {S.~{Ho}},\
  and\ \bibinfo {author} {R.~{Mandelbaum}},\ }\href
  {https://doi.org/10.1093/mnras/stw3056} {\bibfield  {journal} {\bibinfo
  {journal} {\mnras}\ }\textbf {\bibinfo {volume} {465}},\ \bibinfo {pages}
  {4853} (\bibinfo {year} {2017}{\natexlab{b}})},\ \Eprint
  {https://arxiv.org/abs/1610.09410} {arXiv:1610.09410 [astro-ph.CO]}
  \BibitemShut {NoStop}%
\bibitem [{\citenamefont {{Amon}}\ \emph {et~al.}(2018)\citenamefont {{Amon}},
  \citenamefont {{Blake}}, \citenamefont {{Heymans}}, \citenamefont
  {{Leonard}}, \citenamefont {{Asgari}}, \citenamefont {{Bilicki}},
  \citenamefont {{Choi}}, \citenamefont {{Erben}}, \citenamefont
  {{Glazebrook}}, \citenamefont {{Harnois-D{\'e}raps}}, \citenamefont
  {{Hildebrandt}}, \citenamefont {{Hoekstra}}, \citenamefont {{Joachimi}},
  \citenamefont {{Joudaki}}, \citenamefont {{Kuijken}}, \citenamefont
  {{Lidman}}, \citenamefont {{Loveday}}, \citenamefont {{Parkinson}},
  \citenamefont {{Valentijn}},\ and\ \citenamefont {{Wolf}}}]{Amon2018}%
  \BibitemOpen
  \bibfield  {author} {\bibinfo {author} {A.~{Amon}}, \bibinfo {author}
  {C.~{Blake}}, \bibinfo {author} {C.~{Heymans}}, \bibinfo {author} {C.~D.
  {Leonard}}, \bibinfo {author} {M.~{Asgari}}, \bibinfo {author}
  {M.~{Bilicki}}, \bibinfo {author} {A.~{Choi}}, \bibinfo {author}
  {T.~{Erben}}, \bibinfo {author} {K.~{Glazebrook}}, \bibinfo {author}
  {J.~{Harnois-D{\'e}raps}}, et~al.,\ }\href
  {https://doi.org/10.1093/mnras/sty1624} {\bibfield  {journal} {\bibinfo
  {journal} {\mnras}\ }\textbf {\bibinfo {volume} {479}},\ \bibinfo {pages}
  {3422} (\bibinfo {year} {2018})},\ \Eprint {https://arxiv.org/abs/1711.10999}
  {arXiv:1711.10999 [astro-ph.CO]} \BibitemShut {NoStop}%
\bibitem [{\citenamefont {{Singh}}\ \emph {et~al.}(2019)\citenamefont
  {{Singh}}, \citenamefont {{Alam}}, \citenamefont {{Mandelbaum}},
  \citenamefont {{Seljak}}, \citenamefont {{Rodriguez-Torres}},\ and\
  \citenamefont {{Ho}}}]{Singh2019}%
  \BibitemOpen
  \bibfield  {author} {\bibinfo {author} {S.~{Singh}}, \bibinfo {author}
  {S.~{Alam}}, \bibinfo {author} {R.~{Mandelbaum}}, \bibinfo {author}
  {U.~{Seljak}}, \bibinfo {author} {S.~{Rodriguez-Torres}},\ and\ \bibinfo
  {author} {S.~{Ho}},\ }\href {https://doi.org/10.1093/mnras/sty2681}
  {\bibfield  {journal} {\bibinfo  {journal} {\mnras}\ }\textbf {\bibinfo
  {volume} {482}},\ \bibinfo {pages} {785} (\bibinfo {year} {2019})},\ \Eprint
  {https://arxiv.org/abs/1803.08915} {arXiv:1803.08915 [astro-ph.CO]}
  \BibitemShut {NoStop}%
\bibitem [{\citenamefont {{Blake}}\ \emph {et~al.}(2020)\citenamefont
  {{Blake}}, \citenamefont {{Amon}}, \citenamefont {{Asgari}}, \citenamefont
  {{Bilicki}}, \citenamefont {{Dvornik}}, \citenamefont {{Erben}},
  \citenamefont {{Giblin}}, \citenamefont {{Glazebrook}}, \citenamefont
  {{Heymans}}, \citenamefont {{Hildebrandt}}, \citenamefont {{Joachimi}},
  \citenamefont {{Joudaki}}, \citenamefont {{Kannawadi}}, \citenamefont
  {{Kuijken}}, \citenamefont {{Lidman}}, \citenamefont {{Parkinson}},
  \citenamefont {{Shan}}, \citenamefont {{Tr{\"o}ster}}, \citenamefont {{van
  den Busch}}, \citenamefont {{Wolf}},\ and\ \citenamefont
  {{Wright}}}]{Blake2020}%
  \BibitemOpen
  \bibfield  {author} {\bibinfo {author} {C.~{Blake}}, \bibinfo {author}
  {A.~{Amon}}, \bibinfo {author} {M.~{Asgari}}, \bibinfo {author}
  {M.~{Bilicki}}, \bibinfo {author} {A.~{Dvornik}}, \bibinfo {author}
  {T.~{Erben}}, \bibinfo {author} {B.~{Giblin}}, \bibinfo {author}
  {K.~{Glazebrook}}, \bibinfo {author} {C.~{Heymans}}, \bibinfo {author}
  {H.~{Hildebrandt}}, et~al.,\ }\href
  {https://doi.org/10.1051/0004-6361/202038505} {\bibfield  {journal} {\bibinfo
   {journal} {\aap}\ }\textbf {\bibinfo {volume} {642}},\ \bibinfo {eid} {A158}
  (\bibinfo {year} {2020})},\ \Eprint {https://arxiv.org/abs/2005.14351}
  {arXiv:2005.14351 [astro-ph.CO]} \BibitemShut {NoStop}%
\bibitem [{\citenamefont {{Pullen}}\ \emph {et~al.}(2016)\citenamefont
  {{Pullen}}, \citenamefont {{Alam}}, \citenamefont {{He}},\ and\ \citenamefont
  {{Ho}}}]{Pullen2016}%
  \BibitemOpen
  \bibfield  {author} {\bibinfo {author} {A.~R. {Pullen}}, \bibinfo {author}
  {S.~{Alam}}, \bibinfo {author} {S.~{He}},\ and\ \bibinfo {author} {S.~{Ho}},\
  }\href {https://doi.org/10.1093/mnras/stw1249} {\bibfield  {journal}
  {\bibinfo  {journal} {\mnras}\ }\textbf {\bibinfo {volume} {460}},\ \bibinfo
  {pages} {4098} (\bibinfo {year} {2016})},\ \Eprint
  {https://arxiv.org/abs/1511.04457} {arXiv:1511.04457 [astro-ph.CO]}
  \BibitemShut {NoStop}%
\bibitem [{\citenamefont {{Zhang}}\ \emph {et~al.}(2021)\citenamefont
  {{Zhang}}, \citenamefont {{Pullen}}, \citenamefont {{Alam}}, \citenamefont
  {{Singh}}, \citenamefont {{Burtin}}, \citenamefont {{Chuang}}, \citenamefont
  {{Hou}}, \citenamefont {{Lyke}}, \citenamefont {{Myers}}, \citenamefont
  {{Neveux}}, \citenamefont {{Ross}}, \citenamefont {{Rossi}},\ and\
  \citenamefont {{Zhao}}}]{Zhang2021}%
  \BibitemOpen
  \bibfield  {author} {\bibinfo {author} {Y.~{Zhang}}, \bibinfo {author} {A.~R.
  {Pullen}}, \bibinfo {author} {S.~{Alam}}, \bibinfo {author} {S.~{Singh}},
  \bibinfo {author} {E.~{Burtin}}, \bibinfo {author} {C.-H. {Chuang}}, \bibinfo
  {author} {J.~{Hou}}, \bibinfo {author} {B.~W. {Lyke}}, \bibinfo {author}
  {A.~D. {Myers}}, \bibinfo {author} {R.~{Neveux}}, et~al.,\ }\href
  {https://doi.org/10.1093/mnras/staa3672} {\bibfield  {journal} {\bibinfo
  {journal} {\mnras}\ }\textbf {\bibinfo {volume} {501}},\ \bibinfo {pages}
  {1013} (\bibinfo {year} {2021})},\ \Eprint {https://arxiv.org/abs/2007.12607}
  {arXiv:2007.12607 [astro-ph.CO]} \BibitemShut {NoStop}%
\bibitem [{\citenamefont {{Qu}}\ \emph {et~al.}(2023)\citenamefont {{Qu}},
  \citenamefont {{Sherwin}}, \citenamefont {{Madhavacheril}}, \citenamefont
  {{Han}}, \citenamefont {{Crowley}}, \citenamefont {{Abril-Cabezas}},
  \citenamefont {{Ade}}, \citenamefont {{Aiola}}, \citenamefont {{Alford}},
  \citenamefont {{Amiri}}, \citenamefont {{Amodeo}}, \citenamefont {{An}},
  \citenamefont {{Atkins}}, \citenamefont {{Austermann}}, \citenamefont
  {{Battaglia}}, \citenamefont {{Battistelli}}, \citenamefont {{Beall}},
  \citenamefont {{Bean}}, \citenamefont {{Beringue}}, \citenamefont
  {{Bhandarkar}}, \citenamefont {{Biermann}}, \citenamefont {{Bolliet}},
  \citenamefont {{Bond}}, \citenamefont {{Cai}}, \citenamefont {{Calabrese}},
  \citenamefont {{Calafut}}, \citenamefont {{Capalbo}}, \citenamefont
  {{Carrero}}, \citenamefont {{Carron}}, \citenamefont {{Challinor}},
  \citenamefont {{Chesmore}}, \citenamefont {{Cho}}, \citenamefont {{Choi}},
  \citenamefont {{Clark}}, \citenamefont {{C{\'o}rdova Rosado}}, \citenamefont
  {{Cothard}}, \citenamefont {{Coughlin}}, \citenamefont {{Coulton}},
  \citenamefont {{Dalal}}, \citenamefont {{Darwish}}, \citenamefont {{Devlin}},
  \citenamefont {{Dicker}}, \citenamefont {{Doze}}, \citenamefont {{Duell}},
  \citenamefont {{Duff}}, \citenamefont {{Duivenvoorden}}, \citenamefont
  {{Dunkley}}, \citenamefont {{D{\"u}nner}}, \citenamefont {{Fanfani}},
  \citenamefont {{Fankhanel}}, \citenamefont {{Farren}}, \citenamefont
  {{Ferraro}}, \citenamefont {{Freundt}}, \citenamefont {{Fuzia}},
  \citenamefont {{Gallardo}}, \citenamefont {{Garrido}}, \citenamefont
  {{Gluscevic}}, \citenamefont {{Golec}}, \citenamefont {{Guan}}, \citenamefont
  {{Halpern}}, \citenamefont {{Harrison}}, \citenamefont {{Hasselfield}},
  \citenamefont {{Healy}}, \citenamefont {{Henderson}}, \citenamefont
  {{Hensley}}, \citenamefont {{Herv{\'\i}as-Caimapo}}, \citenamefont {{Hill}},
  \citenamefont {{Hilton}}, \citenamefont {{Hilton}}, \citenamefont {{Hincks}},
  \citenamefont {{Hlo{\v{z}}ek}}, \citenamefont {{Ho}}, \citenamefont
  {{Huber}}, \citenamefont {{Hubmayr}}, \citenamefont {{Huffenberger}},
  \citenamefont {{Hughes}}, \citenamefont {{Irwin}}, \citenamefont {{Isopi}},
  \citenamefont {{Jense}}, \citenamefont {{Keller}}, \citenamefont {{Kim}},
  \citenamefont {{Knowles}}, \citenamefont {{Koopman}}, \citenamefont
  {{Kosowsky}}, \citenamefont {{Kramer}}, \citenamefont {{Kusiak}},
  \citenamefont {{La Posta}}, \citenamefont {{Lague}}, \citenamefont {{Lakey}},
  \citenamefont {{Lee}}, \citenamefont {{Li}}, \citenamefont {{Li}},
  \citenamefont {{Limon}}, \citenamefont {{Lokken}}, \citenamefont {{Louis}},
  \citenamefont {{Lungu}}, \citenamefont {{MacCrann}}, \citenamefont
  {{MacInnis}}, \citenamefont {{Maldonado}}, \citenamefont {{Maldonado}},
  \citenamefont {{Mallaby-Kay}}, \citenamefont {{Marques}}, \citenamefont
  {{McMahon}}, \citenamefont {{Mehta}}, \citenamefont {{Menanteau}},
  \citenamefont {{Moodley}}, \citenamefont {{Morris}}, \citenamefont
  {{Mroczkowski}}, \citenamefont {{Naess}}, \citenamefont {{Namikawa}},
  \citenamefont {{Nati}}, \citenamefont {{Newburgh}}, \citenamefont {{Nicola}},
  \citenamefont {{Niemack}}, \citenamefont {{Nolta}}, \citenamefont
  {{Orlowski-Scherer}}, \citenamefont {{Page}}, \citenamefont {{Pandey}},
  \citenamefont {{Partridge}}, \citenamefont {{Prince}}, \citenamefont
  {{Puddu}}, \citenamefont {{Radiconi}}, \citenamefont {{Robertson}},
  \citenamefont {{Rojas}}, \citenamefont {{Sakuma}}, \citenamefont
  {{Salatino}}, \citenamefont {{Schaan}}, \citenamefont {{Schmitt}},
  \citenamefont {{Sehgal}}, \citenamefont {{Shaikh}}, \citenamefont {{Sierra}},
  \citenamefont {{Sievers}}, \citenamefont {{Sif{\'o}n}}, \citenamefont
  {{Simon}}, \citenamefont {{Sonka}}, \citenamefont {{Spergel}}, \citenamefont
  {{Staggs}}, \citenamefont {{Storer}}, \citenamefont {{Switzer}},
  \citenamefont {{Tampier}}, \citenamefont {{Thornton}}, \citenamefont
  {{Trac}}, \citenamefont {{Treu}}, \citenamefont {{Tucker}}, \citenamefont
  {{Ulluom}}, \citenamefont {{Vale}}, \citenamefont {{Van Engelen}},
  \citenamefont {{Van Lanen}}, \citenamefont {{van Marrewijk}}, \citenamefont
  {{Vargas}}, \citenamefont {{Vavagiakis}}, \citenamefont {{Wagoner}},
  \citenamefont {{Wang}}, \citenamefont {{Wenzl}}, \citenamefont {{Wollack}},
  \citenamefont {{Xu}}, \citenamefont {{Zago}},\ and\ \citenamefont
  {{Zhang}}}]{Qu2023}%
  \BibitemOpen
  \bibfield  {author} {\bibinfo {author} {F.~J. {Qu}}, \bibinfo {author} {B.~D.
  {Sherwin}}, \bibinfo {author} {M.~S. {Madhavacheril}}, \bibinfo {author}
  {D.~{Han}}, \bibinfo {author} {K.~T. {Crowley}}, \bibinfo {author}
  {I.~{Abril-Cabezas}}, \bibinfo {author} {P.~A.~R. {Ade}}, \bibinfo {author}
  {S.~{Aiola}}, \bibinfo {author} {T.~{Alford}}, \bibinfo {author}
  {M.~{Amiri}}, et~al.,\ }\href {https://doi.org/10.48550/arXiv.2304.05202}
  {\bibfield  {journal} {\bibinfo  {journal} {arXiv e-prints}\ ,\ \bibinfo
  {eid} {arXiv:2304.05202}} (\bibinfo {year} {2023})},\ \Eprint
  {https://arxiv.org/abs/2304.05202} {arXiv:2304.05202 [astro-ph.CO]}
  \BibitemShut {NoStop}%
\bibitem [{\citenamefont {{Story}}\ \emph {et~al.}(2015)\citenamefont
  {{Story}}, \citenamefont {{Hanson}}, \citenamefont {{Ade}}, \citenamefont
  {{Aird}}, \citenamefont {{Austermann}}, \citenamefont {{Beall}},
  \citenamefont {{Bender}}, \citenamefont {{Benson}}, \citenamefont {{Bleem}},
  \citenamefont {{Carlstrom}}, \citenamefont {{Chang}}, \citenamefont
  {{Chiang}}, \citenamefont {{Cho}}, \citenamefont {{Citron}}, \citenamefont
  {{Crawford}}, \citenamefont {{Crites}}, \citenamefont {{de Haan}},
  \citenamefont {{Dobbs}}, \citenamefont {{Everett}}, \citenamefont
  {{Gallicchio}}, \citenamefont {{Gao}}, \citenamefont {{George}},
  \citenamefont {{Gilbert}}, \citenamefont {{Halverson}}, \citenamefont
  {{Harrington}}, \citenamefont {{Henning}}, \citenamefont {{Hilton}},
  \citenamefont {{Holder}}, \citenamefont {{Holzapfel}}, \citenamefont
  {{Hoover}}, \citenamefont {{Hou}}, \citenamefont {{Hrubes}}, \citenamefont
  {{Huang}}, \citenamefont {{Hubmayr}}, \citenamefont {{Irwin}}, \citenamefont
  {{Keisler}}, \citenamefont {{Knox}}, \citenamefont {{Lee}}, \citenamefont
  {{Leitch}}, \citenamefont {{Li}}, \citenamefont {{Liang}}, \citenamefont
  {{Luong-Van}}, \citenamefont {{McMahon}}, \citenamefont {{Mehl}},
  \citenamefont {{Meyer}}, \citenamefont {{Mocanu}}, \citenamefont {{Montroy}},
  \citenamefont {{Natoli}}, \citenamefont {{Nibarger}}, \citenamefont
  {{Novosad}}, \citenamefont {{Padin}}, \citenamefont {{Pryke}}, \citenamefont
  {{Reichardt}}, \citenamefont {{Ruhl}}, \citenamefont {{Saliwanchik}},
  \citenamefont {{Sayre}}, \citenamefont {{Schaffer}}, \citenamefont
  {{Smecher}}, \citenamefont {{Stark}}, \citenamefont {{Tucker}}, \citenamefont
  {{Vand erlinde}}, \citenamefont {{Vieira}}, \citenamefont {{Wang}},
  \citenamefont {{Whitehorn}}, \citenamefont {{Yefremenko}},\ and\
  \citenamefont {{Zahn}}}]{Story2015}%
  \BibitemOpen
  \bibfield  {author} {\bibinfo {author} {K.~T. {Story}}, \bibinfo {author}
  {D.~{Hanson}}, \bibinfo {author} {P.~A.~R. {Ade}}, \bibinfo {author} {K.~A.
  {Aird}}, \bibinfo {author} {J.~E. {Austermann}}, \bibinfo {author} {J.~A.
  {Beall}}, \bibinfo {author} {A.~N. {Bender}}, \bibinfo {author} {B.~A.
  {Benson}}, \bibinfo {author} {L.~E. {Bleem}}, \bibinfo {author} {J.~E.
  {Carlstrom}}, et~al.,\ }\href {https://doi.org/10.1088/0004-637X/810/1/50}
  {\bibfield  {journal} {\bibinfo  {journal} {\apj}\ }\textbf {\bibinfo
  {volume} {810}},\ \bibinfo {eid} {50} (\bibinfo {year} {2015})},\ \Eprint
  {https://arxiv.org/abs/1412.4760} {arXiv:1412.4760 [astro-ph.CO]}
  \BibitemShut {NoStop}%
\bibitem [{\citenamefont {{Pan}}\ \emph {et~al.}(2023)\citenamefont {{Pan}},
  \citenamefont {{Bianchini}}, \citenamefont {{Wu}}, \citenamefont {{Ade}},
  \citenamefont {{Ahmed}}, \citenamefont {{Anderes}}, \citenamefont
  {{Anderson}}, \citenamefont {{Ansarinejad}}, \citenamefont {{Archipley}},
  \citenamefont {{Aylor}}, \citenamefont {{Balkenhol}}, \citenamefont
  {{Barry}}, \citenamefont {{Basu Thakur}}, \citenamefont {{Benabed}},
  \citenamefont {{Bender}}, \citenamefont {{Benson}}, \citenamefont {{Bleem}},
  \citenamefont {{Bouchet}}, \citenamefont {{Bryant}}, \citenamefont {{Byrum}},
  \citenamefont {{Camphuis}}, \citenamefont {{Carlstrom}}, \citenamefont
  {{Carter}}, \citenamefont {{Cecil}}, \citenamefont {{Chang}}, \citenamefont
  {{Chaubal}}, \citenamefont {{Chen}}, \citenamefont {{Chichura}},
  \citenamefont {{Cho}}, \citenamefont {{Chou}}, \citenamefont {{Cliche}},
  \citenamefont {{Coerver}}, \citenamefont {{Crawford}}, \citenamefont
  {{Cukierman}}, \citenamefont {{Daley}}, \citenamefont {{de Haan}},
  \citenamefont {{Denison}}, \citenamefont {{Dibert}}, \citenamefont {{Ding}},
  \citenamefont {{Dobbs}}, \citenamefont {{Doussot}}, \citenamefont
  {{Dutcher}}, \citenamefont {{Everett}}, \citenamefont {{Feng}}, \citenamefont
  {{Ferguson}}, \citenamefont {{Fichman}}, \citenamefont {{Foster}},
  \citenamefont {{Fu}}, \citenamefont {{Galli}}, \citenamefont {{Gambrel}},
  \citenamefont {{Gardner}}, \citenamefont {{Ge}}, \citenamefont
  {{Goeckner-Wald}}, \citenamefont {{Gualtieri}}, \citenamefont {{Guidi}},
  \citenamefont {{Guns}}, \citenamefont {{Gupta}}, \citenamefont {{Halverson}},
  \citenamefont {{Harke-Hosemann}}, \citenamefont {{Harrington}}, \citenamefont
  {{Henning}}, \citenamefont {{Hilton}}, \citenamefont {{Hivon}}, \citenamefont
  {{Holder}}, \citenamefont {{Holzapfel}}, \citenamefont {{Hood}},
  \citenamefont {{Howe}}, \citenamefont {{Huang}}, \citenamefont {{Irwin}},
  \citenamefont {{Jeong}}, \citenamefont {{Jonas}}, \citenamefont {{Jones}},
  \citenamefont {{K{\'e}ruzor{\'e}}}, \citenamefont {{Khaire}}, \citenamefont
  {{Knox}}, \citenamefont {{Kofman}}, \citenamefont {{Korman}}, \citenamefont
  {{Kubik}}, \citenamefont {{Kuhlmann}}, \citenamefont {{Kuo}}, \citenamefont
  {{Lee}}, \citenamefont {{Leitch}}, \citenamefont {{Levy}}, \citenamefont
  {{Lowitz}}, \citenamefont {{Lu}}, \citenamefont {{Maniyar}}, \citenamefont
  {{Menanteau}}, \citenamefont {{Meyer}}, \citenamefont {{Michalik}},
  \citenamefont {{Millea}}, \citenamefont {{Montgomery}}, \citenamefont
  {{Nadolski}}, \citenamefont {{Nakato}}, \citenamefont {{Natoli}},
  \citenamefont {{Nguyen}}, \citenamefont {{Noble}}, \citenamefont {{Novosad}},
  \citenamefont {{Omori}}, \citenamefont {{Padin}}, \citenamefont {{Paschos}},
  \citenamefont {{Pearson}}, \citenamefont {{Posada}}, \citenamefont
  {{Prabhu}}, \citenamefont {{Quan}}, \citenamefont {{Raghunathan}},
  \citenamefont {{Rahimi}}, \citenamefont {{Rahlin}}, \citenamefont
  {{Reichardt}}, \citenamefont {{Riebel}}, \citenamefont {{Riedel}},
  \citenamefont {{Ruhl}}, \citenamefont {{Sayre}}, \citenamefont
  {{Schiappucci}}, \citenamefont {{Shirokoff}}, \citenamefont {{Smecher}},
  \citenamefont {{Sobrin}}, \citenamefont {{Stark}}, \citenamefont {{Stephen}},
  \citenamefont {{Story}}, \citenamefont {{Suzuki}}, \citenamefont
  {{Takakura}}, \citenamefont {{Tandoi}}, \citenamefont {{Thompson}},
  \citenamefont {{Thorne}}, \citenamefont {{Trendafilova}}, \citenamefont
  {{Tucker}}, \citenamefont {{Umilta}}, \citenamefont {{Vale}}, \citenamefont
  {{Vanderlinde}}, \citenamefont {{Vieira}}, \citenamefont {{Wang}},
  \citenamefont {{Whitehorn}}, \citenamefont {{Yefremenko}}, \citenamefont
  {{Yoon}}, \citenamefont {{Young}},\ and\ \citenamefont
  {{Zebrowski}}}]{Pan2023}%
  \BibitemOpen
  \bibfield  {author} {\bibinfo {author} {Z.~{Pan}}, \bibinfo {author}
  {F.~{Bianchini}}, \bibinfo {author} {W.~L.~K. {Wu}}, \bibinfo {author}
  {P.~A.~R. {Ade}}, \bibinfo {author} {Z.~{Ahmed}}, \bibinfo {author}
  {E.~{Anderes}}, \bibinfo {author} {A.~J. {Anderson}}, \bibinfo {author}
  {B.~{Ansarinejad}}, \bibinfo {author} {M.~{Archipley}}, \bibinfo {author}
  {K.~{Aylor}}, et~al.,\ }\href {https://doi.org/10.1103/PhysRevD.108.122005}
  {\bibfield  {journal} {\bibinfo  {journal} {\prd}\ }\textbf {\bibinfo
  {volume} {108}},\ \bibinfo {eid} {122005} (\bibinfo {year} {2023})},\ \Eprint
  {https://arxiv.org/abs/2308.11608} {arXiv:2308.11608 [astro-ph.CO]}
  \BibitemShut {NoStop}%
\bibitem [{\citenamefont {{Ade}}\ \emph {et~al.}(2019)\citenamefont {{Ade}},
  \citenamefont {{Aguirre}}, \citenamefont {{Ahmed}}, \citenamefont {{Aiola}},
  \citenamefont {{Ali}}, \citenamefont {{Alonso}}, \citenamefont {{Alvarez}},
  \citenamefont {{Arnold}}, \citenamefont {{Ashton}}, \citenamefont
  {{Austermann}}, \citenamefont {{Awan}}, \citenamefont {{Baccigalupi}},
  \citenamefont {{Baildon}}, \citenamefont {{Barron}}, \citenamefont
  {{Battaglia}}, \citenamefont {{Battye}}, \citenamefont {{Baxter}},
  \citenamefont {{Bazarko}}, \citenamefont {{Beall}}, \citenamefont {{Bean}},
  \citenamefont {{Beck}}, \citenamefont {{Beckman}}, \citenamefont
  {{Beringue}}, \citenamefont {{Bianchini}}, \citenamefont {{Boada}},
  \citenamefont {{Boettger}}, \citenamefont {{Bond}}, \citenamefont
  {{Borrill}}, \citenamefont {{Brown}}, \citenamefont {{Bruno}}, \citenamefont
  {{Bryan}}, \citenamefont {{Calabrese}}, \citenamefont {{Calafut}},
  \citenamefont {{Calisse}}, \citenamefont {{Carron}}, \citenamefont
  {{Challinor}}, \citenamefont {{Chesmore}}, \citenamefont {{Chinone}},
  \citenamefont {{Chluba}}, \citenamefont {{Cho}}, \citenamefont {{Choi}},
  \citenamefont {{Coppi}}, \citenamefont {{Cothard}}, \citenamefont
  {{Coughlin}}, \citenamefont {{Crichton}}, \citenamefont {{Crowley}},
  \citenamefont {{Crowley}}, \citenamefont {{Cukierman}}, \citenamefont
  {{D'Ewart}}, \citenamefont {{D{\"u}nner}}, \citenamefont {{de Haan}},
  \citenamefont {{Devlin}}, \citenamefont {{Dicker}}, \citenamefont {{Didier}},
  \citenamefont {{Dobbs}}, \citenamefont {{Dober}}, \citenamefont {{Duell}},
  \citenamefont {{Duff}}, \citenamefont {{Duivenvoorden}}, \citenamefont
  {{Dunkley}}, \citenamefont {{Dusatko}}, \citenamefont {{Errard}},
  \citenamefont {{Fabbian}}, \citenamefont {{Feeney}}, \citenamefont
  {{Ferraro}}, \citenamefont {{Flux{\`a}}}, \citenamefont {{Freese}},
  \citenamefont {{Frisch}}, \citenamefont {{Frolov}}, \citenamefont {{Fuller}},
  \citenamefont {{Fuzia}}, \citenamefont {{Galitzki}}, \citenamefont
  {{Gallardo}}, \citenamefont {{Tomas Galvez Ghersi}}, \citenamefont {{Gao}},
  \citenamefont {{Gawiser}}, \citenamefont {{Gerbino}}, \citenamefont
  {{Gluscevic}}, \citenamefont {{Goeckner-Wald}}, \citenamefont {{Golec}},
  \citenamefont {{Gordon}}, \citenamefont {{Gralla}}, \citenamefont {{Green}},
  \citenamefont {{Grigorian}}, \citenamefont {{Groh}}, \citenamefont
  {{Groppi}}, \citenamefont {{Guan}}, \citenamefont {{Gudmundsson}},
  \citenamefont {{Han}}, \citenamefont {{Hargrave}}, \citenamefont
  {{Hasegawa}}, \citenamefont {{Hasselfield}}, \citenamefont {{Hattori}},
  \citenamefont {{Haynes}}, \citenamefont {{Hazumi}}, \citenamefont {{He}},
  \citenamefont {{Healy}}, \citenamefont {{Henderson}}, \citenamefont
  {{Hervias-Caimapo}}, \citenamefont {{Hill}}, \citenamefont {{Hill}},
  \citenamefont {{Hilton}}, \citenamefont {{Hilton}}, \citenamefont {{Hincks}},
  \citenamefont {{Hinshaw}}, \citenamefont {{Hlo{\v{z}}ek}}, \citenamefont
  {{Ho}}, \citenamefont {{Ho}}, \citenamefont {{Howe}}, \citenamefont
  {{Huang}}, \citenamefont {{Hubmayr}}, \citenamefont {{Huffenberger}},
  \citenamefont {{Hughes}}, \citenamefont {{Ijjas}}, \citenamefont {{Ikape}},
  \citenamefont {{Irwin}}, \citenamefont {{Jaffe}}, \citenamefont {{Jain}},
  \citenamefont {{Jeong}}, \citenamefont {{Kaneko}}, \citenamefont {{Karpel}},
  \citenamefont {{Katayama}}, \citenamefont {{Keating}}, \citenamefont
  {{Kernasovskiy}}, \citenamefont {{Keskitalo}}, \citenamefont {{Kisner}},
  \citenamefont {{Kiuchi}}, \citenamefont {{Klein}}, \citenamefont {{Knowles}},
  \citenamefont {{Koopman}}, \citenamefont {{Kosowsky}}, \citenamefont
  {{Krachmalnicoff}}, \citenamefont {{Kuenstner}}, \citenamefont {{Kuo}},
  \citenamefont {{Kusaka}}, \citenamefont {{Lashner}}, \citenamefont {{Lee}},
  \citenamefont {{Lee}}, \citenamefont {{Leon}}, \citenamefont {{Leung}},
  \citenamefont {{Lewis}}, \citenamefont {{Li}}, \citenamefont {{Li}},
  \citenamefont {{Limon}}, \citenamefont {{Linder}}, \citenamefont
  {{Lopez-Caraballo}}, \citenamefont {{Louis}}, \citenamefont {{Lowry}},
  \citenamefont {{Lungu}}, \citenamefont {{Madhavacheril}}, \citenamefont
  {{Mak}}, \citenamefont {{Maldonado}}, \citenamefont {{Mani}}, \citenamefont
  {{Mates}}, \citenamefont {{Matsuda}}, \citenamefont {{Maurin}}, \citenamefont
  {{Mauskopf}}, \citenamefont {{May}}, \citenamefont {{McCallum}},
  \citenamefont {{McKenney}}, \citenamefont {{McMahon}}, \citenamefont
  {{Meerburg}}, \citenamefont {{Meyers}}, \citenamefont {{Miller}},
  \citenamefont {{Mirmelstein}}, \citenamefont {{Moodley}}, \citenamefont
  {{Munchmeyer}}, \citenamefont {{Munson}}, \citenamefont {{Naess}},
  \citenamefont {{Nati}}, \citenamefont {{Navaroli}}, \citenamefont
  {{Newburgh}}, \citenamefont {{Nguyen}}, \citenamefont {{Niemack}},
  \citenamefont {{Nishino}}, \citenamefont {{Orlowski-Scherer}}, \citenamefont
  {{Page}}, \citenamefont {{Partridge}}, \citenamefont {{Peloton}},
  \citenamefont {{Perrotta}}, \citenamefont {{Piccirillo}}, \citenamefont
  {{Pisano}}, \citenamefont {{Poletti}}, \citenamefont {{Puddu}}, \citenamefont
  {{Puglisi}}, \citenamefont {{Raum}}, \citenamefont {{Reichardt}},
  \citenamefont {{Remazeilles}}, \citenamefont {{Rephaeli}}, \citenamefont
  {{Riechers}}, \citenamefont {{Rojas}}, \citenamefont {{Roy}}, \citenamefont
  {{Sadeh}}, \citenamefont {{Sakurai}}, \citenamefont {{Salatino}},
  \citenamefont {{Sathyanarayana Rao}}, \citenamefont {{Schaan}}, \citenamefont
  {{Schmittfull}}, \citenamefont {{Sehgal}}, \citenamefont {{Seibert}},
  \citenamefont {{Seljak}}, \citenamefont {{Sherwin}}, \citenamefont
  {{Shimon}}, \citenamefont {{Sierra}}, \citenamefont {{Sievers}},
  \citenamefont {{Sikhosana}}, \citenamefont {{Silva-Feaver}}, \citenamefont
  {{Simon}}, \citenamefont {{Sinclair}}, \citenamefont {{Siritanasak}},
  \citenamefont {{Smith}}, \citenamefont {{Smith}}, \citenamefont {{Spergel}},
  \citenamefont {{Staggs}}, \citenamefont {{Stein}}, \citenamefont {{Stevens}},
  \citenamefont {{Stompor}}, \citenamefont {{Suzuki}}, \citenamefont
  {{Tajima}}, \citenamefont {{Takakura}}, \citenamefont {{Teply}},
  \citenamefont {{Thomas}}, \citenamefont {{Thorne}}, \citenamefont
  {{Thornton}}, \citenamefont {{Trac}}, \citenamefont {{Tsai}}, \citenamefont
  {{Tucker}}, \citenamefont {{Ullom}}, \citenamefont {{Vagnozzi}},
  \citenamefont {{van Engelen}}, \citenamefont {{Van Lanen}}, \citenamefont
  {{Van Winkle}}, \citenamefont {{Vavagiakis}}, \citenamefont {{Verg{\`e}s}},
  \citenamefont {{Vissers}}, \citenamefont {{Wagoner}}, \citenamefont
  {{Walker}}, \citenamefont {{Ward}}, \citenamefont {{Westbrook}},
  \citenamefont {{Whitehorn}}, \citenamefont {{Williams}}, \citenamefont
  {{Williams}}, \citenamefont {{Wollack}}, \citenamefont {{Xu}}, \citenamefont
  {{Yu}}, \citenamefont {{Yu}}, \citenamefont {{Zago}}, \citenamefont
  {{Zhang}}, \citenamefont {{Zhu}},\ and\ \citenamefont {{Simons Observatory
  Collaboration}}}]{SimonsObservatory2019}%
  \BibitemOpen
  \bibfield  {author} {\bibinfo {author} {P.~{Ade}}, \bibinfo {author}
  {J.~{Aguirre}}, \bibinfo {author} {Z.~{Ahmed}}, \bibinfo {author}
  {S.~{Aiola}}, \bibinfo {author} {A.~{Ali}}, \bibinfo {author} {D.~{Alonso}},
  \bibinfo {author} {M.~A. {Alvarez}}, \bibinfo {author} {K.~{Arnold}},
  \bibinfo {author} {P.~{Ashton}}, \bibinfo {author} {J.~{Austermann}},
  et~al.,\ }\href {https://doi.org/10.1088/1475-7516/2019/02/056} {\bibfield
  {journal} {\bibinfo  {journal} {\jcap}\ }\textbf {\bibinfo {volume} {2019}},\
  \bibinfo {eid} {056} (\bibinfo {year} {2019})},\ \Eprint
  {https://arxiv.org/abs/1808.07445} {arXiv:1808.07445 [astro-ph.CO]}
  \BibitemShut {NoStop}%
\bibitem [{\citenamefont {{Abazajian}}\ \emph {et~al.}(2016)\citenamefont
  {{Abazajian}}, \citenamefont {{Adshead}}, \citenamefont {{Ahmed}},
  \citenamefont {{Allen}}, \citenamefont {{Alonso}}, \citenamefont {{Arnold}},
  \citenamefont {{Baccigalupi}}, \citenamefont {{Bartlett}}, \citenamefont
  {{Battaglia}}, \citenamefont {{Benson}}, \citenamefont {{Bischoff}},
  \citenamefont {{Borrill}}, \citenamefont {{Buza}}, \citenamefont
  {{Calabrese}}, \citenamefont {{Caldwell}}, \citenamefont {{Carlstrom}},
  \citenamefont {{Chang}}, \citenamefont {{Crawford}}, \citenamefont
  {{Cyr-Racine}}, \citenamefont {{De Bernardis}}, \citenamefont {{de Haan}},
  \citenamefont {{di Serego Alighieri}}, \citenamefont {{Dunkley}},
  \citenamefont {{Dvorkin}}, \citenamefont {{Errard}}, \citenamefont
  {{Fabbian}}, \citenamefont {{Feeney}}, \citenamefont {{Ferraro}},
  \citenamefont {{Filippini}}, \citenamefont {{Flauger}}, \citenamefont
  {{Fuller}}, \citenamefont {{Gluscevic}}, \citenamefont {{Green}},
  \citenamefont {{Grin}}, \citenamefont {{Grohs}}, \citenamefont {{Henning}},
  \citenamefont {{Hill}}, \citenamefont {{Hlozek}}, \citenamefont {{Holder}},
  \citenamefont {{Holzapfel}}, \citenamefont {{Hu}}, \citenamefont
  {{Huffenberger}}, \citenamefont {{Keskitalo}}, \citenamefont {{Knox}},
  \citenamefont {{Kosowsky}}, \citenamefont {{Kovac}}, \citenamefont
  {{Kovetz}}, \citenamefont {{Kuo}}, \citenamefont {{Kusaka}}, \citenamefont
  {{Le Jeune}}, \citenamefont {{Lee}}, \citenamefont {{Lilley}}, \citenamefont
  {{Loverde}}, \citenamefont {{Madhavacheril}}, \citenamefont {{Mantz}},
  \citenamefont {{Marsh}}, \citenamefont {{McMahon}}, \citenamefont
  {{Meerburg}}, \citenamefont {{Meyers}}, \citenamefont {{Miller}},
  \citenamefont {{Munoz}}, \citenamefont {{Nguyen}}, \citenamefont {{Niemack}},
  \citenamefont {{Peloso}}, \citenamefont {{Peloton}}, \citenamefont
  {{Pogosian}}, \citenamefont {{Pryke}}, \citenamefont {{Raveri}},
  \citenamefont {{Reichardt}}, \citenamefont {{Rocha}}, \citenamefont
  {{Rotti}}, \citenamefont {{Schaan}}, \citenamefont {{Schmittfull}},
  \citenamefont {{Scott}}, \citenamefont {{Sehgal}}, \citenamefont
  {{Shandera}}, \citenamefont {{Sherwin}}, \citenamefont {{Smith}},
  \citenamefont {{Sorbo}}, \citenamefont {{Starkman}}, \citenamefont {{Story}},
  \citenamefont {{van Engelen}}, \citenamefont {{Vieira}}, \citenamefont
  {{Watson}}, \citenamefont {{Whitehorn}},\ and\ \citenamefont {{Kimmy
  Wu}}}]{Abazajian2016}%
  \BibitemOpen
  \bibfield  {author} {\bibinfo {author} {K.~N. {Abazajian}}, \bibinfo {author}
  {P.~{Adshead}}, \bibinfo {author} {Z.~{Ahmed}}, \bibinfo {author} {S.~W.
  {Allen}}, \bibinfo {author} {D.~{Alonso}}, \bibinfo {author} {K.~S.
  {Arnold}}, \bibinfo {author} {C.~{Baccigalupi}}, \bibinfo {author} {J.~G.
  {Bartlett}}, \bibinfo {author} {N.~{Battaglia}}, \bibinfo {author} {B.~A.
  {Benson}}, et~al.,\ }\href@noop {} {\bibfield  {journal} {\bibinfo  {journal}
  {arXiv e-prints}\ ,\ \bibinfo {eid} {arXiv:1610.02743}} (\bibinfo {year}
  {2016})},\ \Eprint {https://arxiv.org/abs/1610.02743} {arXiv:1610.02743
  [astro-ph.CO]} \BibitemShut {NoStop}%
\bibitem [{\citenamefont {{DESI Collaboration}}\ \emph
  {et~al.}(2016)\citenamefont {{DESI Collaboration}}, \citenamefont
  {{Aghamousa}}, \citenamefont {{Aguilar}}, \citenamefont {{Ahlen}},
  \citenamefont {{Alam}}, \citenamefont {{Allen}}, \citenamefont {{Allende
  Prieto}}, \citenamefont {{Annis}}, \citenamefont {{Bailey}}, \citenamefont
  {{Balland}}, \citenamefont {{Ballester}}, \citenamefont {{Baltay}},
  \citenamefont {{Beaufore}}, \citenamefont {{Bebek}}, \citenamefont {{Beers}},
  \citenamefont {{Bell}}, \citenamefont {{Bernal}}, \citenamefont {{Besuner}},
  \citenamefont {{Beutler}}, \citenamefont {{Blake}}, \citenamefont
  {{Bleuler}}, \citenamefont {{Blomqvist}}, \citenamefont {{Blum}},
  \citenamefont {{Bolton}}, \citenamefont {{Briceno}}, \citenamefont
  {{Brooks}}, \citenamefont {{Brownstein}}, \citenamefont {{Buckley-Geer}},
  \citenamefont {{Burden}}, \citenamefont {{Burtin}}, \citenamefont {{Busca}},
  \citenamefont {{Cahn}}, \citenamefont {{Cai}}, \citenamefont {{Cardiel-Sas}},
  \citenamefont {{Carlberg}}, \citenamefont {{Carton}}, \citenamefont
  {{Casas}}, \citenamefont {{Castand er}}, \citenamefont {{Cervantes-Cota}},
  \citenamefont {{Claybaugh}}, \citenamefont {{Close}}, \citenamefont
  {{Coker}}, \citenamefont {{Cole}}, \citenamefont {{Comparat}}, \citenamefont
  {{Cooper}}, \citenamefont {{Cousinou}}, \citenamefont {{Crocce}},
  \citenamefont {{Cuby}}, \citenamefont {{Cunningham}}, \citenamefont
  {{Davis}}, \citenamefont {{Dawson}}, \citenamefont {{de la Macorra}},
  \citenamefont {{De Vicente}}, \citenamefont {{Delubac}}, \citenamefont
  {{Derwent}}, \citenamefont {{Dey}}, \citenamefont {{Dhungana}}, \citenamefont
  {{Ding}}, \citenamefont {{Doel}}, \citenamefont {{Duan}}, \citenamefont
  {{Ealet}}, \citenamefont {{Edelstein}}, \citenamefont {{Eftekharzadeh}},
  \citenamefont {{Eisenstein}}, \citenamefont {{Elliott}}, \citenamefont
  {{Escoffier}}, \citenamefont {{Evatt}}, \citenamefont {{Fagrelius}},
  \citenamefont {{Fan}}, \citenamefont {{Fanning}}, \citenamefont {{Farahi}},
  \citenamefont {{Farihi}}, \citenamefont {{Favole}}, \citenamefont {{Feng}},
  \citenamefont {{Fernandez}}, \citenamefont {{Findlay}}, \citenamefont
  {{Finkbeiner}}, \citenamefont {{Fitzpatrick}}, \citenamefont {{Flaugher}},
  \citenamefont {{Flender}}, \citenamefont {{Font-Ribera}}, \citenamefont
  {{Forero-Romero}}, \citenamefont {{Fosalba}}, \citenamefont {{Frenk}},
  \citenamefont {{Fumagalli}}, \citenamefont {{Gaensicke}}, \citenamefont
  {{Gallo}}, \citenamefont {{Garcia-Bellido}}, \citenamefont {{Gaztanaga}},
  \citenamefont {{Pietro Gentile Fusillo}}, \citenamefont {{Gerard}},
  \citenamefont {{Gershkovich}}, \citenamefont {{Giannantonio}}, \citenamefont
  {{Gillet}}, \citenamefont {{Gonzalez-de-Rivera}}, \citenamefont
  {{Gonzalez-Perez}}, \citenamefont {{Gott}}, \citenamefont {{Graur}},
  \citenamefont {{Gutierrez}}, \citenamefont {{Guy}}, \citenamefont {{Habib}},
  \citenamefont {{Heetderks}}, \citenamefont {{Heetderks}}, \citenamefont
  {{Heitmann}}, \citenamefont {{Hellwing}}, \citenamefont {{Herrera}},
  \citenamefont {{Ho}}, \citenamefont {{Holland}}, \citenamefont {{Honscheid}},
  \citenamefont {{Huff}}, \citenamefont {{Hutchinson}}, \citenamefont
  {{Huterer}}, \citenamefont {{Hwang}}, \citenamefont {{Illa Laguna}},
  \citenamefont {{Ishikawa}}, \citenamefont {{Jacobs}}, \citenamefont
  {{Jeffrey}}, \citenamefont {{Jelinsky}}, \citenamefont {{Jennings}},
  \citenamefont {{Jiang}}, \citenamefont {{Jimenez}}, \citenamefont
  {{Johnson}}, \citenamefont {{Joyce}}, \citenamefont {{Jullo}}, \citenamefont
  {{Juneau}}, \citenamefont {{Kama}}, \citenamefont {{Karcher}}, \citenamefont
  {{Karkar}}, \citenamefont {{Kehoe}}, \citenamefont {{Kennamer}},
  \citenamefont {{Kent}}, \citenamefont {{Kilbinger}}, \citenamefont {{Kim}},
  \citenamefont {{Kirkby}}, \citenamefont {{Kisner}}, \citenamefont
  {{Kitanidis}}, \citenamefont {{Kneib}}, \citenamefont {{Koposov}},
  \citenamefont {{Kovacs}}, \citenamefont {{Koyama}}, \citenamefont {{Kremin}},
  \citenamefont {{Kron}}, \citenamefont {{Kronig}}, \citenamefont
  {{Kueter-Young}}, \citenamefont {{Lacey}}, \citenamefont {{Lafever}},
  \citenamefont {{Lahav}}, \citenamefont {{Lambert}}, \citenamefont
  {{Lampton}}, \citenamefont {{Land riau}}, \citenamefont {{Lang}},
  \citenamefont {{Lauer}}, \citenamefont {{Le Goff}}, \citenamefont {{Le
  Guillou}}, \citenamefont {{Le Van Suu}}, \citenamefont {{Lee}}, \citenamefont
  {{Lee}}, \citenamefont {{Leitner}}, \citenamefont {{Lesser}}, \citenamefont
  {{Levi}}, \citenamefont {{L'Huillier}}, \citenamefont {{Li}}, \citenamefont
  {{Liang}}, \citenamefont {{Lin}}, \citenamefont {{Linder}}, \citenamefont
  {{Loebman}}, \citenamefont {{Luki{\'c}}}, \citenamefont {{Ma}}, \citenamefont
  {{MacCrann}}, \citenamefont {{Magneville}}, \citenamefont {{Makarem}},
  \citenamefont {{Manera}}, \citenamefont {{Manser}}, \citenamefont
  {{Marshall}}, \citenamefont {{Martini}}, \citenamefont {{Massey}},
  \citenamefont {{Matheson}}, \citenamefont {{McCauley}}, \citenamefont
  {{McDonald}}, \citenamefont {{McGreer}}, \citenamefont {{Meisner}},
  \citenamefont {{Metcalfe}}, \citenamefont {{Miller}}, \citenamefont
  {{Miquel}}, \citenamefont {{Moustakas}}, \citenamefont {{Myers}},
  \citenamefont {{Naik}}, \citenamefont {{Newman}}, \citenamefont {{Nichol}},
  \citenamefont {{Nicola}}, \citenamefont {{Nicolati da Costa}}, \citenamefont
  {{Nie}}, \citenamefont {{Niz}}, \citenamefont {{Norberg}}, \citenamefont
  {{Nord}}, \citenamefont {{Norman}}, \citenamefont {{Nugent}}, \citenamefont
  {{O'Brien}}, \citenamefont {{Oh}}, \citenamefont {{Olsen}}, \citenamefont
  {{Padilla}}, \citenamefont {{Padmanabhan}}, \citenamefont {{Padmanabhan}},
  \citenamefont {{Palanque-Delabrouille}}, \citenamefont {{Palmese}},
  \citenamefont {{Pappalardo}}, \citenamefont {{P{\^a}ris}}, \citenamefont
  {{Park}}, \citenamefont {{Patej}}, \citenamefont {{Peacock}}, \citenamefont
  {{Peiris}}, \citenamefont {{Peng}}, \citenamefont {{Percival}}, \citenamefont
  {{Perruchot}}, \citenamefont {{Pieri}}, \citenamefont {{Pogge}},
  \citenamefont {{Pollack}}, \citenamefont {{Poppett}}, \citenamefont
  {{Prada}}, \citenamefont {{Prakash}}, \citenamefont {{Probst}}, \citenamefont
  {{Rabinowitz}}, \citenamefont {{Raichoor}}, \citenamefont {{Ree}},
  \citenamefont {{Refregier}}, \citenamefont {{Regal}}, \citenamefont {{Reid}},
  \citenamefont {{Reil}}, \citenamefont {{Rezaie}}, \citenamefont {{Rockosi}},
  \citenamefont {{Roe}}, \citenamefont {{Ronayette}}, \citenamefont
  {{Roodman}}, \citenamefont {{Ross}}, \citenamefont {{Ross}}, \citenamefont
  {{Rossi}}, \citenamefont {{Rozo}}, \citenamefont {{Ruhlmann-Kleider}},
  \citenamefont {{Rykoff}}, \citenamefont {{Sabiu}}, \citenamefont
  {{Samushia}}, \citenamefont {{Sanchez}}, \citenamefont {{Sanchez}},
  \citenamefont {{Schlegel}}, \citenamefont {{Schneider}}, \citenamefont
  {{Schubnell}}, \citenamefont {{Secroun}}, \citenamefont {{Seljak}},
  \citenamefont {{Seo}}, \citenamefont {{Serrano}}, \citenamefont
  {{Shafieloo}}, \citenamefont {{Shan}}, \citenamefont {{Sharples}},
  \citenamefont {{Sholl}}, \citenamefont {{Shourt}}, \citenamefont {{Silber}},
  \citenamefont {{Silva}}, \citenamefont {{Sirk}}, \citenamefont {{Slosar}},
  \citenamefont {{Smith}}, \citenamefont {{Smoot}}, \citenamefont {{Som}},
  \citenamefont {{Song}}, \citenamefont {{Sprayberry}}, \citenamefont
  {{Staten}}, \citenamefont {{Stefanik}}, \citenamefont {{Tarle}},
  \citenamefont {{Sien Tie}}, \citenamefont {{Tinker}}, \citenamefont
  {{Tojeiro}}, \citenamefont {{Valdes}}, \citenamefont {{Valenzuela}},
  \citenamefont {{Valluri}}, \citenamefont {{Vargas-Magana}}, \citenamefont
  {{Verde}}, \citenamefont {{Walker}}, \citenamefont {{Wang}}, \citenamefont
  {{Wang}}, \citenamefont {{Weaver}}, \citenamefont {{Weaverdyck}},
  \citenamefont {{Wechsler}}, \citenamefont {{Weinberg}}, \citenamefont
  {{White}}, \citenamefont {{Yang}}, \citenamefont {{Yeche}}, \citenamefont
  {{Zhang}}, \citenamefont {{Zhao}}, \citenamefont {{Zheng}}, \citenamefont
  {{Zhou}}, \citenamefont {{Zhou}}, \citenamefont {{Zhu}}, \citenamefont
  {{Zou}},\ and\ \citenamefont {{Zu}}}]{DESICollaboration2016}%
  \BibitemOpen
  \bibfield  {author} {\bibinfo {author} {{DESI Collaboration}}, \bibinfo
  {author} {A.~{Aghamousa}}, \bibinfo {author} {J.~{Aguilar}}, \bibinfo
  {author} {S.~{Ahlen}}, \bibinfo {author} {S.~{Alam}}, \bibinfo {author}
  {L.~E. {Allen}}, \bibinfo {author} {C.~{Allende Prieto}}, \bibinfo {author}
  {J.~{Annis}}, \bibinfo {author} {S.~{Bailey}}, \bibinfo {author}
  {C.~{Balland}}, et~al.,\ }\href@noop {} {\bibfield  {journal} {\bibinfo
  {journal} {arXiv e-prints}\ ,\ \bibinfo {eid} {arXiv:1611.00036}} (\bibinfo
  {year} {2016})},\ \Eprint {https://arxiv.org/abs/1611.00036}
  {arXiv:1611.00036 [astro-ph.IM]} \BibitemShut {NoStop}%
\bibitem [{\citenamefont {{Dor{\'e}}}\ \emph {et~al.}(2014)\citenamefont
  {{Dor{\'e}}}, \citenamefont {{Bock}}, \citenamefont {{Ashby}}, \citenamefont
  {{Capak}}, \citenamefont {{Cooray}}, \citenamefont {{de Putter}},
  \citenamefont {{Eifler}}, \citenamefont {{Flagey}}, \citenamefont {{Gong}},
  \citenamefont {{Habib}}, \citenamefont {{Heitmann}}, \citenamefont
  {{Hirata}}, \citenamefont {{Jeong}}, \citenamefont {{Katti}}, \citenamefont
  {{Korngut}}, \citenamefont {{Krause}}, \citenamefont {{Lee}}, \citenamefont
  {{Masters}}, \citenamefont {{Mauskopf}}, \citenamefont {{Melnick}},
  \citenamefont {{Mennesson}}, \citenamefont {{Nguyen}}, \citenamefont
  {{{\"O}berg}}, \citenamefont {{Pullen}}, \citenamefont {{Raccanelli}},
  \citenamefont {{Smith}}, \citenamefont {{Song}}, \citenamefont {{Tolls}},
  \citenamefont {{Unwin}}, \citenamefont {{Venumadhav}}, \citenamefont
  {{Viero}}, \citenamefont {{Werner}},\ and\ \citenamefont
  {{Zemcov}}}]{Dore2014}%
  \BibitemOpen
  \bibfield  {author} {\bibinfo {author} {O.~{Dor{\'e}}}, \bibinfo {author}
  {J.~{Bock}}, \bibinfo {author} {M.~{Ashby}}, \bibinfo {author} {P.~{Capak}},
  \bibinfo {author} {A.~{Cooray}}, \bibinfo {author} {R.~{de Putter}}, \bibinfo
  {author} {T.~{Eifler}}, \bibinfo {author} {N.~{Flagey}}, \bibinfo {author}
  {Y.~{Gong}}, \bibinfo {author} {S.~{Habib}}, et~al.,\ }\href@noop {}
  {\bibfield  {journal} {\bibinfo  {journal} {arXiv e-prints}\ ,\ \bibinfo
  {eid} {arXiv:1412.4872}} (\bibinfo {year} {2014})},\ \Eprint
  {https://arxiv.org/abs/1412.4872} {arXiv:1412.4872 [astro-ph.CO]}
  \BibitemShut {NoStop}%
\bibitem [{\citenamefont {{Doux}}\ \emph {et~al.}(2018)\citenamefont {{Doux}},
  \citenamefont {{Penna-Lima}}, \citenamefont {{Vitenti}}, \citenamefont
  {{Tr{\'e}guer}}, \citenamefont {{Aubourg}},\ and\ \citenamefont
  {{Ganga}}}]{Doux2018}%
  \BibitemOpen
  \bibfield  {author} {\bibinfo {author} {C.~{Doux}}, \bibinfo {author}
  {M.~{Penna-Lima}}, \bibinfo {author} {S.~D.~P. {Vitenti}}, \bibinfo {author}
  {J.~{Tr{\'e}guer}}, \bibinfo {author} {E.~{Aubourg}},\ and\ \bibinfo {author}
  {K.~{Ganga}},\ }\href {https://doi.org/10.1093/mnras/sty2160} {\bibfield
  {journal} {\bibinfo  {journal} {\mnras}\ }\textbf {\bibinfo {volume} {480}},\
  \bibinfo {pages} {5386} (\bibinfo {year} {2018})},\ \Eprint
  {https://arxiv.org/abs/1706.04583} {arXiv:1706.04583 [astro-ph.CO]}
  \BibitemShut {NoStop}%
\bibitem [{\citenamefont {{Chen}}\ \emph {et~al.}(2022)\citenamefont {{Chen}},
  \citenamefont {{White}}, \citenamefont {{DeRose}},\ and\ \citenamefont
  {{Kokron}}}]{Chen2022}%
  \BibitemOpen
  \bibfield  {author} {\bibinfo {author} {S.-F. {Chen}}, \bibinfo {author}
  {M.~{White}}, \bibinfo {author} {J.~{DeRose}},\ and\ \bibinfo {author}
  {N.~{Kokron}},\ }\href {https://doi.org/10.1088/1475-7516/2022/07/041}
  {\bibfield  {journal} {\bibinfo  {journal} {\jcap}\ }\textbf {\bibinfo
  {volume} {2022}},\ \bibinfo {eid} {041} (\bibinfo {year} {2022})},\ \Eprint
  {https://arxiv.org/abs/2204.10392} {arXiv:2204.10392 [astro-ph.CO]}
  \BibitemShut {NoStop}%
\bibitem [{\citenamefont {{Carron}}\ \emph {et~al.}(2022)\citenamefont
  {{Carron}}, \citenamefont {{Mirmelstein}},\ and\ \citenamefont
  {{Lewis}}}]{Carron2022}%
  \BibitemOpen
  \bibfield  {author} {\bibinfo {author} {J.~{Carron}}, \bibinfo {author}
  {M.~{Mirmelstein}},\ and\ \bibinfo {author} {A.~{Lewis}},\ }\href
  {https://doi.org/10.1088/1475-7516/2022/09/039} {\bibfield  {journal}
  {\bibinfo  {journal} {\jcap}\ }\textbf {\bibinfo {volume} {2022}},\ \bibinfo
  {eid} {039} (\bibinfo {year} {2022})},\ \Eprint
  {https://arxiv.org/abs/2206.07773} {arXiv:2206.07773 [astro-ph.CO]}
  \BibitemShut {NoStop}%
\bibitem [{\citenamefont {{Reid}}\ \emph {et~al.}(2016)\citenamefont {{Reid}},
  \citenamefont {{Ho}}, \citenamefont {{Padmanabhan}}, \citenamefont
  {{Percival}}, \citenamefont {{Tinker}}, \citenamefont {{Tojeiro}},
  \citenamefont {{White}}, \citenamefont {{Eisenstein}}, \citenamefont
  {{Maraston}}, \citenamefont {{Ross}}, \citenamefont {{S{\'a}nchez}},
  \citenamefont {{Schlegel}}, \citenamefont {{Sheldon}}, \citenamefont
  {{Strauss}}, \citenamefont {{Thomas}}, \citenamefont {{Wake}}, \citenamefont
  {{Beutler}}, \citenamefont {{Bizyaev}}, \citenamefont {{Bolton}},
  \citenamefont {{Brownstein}}, \citenamefont {{Chuang}}, \citenamefont
  {{Dawson}}, \citenamefont {{Harding}}, \citenamefont {{Kitaura}},
  \citenamefont {{Leauthaud}}, \citenamefont {{Masters}}, \citenamefont
  {{McBride}}, \citenamefont {{More}}, \citenamefont {{Olmstead}},
  \citenamefont {{Oravetz}}, \citenamefont {{Nuza}}, \citenamefont {{Pan}},
  \citenamefont {{Parejko}}, \citenamefont {{Pforr}}, \citenamefont {{Prada}},
  \citenamefont {{Rodr{\'\i}guez-Torres}}, \citenamefont {{Salazar-Albornoz}},
  \citenamefont {{Samushia}}, \citenamefont {{Schneider}}, \citenamefont
  {{Sc{\'o}ccola}}, \citenamefont {{Simmons}},\ and\ \citenamefont
  {{Vargas-Magana}}}]{Reid2016}%
  \BibitemOpen
  \bibfield  {author} {\bibinfo {author} {B.~{Reid}}, \bibinfo {author}
  {S.~{Ho}}, \bibinfo {author} {N.~{Padmanabhan}}, \bibinfo {author} {W.~J.
  {Percival}}, \bibinfo {author} {J.~{Tinker}}, \bibinfo {author}
  {R.~{Tojeiro}}, \bibinfo {author} {M.~{White}}, \bibinfo {author} {D.~J.
  {Eisenstein}}, \bibinfo {author} {C.~{Maraston}}, \bibinfo {author} {A.~J.
  {Ross}}, et~al.,\ }\href {https://doi.org/10.1093/mnras/stv2382} {\bibfield
  {journal} {\bibinfo  {journal} {\mnras}\ }\textbf {\bibinfo {volume} {455}},\
  \bibinfo {pages} {1553} (\bibinfo {year} {2016})},\ \Eprint
  {https://arxiv.org/abs/1509.06529} {arXiv:1509.06529 [astro-ph.CO]}
  \BibitemShut {NoStop}%
\bibitem [{\citenamefont {{Ma}}\ and\ \citenamefont
  {{Bertschinger}}(1995)}]{Ma1995}%
  \BibitemOpen
  \bibfield  {author} {\bibinfo {author} {C.-P. {Ma}}\ and\ \bibinfo {author}
  {E.~{Bertschinger}},\ }\href {https://doi.org/10.1086/176550} {\bibfield
  {journal} {\bibinfo  {journal} {\apj}\ }\textbf {\bibinfo {volume} {455}},\
  \bibinfo {pages} {7} (\bibinfo {year} {1995})},\ \Eprint
  {https://arxiv.org/abs/astro-ph/9506072} {arXiv:astro-ph/9506072 [astro-ph]}
  \BibitemShut {NoStop}%
\bibitem [{\citenamefont {{Silvestri}}\ \emph {et~al.}(2013)\citenamefont
  {{Silvestri}}, \citenamefont {{Pogosian}},\ and\ \citenamefont
  {{Buniy}}}]{Silvestri2013}%
  \BibitemOpen
  \bibfield  {author} {\bibinfo {author} {A.~{Silvestri}}, \bibinfo {author}
  {L.~{Pogosian}},\ and\ \bibinfo {author} {R.~V. {Buniy}},\ }\href
  {https://doi.org/10.1103/PhysRevD.87.104015} {\bibfield  {journal} {\bibinfo
  {journal} {\prd}\ }\textbf {\bibinfo {volume} {87}},\ \bibinfo {eid} {104015}
  (\bibinfo {year} {2013})},\ \Eprint {https://arxiv.org/abs/1302.1193}
  {arXiv:1302.1193 [astro-ph.CO]} \BibitemShut {NoStop}%
\bibitem [{\citenamefont {{Baker}}\ \emph {et~al.}(2014)\citenamefont
  {{Baker}}, \citenamefont {{Ferreira}}, \citenamefont {{Leonard}},\ and\
  \citenamefont {{Motta}}}]{Baker2014}%
  \BibitemOpen
  \bibfield  {author} {\bibinfo {author} {T.~{Baker}}, \bibinfo {author} {P.~G.
  {Ferreira}}, \bibinfo {author} {C.~D. {Leonard}},\ and\ \bibinfo {author}
  {M.~{Motta}},\ }\href {https://doi.org/10.1103/PhysRevD.90.124030} {\bibfield
   {journal} {\bibinfo  {journal} {\prd}\ }\textbf {\bibinfo {volume} {90}},\
  \bibinfo {eid} {124030} (\bibinfo {year} {2014})},\ \Eprint
  {https://arxiv.org/abs/1409.8284} {arXiv:1409.8284 [astro-ph.CO]}
  \BibitemShut {NoStop}%
\bibitem [{\citenamefont {{Wenzl}}\ \emph {et~al.}(2022)\citenamefont
  {{Wenzl}}, \citenamefont {{Doux}}, \citenamefont {{Heinrich}}, \citenamefont
  {{Bean}}, \citenamefont {{Jain}}, \citenamefont {{Dor{\'e}}}, \citenamefont
  {{Eifler}},\ and\ \citenamefont {{Fang}}}]{Wenzl2022}%
  \BibitemOpen
  \bibfield  {author} {\bibinfo {author} {L.~{Wenzl}}, \bibinfo {author}
  {C.~{Doux}}, \bibinfo {author} {C.~{Heinrich}}, \bibinfo {author}
  {R.~{Bean}}, \bibinfo {author} {B.~{Jain}}, \bibinfo {author}
  {O.~{Dor{\'e}}}, \bibinfo {author} {T.~{Eifler}},\ and\ \bibinfo {author}
  {X.~{Fang}},\ }\href {https://doi.org/10.1093/mnras/stac790} {\bibfield
  {journal} {\bibinfo  {journal} {\mnras}\ }\textbf {\bibinfo {volume} {512}},\
  \bibinfo {pages} {5311} (\bibinfo {year} {2022})},\ \Eprint
  {https://arxiv.org/abs/2112.07681} {arXiv:2112.07681 [astro-ph.CO]}
  \BibitemShut {NoStop}%
\bibitem [{\citenamefont {{Linder}}(2008)}]{Linder2008_growth}%
  \BibitemOpen
  \bibfield  {author} {\bibinfo {author} {E.~V. {Linder}},\ }\href
  {https://doi.org/10.1016/j.astropartphys.2008.03.002} {\bibfield  {journal}
  {\bibinfo  {journal} {Astroparticle Physics}\ }\textbf {\bibinfo {volume}
  {29}},\ \bibinfo {pages} {336} (\bibinfo {year} {2008})},\ \Eprint
  {https://arxiv.org/abs/0709.1113} {arXiv:0709.1113 [astro-ph]} \BibitemShut
  {NoStop}%
\bibitem [{\citenamefont {{Planck Collaboration}}\ \emph
  {et~al.}(2020{\natexlab{b}})\citenamefont {{Planck Collaboration}},
  \citenamefont {{Aghanim}}, \citenamefont {{Akrami}}, \citenamefont
  {{Arroja}}, \citenamefont {{Ashdown}}, \citenamefont {{Aumont}},
  \citenamefont {{Baccigalupi}}, \citenamefont {{Ballardini}}, \citenamefont
  {{Banday}}, \citenamefont {{Barreiro}}, \citenamefont {{Bartolo}},
  \citenamefont {{Basak}}, \citenamefont {{Battye}}, \citenamefont {{Benabed}},
  \citenamefont {{Bernard}}, \citenamefont {{Bersanelli}}, \citenamefont
  {{Bielewicz}}, \citenamefont {{Bock}}, \citenamefont {{Bond}}, \citenamefont
  {{Borrill}}, \citenamefont {{Bouchet}}, \citenamefont {{Boulanger}},
  \citenamefont {{Bucher}}, \citenamefont {{Burigana}}, \citenamefont
  {{Butler}}, \citenamefont {{Calabrese}}, \citenamefont {{Cardoso}},
  \citenamefont {{Carron}}, \citenamefont {{Casaponsa}}, \citenamefont
  {{Challinor}}, \citenamefont {{Chiang}}, \citenamefont {{Colombo}},
  \citenamefont {{Combet}}, \citenamefont {{Contreras}}, \citenamefont
  {{Crill}}, \citenamefont {{Cuttaia}}, \citenamefont {{de Bernardis}},
  \citenamefont {{de Zotti}}, \citenamefont {{Delabrouille}}, \citenamefont
  {{Delouis}}, \citenamefont {{D{\'e}sert}}, \citenamefont {{Di Valentino}},
  \citenamefont {{Dickinson}}, \citenamefont {{Diego}}, \citenamefont
  {{Donzelli}}, \citenamefont {{Dor{\'e}}}, \citenamefont {{Douspis}},
  \citenamefont {{Ducout}}, \citenamefont {{Dupac}}, \citenamefont
  {{Efstathiou}}, \citenamefont {{Elsner}}, \citenamefont {{En{\ss}lin}},
  \citenamefont {{Eriksen}}, \citenamefont {{Falgarone}}, \citenamefont
  {{Fantaye}}, \citenamefont {{Fergusson}}, \citenamefont {{Fernandez-Cobos}},
  \citenamefont {{Finelli}}, \citenamefont {{Forastieri}}, \citenamefont
  {{Frailis}}, \citenamefont {{Franceschi}}, \citenamefont {{Frolov}},
  \citenamefont {{Galeotta}}, \citenamefont {{Galli}}, \citenamefont {{Ganga}},
  \citenamefont {{G{\'e}nova-Santos}}, \citenamefont {{Gerbino}}, \citenamefont
  {{Ghosh}}, \citenamefont {{Gonz{\'a}lez-Nuevo}}, \citenamefont
  {{G{\'o}rski}}, \citenamefont {{Gratton}}, \citenamefont {{Gruppuso}},
  \citenamefont {{Gudmundsson}}, \citenamefont {{Hamann}}, \citenamefont
  {{Handley}}, \citenamefont {{Hansen}}, \citenamefont {{Helou}}, \citenamefont
  {{Herranz}}, \citenamefont {{Hildebrandt}}, \citenamefont {{Hivon}},
  \citenamefont {{Huang}}, \citenamefont {{Jaffe}}, \citenamefont {{Jones}},
  \citenamefont {{Karakci}}, \citenamefont {{Keih{\"a}nen}}, \citenamefont
  {{Keskitalo}}, \citenamefont {{Kiiveri}}, \citenamefont {{Kim}},
  \citenamefont {{Kisner}}, \citenamefont {{Knox}}, \citenamefont
  {{Krachmalnicoff}}, \citenamefont {{Kunz}}, \citenamefont {{Kurki-Suonio}},
  \citenamefont {{Lagache}}, \citenamefont {{Lamarre}}, \citenamefont
  {{Langer}}, \citenamefont {{Lasenby}}, \citenamefont {{Lattanzi}},
  \citenamefont {{Lawrence}}, \citenamefont {{Le Jeune}}, \citenamefont
  {{Leahy}}, \citenamefont {{Lesgourgues}}, \citenamefont {{Levrier}},
  \citenamefont {{Lewis}}, \citenamefont {{Liguori}}, \citenamefont {{Lilje}},
  \citenamefont {{Lilley}}, \citenamefont {{Lindholm}}, \citenamefont
  {{L{\'o}pez-Caniego}}, \citenamefont {{Lubin}}, \citenamefont {{Ma}},
  \citenamefont {{Mac{\'\i}as-P{\'e}rez}}, \citenamefont {{Maggio}},
  \citenamefont {{Maino}}, \citenamefont {{Mandolesi}}, \citenamefont
  {{Mangilli}}, \citenamefont {{Marcos-Caballero}}, \citenamefont {{Maris}},
  \citenamefont {{Martin}}, \citenamefont {{Martinelli}}, \citenamefont
  {{Mart{\'\i}nez-Gonz{\'a}lez}}, \citenamefont {{Matarrese}}, \citenamefont
  {{Mauri}}, \citenamefont {{McEwen}}, \citenamefont {{Meerburg}},
  \citenamefont {{Meinhold}}, \citenamefont {{Melchiorri}}, \citenamefont
  {{Mennella}}, \citenamefont {{Migliaccio}}, \citenamefont {{Millea}},
  \citenamefont {{Mitra}}, \citenamefont {{Miville-Desch{\^e}nes}},
  \citenamefont {{Molinari}}, \citenamefont {{Moneti}}, \citenamefont
  {{Montier}}, \citenamefont {{Morgante}}, \citenamefont {{Moss}},
  \citenamefont {{Mottet}}, \citenamefont {{M{\"u}nchmeyer}}, \citenamefont
  {{Natoli}}, \citenamefont {{N{\o}rgaard-Nielsen}}, \citenamefont
  {{Oxborrow}}, \citenamefont {{Pagano}}, \citenamefont {{Paoletti}},
  \citenamefont {{Partridge}}, \citenamefont {{Patanchon}}, \citenamefont
  {{Pearson}}, \citenamefont {{Peel}}, \citenamefont {{Peiris}}, \citenamefont
  {{Perrotta}}, \citenamefont {{Pettorino}}, \citenamefont {{Piacentini}},
  \citenamefont {{Polastri}}, \citenamefont {{Polenta}}, \citenamefont
  {{Puget}}, \citenamefont {{Rachen}}, \citenamefont {{Reinecke}},
  \citenamefont {{Remazeilles}}, \citenamefont {{Renault}}, \citenamefont
  {{Renzi}}, \citenamefont {{Rocha}}, \citenamefont {{Rosset}}, \citenamefont
  {{Roudier}}, \citenamefont {{Rubi{\~n}o-Mart{\'\i}n}}, \citenamefont
  {{Ruiz-Granados}}, \citenamefont {{Salvati}}, \citenamefont {{Sandri}},
  \citenamefont {{Savelainen}}, \citenamefont {{Scott}}, \citenamefont
  {{Shellard}}, \citenamefont {{Shiraishi}}, \citenamefont {{Sirignano}},
  \citenamefont {{Sirri}}, \citenamefont {{Spencer}}, \citenamefont
  {{Sunyaev}}, \citenamefont {{Suur-Uski}}, \citenamefont {{Tauber}},
  \citenamefont {{Tavagnacco}}, \citenamefont {{Tenti}}, \citenamefont
  {{Terenzi}}, \citenamefont {{Toffolatti}}, \citenamefont {{Tomasi}},
  \citenamefont {{Trombetti}}, \citenamefont {{Valiviita}}, \citenamefont {{Van
  Tent}}, \citenamefont {{Vibert}}, \citenamefont {{Vielva}}, \citenamefont
  {{Villa}}, \citenamefont {{Vittorio}}, \citenamefont {{Wandelt}},
  \citenamefont {{Wehus}}, \citenamefont {{White}}, \citenamefont {{White}},
  \citenamefont {{Zacchei}},\ and\ \citenamefont
  {{Zonca}}}]{PlanckCollaboration2018_overview}%
  \BibitemOpen
  \bibfield  {author} {\bibinfo {author} {{Planck Collaboration}}, \bibinfo
  {author} {N.~{Aghanim}}, \bibinfo {author} {Y.~{Akrami}}, \bibinfo {author}
  {F.~{Arroja}}, \bibinfo {author} {M.~{Ashdown}}, \bibinfo {author}
  {J.~{Aumont}}, \bibinfo {author} {C.~{Baccigalupi}}, \bibinfo {author}
  {M.~{Ballardini}}, \bibinfo {author} {A.~J. {Banday}}, \bibinfo {author}
  {R.~B. {Barreiro}}, et~al.,\ }\href
  {https://doi.org/10.1051/0004-6361/201833880} {\bibfield  {journal} {\bibinfo
   {journal} {\aap}\ }\textbf {\bibinfo {volume} {641}},\ \bibinfo {eid} {A1}
  (\bibinfo {year} {2020}{\natexlab{b}})},\ \Eprint
  {https://arxiv.org/abs/1807.06205} {arXiv:1807.06205 [astro-ph.CO]}
  \BibitemShut {NoStop}%
\bibitem [{\citenamefont {{Planck Collaboration}}\ \emph
  {et~al.}(2020{\natexlab{c}})\citenamefont {{Planck Collaboration}},
  \citenamefont {{Aghanim}}, \citenamefont {{Akrami}}, \citenamefont
  {{Ashdown}}, \citenamefont {{Aumont}}, \citenamefont {{Baccigalupi}},
  \citenamefont {{Ballardini}}, \citenamefont {{Banday}}, \citenamefont
  {{Barreiro}}, \citenamefont {{Bartolo}}, \citenamefont {{Basak}},
  \citenamefont {{Benabed}}, \citenamefont {{Bernard}}, \citenamefont
  {{Bersanelli}}, \citenamefont {{Bielewicz}}, \citenamefont {{Bock}},
  \citenamefont {{Bond}}, \citenamefont {{Borrill}}, \citenamefont {{Bouchet}},
  \citenamefont {{Boulanger}}, \citenamefont {{Bucher}}, \citenamefont
  {{Burigana}}, \citenamefont {{Calabrese}}, \citenamefont {{Cardoso}},
  \citenamefont {{Carron}}, \citenamefont {{Challinor}}, \citenamefont
  {{Chiang}}, \citenamefont {{Colombo}}, \citenamefont {{Combet}},
  \citenamefont {{Crill}}, \citenamefont {{Cuttaia}}, \citenamefont {{de
  Bernardis}}, \citenamefont {{de Zotti}}, \citenamefont {{Delabrouille}},
  \citenamefont {{Di Valentino}}, \citenamefont {{Diego}}, \citenamefont
  {{Dor{\'e}}}, \citenamefont {{Douspis}}, \citenamefont {{Ducout}},
  \citenamefont {{Dupac}}, \citenamefont {{Efstathiou}}, \citenamefont
  {{Elsner}}, \citenamefont {{En{\ss}lin}}, \citenamefont {{Eriksen}},
  \citenamefont {{Fantaye}}, \citenamefont {{Fernandez-Cobos}}, \citenamefont
  {{Finelli}}, \citenamefont {{Forastieri}}, \citenamefont {{Frailis}},
  \citenamefont {{Fraisse}}, \citenamefont {{Franceschi}}, \citenamefont
  {{Frolov}}, \citenamefont {{Galeotta}}, \citenamefont {{Galli}},
  \citenamefont {{Ganga}}, \citenamefont {{G{\'e}nova-Santos}}, \citenamefont
  {{Gerbino}}, \citenamefont {{Ghosh}}, \citenamefont {{Gonz{\'a}lez-Nuevo}},
  \citenamefont {{G{\'o}rski}}, \citenamefont {{Gratton}}, \citenamefont
  {{Gruppuso}}, \citenamefont {{Gudmundsson}}, \citenamefont {{Hamann}},
  \citenamefont {{Handley}}, \citenamefont {{Hansen}}, \citenamefont
  {{Herranz}}, \citenamefont {{Hivon}}, \citenamefont {{Huang}}, \citenamefont
  {{Jaffe}}, \citenamefont {{Jones}}, \citenamefont {{Karakci}}, \citenamefont
  {{Keih{\"a}nen}}, \citenamefont {{Keskitalo}}, \citenamefont {{Kiiveri}},
  \citenamefont {{Kim}}, \citenamefont {{Knox}}, \citenamefont
  {{Krachmalnicoff}}, \citenamefont {{Kunz}}, \citenamefont {{Kurki-Suonio}},
  \citenamefont {{Lagache}}, \citenamefont {{Lamarre}}, \citenamefont
  {{Lasenby}}, \citenamefont {{Lattanzi}}, \citenamefont {{Lawrence}},
  \citenamefont {{Le Jeune}}, \citenamefont {{Levrier}}, \citenamefont
  {{Lewis}}, \citenamefont {{Liguori}}, \citenamefont {{Lilje}}, \citenamefont
  {{Lindholm}}, \citenamefont {{L{\'o}pez-Caniego}}, \citenamefont {{Lubin}},
  \citenamefont {{Ma}}, \citenamefont {{Mac{\'\i}as-P{\'e}rez}}, \citenamefont
  {{Maggio}}, \citenamefont {{Maino}}, \citenamefont {{Mandolesi}},
  \citenamefont {{Mangilli}}, \citenamefont {{Marcos-Caballero}}, \citenamefont
  {{Maris}}, \citenamefont {{Martin}}, \citenamefont
  {{Mart{\'\i}nez-Gonz{\'a}lez}}, \citenamefont {{Matarrese}}, \citenamefont
  {{Mauri}}, \citenamefont {{McEwen}}, \citenamefont {{Melchiorri}},
  \citenamefont {{Mennella}}, \citenamefont {{Migliaccio}}, \citenamefont
  {{Miville-Desch{\^e}nes}}, \citenamefont {{Molinari}}, \citenamefont
  {{Moneti}}, \citenamefont {{Montier}}, \citenamefont {{Morgante}},
  \citenamefont {{Moss}}, \citenamefont {{Natoli}}, \citenamefont {{Pagano}},
  \citenamefont {{Paoletti}}, \citenamefont {{Partridge}}, \citenamefont
  {{Patanchon}}, \citenamefont {{Perrotta}}, \citenamefont {{Pettorino}},
  \citenamefont {{Piacentini}}, \citenamefont {{Polastri}}, \citenamefont
  {{Polenta}}, \citenamefont {{Puget}}, \citenamefont {{Rachen}}, \citenamefont
  {{Reinecke}}, \citenamefont {{Remazeilles}}, \citenamefont {{Renzi}},
  \citenamefont {{Rocha}}, \citenamefont {{Rosset}}, \citenamefont {{Roudier}},
  \citenamefont {{Rubi{\~n}o-Mart{\'\i}n}}, \citenamefont {{Ruiz-Granados}},
  \citenamefont {{Salvati}}, \citenamefont {{Sandri}}, \citenamefont
  {{Savelainen}}, \citenamefont {{Scott}}, \citenamefont {{Sirignano}},
  \citenamefont {{Sunyaev}}, \citenamefont {{Suur-Uski}}, \citenamefont
  {{Tauber}}, \citenamefont {{Tavagnacco}}, \citenamefont {{Tenti}},
  \citenamefont {{Toffolatti}}, \citenamefont {{Tomasi}}, \citenamefont
  {{Trombetti}}, \citenamefont {{Valiviita}}, \citenamefont {{Van Tent}},
  \citenamefont {{Vielva}}, \citenamefont {{Villa}}, \citenamefont
  {{Vittorio}}, \citenamefont {{Wandelt}}, \citenamefont {{Wehus}},
  \citenamefont {{White}}, \citenamefont {{White}}, \citenamefont {{Zacchei}},\
  and\ \citenamefont {{Zonca}}}]{PlanckCollaboration2018_lensing}%
  \BibitemOpen
  \bibfield  {author} {\bibinfo {author} {{Planck Collaboration}}, \bibinfo
  {author} {N.~{Aghanim}}, \bibinfo {author} {Y.~{Akrami}}, \bibinfo {author}
  {M.~{Ashdown}}, \bibinfo {author} {J.~{Aumont}}, \bibinfo {author}
  {C.~{Baccigalupi}}, \bibinfo {author} {M.~{Ballardini}}, \bibinfo {author}
  {A.~J. {Banday}}, \bibinfo {author} {R.~B. {Barreiro}}, \bibinfo {author}
  {N.~{Bartolo}}, et~al.,\ }\href {https://doi.org/10.1051/0004-6361/201833886}
  {\bibfield  {journal} {\bibinfo  {journal} {\aap}\ }\textbf {\bibinfo
  {volume} {641}},\ \bibinfo {eid} {A8} (\bibinfo {year}
  {2020}{\natexlab{c}})},\ \Eprint {https://arxiv.org/abs/1807.06210}
  {arXiv:1807.06210 [astro-ph.CO]} \BibitemShut {NoStop}%
\bibitem [{\citenamefont {{Planck Collaboration}}\ \emph
  {et~al.}(2020{\natexlab{d}})\citenamefont {{Planck Collaboration}},
  \citenamefont {{Akrami}}, \citenamefont {{Andersen}}, \citenamefont
  {{Ashdown}}, \citenamefont {{Baccigalupi}}, \citenamefont {{Ballardini}},
  \citenamefont {{Banday}}, \citenamefont {{Barreiro}}, \citenamefont
  {{Bartolo}}, \citenamefont {{Basak}}, \citenamefont {{Benabed}},
  \citenamefont {{Bernard}}, \citenamefont {{Bersanelli}}, \citenamefont
  {{Bielewicz}}, \citenamefont {{Bond}}, \citenamefont {{Borrill}},
  \citenamefont {{Burigana}}, \citenamefont {{Butler}}, \citenamefont
  {{Calabrese}}, \citenamefont {{Casaponsa}}, \citenamefont {{Chiang}},
  \citenamefont {{Colombo}}, \citenamefont {{Combet}}, \citenamefont {{Crill}},
  \citenamefont {{Cuttaia}}, \citenamefont {{de Bernardis}}, \citenamefont {{de
  Rosa}}, \citenamefont {{de Zotti}}, \citenamefont {{Delabrouille}},
  \citenamefont {{Di Valentino}}, \citenamefont {{Diego}}, \citenamefont
  {{Dor{\'e}}}, \citenamefont {{Douspis}}, \citenamefont {{Dupac}},
  \citenamefont {{Eriksen}}, \citenamefont {{Fernandez-Cobos}}, \citenamefont
  {{Finelli}}, \citenamefont {{Frailis}}, \citenamefont {{Fraisse}},
  \citenamefont {{Franceschi}}, \citenamefont {{Frolov}}, \citenamefont
  {{Galeotta}}, \citenamefont {{Galli}}, \citenamefont {{Ganga}}, \citenamefont
  {{Gerbino}}, \citenamefont {{Ghosh}}, \citenamefont {{Gonz{\'a}lez-Nuevo}},
  \citenamefont {{G{\'o}rski}}, \citenamefont {{Gruppuso}}, \citenamefont
  {{Gudmundsson}}, \citenamefont {{Handley}}, \citenamefont {{Helou}},
  \citenamefont {{Herranz}}, \citenamefont {{Hildebrandt}}, \citenamefont
  {{Hivon}}, \citenamefont {{Huang}}, \citenamefont {{Jaffe}}, \citenamefont
  {{Jones}}, \citenamefont {{Keih{\"a}nen}}, \citenamefont {{Keskitalo}},
  \citenamefont {{Kiiveri}}, \citenamefont {{Kim}}, \citenamefont {{Kisner}},
  \citenamefont {{Krachmalnicoff}}, \citenamefont {{Kunz}}, \citenamefont
  {{Kurki-Suonio}}, \citenamefont {{Lasenby}}, \citenamefont {{Lattanzi}},
  \citenamefont {{Lawrence}}, \citenamefont {{Le Jeune}}, \citenamefont
  {{Levrier}}, \citenamefont {{Liguori}}, \citenamefont {{Lilje}},
  \citenamefont {{Lilley}}, \citenamefont {{Lindholm}}, \citenamefont
  {{L{\'o}pez-Caniego}}, \citenamefont {{Lubin}}, \citenamefont
  {{Mac{\'\i}as-P{\'e}rez}}, \citenamefont {{Maino}}, \citenamefont
  {{Mandolesi}}, \citenamefont {{Marcos-Caballero}}, \citenamefont {{Maris}},
  \citenamefont {{Martin}}, \citenamefont {{Mart{\'\i}nez-Gonz{\'a}lez}},
  \citenamefont {{Matarrese}}, \citenamefont {{Mauri}}, \citenamefont
  {{McEwen}}, \citenamefont {{Meinhold}}, \citenamefont {{Mennella}},
  \citenamefont {{Migliaccio}}, \citenamefont {{Mitra}}, \citenamefont
  {{Molinari}}, \citenamefont {{Montier}}, \citenamefont {{Morgante}},
  \citenamefont {{Moss}}, \citenamefont {{Natoli}}, \citenamefont {{Paoletti}},
  \citenamefont {{Partridge}}, \citenamefont {{Patanchon}}, \citenamefont
  {{Pearson}}, \citenamefont {{Pearson}}, \citenamefont {{Perrotta}},
  \citenamefont {{Piacentini}}, \citenamefont {{Polenta}}, \citenamefont
  {{Rachen}}, \citenamefont {{Reinecke}}, \citenamefont {{Remazeilles}},
  \citenamefont {{Renzi}}, \citenamefont {{Rocha}}, \citenamefont {{Rosset}},
  \citenamefont {{Roudier}}, \citenamefont {{Rubi{\~n}o-Mart{\'\i}n}},
  \citenamefont {{Ruiz-Granados}}, \citenamefont {{Salvati}}, \citenamefont
  {{Savelainen}}, \citenamefont {{Scott}}, \citenamefont {{Sirignano}},
  \citenamefont {{Sirri}}, \citenamefont {{Spencer}}, \citenamefont
  {{Suur-Uski}}, \citenamefont {{Svalheim}}, \citenamefont {{Tauber}},
  \citenamefont {{Tavagnacco}}, \citenamefont {{Tenti}}, \citenamefont
  {{Terenzi}}, \citenamefont {{Thommesen}}, \citenamefont {{Toffolatti}},
  \citenamefont {{Tomasi}}, \citenamefont {{Tristram}}, \citenamefont
  {{Trombetti}}, \citenamefont {{Valiviita}}, \citenamefont {{Van Tent}},
  \citenamefont {{Vielva}}, \citenamefont {{Villa}}, \citenamefont
  {{Vittorio}}, \citenamefont {{Wandelt}}, \citenamefont {{Wehus}},
  \citenamefont {{Zacchei}},\ and\ \citenamefont
  {{Zonca}}}]{PlanckCollaboration_PR4_NPIPE}%
  \BibitemOpen
  \bibfield  {author} {\bibinfo {author} {{Planck Collaboration}}, \bibinfo
  {author} {Y.~{Akrami}}, \bibinfo {author} {K.~J. {Andersen}}, \bibinfo
  {author} {M.~{Ashdown}}, \bibinfo {author} {C.~{Baccigalupi}}, \bibinfo
  {author} {M.~{Ballardini}}, \bibinfo {author} {A.~J. {Banday}}, \bibinfo
  {author} {R.~B. {Barreiro}}, \bibinfo {author} {N.~{Bartolo}}, \bibinfo
  {author} {S.~{Basak}}, et~al.,\ }\href
  {https://doi.org/10.1051/0004-6361/202038073} {\bibfield  {journal} {\bibinfo
   {journal} {\aap}\ }\textbf {\bibinfo {volume} {643}},\ \bibinfo {eid} {A42}
  (\bibinfo {year} {2020}{\natexlab{d}})},\ \Eprint
  {https://arxiv.org/abs/2007.04997} {arXiv:2007.04997 [astro-ph.CO]}
  \BibitemShut {NoStop}%
\bibitem [{\citenamefont {{Rosenberg}}\ \emph {et~al.}(2022)\citenamefont
  {{Rosenberg}}, \citenamefont {{Gratton}},\ and\ \citenamefont
  {{Efstathiou}}}]{Rosenberg2022}%
  \BibitemOpen
  \bibfield  {author} {\bibinfo {author} {E.~{Rosenberg}}, \bibinfo {author}
  {S.~{Gratton}},\ and\ \bibinfo {author} {G.~{Efstathiou}},\ }\href
  {https://doi.org/10.1093/mnras/stac2744} {\bibfield  {journal} {\bibinfo
  {journal} {\mnras}\ }\textbf {\bibinfo {volume} {517}},\ \bibinfo {pages}
  {4620} (\bibinfo {year} {2022})},\ \Eprint {https://arxiv.org/abs/2205.10869}
  {arXiv:2205.10869 [astro-ph.CO]} \BibitemShut {NoStop}%
\bibitem [{\citenamefont {{Tristram}}\ \emph {et~al.}(2023)\citenamefont
  {{Tristram}}, \citenamefont {{Banday}}, \citenamefont {{Douspis}},
  \citenamefont {{Garrido}}, \citenamefont {{G{\'o}rski}}, \citenamefont
  {{Henrot-Versill{\'e}}}, \citenamefont {{Hergt}}, \citenamefont {{Ili{\'c}}},
  \citenamefont {{Keskitalo}}, \citenamefont {{Lagache}}, \citenamefont
  {{Lawrence}}, \citenamefont {{Partridge}},\ and\ \citenamefont
  {{Scott}}}]{Tristram2023}%
  \BibitemOpen
  \bibfield  {author} {\bibinfo {author} {M.~{Tristram}}, \bibinfo {author}
  {A.~J. {Banday}}, \bibinfo {author} {M.~{Douspis}}, \bibinfo {author}
  {X.~{Garrido}}, \bibinfo {author} {K.~M. {G{\'o}rski}}, \bibinfo {author}
  {S.~{Henrot-Versill{\'e}}}, \bibinfo {author} {L.~T. {Hergt}}, \bibinfo
  {author} {S.~{Ili{\'c}}}, \bibinfo {author} {R.~{Keskitalo}}, \bibinfo
  {author} {G.~{Lagache}}, et~al.,\ }\href
  {https://doi.org/10.48550/arXiv.2309.10034} {\bibfield  {journal} {\bibinfo
  {journal} {arXiv e-prints}\ ,\ \bibinfo {eid} {arXiv:2309.10034}} (\bibinfo
  {year} {2023})},\ \Eprint {https://arxiv.org/abs/2309.10034}
  {arXiv:2309.10034 [astro-ph.CO]} \BibitemShut {NoStop}%
\bibitem [{\citenamefont {{Krolewski}}\ \emph {et~al.}(2020)\citenamefont
  {{Krolewski}}, \citenamefont {{Ferraro}}, \citenamefont {{Schlafly}},\ and\
  \citenamefont {{White}}}]{Krolewski2020}%
  \BibitemOpen
  \bibfield  {author} {\bibinfo {author} {A.~{Krolewski}}, \bibinfo {author}
  {S.~{Ferraro}}, \bibinfo {author} {E.~F. {Schlafly}},\ and\ \bibinfo {author}
  {M.~{White}},\ }\href {https://doi.org/10.1088/1475-7516/2020/05/047}
  {\bibfield  {journal} {\bibinfo  {journal} {\jcap}\ }\textbf {\bibinfo
  {volume} {2020}},\ \bibinfo {eid} {047} (\bibinfo {year} {2020})},\ \Eprint
  {https://arxiv.org/abs/1909.07412} {arXiv:1909.07412 [astro-ph.CO]}
  \BibitemShut {NoStop}%
\bibitem [{\citenamefont {{White}}\ \emph {et~al.}(2022)\citenamefont
  {{White}}, \citenamefont {{Zhou}}, \citenamefont {{DeRose}}, \citenamefont
  {{Ferraro}}, \citenamefont {{Chen}}, \citenamefont {{Kokron}}, \citenamefont
  {{Bailey}}, \citenamefont {{Brooks}}, \citenamefont {{Garc{\'\i}a-Bellido}},
  \citenamefont {{Guy}}, \citenamefont {{Honscheid}}, \citenamefont {{Kehoe}},
  \citenamefont {{Kremin}}, \citenamefont {{Levi}}, \citenamefont
  {{Palanque-Delabrouille}}, \citenamefont {{Poppett}}, \citenamefont
  {{Schlegel}},\ and\ \citenamefont {{Tarle}}}]{White2022}%
  \BibitemOpen
  \bibfield  {author} {\bibinfo {author} {M.~{White}}, \bibinfo {author}
  {R.~{Zhou}}, \bibinfo {author} {J.~{DeRose}}, \bibinfo {author}
  {S.~{Ferraro}}, \bibinfo {author} {S.-F. {Chen}}, \bibinfo {author}
  {N.~{Kokron}}, \bibinfo {author} {S.~{Bailey}}, \bibinfo {author}
  {D.~{Brooks}}, \bibinfo {author} {J.~{Garc{\'\i}a-Bellido}}, \bibinfo
  {author} {J.~{Guy}}, et~al.,\ }\href
  {https://doi.org/10.1088/1475-7516/2022/02/007} {\bibfield  {journal}
  {\bibinfo  {journal} {\jcap}\ }\textbf {\bibinfo {volume} {2022}},\ \bibinfo
  {eid} {007} (\bibinfo {year} {2022})},\ \Eprint
  {https://arxiv.org/abs/2111.09898} {arXiv:2111.09898 [astro-ph.CO]}
  \BibitemShut {NoStop}%
\bibitem [{\citenamefont {{Dawson}}\ \emph {et~al.}(2013)\citenamefont
  {{Dawson}}, \citenamefont {{Schlegel}}, \citenamefont {{Ahn}}, \citenamefont
  {{Anderson}}, \citenamefont {{Aubourg}}, \citenamefont {{Bailey}},
  \citenamefont {{Barkhouser}}, \citenamefont {{Bautista}}, \citenamefont
  {{Beifiori}}, \citenamefont {{Berlind}}, \citenamefont {{Bhardwaj}},
  \citenamefont {{Bizyaev}}, \citenamefont {{Blake}}, \citenamefont
  {{Blanton}}, \citenamefont {{Blomqvist}}, \citenamefont {{Bolton}},
  \citenamefont {{Borde}}, \citenamefont {{Bovy}}, \citenamefont {{Brandt}},
  \citenamefont {{Brewington}}, \citenamefont {{Brinkmann}}, \citenamefont
  {{Brown}}, \citenamefont {{Brownstein}}, \citenamefont {{Bundy}},
  \citenamefont {{Busca}}, \citenamefont {{Carithers}}, \citenamefont
  {{Carnero}}, \citenamefont {{Carr}}, \citenamefont {{Chen}}, \citenamefont
  {{Comparat}}, \citenamefont {{Connolly}}, \citenamefont {{Cope}},
  \citenamefont {{Croft}}, \citenamefont {{Cuesta}}, \citenamefont {{da
  Costa}}, \citenamefont {{Davenport}}, \citenamefont {{Delubac}},
  \citenamefont {{de Putter}}, \citenamefont {{Dhital}}, \citenamefont
  {{Ealet}}, \citenamefont {{Ebelke}}, \citenamefont {{Eisenstein}},
  \citenamefont {{Escoffier}}, \citenamefont {{Fan}}, \citenamefont {{Filiz
  Ak}}, \citenamefont {{Finley}}, \citenamefont {{Font-Ribera}}, \citenamefont
  {{G{\'e}nova-Santos}}, \citenamefont {{Gunn}}, \citenamefont {{Guo}},
  \citenamefont {{Haggard}}, \citenamefont {{Hall}}, \citenamefont
  {{Hamilton}}, \citenamefont {{Harris}}, \citenamefont {{Harris}},
  \citenamefont {{Ho}}, \citenamefont {{Hogg}}, \citenamefont {{Holder}},
  \citenamefont {{Honscheid}}, \citenamefont {{Huehnerhoff}}, \citenamefont
  {{Jordan}}, \citenamefont {{Jordan}}, \citenamefont {{Kauffmann}},
  \citenamefont {{Kazin}}, \citenamefont {{Kirkby}}, \citenamefont {{Klaene}},
  \citenamefont {{Kneib}}, \citenamefont {{Le Goff}}, \citenamefont {{Lee}},
  \citenamefont {{Long}}, \citenamefont {{Loomis}}, \citenamefont {{Lundgren}},
  \citenamefont {{Lupton}}, \citenamefont {{Maia}}, \citenamefont {{Makler}},
  \citenamefont {{Malanushenko}}, \citenamefont {{Malanushenko}}, \citenamefont
  {{Mandelbaum}}, \citenamefont {{Manera}}, \citenamefont {{Maraston}},
  \citenamefont {{Margala}}, \citenamefont {{Masters}}, \citenamefont
  {{McBride}}, \citenamefont {{McDonald}}, \citenamefont {{McGreer}},
  \citenamefont {{McMahon}}, \citenamefont {{Mena}}, \citenamefont
  {{Miralda-Escud{\'e}}}, \citenamefont {{Montero-Dorta}}, \citenamefont
  {{Montesano}}, \citenamefont {{Muna}}, \citenamefont {{Myers}}, \citenamefont
  {{Naugle}}, \citenamefont {{Nichol}}, \citenamefont {{Noterdaeme}},
  \citenamefont {{Nuza}}, \citenamefont {{Olmstead}}, \citenamefont
  {{Oravetz}}, \citenamefont {{Oravetz}}, \citenamefont {{Owen}}, \citenamefont
  {{Padmanabhan}}, \citenamefont {{Palanque-Delabrouille}}, \citenamefont
  {{Pan}}, \citenamefont {{Parejko}}, \citenamefont {{P{\^a}ris}},
  \citenamefont {{Percival}}, \citenamefont {{P{\'e}rez-Fournon}},
  \citenamefont {{P{\'e}rez-R{\`a}fols}}, \citenamefont {{Petitjean}},
  \citenamefont {{Pfaffenberger}}, \citenamefont {{Pforr}}, \citenamefont
  {{Pieri}}, \citenamefont {{Prada}}, \citenamefont {{Price-Whelan}},
  \citenamefont {{Raddick}}, \citenamefont {{Rebolo}}, \citenamefont {{Rich}},
  \citenamefont {{Richards}}, \citenamefont {{Rockosi}}, \citenamefont {{Roe}},
  \citenamefont {{Ross}}, \citenamefont {{Ross}}, \citenamefont {{Rossi}},
  \citenamefont {{Rubi{\~n}o-Martin}}, \citenamefont {{Samushia}},
  \citenamefont {{S{\'a}nchez}}, \citenamefont {{Sayres}}, \citenamefont
  {{Schmidt}}, \citenamefont {{Schneider}}, \citenamefont {{Sc{\'o}ccola}},
  \citenamefont {{Seo}}, \citenamefont {{Shelden}}, \citenamefont {{Sheldon}},
  \citenamefont {{Shen}}, \citenamefont {{Shu}}, \citenamefont {{Slosar}},
  \citenamefont {{Smee}}, \citenamefont {{Snedden}}, \citenamefont
  {{Stauffer}}, \citenamefont {{Steele}}, \citenamefont {{Strauss}},
  \citenamefont {{Streblyanska}}, \citenamefont {{Suzuki}}, \citenamefont
  {{Swanson}}, \citenamefont {{Tal}}, \citenamefont {{Tanaka}}, \citenamefont
  {{Thomas}}, \citenamefont {{Tinker}}, \citenamefont {{Tojeiro}},
  \citenamefont {{Tremonti}}, \citenamefont {{Vargas Maga{\~n}a}},
  \citenamefont {{Verde}}, \citenamefont {{Viel}}, \citenamefont {{Wake}},
  \citenamefont {{Watson}}, \citenamefont {{Weaver}}, \citenamefont
  {{Weinberg}}, \citenamefont {{Weiner}}, \citenamefont {{West}}, \citenamefont
  {{White}}, \citenamefont {{Wood-Vasey}}, \citenamefont {{Yeche}},
  \citenamefont {{Zehavi}}, \citenamefont {{Zhao}},\ and\ \citenamefont
  {{Zheng}}}]{Dawson2013}%
  \BibitemOpen
  \bibfield  {author} {\bibinfo {author} {K.~S. {Dawson}}, \bibinfo {author}
  {D.~J. {Schlegel}}, \bibinfo {author} {C.~P. {Ahn}}, \bibinfo {author} {S.~F.
  {Anderson}}, \bibinfo {author} {{\'E}.~{Aubourg}}, \bibinfo {author}
  {S.~{Bailey}}, \bibinfo {author} {R.~H. {Barkhouser}}, \bibinfo {author}
  {J.~E. {Bautista}}, \bibinfo {author} {A.~{Beifiori}}, \bibinfo {author}
  {A.~A. {Berlind}}, et~al.,\ }\href
  {https://doi.org/10.1088/0004-6256/145/1/10} {\bibfield  {journal} {\bibinfo
  {journal} {\aj}\ }\textbf {\bibinfo {volume} {145}},\ \bibinfo {eid} {10}
  (\bibinfo {year} {2013})},\ \Eprint {https://arxiv.org/abs/1208.0022}
  {arXiv:1208.0022 [astro-ph.CO]} \BibitemShut {NoStop}%
\bibitem [{\citenamefont {{Eisenstein}}\ \emph {et~al.}(2011)\citenamefont
  {{Eisenstein}}, \citenamefont {{Weinberg}}, \citenamefont {{Agol}},
  \citenamefont {{Aihara}}, \citenamefont {{Allende Prieto}}, \citenamefont
  {{Anderson}}, \citenamefont {{Arns}}, \citenamefont {{Aubourg}},
  \citenamefont {{Bailey}}, \citenamefont {{Balbinot}}, \citenamefont
  {{Barkhouser}}, \citenamefont {{Beers}}, \citenamefont {{Berlind}},
  \citenamefont {{Bickerton}}, \citenamefont {{Bizyaev}}, \citenamefont
  {{Blanton}}, \citenamefont {{Bochanski}}, \citenamefont {{Bolton}},
  \citenamefont {{Bosman}}, \citenamefont {{Bovy}}, \citenamefont {{Brandt}},
  \citenamefont {{Breslauer}}, \citenamefont {{Brewington}}, \citenamefont
  {{Brinkmann}}, \citenamefont {{Brown}}, \citenamefont {{Brownstein}},
  \citenamefont {{Burger}}, \citenamefont {{Busca}}, \citenamefont
  {{Campbell}}, \citenamefont {{Cargile}}, \citenamefont {{Carithers}},
  \citenamefont {{Carlberg}}, \citenamefont {{Carr}}, \citenamefont {{Chang}},
  \citenamefont {{Chen}}, \citenamefont {{Chiappini}}, \citenamefont
  {{Comparat}}, \citenamefont {{Connolly}}, \citenamefont {{Cortes}},
  \citenamefont {{Croft}}, \citenamefont {{Cunha}}, \citenamefont {{da Costa}},
  \citenamefont {{Davenport}}, \citenamefont {{Dawson}}, \citenamefont {{De
  Lee}}, \citenamefont {{Porto de Mello}}, \citenamefont {{de Simoni}},
  \citenamefont {{Dean}}, \citenamefont {{Dhital}}, \citenamefont {{Ealet}},
  \citenamefont {{Ebelke}}, \citenamefont {{Edmondson}}, \citenamefont
  {{Eiting}}, \citenamefont {{Escoffier}}, \citenamefont {{Esposito}},
  \citenamefont {{Evans}}, \citenamefont {{Fan}}, \citenamefont {{Femen{\'\i}a
  Castell{\'a}}}, \citenamefont {{Dutra Ferreira}}, \citenamefont
  {{Fitzgerald}}, \citenamefont {{Fleming}}, \citenamefont {{Font-Ribera}},
  \citenamefont {{Ford}}, \citenamefont {{Frinchaboy}}, \citenamefont
  {{Garc{\'\i}a P{\'e}rez}}, \citenamefont {{Gaudi}}, \citenamefont {{Ge}},
  \citenamefont {{Ghezzi}}, \citenamefont {{Gillespie}}, \citenamefont
  {{Gilmore}}, \citenamefont {{Girardi}}, \citenamefont {{Gott}}, \citenamefont
  {{Gould}}, \citenamefont {{Grebel}}, \citenamefont {{Gunn}}, \citenamefont
  {{Hamilton}}, \citenamefont {{Harding}}, \citenamefont {{Harris}},
  \citenamefont {{Hawley}}, \citenamefont {{Hearty}}, \citenamefont
  {{Hennawi}}, \citenamefont {{Gonz{\'a}lez Hern{\'a}ndez}}, \citenamefont
  {{Ho}}, \citenamefont {{Hogg}}, \citenamefont {{Holtzman}}, \citenamefont
  {{Honscheid}}, \citenamefont {{Inada}}, \citenamefont {{Ivans}},
  \citenamefont {{Jiang}}, \citenamefont {{Jiang}}, \citenamefont {{Johnson}},
  \citenamefont {{Jordan}}, \citenamefont {{Jordan}}, \citenamefont
  {{Kauffmann}}, \citenamefont {{Kazin}}, \citenamefont {{Kirkby}},
  \citenamefont {{Klaene}}, \citenamefont {{Knapp}}, \citenamefont {{Kneib}},
  \citenamefont {{Kochanek}}, \citenamefont {{Koesterke}}, \citenamefont
  {{Kollmeier}}, \citenamefont {{Kron}}, \citenamefont {{Lampeitl}},
  \citenamefont {{Lang}}, \citenamefont {{Lawler}}, \citenamefont {{Le Goff}},
  \citenamefont {{Lee}}, \citenamefont {{Lee}}, \citenamefont {{Leisenring}},
  \citenamefont {{Lin}}, \citenamefont {{Liu}}, \citenamefont {{Long}},
  \citenamefont {{Loomis}}, \citenamefont {{Lucatello}}, \citenamefont
  {{Lundgren}}, \citenamefont {{Lupton}}, \citenamefont {{Ma}}, \citenamefont
  {{Ma}}, \citenamefont {{MacDonald}}, \citenamefont {{Mack}}, \citenamefont
  {{Mahadevan}}, \citenamefont {{Maia}}, \citenamefont {{Majewski}},
  \citenamefont {{Makler}}, \citenamefont {{Malanushenko}}, \citenamefont
  {{Malanushenko}}, \citenamefont {{Mandelbaum}}, \citenamefont {{Maraston}},
  \citenamefont {{Margala}}, \citenamefont {{Maseman}}, \citenamefont
  {{Masters}}, \citenamefont {{McBride}}, \citenamefont {{McDonald}},
  \citenamefont {{McGreer}}, \citenamefont {{McMahon}}, \citenamefont {{Mena
  Requejo}}, \citenamefont {{M{\'e}nard}}, \citenamefont
  {{Miralda-Escud{\'e}}}, \citenamefont {{Morrison}}, \citenamefont
  {{Mullally}}, \citenamefont {{Muna}}, \citenamefont {{Murayama}},
  \citenamefont {{Myers}}, \citenamefont {{Naugle}}, \citenamefont {{Neto}},
  \citenamefont {{Nguyen}}, \citenamefont {{Nichol}}, \citenamefont
  {{Nidever}}, \citenamefont {{O'Connell}}, \citenamefont {{Ogando}},
  \citenamefont {{Olmstead}}, \citenamefont {{Oravetz}}, \citenamefont
  {{Padmanabhan}}, \citenamefont {{Paegert}}, \citenamefont
  {{Palanque-Delabrouille}}, \citenamefont {{Pan}}, \citenamefont {{Pandey}},
  \citenamefont {{Parejko}}, \citenamefont {{P{\^a}ris}}, \citenamefont
  {{Pellegrini}}, \citenamefont {{Pepper}}, \citenamefont {{Percival}},
  \citenamefont {{Petitjean}}, \citenamefont {{Pfaffenberger}}, \citenamefont
  {{Pforr}}, \citenamefont {{Phleps}}, \citenamefont {{Pichon}}, \citenamefont
  {{Pieri}}, \citenamefont {{Prada}}, \citenamefont {{Price-Whelan}},
  \citenamefont {{Raddick}}, \citenamefont {{Ramos}}, \citenamefont {{Reid}},
  \citenamefont {{Reyle}}, \citenamefont {{Rich}}, \citenamefont {{Richards}},
  \citenamefont {{Rieke}}, \citenamefont {{Rieke}}, \citenamefont {{Rix}},
  \citenamefont {{Robin}}, \citenamefont {{Rocha-Pinto}}, \citenamefont
  {{Rockosi}}, \citenamefont {{Roe}}, \citenamefont {{Rollinde}}, \citenamefont
  {{Ross}}, \citenamefont {{Ross}}, \citenamefont {{Rossetto}}, \citenamefont
  {{S{\'a}nchez}}, \citenamefont {{Santiago}}, \citenamefont {{Sayres}},
  \citenamefont {{Schiavon}}, \citenamefont {{Schlegel}}, \citenamefont
  {{Schlesinger}}, \citenamefont {{Schmidt}}, \citenamefont {{Schneider}},
  \citenamefont {{Sellgren}}, \citenamefont {{Shelden}}, \citenamefont
  {{Sheldon}}, \citenamefont {{Shetrone}}, \citenamefont {{Shu}}, \citenamefont
  {{Silverman}}, \citenamefont {{Simmerer}}, \citenamefont {{Simmons}},
  \citenamefont {{Sivarani}}, \citenamefont {{Skrutskie}}, \citenamefont
  {{Slosar}}, \citenamefont {{Smee}}, \citenamefont {{Smith}}, \citenamefont
  {{Snedden}}, \citenamefont {{Stassun}}, \citenamefont {{Steele}},
  \citenamefont {{Steinmetz}}, \citenamefont {{Stockett}}, \citenamefont
  {{Stollberg}}, \citenamefont {{Strauss}}, \citenamefont {{Szalay}},
  \citenamefont {{Tanaka}}, \citenamefont {{Thakar}}, \citenamefont {{Thomas}},
  \citenamefont {{Tinker}}, \citenamefont {{Tofflemire}}, \citenamefont
  {{Tojeiro}}, \citenamefont {{Tremonti}}, \citenamefont {{Vargas Maga{\~n}a}},
  \citenamefont {{Verde}}, \citenamefont {{Vogt}}, \citenamefont {{Wake}},
  \citenamefont {{Wan}}, \citenamefont {{Wang}}, \citenamefont {{Weaver}},
  \citenamefont {{White}}, \citenamefont {{White}}, \citenamefont {{Wilson}},
  \citenamefont {{Wisniewski}}, \citenamefont {{Wood-Vasey}}, \citenamefont
  {{Yanny}}, \citenamefont {{Yasuda}}, \citenamefont {{Y{\`e}che}},
  \citenamefont {{York}}, \citenamefont {{Young}}, \citenamefont {{Zasowski}},
  \citenamefont {{Zehavi}},\ and\ \citenamefont {{Zhao}}}]{Eisenstein2011}%
  \BibitemOpen
  \bibfield  {author} {\bibinfo {author} {D.~J. {Eisenstein}}, \bibinfo
  {author} {D.~H. {Weinberg}}, \bibinfo {author} {E.~{Agol}}, \bibinfo {author}
  {H.~{Aihara}}, \bibinfo {author} {C.~{Allende Prieto}}, \bibinfo {author}
  {S.~F. {Anderson}}, \bibinfo {author} {J.~A. {Arns}}, \bibinfo {author}
  {{\'E}.~{Aubourg}}, \bibinfo {author} {S.~{Bailey}}, \bibinfo {author}
  {E.~{Balbinot}}, et~al.,\ }\href {https://doi.org/10.1088/0004-6256/142/3/72}
  {\bibfield  {journal} {\bibinfo  {journal} {\aj}\ }\textbf {\bibinfo {volume}
  {142}},\ \bibinfo {eid} {72} (\bibinfo {year} {2011})},\ \Eprint
  {https://arxiv.org/abs/1101.1529} {arXiv:1101.1529 [astro-ph.IM]}
  \BibitemShut {NoStop}%
\bibitem [{\citenamefont {{Gil-Mar{\'\i}n}}\ \emph {et~al.}(2016)\citenamefont
  {{Gil-Mar{\'\i}n}}, \citenamefont {{Percival}}, \citenamefont {{Brownstein}},
  \citenamefont {{Chuang}}, \citenamefont {{Grieb}}, \citenamefont {{Ho}},
  \citenamefont {{Kitaura}}, \citenamefont {{Maraston}}, \citenamefont
  {{Prada}}, \citenamefont {{Rodr{\'\i}guez-Torres}}, \citenamefont {{Ross}},
  \citenamefont {{Samushia}}, \citenamefont {{Schlegel}}, \citenamefont
  {{Thomas}}, \citenamefont {{Tinker}},\ and\ \citenamefont
  {{Zhao}}}]{GilMarin2016}%
  \BibitemOpen
  \bibfield  {author} {\bibinfo {author} {H.~{Gil-Mar{\'\i}n}}, \bibinfo
  {author} {W.~J. {Percival}}, \bibinfo {author} {J.~R. {Brownstein}}, \bibinfo
  {author} {C.-H. {Chuang}}, \bibinfo {author} {J.~N. {Grieb}}, \bibinfo
  {author} {S.~{Ho}}, \bibinfo {author} {F.-S. {Kitaura}}, \bibinfo {author}
  {C.~{Maraston}}, \bibinfo {author} {F.~{Prada}}, \bibinfo {author}
  {S.~{Rodr{\'\i}guez-Torres}}, et~al.,\ }\href
  {https://doi.org/10.1093/mnras/stw1096} {\bibfield  {journal} {\bibinfo
  {journal} {\mnras}\ }\textbf {\bibinfo {volume} {460}},\ \bibinfo {pages}
  {4188} (\bibinfo {year} {2016})},\ \Eprint {https://arxiv.org/abs/1509.06386}
  {arXiv:1509.06386 [astro-ph.CO]} \BibitemShut {NoStop}%
\bibitem [{\citenamefont {{Chuang}}\ \emph {et~al.}(2017)\citenamefont
  {{Chuang}}, \citenamefont {{Pellejero-Ibanez}}, \citenamefont
  {{Rodr{\'\i}guez-Torres}}, \citenamefont {{Ross}}, \citenamefont {{Zhao}},
  \citenamefont {{Wang}}, \citenamefont {{Cuesta}}, \citenamefont
  {{Rubi{\~n}o-Mart{\'\i}n}}, \citenamefont {{Prada}}, \citenamefont {{Alam}},
  \citenamefont {{Beutler}}, \citenamefont {{Eisenstein}}, \citenamefont
  {{Gil-Mar{\'\i}n}}, \citenamefont {{Grieb}}, \citenamefont {{Ho}},
  \citenamefont {{Kitaura}}, \citenamefont {{Percival}}, \citenamefont
  {{Rossi}}, \citenamefont {{Salazar-Albornoz}}, \citenamefont {{Samushia}},
  \citenamefont {{S{\'a}nchez}}, \citenamefont {{Satpathy}}, \citenamefont
  {{Slosar}}, \citenamefont {{Thomas}}, \citenamefont {{Tinker}}, \citenamefont
  {{Tojeiro}}, \citenamefont {{Vargas-Maga{\~n}a}}, \citenamefont {{Vazquez}},
  \citenamefont {{Brownstein}}, \citenamefont {{Nichol}},\ and\ \citenamefont
  {{Olmstead}}}]{Chuang2017}%
  \BibitemOpen
  \bibfield  {author} {\bibinfo {author} {C.-H. {Chuang}}, \bibinfo {author}
  {M.~{Pellejero-Ibanez}}, \bibinfo {author} {S.~{Rodr{\'\i}guez-Torres}},
  \bibinfo {author} {A.~J. {Ross}}, \bibinfo {author} {G.-b. {Zhao}}, \bibinfo
  {author} {Y.~{Wang}}, \bibinfo {author} {A.~J. {Cuesta}}, \bibinfo {author}
  {J.~A. {Rubi{\~n}o-Mart{\'\i}n}}, \bibinfo {author} {F.~{Prada}}, \bibinfo
  {author} {S.~{Alam}}, et~al.,\ }\href {https://doi.org/10.1093/mnras/stx1641}
  {\bibfield  {journal} {\bibinfo  {journal} {\mnras}\ }\textbf {\bibinfo
  {volume} {471}},\ \bibinfo {pages} {2370} (\bibinfo {year} {2017})},\ \Eprint
  {https://arxiv.org/abs/1607.03151} {arXiv:1607.03151 [astro-ph.CO]}
  \BibitemShut {NoStop}%
\bibitem [{\citenamefont {{Anderson}}\ \emph {et~al.}(2014)\citenamefont
  {{Anderson}}, \citenamefont {{Aubourg}}, \citenamefont {{Bailey}},
  \citenamefont {{Beutler}}, \citenamefont {{Bhardwaj}}, \citenamefont
  {{Blanton}}, \citenamefont {{Bolton}}, \citenamefont {{Brinkmann}},
  \citenamefont {{Brownstein}}, \citenamefont {{Burden}}, \citenamefont
  {{Chuang}}, \citenamefont {{Cuesta}}, \citenamefont {{Dawson}}, \citenamefont
  {{Eisenstein}}, \citenamefont {{Escoffier}}, \citenamefont {{Gunn}},
  \citenamefont {{Guo}}, \citenamefont {{Ho}}, \citenamefont {{Honscheid}},
  \citenamefont {{Howlett}}, \citenamefont {{Kirkby}}, \citenamefont
  {{Lupton}}, \citenamefont {{Manera}}, \citenamefont {{Maraston}},
  \citenamefont {{McBride}}, \citenamefont {{Mena}}, \citenamefont
  {{Montesano}}, \citenamefont {{Nichol}}, \citenamefont {{Nuza}},
  \citenamefont {{Olmstead}}, \citenamefont {{Padmanabhan}}, \citenamefont
  {{Palanque-Delabrouille}}, \citenamefont {{Parejko}}, \citenamefont
  {{Percival}}, \citenamefont {{Petitjean}}, \citenamefont {{Prada}},
  \citenamefont {{Price-Whelan}}, \citenamefont {{Reid}}, \citenamefont
  {{Roe}}, \citenamefont {{Ross}}, \citenamefont {{Ross}}, \citenamefont
  {{Sabiu}}, \citenamefont {{Saito}}, \citenamefont {{Samushia}}, \citenamefont
  {{S{\'a}nchez}}, \citenamefont {{Schlegel}}, \citenamefont {{Schneider}},
  \citenamefont {{Scoccola}}, \citenamefont {{Seo}}, \citenamefont {{Skibba}},
  \citenamefont {{Strauss}}, \citenamefont {{Swanson}}, \citenamefont
  {{Thomas}}, \citenamefont {{Tinker}}, \citenamefont {{Tojeiro}},
  \citenamefont {{Maga{\~n}a}}, \citenamefont {{Verde}}, \citenamefont
  {{Wake}}, \citenamefont {{Weaver}}, \citenamefont {{Weinberg}}, \citenamefont
  {{White}}, \citenamefont {{Xu}}, \citenamefont {{Y{\`e}che}}, \citenamefont
  {{Zehavi}},\ and\ \citenamefont {{Zhao}}}]{Anderson2014}%
  \BibitemOpen
  \bibfield  {author} {\bibinfo {author} {L.~{Anderson}}, \bibinfo {author}
  {{\'E}.~{Aubourg}}, \bibinfo {author} {S.~{Bailey}}, \bibinfo {author}
  {F.~{Beutler}}, \bibinfo {author} {V.~{Bhardwaj}}, \bibinfo {author}
  {M.~{Blanton}}, \bibinfo {author} {A.~S. {Bolton}}, \bibinfo {author}
  {J.~{Brinkmann}}, \bibinfo {author} {J.~R. {Brownstein}}, \bibinfo {author}
  {A.~{Burden}}, et~al.,\ }\href {https://doi.org/10.1093/mnras/stu523}
  {\bibfield  {journal} {\bibinfo  {journal} {\mnras}\ }\textbf {\bibinfo
  {volume} {441}},\ \bibinfo {pages} {24} (\bibinfo {year} {2014})},\ \Eprint
  {https://arxiv.org/abs/1312.4877} {arXiv:1312.4877 [astro-ph.CO]}
  \BibitemShut {NoStop}%
\bibitem [{\citenamefont {{Nicola}}\ \emph {et~al.}(2021)\citenamefont
  {{Nicola}}, \citenamefont {{Garc{\'\i}a-Garc{\'\i}a}}, \citenamefont
  {{Alonso}}, \citenamefont {{Dunkley}}, \citenamefont {{Ferreira}},
  \citenamefont {{Slosar}},\ and\ \citenamefont {{Spergel}}}]{Nicola2021}%
  \BibitemOpen
  \bibfield  {author} {\bibinfo {author} {A.~{Nicola}}, \bibinfo {author}
  {C.~{Garc{\'\i}a-Garc{\'\i}a}}, \bibinfo {author} {D.~{Alonso}}, \bibinfo
  {author} {J.~{Dunkley}}, \bibinfo {author} {P.~G. {Ferreira}}, \bibinfo
  {author} {A.~{Slosar}},\ and\ \bibinfo {author} {D.~N. {Spergel}},\ }\href
  {https://doi.org/10.1088/1475-7516/2021/03/067} {\bibfield  {journal}
  {\bibinfo  {journal} {\jcap}\ }\textbf {\bibinfo {volume} {2021}},\ \bibinfo
  {eid} {067} (\bibinfo {year} {2021})},\ \Eprint
  {https://arxiv.org/abs/2010.09717} {arXiv:2010.09717 [astro-ph.CO]}
  \BibitemShut {NoStop}%
\bibitem [{\citenamefont {{Marques}}\ \emph {et~al.}(2023)\citenamefont
  {{Marques}}, \citenamefont {{Madhavacheril}}, \citenamefont {{Darwish}},
  \citenamefont {{Shaikh}}, \citenamefont {{Aguena}}, \citenamefont {{Alves}},
  \citenamefont {{Avila}}, \citenamefont {{Bacon}}, \citenamefont {{Baxter}},
  \citenamefont {{Bechtol}}, \citenamefont {{Becker}}, \citenamefont
  {{Bertin}}, \citenamefont {{Blazek}}, \citenamefont {{Bond}}, \citenamefont
  {{Brooks}}, \citenamefont {{Cai}}, \citenamefont {{Calabrese}}, \citenamefont
  {{Carnero Rosell}}, \citenamefont {{Carretero}}, \citenamefont {{Cawthon}},
  \citenamefont {{Crocce}}, \citenamefont {{da Costa}}, \citenamefont
  {{Pereira}}, \citenamefont {{De Vicente}}, \citenamefont {{Desai}},
  \citenamefont {{Diehl}}, \citenamefont {{Doel}}, \citenamefont {{Doux}},
  \citenamefont {{Drlica-Wagner}}, \citenamefont {{Dunkley}}, \citenamefont
  {{Elvin-Poole}}, \citenamefont {{Everett}}, \citenamefont {{Ferraro}},
  \citenamefont {{Ferrero}}, \citenamefont {{Flaugher}}, \citenamefont
  {{Fosalba}}, \citenamefont {{Garc{\'\i}a-Bellido}}, \citenamefont {{Gatti}},
  \citenamefont {{Giannini}}, \citenamefont {{Gluscevic}}, \citenamefont
  {{Gruen}}, \citenamefont {{Gruendl}}, \citenamefont {{Gutierrez}},
  \citenamefont {{Harrison}}, \citenamefont {{Hill}}, \citenamefont {{Hinton}},
  \citenamefont {{Hollowood}}, \citenamefont {{Honscheid}}, \citenamefont
  {{Huterer}}, \citenamefont {{Jeffrey}}, \citenamefont {{Kim}}, \citenamefont
  {{Kuehn}}, \citenamefont {{Lahav}}, \citenamefont {{Lemos}}, \citenamefont
  {{Lima}}, \citenamefont {{Huffenberger}}, \citenamefont {{MacCrann}},
  \citenamefont {{Marshall}}, \citenamefont {{Mena-Fern{\'a}ndez}},
  \citenamefont {{Miquel}}, \citenamefont {{Mohr}}, \citenamefont {{Moodley}},
  \citenamefont {{Muir}}, \citenamefont {{Naess}}, \citenamefont {{Nati}},
  \citenamefont {{Page}}, \citenamefont {{Palmese}}, \citenamefont {{Plazas
  Malag{\'o}n}}, \citenamefont {{Porredon}}, \citenamefont {{Prat}},
  \citenamefont {{Qu}}, \citenamefont {{Raveri}}, \citenamefont {{Ross}},
  \citenamefont {{Rykoff}}, \citenamefont {{Farren}}, \citenamefont
  {{Samuroff}}, \citenamefont {{Sanchez}}, \citenamefont {{Schubnell}},
  \citenamefont {{Sevilla-Noarbe}}, \citenamefont {{Sheldon}}, \citenamefont
  {{Sherwin}}, \citenamefont {{Sif{\'o}n}}, \citenamefont {{Smith}},
  \citenamefont {{Spergel}}, \citenamefont {{Staggs}}, \citenamefont
  {{Suchyta}}, \citenamefont {{Tarle}}, \citenamefont {{To}}, \citenamefont
  {{Van Engelen}}, \citenamefont {{Weaverdyck}}, \citenamefont {{Weller}},
  \citenamefont {{Wenzl}}, \citenamefont {{Wiseman}}, \citenamefont
  {{Wollack}},\ and\ \citenamefont {{Yanny}}}]{Marques2023}%
  \BibitemOpen
  \bibfield  {author} {\bibinfo {author} {G.~A. {Marques}}, \bibinfo {author}
  {M.~S. {Madhavacheril}}, \bibinfo {author} {O.~{Darwish}}, \bibinfo {author}
  {S.~{Shaikh}}, \bibinfo {author} {M.~{Aguena}}, \bibinfo {author}
  {O.~{Alves}}, \bibinfo {author} {S.~{Avila}}, \bibinfo {author} {D.~{Bacon}},
  \bibinfo {author} {E.~J. {Baxter}}, \bibinfo {author} {K.~{Bechtol}},
  et~al.,\ }\href {https://doi.org/10.48550/arXiv.2306.17268} {\bibfield
  {journal} {\bibinfo  {journal} {arXiv e-prints}\ ,\ \bibinfo {eid}
  {arXiv:2306.17268}} (\bibinfo {year} {2023})},\ \Eprint
  {https://arxiv.org/abs/2306.17268} {arXiv:2306.17268 [astro-ph.CO]}
  \BibitemShut {NoStop}%
\bibitem [{\citenamefont {{Chisari}}\ \emph {et~al.}(2019)\citenamefont
  {{Chisari}}, \citenamefont {{Alonso}}, \citenamefont {{Krause}},
  \citenamefont {{Leonard}}, \citenamefont {{Bull}}, \citenamefont {{Neveu}},
  \citenamefont {{Villarreal}}, \citenamefont {{Singh}}, \citenamefont
  {{McClintock}}, \citenamefont {{Ellison}}, \citenamefont {{Du}},
  \citenamefont {{Zuntz}}, \citenamefont {{Mead}}, \citenamefont {{Joudaki}},
  \citenamefont {{Lorenz}}, \citenamefont {{Tr{\"o}ster}}, \citenamefont
  {{Sanchez}}, \citenamefont {{Lanusse}}, \citenamefont {{Ishak}},
  \citenamefont {{Hlozek}}, \citenamefont {{Blazek}}, \citenamefont
  {{Campagne}}, \citenamefont {{Almoubayyed}}, \citenamefont {{Eifler}},
  \citenamefont {{Kirby}}, \citenamefont {{Kirkby}}, \citenamefont
  {{Plaszczynski}}, \citenamefont {{Slosar}}, \citenamefont {{Vrastil}},
  \citenamefont {{Wagoner}},\ and\ \citenamefont {{LSST Dark Energy Science
  Collaboration}}}]{Chisari2019}%
  \BibitemOpen
  \bibfield  {author} {\bibinfo {author} {N.~E. {Chisari}}, \bibinfo {author}
  {D.~{Alonso}}, \bibinfo {author} {E.~{Krause}}, \bibinfo {author} {C.~D.
  {Leonard}}, \bibinfo {author} {P.~{Bull}}, \bibinfo {author} {J.~{Neveu}},
  \bibinfo {author} {A.~S. {Villarreal}}, \bibinfo {author} {S.~{Singh}},
  \bibinfo {author} {T.~{McClintock}}, \bibinfo {author} {J.~{Ellison}},
  et~al.,\ }\href {https://doi.org/10.3847/1538-4365/ab1658} {\bibfield
  {journal} {\bibinfo  {journal} {\apjs}\ }\textbf {\bibinfo {volume} {242}},\
  \bibinfo {eid} {2} (\bibinfo {year} {2019})},\ \Eprint
  {https://arxiv.org/abs/1812.05995} {arXiv:1812.05995 [astro-ph.CO]}
  \BibitemShut {NoStop}%
\bibitem [{\citenamefont {{Lewis}}\ \emph {et~al.}(2000)\citenamefont
  {{Lewis}}, \citenamefont {{Challinor}},\ and\ \citenamefont
  {{Lasenby}}}]{Lewis2000}%
  \BibitemOpen
  \bibfield  {author} {\bibinfo {author} {A.~{Lewis}}, \bibinfo {author}
  {A.~{Challinor}},\ and\ \bibinfo {author} {A.~{Lasenby}},\ }\href
  {https://doi.org/10.1086/309179} {\bibfield  {journal} {\bibinfo  {journal}
  {\apj}\ }\textbf {\bibinfo {volume} {538}},\ \bibinfo {pages} {473} (\bibinfo
  {year} {2000})},\ \Eprint {https://arxiv.org/abs/astro-ph/9911177}
  {arXiv:astro-ph/9911177 [astro-ph]} \BibitemShut {NoStop}%
\bibitem [{\citenamefont {{Howlett}}\ \emph {et~al.}(2012)\citenamefont
  {{Howlett}}, \citenamefont {{Lewis}}, \citenamefont {{Hall}},\ and\
  \citenamefont {{Challinor}}}]{Howlett2012}%
  \BibitemOpen
  \bibfield  {author} {\bibinfo {author} {C.~{Howlett}}, \bibinfo {author}
  {A.~{Lewis}}, \bibinfo {author} {A.~{Hall}},\ and\ \bibinfo {author}
  {A.~{Challinor}},\ }\href {https://doi.org/10.1088/1475-7516/2012/04/027}
  {\bibfield  {journal} {\bibinfo  {journal} {\jcap}\ }\textbf {\bibinfo
  {volume} {2012}},\ \bibinfo {eid} {027} (\bibinfo {year} {2012})},\ \Eprint
  {https://arxiv.org/abs/1201.3654} {arXiv:1201.3654 [astro-ph.CO]}
  \BibitemShut {NoStop}%
\bibitem [{\citenamefont {{Farren}}\ \emph {et~al.}(2023)\citenamefont
  {{Farren}}, \citenamefont {{Krolewski}}, \citenamefont {{MacCrann}},
  \citenamefont {{Ferraro}}, \citenamefont {{Abril-Cabezas}}, \citenamefont
  {{An}}, \citenamefont {{Atkins}}, \citenamefont {{Battaglia}}, \citenamefont
  {{Bond}}, \citenamefont {{Calabrese}}, \citenamefont {{Choi}}, \citenamefont
  {{Darwish}}, \citenamefont {{Devlin}}, \citenamefont {{Duivenvoorden}},
  \citenamefont {{Dunkley}}, \citenamefont {{Hill}}, \citenamefont {{Hilton}},
  \citenamefont {{Huffenberger}}, \citenamefont {{Kim}}, \citenamefont
  {{Louis}}, \citenamefont {{Madhavacheril}}, \citenamefont {{Marques}},
  \citenamefont {{Moodley}}, \citenamefont {{Page}}, \citenamefont
  {{Partridge}}, \citenamefont {{Qu}}, \citenamefont {{Sehgal}}, \citenamefont
  {{Sherwin}}, \citenamefont {{Sif{\'o}n}}, \citenamefont {{Staggs}},
  \citenamefont {{Van Engelen}}, \citenamefont {{Vargas}}, \citenamefont
  {{Wenzl}}, \citenamefont {{White}},\ and\ \citenamefont
  {{Wollack}}}]{Farren2023}%
  \BibitemOpen
  \bibfield  {author} {\bibinfo {author} {G.~S. {Farren}}, \bibinfo {author}
  {A.~{Krolewski}}, \bibinfo {author} {N.~{MacCrann}}, \bibinfo {author}
  {S.~{Ferraro}}, \bibinfo {author} {I.~{Abril-Cabezas}}, \bibinfo {author}
  {R.~{An}}, \bibinfo {author} {Z.~{Atkins}}, \bibinfo {author}
  {N.~{Battaglia}}, \bibinfo {author} {J.~R. {Bond}}, \bibinfo {author}
  {E.~{Calabrese}}, et~al.,\ }\href {https://doi.org/10.48550/arXiv.2309.05659}
  {\bibfield  {journal} {\bibinfo  {journal} {arXiv e-prints}\ ,\ \bibinfo
  {eid} {arXiv:2309.05659}} (\bibinfo {year} {2023})},\ \Eprint
  {https://arxiv.org/abs/2309.05659} {arXiv:2309.05659 [astro-ph.CO]}
  \BibitemShut {NoStop}%
\bibitem [{\citenamefont {{Hildebrandt}}\ \emph {et~al.}(2009)\citenamefont
  {{Hildebrandt}}, \citenamefont {{van Waerbeke}},\ and\ \citenamefont
  {{Erben}}}]{Hildebrandt2009}%
  \BibitemOpen
  \bibfield  {author} {\bibinfo {author} {H.~{Hildebrandt}}, \bibinfo {author}
  {L.~{van Waerbeke}},\ and\ \bibinfo {author} {T.~{Erben}},\ }\href
  {https://doi.org/10.1051/0004-6361/200912655} {\bibfield  {journal} {\bibinfo
   {journal} {\aap}\ }\textbf {\bibinfo {volume} {507}},\ \bibinfo {pages}
  {683} (\bibinfo {year} {2009})},\ \Eprint {https://arxiv.org/abs/0906.1580}
  {arXiv:0906.1580 [astro-ph.CO]} \BibitemShut {NoStop}%
\bibitem [{\citenamefont {Schmidt}\ \emph {et~al.}(2012)\citenamefont
  {Schmidt}, \citenamefont {Leauthaud}, \citenamefont {Massey}, \citenamefont
  {Rhodes}, \citenamefont {George}, \citenamefont {Koekemoer}, \citenamefont
  {Finoguenov},\ and\ \citenamefont {Tanaka}}]{Schmidt2012}%
  \BibitemOpen
  \bibfield  {author} {\bibinfo {author} {F.~Schmidt}, \bibinfo {author}
  {A.~Leauthaud}, \bibinfo {author} {R.~Massey}, \bibinfo {author} {J.~Rhodes},
  \bibinfo {author} {M.~R. George}, \bibinfo {author} {A.~M. Koekemoer},
  \bibinfo {author} {A.~Finoguenov},\ and\ \bibinfo {author} {M.~Tanaka},\
  }\href {https://doi.org/10.1088/2041-8205/744/2/L22} {\bibfield  {journal}
  {\bibinfo  {journal} {Astrophys. J. Lett.}\ }\textbf {\bibinfo {volume}
  {744}},\ \bibinfo {pages} {L22} (\bibinfo {year} {2012})},\ \Eprint
  {https://arxiv.org/abs/1111.3679} {arXiv:1111.3679 [astro-ph.CO]}
  \BibitemShut {NoStop}%
\bibitem [{\citenamefont {Duncan}\ \emph {et~al.}(2014)\citenamefont {Duncan},
  \citenamefont {Joachimi}, \citenamefont {Heavens}, \citenamefont {Heymans},\
  and\ \citenamefont {Hildebrandt}}]{Duncan2014}%
  \BibitemOpen
  \bibfield  {author} {\bibinfo {author} {C.~Duncan}, \bibinfo {author}
  {B.~Joachimi}, \bibinfo {author} {A.~Heavens}, \bibinfo {author}
  {C.~Heymans},\ and\ \bibinfo {author} {H.~Hildebrandt},\ }\href
  {https://doi.org/10.1093/mnras/stt2060} {\bibfield  {journal} {\bibinfo
  {journal} {Mon. Not. Roy. Astron. Soc.}\ }\textbf {\bibinfo {volume} {437}},\
  \bibinfo {pages} {2471} (\bibinfo {year} {2014})},\ \Eprint
  {https://arxiv.org/abs/1306.6870} {arXiv:1306.6870 [astro-ph.CO]}
  \BibitemShut {NoStop}%
\bibitem [{\citenamefont {Hildebrandt}(2016)}]{Hildebrandt2016}%
  \BibitemOpen
  \bibfield  {author} {\bibinfo {author} {H.~Hildebrandt},\ }\href
  {https://doi.org/10.1093/mnras/stv2575} {\bibfield  {journal} {\bibinfo
  {journal} {Mon. Not. Roy. Astron. Soc.}\ }\textbf {\bibinfo {volume} {455}},\
  \bibinfo {pages} {3943} (\bibinfo {year} {2016})},\ \Eprint
  {https://arxiv.org/abs/1511.01352} {arXiv:1511.01352 [astro-ph.GA]}
  \BibitemShut {NoStop}%
\bibitem [{\citenamefont {Thiele}\ \emph {et~al.}(2020)\citenamefont {Thiele},
  \citenamefont {Duncan},\ and\ \citenamefont {Alonso}}]{Thiele2020}%
  \BibitemOpen
  \bibfield  {author} {\bibinfo {author} {L.~Thiele}, \bibinfo {author}
  {C.~A.~J. Duncan},\ and\ \bibinfo {author} {D.~Alonso},\ }\href
  {https://doi.org/10.1093/mnras/stz3103} {\bibfield  {journal} {\bibinfo
  {journal} {Mon. Not. Roy. Astron. Soc.}\ }\textbf {\bibinfo {volume} {491}},\
  \bibinfo {pages} {1746} (\bibinfo {year} {2020})},\ \Eprint
  {https://arxiv.org/abs/1907.13205} {arXiv:1907.13205 [astro-ph.CO]}
  \BibitemShut {NoStop}%
\bibitem [{\citenamefont {{Unruh}}\ \emph {et~al.}(2020)\citenamefont
  {{Unruh}}, \citenamefont {{Schneider}}, \citenamefont {{Hilbert}},
  \citenamefont {{Simon}}, \citenamefont {{Martin}},\ and\ \citenamefont
  {{Puertas}}}]{Unruh2020}%
  \BibitemOpen
  \bibfield  {author} {\bibinfo {author} {S.~{Unruh}}, \bibinfo {author}
  {P.~{Schneider}}, \bibinfo {author} {S.~{Hilbert}}, \bibinfo {author}
  {P.~{Simon}}, \bibinfo {author} {S.~{Martin}},\ and\ \bibinfo {author} {J.~C.
  {Puertas}},\ }\href {https://doi.org/10.1051/0004-6361/201936915} {\bibfield
  {journal} {\bibinfo  {journal} {\aap}\ }\textbf {\bibinfo {volume} {638}},\
  \bibinfo {eid} {A96} (\bibinfo {year} {2020})},\ \Eprint
  {https://arxiv.org/abs/1910.06400} {arXiv:1910.06400 [astro-ph.CO]}
  \BibitemShut {NoStop}%
\bibitem [{\citenamefont {{Maartens}}\ \emph {et~al.}(2021)\citenamefont
  {{Maartens}}, \citenamefont {{Fonseca}}, \citenamefont {{Camera}},
  \citenamefont {{Jolicoeur}}, \citenamefont {{Viljoen}},\ and\ \citenamefont
  {{Clarkson}}}]{Maartens2021}%
  \BibitemOpen
  \bibfield  {author} {\bibinfo {author} {R.~{Maartens}}, \bibinfo {author}
  {J.~{Fonseca}}, \bibinfo {author} {S.~{Camera}}, \bibinfo {author}
  {S.~{Jolicoeur}}, \bibinfo {author} {J.-A. {Viljoen}},\ and\ \bibinfo
  {author} {C.~{Clarkson}},\ }\href
  {https://doi.org/10.1088/1475-7516/2021/12/009} {\bibfield  {journal}
  {\bibinfo  {journal} {\jcap}\ }\textbf {\bibinfo {volume} {2021}},\ \bibinfo
  {eid} {009} (\bibinfo {year} {2021})},\ \Eprint
  {https://arxiv.org/abs/2107.13401} {arXiv:2107.13401 [astro-ph.CO]}
  \BibitemShut {NoStop}%
\bibitem [{\citenamefont {{von Wietersheim-Kramsta}}\ \emph
  {et~al.}(2021)\citenamefont {{von Wietersheim-Kramsta}}, \citenamefont
  {{Joachimi}}, \citenamefont {{van den Busch}}, \citenamefont {{Heymans}},
  \citenamefont {{Hildebrandt}}, \citenamefont {{Asgari}}, \citenamefont
  {{Tr{\"o}ster}}, \citenamefont {{Unruh}},\ and\ \citenamefont
  {{Wright}}}]{vonWietersheimKramsta2021}%
  \BibitemOpen
  \bibfield  {author} {\bibinfo {author} {M.~{von Wietersheim-Kramsta}},
  \bibinfo {author} {B.~{Joachimi}}, \bibinfo {author} {J.~L. {van den Busch}},
  \bibinfo {author} {C.~{Heymans}}, \bibinfo {author} {H.~{Hildebrandt}},
  \bibinfo {author} {M.~{Asgari}}, \bibinfo {author} {T.~{Tr{\"o}ster}},
  \bibinfo {author} {S.~{Unruh}},\ and\ \bibinfo {author} {A.~H. {Wright}},\
  }\href {https://doi.org/10.1093/mnras/stab1000} {\bibfield  {journal}
  {\bibinfo  {journal} {\mnras}\ }\textbf {\bibinfo {volume} {504}},\ \bibinfo
  {pages} {1452} (\bibinfo {year} {2021})},\ \Eprint
  {https://arxiv.org/abs/2101.05261} {arXiv:2101.05261 [astro-ph.CO]}
  \BibitemShut {NoStop}%
\bibitem [{\citenamefont {{Duncan}}\ \emph {et~al.}(2022)\citenamefont
  {{Duncan}}, \citenamefont {{Harnois-D{\'e}raps}}, \citenamefont {{Miller}},\
  and\ \citenamefont {{Langedijk}}}]{Duncan2022}%
  \BibitemOpen
  \bibfield  {author} {\bibinfo {author} {C.~A.~J. {Duncan}}, \bibinfo {author}
  {J.~{Harnois-D{\'e}raps}}, \bibinfo {author} {L.~{Miller}},\ and\ \bibinfo
  {author} {A.~{Langedijk}},\ }\href {https://doi.org/10.1093/mnras/stac1809}
  {\bibfield  {journal} {\bibinfo  {journal} {\mnras}\ }\textbf {\bibinfo
  {volume} {515}},\ \bibinfo {pages} {1130} (\bibinfo {year} {2022})},\ \Eprint
  {https://arxiv.org/abs/2111.09867} {arXiv:2111.09867 [astro-ph.CO]}
  \BibitemShut {NoStop}%
\bibitem [{\citenamefont {{Euclid Collaboration}}\ \emph
  {et~al.}(2022)\citenamefont {{Euclid Collaboration}}, \citenamefont
  {{Lepori}}, \citenamefont {{Tutusaus}}, \citenamefont {{Viglione}},
  \citenamefont {{Bonvin}}, \citenamefont {{Camera}}, \citenamefont
  {{Castander}}, \citenamefont {{Durrer}}, \citenamefont {{Fosalba}},
  \citenamefont {{Jelic-Cizmek}}, \citenamefont {{Kunz}}, \citenamefont
  {{Adamek}}, \citenamefont {{Casas}}, \citenamefont {{Martinelli}},
  \citenamefont {{Sakr}}, \citenamefont {{Sapone}}, \citenamefont {{Amara}},
  \citenamefont {{Auricchio}}, \citenamefont {{Bodendorf}}, \citenamefont
  {{Bonino}}, \citenamefont {{Branchini}}, \citenamefont {{Brescia}},
  \citenamefont {{Brinchmann}}, \citenamefont {{Capobianco}}, \citenamefont
  {{Carbone}}, \citenamefont {{Carretero}}, \citenamefont {{Castellano}},
  \citenamefont {{Cavuoti}}, \citenamefont {{Cimatti}}, \citenamefont
  {{Cledassou}}, \citenamefont {{Congedo}}, \citenamefont {{Conselice}},
  \citenamefont {{Conversi}}, \citenamefont {{Copin}}, \citenamefont
  {{Corcione}}, \citenamefont {{Courbin}}, \citenamefont {{Da Silva}},
  \citenamefont {{Degaudenzi}}, \citenamefont {{Douspis}}, \citenamefont
  {{Dubath}}, \citenamefont {{Dupac}}, \citenamefont {{Dusini}}, \citenamefont
  {{Ealet}}, \citenamefont {{Farrens}}, \citenamefont {{Ferriol}},
  \citenamefont {{Franceschi}}, \citenamefont {{Fumana}}, \citenamefont
  {{Garilli}}, \citenamefont {{Gillard}}, \citenamefont {{Gillis}},
  \citenamefont {{Giocoli}}, \citenamefont {{Grazian}}, \citenamefont
  {{Grupp}}, \citenamefont {{Guzzo}}, \citenamefont {{Haugan}}, \citenamefont
  {{Holmes}}, \citenamefont {{Hormuth}}, \citenamefont {{Hudelot}},
  \citenamefont {{Jahnke}}, \citenamefont {{Kermiche}}, \citenamefont
  {{Kiessling}}, \citenamefont {{Kilbinger}}, \citenamefont {{Kitching}},
  \citenamefont {{K{\"u}mmel}}, \citenamefont {{Kurki-Suonio}}, \citenamefont
  {{Ligori}}, \citenamefont {{Lilje}}, \citenamefont {{Lloro}}, \citenamefont
  {{Mansutti}}, \citenamefont {{Marggraf}}, \citenamefont {{Markovic}},
  \citenamefont {{Marulli}}, \citenamefont {{Massey}}, \citenamefont
  {{Maurogordato}}, \citenamefont {{Melchior}}, \citenamefont {{Meneghetti}},
  \citenamefont {{Merlin}}, \citenamefont {{Meylan}}, \citenamefont
  {{Moresco}}, \citenamefont {{Moscardini}}, \citenamefont {{Munari}},
  \citenamefont {{Nakajima}}, \citenamefont {{Niemi}}, \citenamefont
  {{Padilla}}, \citenamefont {{Paltani}}, \citenamefont {{Pasian}},
  \citenamefont {{Pedersen}}, \citenamefont {{Percival}}, \citenamefont
  {{Pettorino}}, \citenamefont {{Pires}}, \citenamefont {{Poncet}},
  \citenamefont {{Popa}}, \citenamefont {{Pozzetti}}, \citenamefont {{Raison}},
  \citenamefont {{Rhodes}}, \citenamefont {{Roncarelli}}, \citenamefont
  {{Rossetti}}, \citenamefont {{Saglia}}, \citenamefont {{Schneider}},
  \citenamefont {{Secroun}}, \citenamefont {{Seidel}}, \citenamefont
  {{Serrano}}, \citenamefont {{Sirignano}}, \citenamefont {{Sirri}},
  \citenamefont {{Stanco}}, \citenamefont {{Starck}}, \citenamefont
  {{Tallada-Cresp{\'\i}}}, \citenamefont {{Taylor}}, \citenamefont {{Tereno}},
  \citenamefont {{Toledo-Moreo}}, \citenamefont {{Torradeflot}}, \citenamefont
  {{Valentijn}}, \citenamefont {{Valenziano}}, \citenamefont {{Wang}},
  \citenamefont {{Weller}}, \citenamefont {{Zamorani}}, \citenamefont
  {{Zoubian}}, \citenamefont {{Andreon}}, \citenamefont {{Bardelli}},
  \citenamefont {{Fabbian}}, \citenamefont {{Graci{\'a}-Carpio}}, \citenamefont
  {{Maino}}, \citenamefont {{Medinaceli}}, \citenamefont {{Mei}}, \citenamefont
  {{Renzi}}, \citenamefont {{Romelli}}, \citenamefont {{Sureau}}, \citenamefont
  {{Vassallo}}, \citenamefont {{Zacchei}}, \citenamefont {{Zucca}},
  \citenamefont {{Baccigalupi}}, \citenamefont {{Balaguera-Antol{\'\i}nez}},
  \citenamefont {{Bernardeau}}, \citenamefont {{Biviano}}, \citenamefont
  {{Blanchard}}, \citenamefont {{Bolzonella}}, \citenamefont {{Borgani}},
  \citenamefont {{Bozzo}}, \citenamefont {{Burigana}}, \citenamefont
  {{Cabanac}}, \citenamefont {{Cappi}}, \citenamefont {{Carvalho}},
  \citenamefont {{Castignani}}, \citenamefont {{Colodro-Conde}}, \citenamefont
  {{Coupon}}, \citenamefont {{Courtois}}, \citenamefont {{Cuby}}, \citenamefont
  {{Davini}}, \citenamefont {{de la Torre}}, \citenamefont {{Di Ferdinando}},
  \citenamefont {{Farina}}, \citenamefont {{Ferreira}}, \citenamefont
  {{Finelli}}, \citenamefont {{Galeotta}}, \citenamefont {{Ganga}},
  \citenamefont {{Garcia-Bellido}}, \citenamefont {{Gaztanaga}}, \citenamefont
  {{Gozaliasl}}, \citenamefont {{Hook}}, \citenamefont {{Ili{\'c}}},
  \citenamefont {{Joachimi}}, \citenamefont {{Kansal}}, \citenamefont
  {{Keihanen}}, \citenamefont {{Kirkpatrick}}, \citenamefont {{Lindholm}},
  \citenamefont {{Mainetti}}, \citenamefont {{Maoli}}, \citenamefont
  {{Martinet}}, \citenamefont {{Maturi}}, \citenamefont {{Metcalf}},
  \citenamefont {{Monaco}}, \citenamefont {{Morgante}}, \citenamefont
  {{Nightingale}}, \citenamefont {{Nucita}}, \citenamefont {{Patrizii}},
  \citenamefont {{Popa}}, \citenamefont {{Potter}}, \citenamefont {{Riccio}},
  \citenamefont {{S{\'a}nchez}}, \citenamefont {{Schirmer}}, \citenamefont
  {{Schultheis}}, \citenamefont {{Scottez}}, \citenamefont {{Sefusatti}},
  \citenamefont {{Tramacere}}, \citenamefont {{Valiviita}}, \citenamefont
  {{Viel}},\ and\ \citenamefont
  {{Hildebrandt}}}]{EuclidCollaboration2022_magnification_bias}%
  \BibitemOpen
  \bibfield  {author} {\bibinfo {author} {{Euclid Collaboration}}, \bibinfo
  {author} {F.~{Lepori}}, \bibinfo {author} {I.~{Tutusaus}}, \bibinfo {author}
  {C.~{Viglione}}, \bibinfo {author} {C.~{Bonvin}}, \bibinfo {author}
  {S.~{Camera}}, \bibinfo {author} {F.~J. {Castander}}, \bibinfo {author}
  {R.~{Durrer}}, \bibinfo {author} {P.~{Fosalba}}, \bibinfo {author}
  {G.~{Jelic-Cizmek}}, et~al.,\ }\href
  {https://doi.org/10.1051/0004-6361/202142419} {\bibfield  {journal} {\bibinfo
   {journal} {\aap}\ }\textbf {\bibinfo {volume} {662}},\ \bibinfo {eid} {A93}
  (\bibinfo {year} {2022})},\ \Eprint {https://arxiv.org/abs/2110.05435}
  {arXiv:2110.05435 [astro-ph.CO]} \BibitemShut {NoStop}%
\bibitem [{\citenamefont {{Elvin-Poole}}\ \emph {et~al.}(2023)\citenamefont
  {{Elvin-Poole}}, \citenamefont {{MacCrann}}, \citenamefont {{Everett}},
  \citenamefont {{Prat}}, \citenamefont {{Rykoff}}, \citenamefont {{De
  Vicente}}, \citenamefont {{Yanny}}, \citenamefont {{Herner}}, \citenamefont
  {{Fert{\'e}}}, \citenamefont {{Di Valentino}}, \citenamefont {{Choi}},
  \citenamefont {{Burke}}, \citenamefont {{Sevilla-Noarbe}}, \citenamefont
  {{Alarcon}}, \citenamefont {{Alves}}, \citenamefont {{Amon}}, \citenamefont
  {{Andrade-Oliveira}}, \citenamefont {{Baxter}}, \citenamefont {{Bechtol}},
  \citenamefont {{Becker}}, \citenamefont {{Bernstein}}, \citenamefont
  {{Blazek}}, \citenamefont {{Camacho}}, \citenamefont {{Campos}},
  \citenamefont {{Carnero Rosell}}, \citenamefont {{Carrasco Kind}},
  \citenamefont {{Cawthon}}, \citenamefont {{Chang}}, \citenamefont {{Chen}},
  \citenamefont {{Cordero}}, \citenamefont {{Crocce}}, \citenamefont {{Davis}},
  \citenamefont {{DeRose}}, \citenamefont {{Diehl}}, \citenamefont
  {{Dodelson}}, \citenamefont {{Doux}}, \citenamefont {{Drlica-Wagner}},
  \citenamefont {{Eckert}}, \citenamefont {{Eifler}}, \citenamefont {{Elsner}},
  \citenamefont {{Fang}}, \citenamefont {{Fosalba}}, \citenamefont
  {{Friedrich}}, \citenamefont {{Gatti}}, \citenamefont {{Giannini}},
  \citenamefont {{Gruen}}, \citenamefont {{Gruendl}}, \citenamefont
  {{Harrison}}, \citenamefont {{Hartley}}, \citenamefont {{Huang}},
  \citenamefont {{Huff}}, \citenamefont {{Huterer}}, \citenamefont {{Krause}},
  \citenamefont {{Kuropatkin}}, \citenamefont {{Leget}}, \citenamefont
  {{Lemos}}, \citenamefont {{Liddle}}, \citenamefont {{McCullough}},
  \citenamefont {{Muir}}, \citenamefont {{Myles}}, \citenamefont
  {{Navarro-Alsina}}, \citenamefont {{Pandey}}, \citenamefont {{Park}},
  \citenamefont {{Porredon}}, \citenamefont {{Raveri}}, \citenamefont
  {{Rodriguez-Monroy}}, \citenamefont {{Rollins}}, \citenamefont {{Roodman}},
  \citenamefont {{Rosenfeld}}, \citenamefont {{Ross}}, \citenamefont
  {{S{\'a}nchez}}, \citenamefont {{Sanchez}}, \citenamefont {{Secco}},
  \citenamefont {{Sheldon}}, \citenamefont {{Shin}}, \citenamefont {{Troxel}},
  \citenamefont {{Tutusaus}}, \citenamefont {{Varga}}, \citenamefont
  {{Weaverdyck}}, \citenamefont {{Wechsler}}, \citenamefont {{Yin}},
  \citenamefont {{Zhang}}, \citenamefont {{Zuntz}}, \citenamefont {{Aguena}},
  \citenamefont {{Avila}}, \citenamefont {{Bacon}}, \citenamefont {{Bertin}},
  \citenamefont {{Bocquet}}, \citenamefont {{Brooks}}, \citenamefont
  {{Garc{\'\i}a-Bellido}}, \citenamefont {{Honscheid}}, \citenamefont
  {{Jarvis}}, \citenamefont {{Li}}, \citenamefont {{Mena-Fern{\'a}ndez}},
  \citenamefont {{To}}, \citenamefont {{Wilkinson}},\ and\ \citenamefont {{DES
  Collaboration}}}]{ElvinPoole2023}%
  \BibitemOpen
  \bibfield  {author} {\bibinfo {author} {J.~{Elvin-Poole}}, \bibinfo {author}
  {N.~{MacCrann}}, \bibinfo {author} {S.~{Everett}}, \bibinfo {author}
  {J.~{Prat}}, \bibinfo {author} {E.~S. {Rykoff}}, \bibinfo {author} {J.~{De
  Vicente}}, \bibinfo {author} {B.~{Yanny}}, \bibinfo {author} {K.~{Herner}},
  \bibinfo {author} {A.~{Fert{\'e}}}, \bibinfo {author} {E.~{Di Valentino}},
  et~al.,\ }\href {https://doi.org/10.1093/mnras/stad1594} {\bibfield
  {journal} {\bibinfo  {journal} {\mnras}\ }\textbf {\bibinfo {volume} {523}},\
  \bibinfo {pages} {3649} (\bibinfo {year} {2023})},\ \Eprint
  {https://arxiv.org/abs/2209.09782} {arXiv:2209.09782 [astro-ph.CO]}
  \BibitemShut {NoStop}%
\bibitem [{\citenamefont {{Wenzl}}\ \emph {et~al.}(2023)\citenamefont
  {{Wenzl}}, \citenamefont {{Chen}},\ and\ \citenamefont
  {{Bean}}}]{Wenzl2023magbias}%
  \BibitemOpen
  \bibfield  {author} {\bibinfo {author} {L.~{Wenzl}}, \bibinfo {author} {S.-F.
  {Chen}},\ and\ \bibinfo {author} {R.~{Bean}},\ }\bibfield  {journal}
  {\bibinfo  {journal} {\mnras}\ }\href
  {https://doi.org/10.1093/mnras/stad3314} {10.1093/mnras/stad3314} (\bibinfo
  {year} {2023}),\ \Eprint {https://arxiv.org/abs/2308.05892} {arXiv:2308.05892
  [astro-ph.CO]} \BibitemShut {NoStop}%
\bibitem [{\citenamefont {{Moradinezhad Dizgah}}\ and\ \citenamefont
  {{Durrer}}(2016)}]{MoradinezhadDizgah2016}%
  \BibitemOpen
  \bibfield  {author} {\bibinfo {author} {A.~{Moradinezhad Dizgah}}\ and\
  \bibinfo {author} {R.~{Durrer}},\ }\href
  {https://doi.org/10.1088/1475-7516/2016/09/035} {\bibfield  {journal}
  {\bibinfo  {journal} {\jcap}\ }\textbf {\bibinfo {volume} {2016}},\ \bibinfo
  {eid} {035} (\bibinfo {year} {2016})},\ \Eprint
  {https://arxiv.org/abs/1604.08914} {arXiv:1604.08914 [astro-ph.CO]}
  \BibitemShut {NoStop}%
\bibitem [{\citenamefont {{Yang}}\ and\ \citenamefont
  {{Pullen}}(2018)}]{Yang2018}%
  \BibitemOpen
  \bibfield  {author} {\bibinfo {author} {S.~{Yang}}\ and\ \bibinfo {author}
  {A.~R. {Pullen}},\ }\href {https://doi.org/10.1093/mnras/sty2353} {\bibfield
  {journal} {\bibinfo  {journal} {\mnras}\ }\textbf {\bibinfo {volume} {481}},\
  \bibinfo {pages} {1441} (\bibinfo {year} {2018})},\ \Eprint
  {https://arxiv.org/abs/1807.05639} {arXiv:1807.05639 [astro-ph.CO]}
  \BibitemShut {NoStop}%
\bibitem [{\citenamefont {{Ghosh}}\ and\ \citenamefont
  {{Durrer}}(2019)}]{Ghosh2019}%
  \BibitemOpen
  \bibfield  {author} {\bibinfo {author} {B.~{Ghosh}}\ and\ \bibinfo {author}
  {R.~{Durrer}},\ }\href {https://doi.org/10.1088/1475-7516/2019/06/010}
  {\bibfield  {journal} {\bibinfo  {journal} {\jcap}\ }\textbf {\bibinfo
  {volume} {2019}},\ \bibinfo {eid} {010} (\bibinfo {year} {2019})},\ \Eprint
  {https://arxiv.org/abs/1812.09546} {arXiv:1812.09546 [astro-ph.CO]}
  \BibitemShut {NoStop}%
\bibitem [{\citenamefont {{Alonso}}\ \emph {et~al.}(2019)\citenamefont
  {{Alonso}}, \citenamefont {{Sanchez}}, \citenamefont {{Slosar}},\ and\
  \citenamefont {{LSST Dark Energy Science Collaboration}}}]{Alonso2019}%
  \BibitemOpen
  \bibfield  {author} {\bibinfo {author} {D.~{Alonso}}, \bibinfo {author}
  {J.~{Sanchez}}, \bibinfo {author} {A.~{Slosar}},\ and\ \bibinfo {author}
  {{LSST Dark Energy Science Collaboration}},\ }\href
  {https://doi.org/10.1093/mnras/stz093} {\bibfield  {journal} {\bibinfo
  {journal} {\mnras}\ }\textbf {\bibinfo {volume} {484}},\ \bibinfo {pages}
  {4127} (\bibinfo {year} {2019})},\ \Eprint {https://arxiv.org/abs/1809.09603}
  {arXiv:1809.09603 [astro-ph.CO]} \BibitemShut {NoStop}%
\bibitem [{\citenamefont {{Benoit-L{\'e}vy}}\ \emph {et~al.}(2013)\citenamefont
  {{Benoit-L{\'e}vy}}, \citenamefont {{D{\'e}chelette}}, \citenamefont
  {{Benabed}}, \citenamefont {{Cardoso}}, \citenamefont {{Hanson}},\ and\
  \citenamefont {{Prunet}}}]{BenoitLevy2013}%
  \BibitemOpen
  \bibfield  {author} {\bibinfo {author} {A.~{Benoit-L{\'e}vy}}, \bibinfo
  {author} {T.~{D{\'e}chelette}}, \bibinfo {author} {K.~{Benabed}}, \bibinfo
  {author} {J.~F. {Cardoso}}, \bibinfo {author} {D.~{Hanson}},\ and\ \bibinfo
  {author} {S.~{Prunet}},\ }\href {https://doi.org/10.1051/0004-6361/201321048}
  {\bibfield  {journal} {\bibinfo  {journal} {\aap}\ }\textbf {\bibinfo
  {volume} {555}},\ \bibinfo {eid} {A37} (\bibinfo {year} {2013})},\ \Eprint
  {https://arxiv.org/abs/1301.4145} {arXiv:1301.4145 [astro-ph.CO]}
  \BibitemShut {NoStop}%
\bibitem [{\citenamefont {{Carron}}(2023)}]{Carron2023}%
  \BibitemOpen
  \bibfield  {author} {\bibinfo {author} {J.~{Carron}},\ }\href
  {https://doi.org/10.1088/1475-7516/2023/02/057} {\bibfield  {journal}
  {\bibinfo  {journal} {\jcap}\ }\textbf {\bibinfo {volume} {2023}},\ \bibinfo
  {eid} {057} (\bibinfo {year} {2023})},\ \Eprint
  {https://arxiv.org/abs/2210.05449} {arXiv:2210.05449 [astro-ph.CO]}
  \BibitemShut {NoStop}%
\bibitem [{\citenamefont {{Krolewski}}\ \emph {et~al.}(2023)\citenamefont
  {{Krolewski}}, \citenamefont {{Percival}}, \citenamefont {{Ferraro}},
  \citenamefont {{Chaussidon}}, \citenamefont {{Rezaie}}, \citenamefont
  {{Aguilar}}, \citenamefont {{Ahlen}}, \citenamefont {{Brooks}}, \citenamefont
  {{Dawson}}, \citenamefont {{de la Macorra}}, \citenamefont {{Doel}},
  \citenamefont {{Fanning}}, \citenamefont {{Font-Ribera}}, \citenamefont
  {{Gontcho}}, \citenamefont {{Guy}}, \citenamefont {{Honscheid}},
  \citenamefont {{Kehoe}}, \citenamefont {{Kisner}}, \citenamefont {{Kremin}},
  \citenamefont {{Landriau}}, \citenamefont {{Levi}}, \citenamefont
  {{Martini}}, \citenamefont {{Meisner}}, \citenamefont {{Miquel}},
  \citenamefont {{Nie}}, \citenamefont {{Poppett}}, \citenamefont {{Ross}},
  \citenamefont {{Rossi}}, \citenamefont {{Schubnell}}, \citenamefont {{Seo}},
  \citenamefont {{Tarle}}, \citenamefont {{Vargas-Magana}}, \citenamefont
  {{Weaver}}, \citenamefont {{Yeche}}, \citenamefont {{Zhou}},\ and\
  \citenamefont {{Zhou}}}]{Krolewski2023}%
  \BibitemOpen
  \bibfield  {author} {\bibinfo {author} {A.~{Krolewski}}, \bibinfo {author}
  {W.~J. {Percival}}, \bibinfo {author} {S.~{Ferraro}}, \bibinfo {author}
  {E.~{Chaussidon}}, \bibinfo {author} {M.~{Rezaie}}, \bibinfo {author} {J.~N.
  {Aguilar}}, \bibinfo {author} {S.~{Ahlen}}, \bibinfo {author} {D.~{Brooks}},
  \bibinfo {author} {K.~{Dawson}}, \bibinfo {author} {A.~{de la Macorra}},
  et~al.,\ }\href {https://doi.org/10.48550/arXiv.2305.07650} {\bibfield
  {journal} {\bibinfo  {journal} {arXiv e-prints}\ ,\ \bibinfo {eid}
  {arXiv:2305.07650}} (\bibinfo {year} {2023})},\ \Eprint
  {https://arxiv.org/abs/2305.07650} {arXiv:2305.07650 [astro-ph.CO]}
  \BibitemShut {NoStop}%
\bibitem [{\citenamefont {{G{\'o}rski}}\ \emph {et~al.}(2005)\citenamefont
  {{G{\'o}rski}}, \citenamefont {{Hivon}}, \citenamefont {{Banday}},
  \citenamefont {{Wandelt}}, \citenamefont {{Hansen}}, \citenamefont
  {{Reinecke}},\ and\ \citenamefont {{Bartelmann}}}]{Gorski2005}%
  \BibitemOpen
  \bibfield  {author} {\bibinfo {author} {K.~M. {G{\'o}rski}}, \bibinfo
  {author} {E.~{Hivon}}, \bibinfo {author} {A.~J. {Banday}}, \bibinfo {author}
  {B.~D. {Wandelt}}, \bibinfo {author} {F.~K. {Hansen}}, \bibinfo {author}
  {M.~{Reinecke}},\ and\ \bibinfo {author} {M.~{Bartelmann}},\ }\href
  {https://doi.org/10.1086/427976} {\bibfield  {journal} {\bibinfo  {journal}
  {\apj}\ }\textbf {\bibinfo {volume} {622}},\ \bibinfo {pages} {759} (\bibinfo
  {year} {2005})},\ \Eprint {https://arxiv.org/abs/astro-ph/0409513}
  {arXiv:astro-ph/0409513 [astro-ph]} \BibitemShut {NoStop}%
\bibitem [{\citenamefont {Zonca}\ \emph {et~al.}(2019)\citenamefont {Zonca},
  \citenamefont {Singer}, \citenamefont {Lenz}, \citenamefont {Reinecke},
  \citenamefont {Rosset}, \citenamefont {Hivon},\ and\ \citenamefont
  {Gorski}}]{Zonca2019}%
  \BibitemOpen
  \bibfield  {author} {\bibinfo {author} {A.~Zonca}, \bibinfo {author}
  {L.~Singer}, \bibinfo {author} {D.~Lenz}, \bibinfo {author} {M.~Reinecke},
  \bibinfo {author} {C.~Rosset}, \bibinfo {author} {E.~Hivon},\ and\ \bibinfo
  {author} {K.~Gorski},\ }\href {https://doi.org/10.21105/joss.01298}
  {\bibfield  {journal} {\bibinfo  {journal} {Journal of Open Source Software}\
  }\textbf {\bibinfo {volume} {4}},\ \bibinfo {pages} {1298} (\bibinfo {year}
  {2019})}\BibitemShut {NoStop}%
\bibitem [{\citenamefont {{White}}\ \emph {et~al.}(2011)\citenamefont
  {{White}}, \citenamefont {{Blanton}}, \citenamefont {{Bolton}}, \citenamefont
  {{Schlegel}}, \citenamefont {{Tinker}}, \citenamefont {{Berlind}},
  \citenamefont {{da Costa}}, \citenamefont {{Kazin}}, \citenamefont {{Lin}},
  \citenamefont {{Maia}}, \citenamefont {{McBride}}, \citenamefont
  {{Padmanabhan}}, \citenamefont {{Parejko}}, \citenamefont {{Percival}},
  \citenamefont {{Prada}}, \citenamefont {{Ramos}}, \citenamefont {{Sheldon}},
  \citenamefont {{de Simoni}}, \citenamefont {{Skibba}}, \citenamefont
  {{Thomas}}, \citenamefont {{Wake}}, \citenamefont {{Zehavi}}, \citenamefont
  {{Zheng}}, \citenamefont {{Nichol}}, \citenamefont {{Schneider}},
  \citenamefont {{Strauss}}, \citenamefont {{Weaver}},\ and\ \citenamefont
  {{Weinberg}}}]{White2011}%
  \BibitemOpen
  \bibfield  {author} {\bibinfo {author} {M.~{White}}, \bibinfo {author}
  {M.~{Blanton}}, \bibinfo {author} {A.~{Bolton}}, \bibinfo {author}
  {D.~{Schlegel}}, \bibinfo {author} {J.~{Tinker}}, \bibinfo {author}
  {A.~{Berlind}}, \bibinfo {author} {L.~{da Costa}}, \bibinfo {author}
  {E.~{Kazin}}, \bibinfo {author} {Y.~T. {Lin}}, \bibinfo {author} {M.~{Maia}},
  et~al.,\ }\href {https://doi.org/10.1088/0004-637X/728/2/126} {\bibfield
  {journal} {\bibinfo  {journal} {\apj}\ }\textbf {\bibinfo {volume} {728}},\
  \bibinfo {eid} {126} (\bibinfo {year} {2011})},\ \Eprint
  {https://arxiv.org/abs/1010.4915} {arXiv:1010.4915 [astro-ph.CO]}
  \BibitemShut {NoStop}%
\bibitem [{\citenamefont {{Parejko}}\ \emph {et~al.}(2013)\citenamefont
  {{Parejko}}, \citenamefont {{Sunayama}}, \citenamefont {{Padmanabhan}},
  \citenamefont {{Wake}}, \citenamefont {{Berlind}}, \citenamefont {{Bizyaev}},
  \citenamefont {{Blanton}}, \citenamefont {{Bolton}}, \citenamefont {{van den
  Bosch}}, \citenamefont {{Brinkmann}}, \citenamefont {{Brownstein}},
  \citenamefont {{da Costa}}, \citenamefont {{Eisenstein}}, \citenamefont
  {{Guo}}, \citenamefont {{Kazin}}, \citenamefont {{Maia}}, \citenamefont
  {{Malanushenko}}, \citenamefont {{Maraston}}, \citenamefont {{McBride}},
  \citenamefont {{Nichol}}, \citenamefont {{Oravetz}}, \citenamefont {{Pan}},
  \citenamefont {{Percival}}, \citenamefont {{Prada}}, \citenamefont {{Ross}},
  \citenamefont {{Ross}}, \citenamefont {{Schlegel}}, \citenamefont
  {{Schneider}}, \citenamefont {{Simmons}}, \citenamefont {{Skibba}},
  \citenamefont {{Tinker}}, \citenamefont {{Tojeiro}}, \citenamefont
  {{Weaver}}, \citenamefont {{Wetzel}}, \citenamefont {{White}}, \citenamefont
  {{Weinberg}}, \citenamefont {{Thomas}}, \citenamefont {{Zehavi}},\ and\
  \citenamefont {{Zheng}}}]{Parejko2013}%
  \BibitemOpen
  \bibfield  {author} {\bibinfo {author} {J.~K. {Parejko}}, \bibinfo {author}
  {T.~{Sunayama}}, \bibinfo {author} {N.~{Padmanabhan}}, \bibinfo {author}
  {D.~A. {Wake}}, \bibinfo {author} {A.~A. {Berlind}}, \bibinfo {author}
  {D.~{Bizyaev}}, \bibinfo {author} {M.~{Blanton}}, \bibinfo {author} {A.~S.
  {Bolton}}, \bibinfo {author} {F.~{van den Bosch}}, \bibinfo {author}
  {J.~{Brinkmann}}, et~al.,\ }\href {https://doi.org/10.1093/mnras/sts314}
  {\bibfield  {journal} {\bibinfo  {journal} {\mnras}\ }\textbf {\bibinfo
  {volume} {429}},\ \bibinfo {pages} {98} (\bibinfo {year} {2013})},\ \Eprint
  {https://arxiv.org/abs/1211.3976} {arXiv:1211.3976 [astro-ph.CO]}
  \BibitemShut {NoStop}%
\bibitem [{\citenamefont {{Kamionkowski}}\ \emph {et~al.}(1997)\citenamefont
  {{Kamionkowski}}, \citenamefont {{Kosowsky}},\ and\ \citenamefont
  {{Stebbins}}}]{Kamionkowski1997}%
  \BibitemOpen
  \bibfield  {author} {\bibinfo {author} {M.~{Kamionkowski}}, \bibinfo {author}
  {A.~{Kosowsky}},\ and\ \bibinfo {author} {A.~{Stebbins}},\ }\href
  {https://doi.org/10.1103/PhysRevD.55.7368} {\bibfield  {journal} {\bibinfo
  {journal} {\prd}\ }\textbf {\bibinfo {volume} {55}},\ \bibinfo {pages} {7368}
  (\bibinfo {year} {1997})},\ \Eprint {https://arxiv.org/abs/astro-ph/9611125}
  {arXiv:astro-ph/9611125 [astro-ph]} \BibitemShut {NoStop}%
\bibitem [{\citenamefont {{Hartlap}}\ \emph {et~al.}(2007)\citenamefont
  {{Hartlap}}, \citenamefont {{Simon}},\ and\ \citenamefont
  {{Schneider}}}]{Hartlap2007}%
  \BibitemOpen
  \bibfield  {author} {\bibinfo {author} {J.~{Hartlap}}, \bibinfo {author}
  {P.~{Simon}},\ and\ \bibinfo {author} {P.~{Schneider}},\ }\href
  {https://doi.org/10.1051/0004-6361:20066170} {\bibfield  {journal} {\bibinfo
  {journal} {\aap}\ }\textbf {\bibinfo {volume} {464}},\ \bibinfo {pages} {399}
  (\bibinfo {year} {2007})},\ \Eprint {https://arxiv.org/abs/astro-ph/0608064}
  {arXiv:astro-ph/0608064 [astro-ph]} \BibitemShut {NoStop}%
\bibitem [{\citenamefont {{Efstathiou}}(2004)}]{Efstathiou2004}%
  \BibitemOpen
  \bibfield  {author} {\bibinfo {author} {G.~{Efstathiou}},\ }\href
  {https://doi.org/10.1111/j.1365-2966.2004.07530.x} {\bibfield  {journal}
  {\bibinfo  {journal} {\mnras}\ }\textbf {\bibinfo {volume} {349}},\ \bibinfo
  {pages} {603} (\bibinfo {year} {2004})},\ \Eprint
  {https://arxiv.org/abs/astro-ph/0307515} {arXiv:astro-ph/0307515 [astro-ph]}
  \BibitemShut {NoStop}%
\bibitem [{\citenamefont {{Couchot}}\ \emph {et~al.}(2017)\citenamefont
  {{Couchot}}, \citenamefont {{Henrot-Versill{\'e}}}, \citenamefont
  {{Perdereau}}, \citenamefont {{Plaszczynski}}, \citenamefont {{Rouill{\'e}
  d'Orfeuil}}, \citenamefont {{Spinelli}},\ and\ \citenamefont
  {{Tristram}}}]{Couchot2017}%
  \BibitemOpen
  \bibfield  {author} {\bibinfo {author} {F.~{Couchot}}, \bibinfo {author}
  {S.~{Henrot-Versill{\'e}}}, \bibinfo {author} {O.~{Perdereau}}, \bibinfo
  {author} {S.~{Plaszczynski}}, \bibinfo {author} {B.~{Rouill{\'e} d'Orfeuil}},
  \bibinfo {author} {M.~{Spinelli}},\ and\ \bibinfo {author} {M.~{Tristram}},\
  }\href {https://doi.org/10.1051/0004-6361/201629815} {\bibfield  {journal}
  {\bibinfo  {journal} {\aap}\ }\textbf {\bibinfo {volume} {602}},\ \bibinfo
  {eid} {A41} (\bibinfo {year} {2017})},\ \Eprint
  {https://arxiv.org/abs/1609.09730} {arXiv:1609.09730 [astro-ph.CO]}
  \BibitemShut {NoStop}%
\bibitem [{\citenamefont {{Garc{\'\i}a-Garc{\'\i}a}}\ \emph
  {et~al.}(2019)\citenamefont {{Garc{\'\i}a-Garc{\'\i}a}}, \citenamefont
  {{Alonso}},\ and\ \citenamefont {{Bellini}}}]{GarciaGarcia2019}%
  \BibitemOpen
  \bibfield  {author} {\bibinfo {author} {C.~{Garc{\'\i}a-Garc{\'\i}a}},
  \bibinfo {author} {D.~{Alonso}},\ and\ \bibinfo {author} {E.~{Bellini}},\
  }\href {https://doi.org/10.1088/1475-7516/2019/11/043} {\bibfield  {journal}
  {\bibinfo  {journal} {\jcap}\ }\textbf {\bibinfo {volume} {2019}},\ \bibinfo
  {eid} {043} (\bibinfo {year} {2019})},\ \Eprint
  {https://arxiv.org/abs/1906.11765} {arXiv:1906.11765 [astro-ph.CO]}
  \BibitemShut {NoStop}%
\bibitem [{\citenamefont {{Norberg}}\ \emph {et~al.}(2009)\citenamefont
  {{Norberg}}, \citenamefont {{Baugh}}, \citenamefont {{Gazta{\~n}aga}},\ and\
  \citenamefont {{Croton}}}]{Norberg2009}%
  \BibitemOpen
  \bibfield  {author} {\bibinfo {author} {P.~{Norberg}}, \bibinfo {author}
  {C.~M. {Baugh}}, \bibinfo {author} {E.~{Gazta{\~n}aga}},\ and\ \bibinfo
  {author} {D.~J. {Croton}},\ }\href
  {https://doi.org/10.1111/j.1365-2966.2009.14389.x} {\bibfield  {journal}
  {\bibinfo  {journal} {\mnras}\ }\textbf {\bibinfo {volume} {396}},\ \bibinfo
  {pages} {19} (\bibinfo {year} {2009})},\ \Eprint
  {https://arxiv.org/abs/0810.1885} {arXiv:0810.1885 [astro-ph]} \BibitemShut
  {NoStop}%
\bibitem [{\citenamefont {{Marques}}\ \emph {et~al.}(2020)\citenamefont
  {{Marques}}, \citenamefont {{Liu}}, \citenamefont {{Huffenberger}},\ and\
  \citenamefont {{Colin Hill}}}]{Marques2020}%
  \BibitemOpen
  \bibfield  {author} {\bibinfo {author} {G.~A. {Marques}}, \bibinfo {author}
  {J.~{Liu}}, \bibinfo {author} {K.~M. {Huffenberger}},\ and\ \bibinfo {author}
  {J.~{Colin Hill}},\ }\href {https://doi.org/10.3847/1538-4357/abc003}
  {\bibfield  {journal} {\bibinfo  {journal} {\apj}\ }\textbf {\bibinfo
  {volume} {904}},\ \bibinfo {eid} {182} (\bibinfo {year} {2020})},\ \Eprint
  {https://arxiv.org/abs/2008.04369} {arXiv:2008.04369 [astro-ph.CO]}
  \BibitemShut {NoStop}%
\bibitem [{\citenamefont {{Mohammad}}\ and\ \citenamefont
  {{Percival}}(2022)}]{Mohammad2022}%
  \BibitemOpen
  \bibfield  {author} {\bibinfo {author} {F.~G. {Mohammad}}\ and\ \bibinfo
  {author} {W.~J. {Percival}},\ }\href {https://doi.org/10.1093/mnras/stac1458}
  {\bibfield  {journal} {\bibinfo  {journal} {\mnras}\ }\textbf {\bibinfo
  {volume} {514}},\ \bibinfo {pages} {1289} (\bibinfo {year} {2022})},\ \Eprint
  {https://arxiv.org/abs/2109.07071} {arXiv:2109.07071 [astro-ph.CO]}
  \BibitemShut {NoStop}%
\bibitem [{\citenamefont {{Nicola}}\ \emph {et~al.}(2020)\citenamefont
  {{Nicola}}, \citenamefont {{Alonso}}, \citenamefont {{S{\'a}nchez}},
  \citenamefont {{Slosar}}, \citenamefont {{Awan}}, \citenamefont
  {{Broussard}}, \citenamefont {{Dunkley}}, \citenamefont {{Gawiser}},
  \citenamefont {{Gomes}}, \citenamefont {{Mandelbaum}}, \citenamefont
  {{Miyatake}}, \citenamefont {{Newman}}, \citenamefont {{Sevilla-Noarbe}},
  \citenamefont {{Skinner}},\ and\ \citenamefont {{Wagoner}}}]{Nicola2020}%
  \BibitemOpen
  \bibfield  {author} {\bibinfo {author} {A.~{Nicola}}, \bibinfo {author}
  {D.~{Alonso}}, \bibinfo {author} {J.~{S{\'a}nchez}}, \bibinfo {author}
  {A.~{Slosar}}, \bibinfo {author} {H.~{Awan}}, \bibinfo {author}
  {A.~{Broussard}}, \bibinfo {author} {J.~{Dunkley}}, \bibinfo {author}
  {E.~{Gawiser}}, \bibinfo {author} {Z.~{Gomes}}, \bibinfo {author}
  {R.~{Mandelbaum}}, et~al.,\ }\href
  {https://doi.org/10.1088/1475-7516/2020/03/044} {\bibfield  {journal}
  {\bibinfo  {journal} {\jcap}\ }\textbf {\bibinfo {volume} {2020}},\ \bibinfo
  {eid} {044} (\bibinfo {year} {2020})},\ \Eprint
  {https://arxiv.org/abs/1912.08209} {arXiv:1912.08209 [astro-ph.CO]}
  \BibitemShut {NoStop}%
\bibitem [{\citenamefont {{Maus}}\ \emph {et~al.}(2023)\citenamefont {{Maus}},
  \citenamefont {{Chen}},\ and\ \citenamefont {{White}}}]{Maus2023}%
  \BibitemOpen
  \bibfield  {author} {\bibinfo {author} {M.~{Maus}}, \bibinfo {author} {S.-F.
  {Chen}},\ and\ \bibinfo {author} {M.~{White}},\ }\href
  {https://doi.org/10.1088/1475-7516/2023/06/005} {\bibfield  {journal}
  {\bibinfo  {journal} {\jcap}\ }\textbf {\bibinfo {volume} {2023}},\ \bibinfo
  {eid} {005} (\bibinfo {year} {2023})},\ \Eprint
  {https://arxiv.org/abs/2302.07430} {arXiv:2302.07430 [astro-ph.CO]}
  \BibitemShut {NoStop}%
\bibitem [{\citenamefont {{Chen}}\ \emph {et~al.}(2020)\citenamefont {{Chen}},
  \citenamefont {{Vlah}},\ and\ \citenamefont {{White}}}]{Chen2020}%
  \BibitemOpen
  \bibfield  {author} {\bibinfo {author} {S.-F. {Chen}}, \bibinfo {author}
  {Z.~{Vlah}},\ and\ \bibinfo {author} {M.~{White}},\ }\href
  {https://doi.org/10.1088/1475-7516/2020/07/062} {\bibfield  {journal}
  {\bibinfo  {journal} {\jcap}\ }\textbf {\bibinfo {volume} {2020}},\ \bibinfo
  {eid} {062} (\bibinfo {year} {2020})},\ \Eprint
  {https://arxiv.org/abs/2005.00523} {arXiv:2005.00523 [astro-ph.CO]}
  \BibitemShut {NoStop}%
\bibitem [{\citenamefont {{Chen}}\ \emph {et~al.}(2021)\citenamefont {{Chen}},
  \citenamefont {{Vlah}}, \citenamefont {{Castorina}},\ and\ \citenamefont
  {{White}}}]{Chen2021}%
  \BibitemOpen
  \bibfield  {author} {\bibinfo {author} {S.-F. {Chen}}, \bibinfo {author}
  {Z.~{Vlah}}, \bibinfo {author} {E.~{Castorina}},\ and\ \bibinfo {author}
  {M.~{White}},\ }\href {https://doi.org/10.1088/1475-7516/2021/03/100}
  {\bibfield  {journal} {\bibinfo  {journal} {\jcap}\ }\textbf {\bibinfo
  {volume} {2021}},\ \bibinfo {eid} {100} (\bibinfo {year} {2021})},\ \Eprint
  {https://arxiv.org/abs/2012.04636} {arXiv:2012.04636 [astro-ph.CO]}
  \BibitemShut {NoStop}%
\bibitem [{\citenamefont {{Lewis}}(2019)}]{Lewis2019getdist}%
  \BibitemOpen
  \bibfield  {author} {\bibinfo {author} {A.~{Lewis}},\ }\href@noop {}
  {\bibfield  {journal} {\bibinfo  {journal} {arXiv e-prints}\ ,\ \bibinfo
  {eid} {arXiv:1910.13970}} (\bibinfo {year} {2019})},\ \Eprint
  {https://arxiv.org/abs/1910.13970} {arXiv:1910.13970 [astro-ph.IM]}
  \BibitemShut {NoStop}%
\bibitem [{\citenamefont {{Sun}}\ \emph {et~al.}(2023)\citenamefont {{Sun}},
  \citenamefont {{Zhang}}, \citenamefont {{Dong}}, \citenamefont {{Yao}},
  \citenamefont {{Shan}}, \citenamefont {{Jullo}}, \citenamefont {{Kneib}},\
  and\ \citenamefont {{Yin}}}]{Sun2023}%
  \BibitemOpen
  \bibfield  {author} {\bibinfo {author} {Z.~{Sun}}, \bibinfo {author}
  {P.~{Zhang}}, \bibinfo {author} {F.~{Dong}}, \bibinfo {author} {J.~{Yao}},
  \bibinfo {author} {H.~{Shan}}, \bibinfo {author} {E.~{Jullo}}, \bibinfo
  {author} {J.-P. {Kneib}},\ and\ \bibinfo {author} {B.~{Yin}},\ }\href
  {https://doi.org/10.3847/1538-4365/acda2a} {\bibfield  {journal} {\bibinfo
  {journal} {\apjs}\ }\textbf {\bibinfo {volume} {267}},\ \bibinfo {eid} {21}
  (\bibinfo {year} {2023})},\ \Eprint {https://arxiv.org/abs/2210.13717}
  {arXiv:2210.13717 [astro-ph.CO]} \BibitemShut {NoStop}%
\bibitem [{\citenamefont {Curtiss}(1941)}]{Curtiss1941}%
  \BibitemOpen
  \bibfield  {author} {\bibinfo {author} {J.~H. Curtiss},\ }\href
  {https://doi.org/10.1214/aoms/1177731679} {\bibfield  {journal} {\bibinfo
  {journal} {The Annals of Mathematical Statistics}\ }\textbf {\bibinfo
  {volume} {12}},\ \bibinfo {pages} {409 } (\bibinfo {year}
  {1941})}\BibitemShut {NoStop}%
\bibitem [{\citenamefont {{Miyatake}}\ \emph {et~al.}(2017)\citenamefont
  {{Miyatake}}, \citenamefont {{Madhavacheril}}, \citenamefont {{Sehgal}},
  \citenamefont {{Slosar}}, \citenamefont {{Spergel}}, \citenamefont
  {{Sherwin}},\ and\ \citenamefont {{van Engelen}}}]{Miyatake2017}%
  \BibitemOpen
  \bibfield  {author} {\bibinfo {author} {H.~{Miyatake}}, \bibinfo {author}
  {M.~S. {Madhavacheril}}, \bibinfo {author} {N.~{Sehgal}}, \bibinfo {author}
  {A.~{Slosar}}, \bibinfo {author} {D.~N. {Spergel}}, \bibinfo {author}
  {B.~{Sherwin}},\ and\ \bibinfo {author} {A.~{van Engelen}},\ }\href
  {https://doi.org/10.1103/PhysRevLett.118.161301} {\bibfield  {journal}
  {\bibinfo  {journal} {\prl}\ }\textbf {\bibinfo {volume} {118}},\ \bibinfo
  {eid} {161301} (\bibinfo {year} {2017})},\ \Eprint
  {https://arxiv.org/abs/1605.05337} {arXiv:1605.05337 [astro-ph.CO]}
  \BibitemShut {NoStop}%
\bibitem [{\citenamefont {{Planck Collaboration}}\ \emph
  {et~al.}(2016)\citenamefont {{Planck Collaboration}}, \citenamefont {{Ade}},
  \citenamefont {{Aghanim}}, \citenamefont {{Arnaud}}, \citenamefont
  {{Ashdown}}, \citenamefont {{Aumont}}, \citenamefont {{Baccigalupi}},
  \citenamefont {{Banday}}, \citenamefont {{Barreiro}}, \citenamefont
  {{Bartlett}}, \citenamefont {{Bartolo}}, \citenamefont {{Battaner}},
  \citenamefont {{Battye}}, \citenamefont {{Benabed}}, \citenamefont
  {{Beno{\^\i}t}}, \citenamefont {{Benoit-L{\'e}vy}}, \citenamefont
  {{Bernard}}, \citenamefont {{Bersanelli}}, \citenamefont {{Bielewicz}},
  \citenamefont {{Bock}}, \citenamefont {{Bonaldi}}, \citenamefont
  {{Bonavera}}, \citenamefont {{Bond}}, \citenamefont {{Borrill}},
  \citenamefont {{Bouchet}}, \citenamefont {{Boulanger}}, \citenamefont
  {{Bucher}}, \citenamefont {{Burigana}}, \citenamefont {{Butler}},
  \citenamefont {{Calabrese}}, \citenamefont {{Cardoso}}, \citenamefont
  {{Catalano}}, \citenamefont {{Challinor}}, \citenamefont {{Chamballu}},
  \citenamefont {{Chary}}, \citenamefont {{Chiang}}, \citenamefont {{Chluba}},
  \citenamefont {{Christensen}}, \citenamefont {{Church}}, \citenamefont
  {{Clements}}, \citenamefont {{Colombi}}, \citenamefont {{Colombo}},
  \citenamefont {{Combet}}, \citenamefont {{Coulais}}, \citenamefont {{Crill}},
  \citenamefont {{Curto}}, \citenamefont {{Cuttaia}}, \citenamefont {{Danese}},
  \citenamefont {{Davies}}, \citenamefont {{Davis}}, \citenamefont {{de
  Bernardis}}, \citenamefont {{de Rosa}}, \citenamefont {{de Zotti}},
  \citenamefont {{Delabrouille}}, \citenamefont {{D{\'e}sert}}, \citenamefont
  {{Di Valentino}}, \citenamefont {{Dickinson}}, \citenamefont {{Diego}},
  \citenamefont {{Dolag}}, \citenamefont {{Dole}}, \citenamefont {{Donzelli}},
  \citenamefont {{Dor{\'e}}}, \citenamefont {{Douspis}}, \citenamefont
  {{Ducout}}, \citenamefont {{Dunkley}}, \citenamefont {{Dupac}}, \citenamefont
  {{Efstathiou}}, \citenamefont {{Elsner}}, \citenamefont {{En{\ss}lin}},
  \citenamefont {{Eriksen}}, \citenamefont {{Farhang}}, \citenamefont
  {{Fergusson}}, \citenamefont {{Finelli}}, \citenamefont {{Forni}},
  \citenamefont {{Frailis}}, \citenamefont {{Fraisse}}, \citenamefont
  {{Franceschi}}, \citenamefont {{Frejsel}}, \citenamefont {{Galeotta}},
  \citenamefont {{Galli}}, \citenamefont {{Ganga}}, \citenamefont {{Gauthier}},
  \citenamefont {{Gerbino}}, \citenamefont {{Ghosh}}, \citenamefont {{Giard}},
  \citenamefont {{Giraud-H{\'e}raud}}, \citenamefont {{Giusarma}},
  \citenamefont {{Gjerl{\o}w}}, \citenamefont {{Gonz{\'a}lez-Nuevo}},
  \citenamefont {{G{\'o}rski}}, \citenamefont {{Gratton}}, \citenamefont
  {{Gregorio}}, \citenamefont {{Gruppuso}}, \citenamefont {{Gudmundsson}},
  \citenamefont {{Hamann}}, \citenamefont {{Hansen}}, \citenamefont {{Hanson}},
  \citenamefont {{Harrison}}, \citenamefont {{Helou}}, \citenamefont
  {{Henrot-Versill{\'e}}}, \citenamefont {{Hern{\'a}ndez-Monteagudo}},
  \citenamefont {{Herranz}}, \citenamefont {{Hildebrand t}}, \citenamefont
  {{Hivon}}, \citenamefont {{Hobson}}, \citenamefont {{Holmes}}, \citenamefont
  {{Hornstrup}}, \citenamefont {{Hovest}}, \citenamefont {{Huang}},
  \citenamefont {{Huffenberger}}, \citenamefont {{Hurier}}, \citenamefont
  {{Jaffe}}, \citenamefont {{Jaffe}}, \citenamefont {{Jones}}, \citenamefont
  {{Juvela}}, \citenamefont {{Keih{\"a}nen}}, \citenamefont {{Keskitalo}},
  \citenamefont {{Kisner}}, \citenamefont {{Kneissl}}, \citenamefont
  {{Knoche}}, \citenamefont {{Knox}}, \citenamefont {{Kunz}}, \citenamefont
  {{Kurki-Suonio}}, \citenamefont {{Lagache}}, \citenamefont
  {{L{\"a}hteenm{\"a}ki}}, \citenamefont {{Lamarre}}, \citenamefont
  {{Lasenby}}, \citenamefont {{Lattanzi}}, \citenamefont {{Lawrence}},
  \citenamefont {{Leahy}}, \citenamefont {{Leonardi}}, \citenamefont
  {{Lesgourgues}}, \citenamefont {{Levrier}}, \citenamefont {{Lewis}},
  \citenamefont {{Liguori}}, \citenamefont {{Lilje}}, \citenamefont
  {{Linden-V{\o}rnle}}, \citenamefont {{L{\'o}pez-Caniego}}, \citenamefont
  {{Lubin}}, \citenamefont {{Mac{\'\i}as-P{\'e}rez}}, \citenamefont {{Maggio}},
  \citenamefont {{Maino}}, \citenamefont {{Mandolesi}}, \citenamefont
  {{Mangilli}}, \citenamefont {{Marchini}}, \citenamefont {{Maris}},
  \citenamefont {{Martin}}, \citenamefont {{Martinelli}}, \citenamefont
  {{Mart{\'\i}nez-Gonz{\'a}lez}}, \citenamefont {{Masi}}, \citenamefont
  {{Matarrese}}, \citenamefont {{McGehee}}, \citenamefont {{Meinhold}},
  \citenamefont {{Melchiorri}}, \citenamefont {{Melin}}, \citenamefont
  {{Mendes}}, \citenamefont {{Mennella}}, \citenamefont {{Migliaccio}},
  \citenamefont {{Millea}}, \citenamefont {{Mitra}}, \citenamefont
  {{Miville-Desch{\^e}nes}}, \citenamefont {{Moneti}}, \citenamefont
  {{Montier}}, \citenamefont {{Morgante}}, \citenamefont {{Mortlock}},
  \citenamefont {{Moss}}, \citenamefont {{Munshi}}, \citenamefont {{Murphy}},
  \citenamefont {{Naselsky}}, \citenamefont {{Nati}}, \citenamefont {{Natoli}},
  \citenamefont {{Netterfield}}, \citenamefont {{N{\o}rgaard-Nielsen}},
  \citenamefont {{Noviello}}, \citenamefont {{Novikov}}, \citenamefont
  {{Novikov}}, \citenamefont {{Oxborrow}}, \citenamefont {{Paci}},
  \citenamefont {{Pagano}}, \citenamefont {{Pajot}}, \citenamefont
  {{Paladini}}, \citenamefont {{Paoletti}}, \citenamefont {{Partridge}},
  \citenamefont {{Pasian}}, \citenamefont {{Patanchon}}, \citenamefont
  {{Pearson}}, \citenamefont {{Perdereau}}, \citenamefont {{Perotto}},
  \citenamefont {{Perrotta}}, \citenamefont {{Pettorino}}, \citenamefont
  {{Piacentini}}, \citenamefont {{Piat}}, \citenamefont {{Pierpaoli}},
  \citenamefont {{Pietrobon}}, \citenamefont {{Plaszczynski}}, \citenamefont
  {{Pointecouteau}}, \citenamefont {{Polenta}}, \citenamefont {{Popa}},
  \citenamefont {{Pratt}}, \citenamefont {{Pr{\'e}zeau}}, \citenamefont
  {{Prunet}}, \citenamefont {{Puget}}, \citenamefont {{Rachen}}, \citenamefont
  {{Reach}}, \citenamefont {{Rebolo}}, \citenamefont {{Reinecke}},
  \citenamefont {{Remazeilles}}, \citenamefont {{Renault}}, \citenamefont
  {{Renzi}}, \citenamefont {{Ristorcelli}}, \citenamefont {{Rocha}},
  \citenamefont {{Rosset}}, \citenamefont {{Rossetti}}, \citenamefont
  {{Roudier}}, \citenamefont {{Rouill{\'e} d'Orfeuil}}, \citenamefont
  {{Rowan-Robinson}}, \citenamefont {{Rubi{\~n}o-Mart{\'\i}n}}, \citenamefont
  {{Rusholme}}, \citenamefont {{Said}}, \citenamefont {{Salvatelli}},
  \citenamefont {{Salvati}}, \citenamefont {{Sandri}}, \citenamefont
  {{Santos}}, \citenamefont {{Savelainen}}, \citenamefont {{Savini}},
  \citenamefont {{Scott}}, \citenamefont {{Seiffert}}, \citenamefont {{Serra}},
  \citenamefont {{Shellard}}, \citenamefont {{Spencer}}, \citenamefont
  {{Spinelli}}, \citenamefont {{Stolyarov}}, \citenamefont {{Stompor}},
  \citenamefont {{Sudiwala}}, \citenamefont {{Sunyaev}}, \citenamefont
  {{Sutton}}, \citenamefont {{Suur-Uski}}, \citenamefont {{Sygnet}},
  \citenamefont {{Tauber}}, \citenamefont {{Terenzi}}, \citenamefont
  {{Toffolatti}}, \citenamefont {{Tomasi}}, \citenamefont {{Tristram}},
  \citenamefont {{Trombetti}}, \citenamefont {{Tucci}}, \citenamefont
  {{Tuovinen}}, \citenamefont {{T{\"u}rler}}, \citenamefont {{Umana}},
  \citenamefont {{Valenziano}}, \citenamefont {{Valiviita}}, \citenamefont
  {{Van Tent}}, \citenamefont {{Vielva}}, \citenamefont {{Villa}},
  \citenamefont {{Wade}}, \citenamefont {{Wandelt}}, \citenamefont {{Wehus}},
  \citenamefont {{White}}, \citenamefont {{White}}, \citenamefont
  {{Wilkinson}}, \citenamefont {{Yvon}}, \citenamefont {{Zacchei}},\ and\
  \citenamefont {{Zonca}}}]{PlanckCollaboration2016}%
  \BibitemOpen
  \bibfield  {author} {\bibinfo {author} {{Planck Collaboration}}, \bibinfo
  {author} {P.~A.~R. {Ade}}, \bibinfo {author} {N.~{Aghanim}}, \bibinfo
  {author} {M.~{Arnaud}}, \bibinfo {author} {M.~{Ashdown}}, \bibinfo {author}
  {J.~{Aumont}}, \bibinfo {author} {C.~{Baccigalupi}}, \bibinfo {author} {A.~J.
  {Banday}}, \bibinfo {author} {R.~B. {Barreiro}}, \bibinfo {author} {J.~G.
  {Bartlett}}, et~al.,\ }\href {https://doi.org/10.1051/0004-6361/201525830}
  {\bibfield  {journal} {\bibinfo  {journal} {\aap}\ }\textbf {\bibinfo
  {volume} {594}},\ \bibinfo {eid} {A13} (\bibinfo {year} {2016})},\ \Eprint
  {https://arxiv.org/abs/1502.01589} {arXiv:1502.01589 [astro-ph.CO]}
  \BibitemShut {NoStop}%
\bibitem [{\citenamefont {{Ivezi{\'c}}}\ \emph {et~al.}(2019)\citenamefont
  {{Ivezi{\'c}}}, \citenamefont {{Kahn}}, \citenamefont {{Tyson}},
  \citenamefont {{Abel}}, \citenamefont {{Acosta}}, \citenamefont {{Allsman}},
  \citenamefont {{Alonso}}, \citenamefont {{AlSayyad}}, \citenamefont
  {{Anderson}}, \citenamefont {{Andrew}}, \citenamefont {{Angel}},
  \citenamefont {{Angeli}}, \citenamefont {{Ansari}}, \citenamefont
  {{Antilogus}}, \citenamefont {{Araujo}}, \citenamefont {{Armstrong}},
  \citenamefont {{Arndt}}, \citenamefont {{Astier}}, \citenamefont {{Aubourg}},
  \citenamefont {{Auza}}, \citenamefont {{Axelrod}}, \citenamefont {{Bard}},
  \citenamefont {{Barr}}, \citenamefont {{Barrau}}, \citenamefont {{Bartlett}},
  \citenamefont {{Bauer}}, \citenamefont {{Bauman}}, \citenamefont {{Baumont}},
  \citenamefont {{Bechtol}}, \citenamefont {{Bechtol}}, \citenamefont
  {{Becker}}, \citenamefont {{Becla}}, \citenamefont {{Beldica}}, \citenamefont
  {{Bellavia}}, \citenamefont {{Bianco}}, \citenamefont {{Biswas}},
  \citenamefont {{Blanc}}, \citenamefont {{Blazek}}, \citenamefont {{Bland
  ford}}, \citenamefont {{Bloom}}, \citenamefont {{Bogart}}, \citenamefont
  {{Bond}}, \citenamefont {{Booth}}, \citenamefont {{Borgland}}, \citenamefont
  {{Borne}}, \citenamefont {{Bosch}}, \citenamefont {{Boutigny}}, \citenamefont
  {{Brackett}}, \citenamefont {{Bradshaw}}, \citenamefont {{Brand t}},
  \citenamefont {{Brown}}, \citenamefont {{Bullock}}, \citenamefont
  {{Burchat}}, \citenamefont {{Burke}}, \citenamefont {{Cagnoli}},
  \citenamefont {{Calabrese}}, \citenamefont {{Callahan}}, \citenamefont
  {{Callen}}, \citenamefont {{Carlin}}, \citenamefont {{Carlson}},
  \citenamefont {{Chand rasekharan}}, \citenamefont {{Charles-Emerson}},
  \citenamefont {{Chesley}}, \citenamefont {{Cheu}}, \citenamefont {{Chiang}},
  \citenamefont {{Chiang}}, \citenamefont {{Chirino}}, \citenamefont {{Chow}},
  \citenamefont {{Ciardi}}, \citenamefont {{Claver}}, \citenamefont
  {{Cohen-Tanugi}}, \citenamefont {{Cockrum}}, \citenamefont {{Coles}},
  \citenamefont {{Connolly}}, \citenamefont {{Cook}}, \citenamefont {{Cooray}},
  \citenamefont {{Covey}}, \citenamefont {{Cribbs}}, \citenamefont {{Cui}},
  \citenamefont {{Cutri}}, \citenamefont {{Daly}}, \citenamefont {{Daniel}},
  \citenamefont {{Daruich}}, \citenamefont {{Daubard}}, \citenamefont
  {{Daues}}, \citenamefont {{Dawson}}, \citenamefont {{Delgado}}, \citenamefont
  {{Dellapenna}}, \citenamefont {{de Peyster}}, \citenamefont {{de Val-Borro}},
  \citenamefont {{Digel}}, \citenamefont {{Doherty}}, \citenamefont {{Dubois}},
  \citenamefont {{Dubois-Felsmann}}, \citenamefont {{Durech}}, \citenamefont
  {{Economou}}, \citenamefont {{Eifler}}, \citenamefont {{Eracleous}},
  \citenamefont {{Emmons}}, \citenamefont {{Fausti Neto}}, \citenamefont
  {{Ferguson}}, \citenamefont {{Figueroa}}, \citenamefont {{Fisher-Levine}},
  \citenamefont {{Focke}}, \citenamefont {{Foss}}, \citenamefont {{Frank}},
  \citenamefont {{Freemon}}, \citenamefont {{Gangler}}, \citenamefont
  {{Gawiser}}, \citenamefont {{Geary}}, \citenamefont {{Gee}}, \citenamefont
  {{Geha}}, \citenamefont {{Gessner}}, \citenamefont {{Gibson}}, \citenamefont
  {{Gilmore}}, \citenamefont {{Glanzman}}, \citenamefont {{Glick}},
  \citenamefont {{Goldina}}, \citenamefont {{Goldstein}}, \citenamefont
  {{Goodenow}}, \citenamefont {{Graham}}, \citenamefont {{Gressler}},
  \citenamefont {{Gris}}, \citenamefont {{Guy}}, \citenamefont {{Guyonnet}},
  \citenamefont {{Haller}}, \citenamefont {{Harris}}, \citenamefont
  {{Hascall}}, \citenamefont {{Haupt}}, \citenamefont {{Hernand ez}},
  \citenamefont {{Herrmann}}, \citenamefont {{Hileman}}, \citenamefont
  {{Hoblitt}}, \citenamefont {{Hodgson}}, \citenamefont {{Hogan}},
  \citenamefont {{Howard}}, \citenamefont {{Huang}}, \citenamefont {{Huffer}},
  \citenamefont {{Ingraham}}, \citenamefont {{Innes}}, \citenamefont
  {{Jacoby}}, \citenamefont {{Jain}}, \citenamefont {{Jammes}}, \citenamefont
  {{Jee}}, \citenamefont {{Jenness}}, \citenamefont {{Jernigan}}, \citenamefont
  {{Jevremovi{\'c}}}, \citenamefont {{Johns}}, \citenamefont {{Johnson}},
  \citenamefont {{Johnson}}, \citenamefont {{Jones}}, \citenamefont
  {{Juramy-Gilles}}, \citenamefont {{Juri{\'c}}}, \citenamefont {{Kalirai}},
  \citenamefont {{Kallivayalil}}, \citenamefont {{Kalmbach}}, \citenamefont
  {{Kantor}}, \citenamefont {{Karst}}, \citenamefont {{Kasliwal}},
  \citenamefont {{Kelly}}, \citenamefont {{Kessler}}, \citenamefont
  {{Kinnison}}, \citenamefont {{Kirkby}}, \citenamefont {{Knox}}, \citenamefont
  {{Kotov}}, \citenamefont {{Krabbendam}}, \citenamefont {{Krughoff}},
  \citenamefont {{Kub{\'a}nek}}, \citenamefont {{Kuczewski}}, \citenamefont
  {{Kulkarni}}, \citenamefont {{Ku}}, \citenamefont {{Kurita}}, \citenamefont
  {{Lage}}, \citenamefont {{Lambert}}, \citenamefont {{Lange}}, \citenamefont
  {{Langton}}, \citenamefont {{Le Guillou}}, \citenamefont {{Levine}},
  \citenamefont {{Liang}}, \citenamefont {{Lim}}, \citenamefont {{Lintott}},
  \citenamefont {{Long}}, \citenamefont {{Lopez}}, \citenamefont {{Lotz}},
  \citenamefont {{Lupton}}, \citenamefont {{Lust}}, \citenamefont
  {{MacArthur}}, \citenamefont {{Mahabal}}, \citenamefont {{Mand elbaum}},
  \citenamefont {{Markiewicz}}, \citenamefont {{Marsh}}, \citenamefont
  {{Marshall}}, \citenamefont {{Marshall}}, \citenamefont {{May}},
  \citenamefont {{McKercher}}, \citenamefont {{McQueen}}, \citenamefont
  {{Meyers}}, \citenamefont {{Migliore}}, \citenamefont {{Miller}},
  \citenamefont {{Mills}}, \citenamefont {{Miraval}}, \citenamefont
  {{Moeyens}}, \citenamefont {{Moolekamp}}, \citenamefont {{Monet}},
  \citenamefont {{Moniez}}, \citenamefont {{Monkewitz}}, \citenamefont
  {{Montgomery}}, \citenamefont {{Morrison}}, \citenamefont {{Mueller}},
  \citenamefont {{Muller}}, \citenamefont {{Mu{\~n}oz Arancibia}},
  \citenamefont {{Neill}}, \citenamefont {{Newbry}}, \citenamefont {{Nief}},
  \citenamefont {{Nomerotski}}, \citenamefont {{Nordby}}, \citenamefont
  {{O'Connor}}, \citenamefont {{Oliver}}, \citenamefont {{Olivier}},
  \citenamefont {{Olsen}}, \citenamefont {{O'Mullane}}, \citenamefont
  {{Ortiz}}, \citenamefont {{Osier}}, \citenamefont {{Owen}}, \citenamefont
  {{Pain}}, \citenamefont {{Palecek}}, \citenamefont {{Parejko}}, \citenamefont
  {{Parsons}}, \citenamefont {{Pease}}, \citenamefont {{Peterson}},
  \citenamefont {{Peterson}}, \citenamefont {{Petravick}}, \citenamefont
  {{Libby Petrick}}, \citenamefont {{Petry}}, \citenamefont {{Pierfederici}},
  \citenamefont {{Pietrowicz}}, \citenamefont {{Pike}}, \citenamefont
  {{Pinto}}, \citenamefont {{Plante}}, \citenamefont {{Plate}}, \citenamefont
  {{Plutchak}}, \citenamefont {{Price}}, \citenamefont {{Prouza}},
  \citenamefont {{Radeka}}, \citenamefont {{Rajagopal}}, \citenamefont
  {{Rasmussen}}, \citenamefont {{Regnault}}, \citenamefont {{Reil}},
  \citenamefont {{Reiss}}, \citenamefont {{Reuter}}, \citenamefont {{Ridgway}},
  \citenamefont {{Riot}}, \citenamefont {{Ritz}}, \citenamefont {{Robinson}},
  \citenamefont {{Roby}}, \citenamefont {{Roodman}}, \citenamefont {{Rosing}},
  \citenamefont {{Roucelle}}, \citenamefont {{Rumore}}, \citenamefont
  {{Russo}}, \citenamefont {{Saha}}, \citenamefont {{Sassolas}}, \citenamefont
  {{Schalk}}, \citenamefont {{Schellart}}, \citenamefont {{Schindler}},
  \citenamefont {{Schmidt}}, \citenamefont {{Schneider}}, \citenamefont
  {{Schneider}}, \citenamefont {{Schoening}}, \citenamefont {{Schumacher}},
  \citenamefont {{Schwamb}}, \citenamefont {{Sebag}}, \citenamefont {{Selvy}},
  \citenamefont {{Sembroski}}, \citenamefont {{Seppala}}, \citenamefont
  {{Serio}}, \citenamefont {{Serrano}}, \citenamefont {{Shaw}}, \citenamefont
  {{Shipsey}}, \citenamefont {{Sick}}, \citenamefont {{Silvestri}},
  \citenamefont {{Slater}}, \citenamefont {{Smith}}, \citenamefont {{Smith}},
  \citenamefont {{Sobhani}}, \citenamefont {{Soldahl}}, \citenamefont
  {{Storrie-Lombardi}}, \citenamefont {{Stover}}, \citenamefont {{Strauss}},
  \citenamefont {{Street}}, \citenamefont {{Stubbs}}, \citenamefont
  {{Sullivan}}, \citenamefont {{Sweeney}}, \citenamefont {{Swinbank}},
  \citenamefont {{Szalay}}, \citenamefont {{Takacs}}, \citenamefont {{Tether}},
  \citenamefont {{Thaler}}, \citenamefont {{Thayer}}, \citenamefont {{Thomas}},
  \citenamefont {{Thornton}}, \citenamefont {{Thukral}}, \citenamefont
  {{Tice}}, \citenamefont {{Trilling}}, \citenamefont {{Turri}}, \citenamefont
  {{Van Berg}}, \citenamefont {{Vanden Berk}}, \citenamefont {{Vetter}},
  \citenamefont {{Virieux}}, \citenamefont {{Vucina}}, \citenamefont {{Wahl}},
  \citenamefont {{Walkowicz}}, \citenamefont {{Walsh}}, \citenamefont
  {{Walter}}, \citenamefont {{Wang}}, \citenamefont {{Wang}}, \citenamefont
  {{Warner}}, \citenamefont {{Wiecha}}, \citenamefont {{Willman}},
  \citenamefont {{Winters}}, \citenamefont {{Wittman}}, \citenamefont
  {{Wolff}}, \citenamefont {{Wood-Vasey}}, \citenamefont {{Wu}}, \citenamefont
  {{Xin}}, \citenamefont {{Yoachim}},\ and\ \citenamefont
  {{Zhan}}}]{Ivezic2019}%
  \BibitemOpen
  \bibfield  {author} {\bibinfo {author} {{\v{Z}}.~{Ivezi{\'c}}}, \bibinfo
  {author} {S.~M. {Kahn}}, \bibinfo {author} {J.~A. {Tyson}}, \bibinfo {author}
  {B.~{Abel}}, \bibinfo {author} {E.~{Acosta}}, \bibinfo {author}
  {R.~{Allsman}}, \bibinfo {author} {D.~{Alonso}}, \bibinfo {author}
  {Y.~{AlSayyad}}, \bibinfo {author} {S.~F. {Anderson}}, \bibinfo {author}
  {J.~{Andrew}}, et~al.,\ }\href {https://doi.org/10.3847/1538-4357/ab042c}
  {\bibfield  {journal} {\bibinfo  {journal} {\apj}\ }\textbf {\bibinfo
  {volume} {873}},\ \bibinfo {eid} {111} (\bibinfo {year} {2019})},\ \Eprint
  {https://arxiv.org/abs/0805.2366} {arXiv:0805.2366 [astro-ph]} \BibitemShut
  {NoStop}%
\bibitem [{\citenamefont {{Laureijs}}\ \emph {et~al.}(2011)\citenamefont
  {{Laureijs}}, \citenamefont {{Amiaux}}, \citenamefont {{Arduini}},
  \citenamefont {{Augu{\`e}res}}, \citenamefont {{Brinchmann}}, \citenamefont
  {{Cole}}, \citenamefont {{Cropper}}, \citenamefont {{Dabin}}, \citenamefont
  {{Duvet}}, \citenamefont {{Ealet}}, \citenamefont {{Garilli}}, \citenamefont
  {{Gondoin}}, \citenamefont {{Guzzo}}, \citenamefont {{Hoar}}, \citenamefont
  {{Hoekstra}}, \citenamefont {{Holmes}}, \citenamefont {{Kitching}},
  \citenamefont {{Maciaszek}}, \citenamefont {{Mellier}}, \citenamefont
  {{Pasian}}, \citenamefont {{Percival}}, \citenamefont {{Rhodes}},
  \citenamefont {{Saavedra Criado}}, \citenamefont {{Sauvage}}, \citenamefont
  {{Scaramella}}, \citenamefont {{Valenziano}}, \citenamefont {{Warren}},
  \citenamefont {{Bender}}, \citenamefont {{Castander}}, \citenamefont
  {{Cimatti}}, \citenamefont {{Le F{\`e}vre}}, \citenamefont {{Kurki-Suonio}},
  \citenamefont {{Levi}}, \citenamefont {{Lilje}}, \citenamefont {{Meylan}},
  \citenamefont {{Nichol}}, \citenamefont {{Pedersen}}, \citenamefont {{Popa}},
  \citenamefont {{Rebolo Lopez}}, \citenamefont {{Rix}}, \citenamefont
  {{Rottgering}}, \citenamefont {{Zeilinger}}, \citenamefont {{Grupp}},
  \citenamefont {{Hudelot}}, \citenamefont {{Massey}}, \citenamefont
  {{Meneghetti}}, \citenamefont {{Miller}}, \citenamefont {{Paltani}},
  \citenamefont {{Paulin-Henriksson}}, \citenamefont {{Pires}}, \citenamefont
  {{Saxton}}, \citenamefont {{Schrabback}}, \citenamefont {{Seidel}},
  \citenamefont {{Walsh}}, \citenamefont {{Aghanim}}, \citenamefont
  {{Amendola}}, \citenamefont {{Bartlett}}, \citenamefont {{Baccigalupi}},
  \citenamefont {{Beaulieu}}, \citenamefont {{Benabed}}, \citenamefont
  {{Cuby}}, \citenamefont {{Elbaz}}, \citenamefont {{Fosalba}}, \citenamefont
  {{Gavazzi}}, \citenamefont {{Helmi}}, \citenamefont {{Hook}}, \citenamefont
  {{Irwin}}, \citenamefont {{Kneib}}, \citenamefont {{Kunz}}, \citenamefont
  {{Mannucci}}, \citenamefont {{Moscardini}}, \citenamefont {{Tao}},
  \citenamefont {{Teyssier}}, \citenamefont {{Weller}}, \citenamefont
  {{Zamorani}}, \citenamefont {{Zapatero Osorio}}, \citenamefont {{Boulade}},
  \citenamefont {{Foumond}}, \citenamefont {{Di Giorgio}}, \citenamefont
  {{Guttridge}}, \citenamefont {{James}}, \citenamefont {{Kemp}}, \citenamefont
  {{Martignac}}, \citenamefont {{Spencer}}, \citenamefont {{Walton}},
  \citenamefont {{Bl{\"u}mchen}}, \citenamefont {{Bonoli}}, \citenamefont
  {{Bortoletto}}, \citenamefont {{Cerna}}, \citenamefont {{Corcione}},
  \citenamefont {{Fabron}}, \citenamefont {{Jahnke}}, \citenamefont {{Ligori}},
  \citenamefont {{Madrid}}, \citenamefont {{Martin}}, \citenamefont
  {{Morgante}}, \citenamefont {{Pamplona}}, \citenamefont {{Prieto}},
  \citenamefont {{Riva}}, \citenamefont {{Toledo}}, \citenamefont
  {{Trifoglio}}, \citenamefont {{Zerbi}}, \citenamefont {{Abdalla}},
  \citenamefont {{Douspis}}, \citenamefont {{Grenet}}, \citenamefont
  {{Borgani}}, \citenamefont {{Bouwens}}, \citenamefont {{Courbin}},
  \citenamefont {{Delouis}}, \citenamefont {{Dubath}}, \citenamefont
  {{Fontana}}, \citenamefont {{Frailis}}, \citenamefont {{Grazian}},
  \citenamefont {{Koppenh{\"o}fer}}, \citenamefont {{Mansutti}}, \citenamefont
  {{Melchior}}, \citenamefont {{Mignoli}}, \citenamefont {{Mohr}},
  \citenamefont {{Neissner}}, \citenamefont {{Noddle}}, \citenamefont
  {{Poncet}}, \citenamefont {{Scodeggio}}, \citenamefont {{Serrano}},
  \citenamefont {{Shane}}, \citenamefont {{Starck}}, \citenamefont {{Surace}},
  \citenamefont {{Taylor}}, \citenamefont {{Verdoes-Kleijn}}, \citenamefont
  {{Vuerli}}, \citenamefont {{Williams}}, \citenamefont {{Zacchei}},
  \citenamefont {{Altieri}}, \citenamefont {{Escudero Sanz}}, \citenamefont
  {{Kohley}}, \citenamefont {{Oosterbroek}}, \citenamefont {{Astier}},
  \citenamefont {{Bacon}}, \citenamefont {{Bardelli}}, \citenamefont {{Baugh}},
  \citenamefont {{Bellagamba}}, \citenamefont {{Benoist}}, \citenamefont
  {{Bianchi}}, \citenamefont {{Biviano}}, \citenamefont {{Branchini}},
  \citenamefont {{Carbone}}, \citenamefont {{Cardone}}, \citenamefont
  {{Clements}}, \citenamefont {{Colombi}}, \citenamefont {{Conselice}},
  \citenamefont {{Cresci}}, \citenamefont {{Deacon}}, \citenamefont {{Dunlop}},
  \citenamefont {{Fedeli}}, \citenamefont {{Fontanot}}, \citenamefont
  {{Franzetti}}, \citenamefont {{Giocoli}}, \citenamefont {{Garcia-Bellido}},
  \citenamefont {{Gow}}, \citenamefont {{Heavens}}, \citenamefont {{Hewett}},
  \citenamefont {{Heymans}}, \citenamefont {{Holland}}, \citenamefont
  {{Huang}}, \citenamefont {{Ilbert}}, \citenamefont {{Joachimi}},
  \citenamefont {{Jennins}}, \citenamefont {{Kerins}}, \citenamefont
  {{Kiessling}}, \citenamefont {{Kirk}}, \citenamefont {{Kotak}}, \citenamefont
  {{Krause}}, \citenamefont {{Lahav}}, \citenamefont {{van Leeuwen}},
  \citenamefont {{Lesgourgues}}, \citenamefont {{Lombardi}}, \citenamefont
  {{Magliocchetti}}, \citenamefont {{Maguire}}, \citenamefont {{Majerotto}},
  \citenamefont {{Maoli}}, \citenamefont {{Marulli}}, \citenamefont
  {{Maurogordato}}, \citenamefont {{McCracken}}, \citenamefont {{McLure}},
  \citenamefont {{Melchiorri}}, \citenamefont {{Merson}}, \citenamefont
  {{Moresco}}, \citenamefont {{Nonino}}, \citenamefont {{Norberg}},
  \citenamefont {{Peacock}}, \citenamefont {{Pello}}, \citenamefont {{Penny}},
  \citenamefont {{Pettorino}}, \citenamefont {{Di Porto}}, \citenamefont
  {{Pozzetti}}, \citenamefont {{Quercellini}}, \citenamefont {{Radovich}},
  \citenamefont {{Rassat}}, \citenamefont {{Roche}}, \citenamefont
  {{Ronayette}}, \citenamefont {{Rossetti}}, \citenamefont {{Sartoris}},
  \citenamefont {{Schneider}}, \citenamefont {{Semboloni}}, \citenamefont
  {{Serjeant}}, \citenamefont {{Simpson}}, \citenamefont {{Skordis}},
  \citenamefont {{Smadja}}, \citenamefont {{Smartt}}, \citenamefont {{Spano}},
  \citenamefont {{Spiro}}, \citenamefont {{Sullivan}}, \citenamefont
  {{Tilquin}}, \citenamefont {{Trotta}}, \citenamefont {{Verde}}, \citenamefont
  {{Wang}}, \citenamefont {{Williger}}, \citenamefont {{Zhao}}, \citenamefont
  {{Zoubian}},\ and\ \citenamefont {{Zucca}}}]{Laureijs2011}%
  \BibitemOpen
  \bibfield  {author} {\bibinfo {author} {R.~{Laureijs}}, \bibinfo {author}
  {J.~{Amiaux}}, \bibinfo {author} {S.~{Arduini}}, \bibinfo {author} {J.~L.
  {Augu{\`e}res}}, \bibinfo {author} {J.~{Brinchmann}}, \bibinfo {author}
  {R.~{Cole}}, \bibinfo {author} {M.~{Cropper}}, \bibinfo {author}
  {C.~{Dabin}}, \bibinfo {author} {L.~{Duvet}}, \bibinfo {author} {A.~{Ealet}},
  et~al.,\ }\href@noop {} {\bibfield  {journal} {\bibinfo  {journal} {arXiv
  e-prints}\ ,\ \bibinfo {eid} {arXiv:1110.3193}} (\bibinfo {year} {2011})},\
  \Eprint {https://arxiv.org/abs/1110.3193} {arXiv:1110.3193 [astro-ph.CO]}
  \BibitemShut {NoStop}%
\bibitem [{\citenamefont {{Spergel}}\ \emph {et~al.}(2015)\citenamefont
  {{Spergel}}, \citenamefont {{Gehrels}}, \citenamefont {{Baltay}},
  \citenamefont {{Bennett}}, \citenamefont {{Breckinridge}}, \citenamefont
  {{Donahue}}, \citenamefont {{Dressler}}, \citenamefont {{Gaudi}},
  \citenamefont {{Greene}}, \citenamefont {{Guyon}}, \citenamefont {{Hirata}},
  \citenamefont {{Kalirai}}, \citenamefont {{Kasdin}}, \citenamefont
  {{Macintosh}}, \citenamefont {{Moos}}, \citenamefont {{Perlmutter}},
  \citenamefont {{Postman}}, \citenamefont {{Rauscher}}, \citenamefont
  {{Rhodes}}, \citenamefont {{Wang}}, \citenamefont {{Weinberg}}, \citenamefont
  {{Benford}}, \citenamefont {{Hudson}}, \citenamefont {{Jeong}}, \citenamefont
  {{Mellier}}, \citenamefont {{Traub}}, \citenamefont {{Yamada}}, \citenamefont
  {{Capak}}, \citenamefont {{Colbert}}, \citenamefont {{Masters}},
  \citenamefont {{Penny}}, \citenamefont {{Savransky}}, \citenamefont
  {{Stern}}, \citenamefont {{Zimmerman}}, \citenamefont {{Barry}},
  \citenamefont {{Bartusek}}, \citenamefont {{Carpenter}}, \citenamefont
  {{Cheng}}, \citenamefont {{Content}}, \citenamefont {{Dekens}}, \citenamefont
  {{Demers}}, \citenamefont {{Grady}}, \citenamefont {{Jackson}}, \citenamefont
  {{Kuan}}, \citenamefont {{Kruk}}, \citenamefont {{Melton}}, \citenamefont
  {{Nemati}}, \citenamefont {{Parvin}}, \citenamefont {{Poberezhskiy}},
  \citenamefont {{Peddie}}, \citenamefont {{Ruffa}}, \citenamefont {{Wallace}},
  \citenamefont {{Whipple}}, \citenamefont {{Wollack}},\ and\ \citenamefont
  {{Zhao}}}]{Spergel2015}%
  \BibitemOpen
  \bibfield  {author} {\bibinfo {author} {D.~{Spergel}}, \bibinfo {author}
  {N.~{Gehrels}}, \bibinfo {author} {C.~{Baltay}}, \bibinfo {author}
  {D.~{Bennett}}, \bibinfo {author} {J.~{Breckinridge}}, \bibinfo {author}
  {M.~{Donahue}}, \bibinfo {author} {A.~{Dressler}}, \bibinfo {author} {B.~S.
  {Gaudi}}, \bibinfo {author} {T.~{Greene}}, \bibinfo {author} {O.~{Guyon}},
  et~al.,\ }\href@noop {} {\bibfield  {journal} {\bibinfo  {journal} {arXiv
  e-prints}\ ,\ \bibinfo {eid} {arXiv:1503.03757}} (\bibinfo {year} {2015})},\
  \Eprint {https://arxiv.org/abs/1503.03757} {arXiv:1503.03757 [astro-ph.IM]}
  \BibitemShut {NoStop}%
\end{thebibliography}%

\end{document}